\documentclass[nofootinbib,twocolumn,pra,floatfix,showpacs]{revtex4-1} 

\usepackage{amsmath,amssymb}

\newcommand{\out}[1]{ {}}

\newcommand{\udl}[1]{{#1}}
\newcommand{\cbp}{Ce$_3$Bi$_4$Pt$_3$}
\newcommand{\feas}{FeAs$_2$}
\newcommand{\fesb}{FeSb$_2$}
\newcommand{\cm}{\hbox{cm}}
\newcommand{\fval}{Fe$_2$VAl}
\newcommand{\oo}[1]{ {\frac{1}{#1}}}
\newcommand{\vek}[1]{ {\mathbf{#1}}}
\newcommand{\svek}[1]{ {\mathbf{#1}}}
\newcommand{\cc}{%
        \ensuremath{c^\dag}   } 
\newcommand{\ca}{%
        \ensuremath{c^{\phantom{\dag}}}}

\newcommand{\fc}{%
        \ensuremath{f^\dag}}    
\newcommand{\fa}{%
        \ensuremath{f^{\phantom{\dag}}}}                   

\newcommand{\pr}{%
        \ensuremath{^\prime}}

\newcommand{\etal}{{\it et al.}}
\newcommand{\eref}[1]{Eq.\ (\ref{#1})}
\newcommand{\Eref}[1]{Eq.\ (\ref{#1})}

\newcommand{\fref}[1]{Fig.\ (\ref{#1})}
\newcommand{\Fref}[1]{Fig.\ (\ref{#1})}

\newcommand{\sref}[1]{Section (\ref{#1})}
\newcommand{\Sref}[1]{section (\ref{#1})}

\newcommand{\tref}[1]{table (\ref{#1})}
\newcommand{\Tref}[1]{Table (\ref{#1})}

\DeclareTextFontCommand{\textbfit}{%
  \fontseries\bfdefault 
  \itshape
}

\usepackage{xcolor,colortbl}
\definecolor{Gray}{gray}{0.85}
\definecolor{lblue}{rgb}{0.8,0.8,1}  
\definecolor{blue}{rgb}{0.6,0.6,0.9}  

\usepackage{hyperref}

\usepackage{url}

\usepackage{framed}

\usepackage{graphicx}

\usepackage[caption=false]{subfig}

\begin{document}

\title[Correlated narrow-gap semiconductors]{Thermoelectricity in\\ Correlated narrow-gap semiconductors}

\author{Jan M.\ Tomczak}

\affiliation{Institute of Solid State Physics,  TU Wien, A-1040 Vienna, Austria}

\begin{abstract}
We review 
many-body effects,
their microscopic origin, as well as their impact onto thermoelectricity in correlated narrow-gap semiconductors.
Members of this class---such as FeSi and FeSb$_2$---display an unusual temperature dependence in various observables:
insulating with large thermopowers at low temperatures, they turn bad metals at temperatures much smaller than the size of their gaps.
This insulator-to-metal crossover is accompanied by spectral weight-transfers over large energies in the optical conductivity and 
by a gradual transition from activated to Curie-Weiss-like behaviour in the magnetic susceptibility. 
We show a retrospective of the understanding
of these phenomena, discuss the relation to heavy-fermion Kondo insulators---such as Ce$_3$Bi$_4$Pt$_3$ for which we present
new results---and propose a general classification of paramagnetic insulators.
From the latter FeSi emerges as an orbital-selective Kondo insulator.
Focussing on intermetallics such as silicides, antimonides, skutterudites, and Heusler compounds we showcase successes and
challenges for the realistic simulation of transport properties in the presence of electronic correlations.
Further, we advert to new avenues in which electronic correlations may contribute to the improvement of thermoelectric performance.
\end{abstract}

%
%
%
\maketitle

\tableofcontents

\section*{Avant-propos}
\label{avant}

Thermoelectricity is the striking ability of some materials to generate an electrical voltage when subjected to a temperature gradient (and vice versa). Prospective applications are multifarious, ranging from radiation and temperature sensors, vacuum gauges, to Peltier coolers and power generators.
However, the current efficiency of thermoelectric conversion is insufficient for commercially viable coolers and generators on a larger scale. 
Indeed, for these applications, thermoelectric devices are so far limited to niche products for which flexibility in design, reliability, and maintenance-free operation trump aspects of cost.
Examples are low-performance fridges, waste-heat recovery modules and power supplies for space-probes.%
\footnote{The prime example are the radioisotope thermoelectric generators aboard Voyager 1 that have been operating since 5-Sep-1977 and will continue to power the spacecraft---now well into interstellar space---until at least 2025.}

Nowadays, most commercially used thermoelectrics are doped intermetallic {\it narrow-gap semiconductors}, built from
post-transition metals, pnictogens, and chalcogens, e.g., Pb(Se,Te) and Bi$_2$Te$_3$,
while at high temperatures group-IV element alloys, Si$_{1-x}$Ge$_x$, are deployed.
These materials combine good powerfactors $S^2/\rho$ (determined by the thermopower $S$ and the resistivity $\rho$) with reasonably low thermal conductivities $\kappa$ 
 to reach a favourable thermoelectric conversion efficiency (measured by the figure of merit $ZT=S^2T/(\kappa \rho)$).
Indeed doped semiconductors present the {optimal compromise} between good conduction ($\rho$ small for high carrier density) and a high thermopower ($S$ large for low carrier densities), see \fref{ZT}.
Much recent thermoelectric research has been devoted to reducing contributions from phonons to the thermal conduction {via} micro-/nano-/super-structuring or by finding crystal structures with intrinsically low $\kappa$. 
Here, we will instead focus on the powerfactor $S^2/\rho$, which is controlled by the electronic structure of a material (see however Sec.~\ref{pdrag}).

One class of materials whose electronic structures give rise to a particularly extensive---and so far largely untapped---panoply of functionalities are {\it correlated materials}.
In these
materials electrons interact strongly with each other, causing a high sensitivity of even fundamental properties
to perturbations,  
such as changes in temperature, pressure, doping, and applied external fields.
These stimuli can induce unusual behaviours, e.g., metal-insulator transitions, various types of long-range order, and, in particular, large response functions. 
These phenomena are, both, a great challenge for fundamental research and a harbinger for technical developments in sensors, switches, transistors and, indeed, thermoelectric devices.

Electronic correlation effects occur in systems in which orbitals of small radial extent---thus particularly those with maximal angular quantum number: $3d$, $4f$---are partially filled and thus strongly affect low-energy properties. In consequence, correlations effects are ubiquitous among many families of compounds, such as transition metals, their oxides, silicides, pnictides, chalcogenides, as well as lanthanide and actinide based materials.
Exploiting various aspects of correlation effects numerous new functionalities were realized or have been proposed. Among them are 
new transparent conductors\cite{Zhang2015}, oxide electronics\cite{doi:10.1146/annurev-matsci-070813-113248},
 such as the Mott transistor\cite{ADMA:ADMA201003241,PhysRevLett.114.246401}, electrode materials\cite{ADMA:ADMA201300900},
eco-friendly pigments\cite{jmt_cesf},  or
intelligent window coatings\cite{Li20123823,me_psik}.

Naturally the question arises, whether correlation effects could also be used to overcome limitations of present thermoelectrics, 
all of which are (see above) uncorrelated narrow-gap semiconductors.
This question constitutes one of the interests in {\it correlated narrow-gap semiconductors}---the subject of this article.
The review is organized as follows:

{\it 1.\ Introduction.}
In \sref{intro} we give an overview of characteristic signatures of electronic correlation effects
in narrow-gap semiconductors and detail in which material classes they can be found.
Examples include transition-metal-based intermetallics, Skutterudites, and Heusler compounds.  
Since basic observables in some of the identified compounds bear a cunning empirical resemblance to that of heavy-fermion Kondo insulators,
we give in \sref{KI} a quite extensive comparison of both types of materials, focusing on a contrasting juxtaposition
of the prototype compounds FeSi and \cbp.

{\it 2.\ Theories of correlated narrow-gap semiconductors.}

A key part of this review is dedicated to the description and microscopic understanding of the electronic structure of correlated narrow-gap
semiconductors.
We begin \sref{theo} with a review of early phenomenological proposals that advocate the importance of electronic correlation effects.
Again, we focus the discussion on protagonist materials, such as FeSi, FeSb$_2$ and \cbp. 
\Sref{band} presents the band-structure point of view for these materials. While highlighting the deficiencies
of effective one-particle theories in describing the physics of our materials of interest, these first principle techniques provide key insights into the chemistry
of bonding, and complexities related to crystal fields and other multi-orbital effects.
\Sref{Ce3dft} includes new material on the Kondo insulator \cbp.
Guided, or at least inspired, by realistic band-structure calculations, the physics of electronic correlations was investigated within many-body {\it models},
such as that due to Hubbard, Kanamori and Gutzwiller, as well as the periodic Anderson model. Works that are directly
linked to our compounds of interest are reviewed in \sref{model}.
Next, we discuss in \sref{real} the current understanding of the physics of correlated narrow-gap semiconductors from the perspective
or {\it realistic} many-body simulations.
In \sref{pnas}, we  first showcase results from state-of-the-art dynamical mean-field methodologies
 for the example of FeSi. Indeed, a vast array of observables
can be simulated in quantitative congruence with experiment:
photoemission spectra, optical conductivities, neutron spectra, and---as discussed in \sref{silicides}---transport properties, including the thermopower.
After such validation of the theoretical framework, we discuss in \sref{micro} the microscopic insights that have been extracted from the numerical calculations.
In \sref{Ce3DMFT} we further include new realistic many-body simulations for the Kondo insulator \cbp.
For paramagnetic insulators with partially filled $d$ or $f$-shells we then
(i) introduce a microscopic classification from the dynamical mean-field perspective (\sref{dmftins}), and
(ii) propose a simple measure for the occurrence of Kondo-characteristics in said systems (\sref{proxy}).
Equipped with these tool, we then compare the {\it ab initio} results for FeSi and \cbp\ (\sref{abinitiohyb}) and propose an orbital-selective Kondo-scenario for FeSi.
Finally, \sref{lattice} discusses the potential influence of the lattice degrees of freedom onto the electronic structure.

{\it 3.\ Thermoelectricity.}

After a brief introduction to thermoelectricity in \sref{thermointro}, we review in 
\sref{correlS} the potential influence of electronic correlations onto thermoelectric performance.
In \sref{limit} we investigate the impact of basic many-body renormalizations in a semiconductors and
motivate that the thermopower has an upper bound if excitations are coherent.
Effects of incoherence (finite lifetimes) are studied in \sref{KubBoltz} and demonstrated to be beyond
semi-classical Boltzmann approaches.
Then, we review experimental findings and  theoretical simulations for transport 
and thermoelectric properties for representative 
materials among silicides (\sref{silicides}), antimonides (\sref{dianti}), skutterudites (\sref{skutts}), and
Heusler compounds (\sref{heuslers}).
In particular, we discuss the impact of finite lifetimes and effective mass renormalizations for FeSi,
the proposal that the colossal thermopower of FeSb$_2$ is driven by the so-called phonon-drag effect,
as well as the pitfalls of electronic structure theory and the influence of defects and disorder
in CoSb$_3$ and Fe$_2$VAl.

{\it 4.\ Conclusions and outlook.}
Finally, we summarize the covered material, point out open challenges,
and identify possible directions of future research.

\begin{table*}[p]%
\rotatebox{90}{
\renewcommand{\arraystretch}{1.2}
\begin{tabular}{|l|l|l|l|l|l|l|}
\hline
\rowcolor{blue}
structure/ & $\Delta_{dir}$        & $\Delta_{indir}$ & $\Delta_s$ & $S_{max}$ & $(S^2/\rho)_{max}$  & see\\
\rowcolor{blue}
material &  [meV]                 & [meV]            &  [meV]     & [mV/K]    & [$\mu$W/K$^2$cm] & section\\
\hline 
\rowcolor{lblue}
\multicolumn{7}{|l|}{B20  structure }  \\ 
\hline
FeSi     & 60-73 \cite{PhysRevLett.71.1748}             &  50-70\cite{Bocelli_FeSi,PhysRevB.56.12916}         &     50\cite{PhysRev.160.476,JPSJ.50.2539}         &  0.5@50K\cite{Wolfe1965449}  & 42.1@62K\cite{Wolfe1965449} &  \ref{overview} \ref{bandFeSi} \ref{KI} \\
         & \cite{Damascelli1997787,PhysRevB.79.165111}                 &           &                 &     1.2@32K\cite{PhysRevB.83.125209}           &                  & \ref{pnas} \ref{silicides} \\
\hline
RuSi     & 200-400           &    310\cite{Buschinger1997238}       &      not act.\cite{BUSCHINGER199757}           &                &                    &   \ref{overview} \ref{bandFeSi} \\
         & \cite{Buschinger1997238,Vescoli1998367}     & 260\cite{Hohl199839}   &                 &    0.27@286K\cite{Hohl199839}            &      8.7@347K\cite{Hohl199839}     &  \ref{silicides} \\
\hline
\rowcolor{lblue}
\multicolumn{7}{|l|}{marcasites} \\
\hline
FeSb$_2$    & 130\cite{PhysRevB.82.245205}                       &  30\cite{0295-5075-80-1-17008}         &  32\cite{PhysRev.160.476}               & -45\cite{0295-5075-80-1-17008}        & 2300@12K\cite{0295-5075-80-1-17008,APEX.2.091102} & \ref{overview} \ref{marcasites} \\
         &                                                &           &  37-48\cite{PhysRevB.74.195130}               & -18\cite{PhysRevB.88.245203} -25\cite{Takahashi2016}   & 8000@28K\cite{PhysRevB.86.115121}    &  \ref{dianti}\\
\hline
RuSb$_2$	   &  790\cite{PhysRevB.82.245205}        & 290\cite{APEX.2.091102}    330\cite{Harada2004200} &   not act.\cite{APEX.2.091102}  & -3.1@15K\cite{APEX.2.091102}               &   40@32K\cite{APEX.2.091102}                &  \ref{overview} \ref{marcasites} \\
\hline
FeAs$_2$    &            &   200-220\cite{Fan1972136,APEX.2.091102}         &   200\cite{APEX.2.091102}              &   -7@12K\cite{APEX.2.091102}             &  153@30K\cite{APEX.2.091102}                 &   \ref{overview} \ref{marcasites} \\
\hline
FeP$_2$     &              &  370-400\cite{Hulliger1959,1402-4896-4-3-010}         &   not act.\cite{1402-4896-4-3-010}              &               &                    & \\
\hline
CrSb$_2$$^*$    &                                           &   50\cite{PhysRevB.86.235136},  70\cite{Harada2004200}        &    50\cite{PhysRevB.86.235136}             &    -4.5@18K\cite{PhysRevB.86.235136}            & 46@24K\cite{PhysRevB.86.235136}                  &  \ref{dianti} \\
\hline
\rowcolor{lblue}
\multicolumn{7}{|l|}{CoGa3-type structure}  \\
\hline
FeGa3    &  600\cite{Knyazev2017} &  260\cite{doi:10.1063/1.1803947}   &  300-400\cite{monika_fega3}      &    -0.35@350K\cite{JPSJ.78.013702}            &     0.003@100K\cite{JPSJ.78.013702}              &  \ref{overview} \ref{dianti} \\
    &   &    400-500\cite{JPSJ.78.013702, PhysRevB.83.245116}                  &  290-450\cite{doi:10.1143/JPSJ.77.024705}              &    -0.45@300K  \cite{haldolaarachchige:103712}            &       0.2@400K\cite{haldolaarachchige:103712}              &  \\
		& &   \cite{doi:10.1143/JPSJ.77.024705,monika_fega3,PhysRevB.90.195206}               &                &        -15@12K\cite{PhysRevB.90.195206}       &                    & \\
\hline
RuGa3    &   650\cite{Knyazev2017}          &    320\cite{doi:10.1063/1.1803947}         &                 &    0.3@566K\cite{doi:10.1063/1.1803947}           &     3.8@920K\cite{doi:10.1063/1.1803947}            & \ref{overview}\\
\hline
\rowcolor{lblue}
\multicolumn{7}{|l|}{skutterudites}   \\
\hline
CoSb$_3$    &   230\cite{Tang2015}               &    230-700\cite{doi:10.1063/1.360402,Tang2015}      &     not act.\cite{PhysRevB.63.014410} &     0.2@300K\cite{PhysRevB.52.4926} &  9.5@625K\cite{doi:10.1063/1.3557068}            &  \ref{overview}\ref{skutts} \\
            &                                    &    \cite{KAWAHARADA2001193,PhysRevB.52.4926}       &               &      -0.4@450K\cite{0953-8984-15-29-315}     &                       &   \\
\hline
\rowcolor{lblue}
\multicolumn{7}{|l|}{Heuslers}  \\   
\hline
Fe$_2$VAl   & 100-200    &    90-130\cite{0953-8984-12-8-318,doi:10.1143/JPSJ.74.1378}       &   270\cite{PhysRevB.58.9763}              & $-0.14@200$K\cite{doi:10.1143/JPSJ.74.1378}                &       3.3@300K\cite{PhysRevB.96.045204}             & \ref{overview} \ref{heuslers} \\
        &    \cite{PhysRevLett.84.3674,Shreder2015}              &           &                 &  doped:68@300K\cite{2053-1591-1-1-015901}              &                    & \\
\hline
\rowcolor{lblue}
\multicolumn{7}{|l|}{Oxides}  \\   
\hline
LaCoO$_3$   & 100\cite{PhysRevB.53.R2926,PhysRevB.58.R1699}    &    100\cite{PhysRevB.53.R2926,PhysRevB.58.R1699}       &   30\cite{PhysRevB.53.R2926}              & 1.2\cite{SENARISRODRIGUEZ1995224}, 33@80K\cite{BENEDICT2016145}                &       0.136@80K\cite{BENEDICT2016145}            & \ref{bandFeSi} \ref{micro} \ref{sec:covalent} \\
\hline
\rowcolor{lblue}
\multicolumn{7}{|l}{Kondo insulators} \\
\hline
\cbp   &    37\cite{PhysRevLett.72.522}           &  8-12\cite{PhysRevB.42.6842,PhysRevLett.118.246601}            &    12\cite{PhysRevB.44.6832}             &    0.055@23K\cite{Katoh199822}            &        1.67@30K\cite{Katoh199822}           &  \ref{KI} \ref{Ce3dft}\\
  &               &              &                 &                &                &  \ref{Ce3DMFT}\\ 
\hline
\end{tabular}  
}
\caption{{\bf Collection of experimental data of prototypical narrow-gap semiconductors.}
$\Delta_{dir}$, $\Delta_{indir}$, $\Delta_{s}$: direct (optical), indirect (transport), and spin gap; $S_{max}$, $(S^2/\rho)_{max}$: peak values of the thermopower and the powerfactor at indicated temperatures. ``not act.'' = no activated behaviour up to largest $T$ measured.
$^*$CrSb$_2$ is antiferromagnetic below $T_N=273$K\cite{Holseth1970,PhysRevB.76.115105}.
}
\label{table1}
\renewcommand{\arraystretch}{1.0}
\end{table*}

\section{Introduction}
\label{intro}

Correlated narrow-gap semiconductors are systems with small band-gaps in which many-body effects play an important role.
In the following \sref{overview} we will give an overview over what typical signatures of such correlation effects are and in which kind of materials one can expect to find them.
In \sref{KI}  we will give a more detailed account of the intriguing physical observations, but focus the discussion on a
one-to-one comparison of a prototypical correlated narrow-gap semiconductor, FeSi, and the archetypical Kondo insulator \cbp, with which it shares
many empirical similarities.

\subsection{Correlated narrow-gap semiconductors: a brief overview}
\label{overview}


A paradigmatic class of materials with strong electronic correlations are $3d$ transition metal oxides\cite{Imada}.
They boast extremely rich phase diagrams, indicative of competing energy scales.
Indeed, the electronic structure of these oxides is dominated by the interplay of a relatively strong ionic character of bonding (causing large crystal-field splittings),
and strong electron-electron interactions (Mott \& Hund physics).
As a result---while they can be induced by small changes in external conditions---insulating phases in transition metal oxides have comparatively large gaps ($\Delta\gtrsim 0.5$eV).
Therefore, this review does not mention oxides, with the notable exception of LaCoO$_3$.
In transition metal {\it intermetallics}, on the other hand, the bonding has a much stronger covalent character 
(see \sref{bandFeSi} and \sref{sec:covalent} for a discussion of ionic vs.\ covalent compounds). 
Also, in most insulating intermetallics the gap 
is determined by  bonding and/or anti-bondig states formed by the $d$-orbitals of a transition metal and the $p$-orbitals of a group 13-15 element.
Such hybridization gaps in covalent insulators are typically  rather narrow ($\Delta\lesssim 0.5$eV, see \tref{table1} for examples). 
Insulating intermetallic compounds that contain transition metal atoms with partially filled $d$-shells constitute the pool of materials that we focus on in this review.
We will group materials into classes according to their crystal-structure and discuss 
the electronic structure and thermoelectric effects of
representative members in \sref{theo} and \sref{thermo}, respectively. To be specific, the chosen prototypical classes and representative compounds are:
\begin{itemize}
	\item cubic B20-structure: FeSi, RuSi \hfill [see \sref{KI}, \sref{bandFeSi}, \sref{silicides}]
	\item marcasite structure: FeSb$_2$, CrSb$_2$, RuSb$_2$, FeAs$_2$ \hfill [see \sref{marcasites}, \sref{dianti}]
	\item CoGa$_3$-structure: FeGa$_3$, RuGa$_3$ \hfill[see \sref{pdrago}]
	\item skutterudites: CoSb$_3$ \hfill[see \sref{skutts}]
	\item Heusler compounds: Fe$_2$VAl \hfill[see \sref{heuslers}]  
\end{itemize}
The following gives a brief overview of basic experimental findings for the above classes of correlated narrow-gap semiconductors.
Besides, for the above materials, we collected in \tref{table1} 
a list of experimental data, such as the direct (optical), indirect, and spin gaps,
as well as peak amplitudes of thermopowers and powerfactors.
In that list, we have further included a single oxide: LaCoO$_3$. While exhibiting similarities in some physical observables
(see \fref{overviewrho}, \fref{overviewchi}), the microscopic physics underlying them are pronouncedly different than in all other compounds listed here.
These differences will be discussed in \sref{bandFeSi}, \sref{micro} as well as \sref{sec:covalent}.

\subsubsection{Signatures of correlation effects.} 
\label{signature}

\begin{figure*}[th]
  \begin{center}
	{\includegraphics[angle=0,width=.45\textwidth]{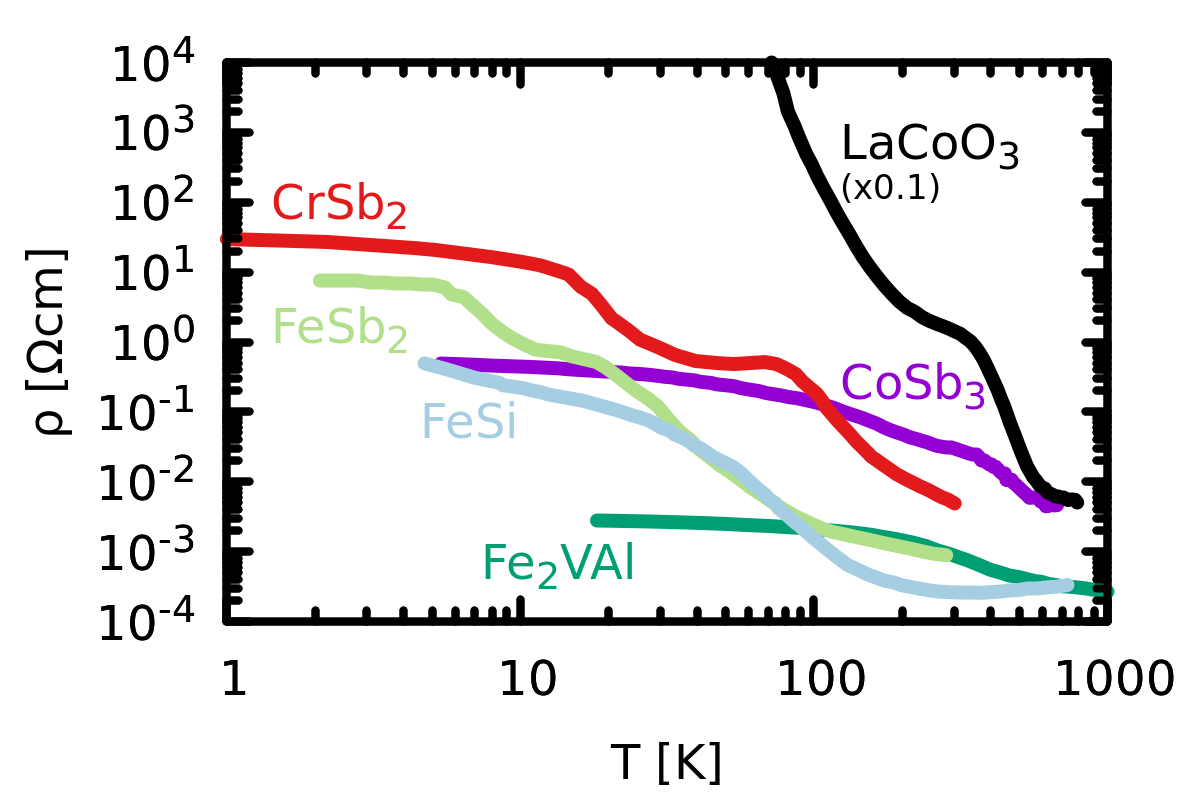}}
	{\includegraphics[angle=0,width=.45\textwidth]{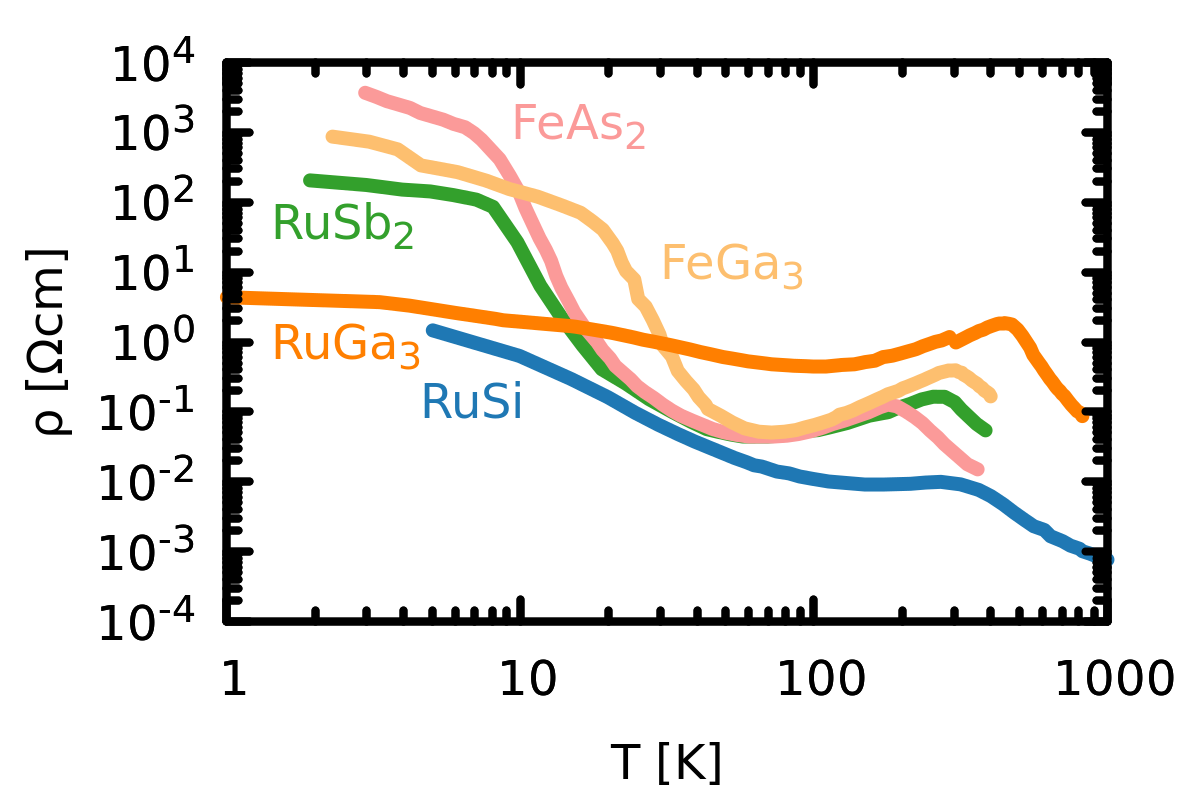}}
	      \caption{{\bf Overview: resistivities.}  Shown is a selection of resistivities
				for some materials from \tref{table1}.
				FeSi: Bocelli \etal\ \cite{Bocelli_FeSi},
				RuSi: Hohl \etal\ \cite{Hohl199839},
				FeSb$_2$, RuSb$_2$, FeAs$_2$: Sun \etal\ \cite{APEX.2.091102},
				CrSb$_2$: Sales \etal\ \cite{PhysRevB.86.235136},     
				FeGa$_3$: Gamza \etal\ \cite{monika_fega3},
				RuGa$_3$: Wagner-Reetz \etal\ \cite{PhysRevB.90.195206},
        CoSb$_3$: Mandrus \etal\ \cite{PhysRevB.52.4926},
				Fe$_2$VAl: Nishino \etal\ \cite{PhysRevLett.79.1909},
				LaCoO$_3$: Yamaguchi \etal\ \cite{PhysRevB.53.R2926}.
				For more experimental data see \fref{FeSirhoPF} for FeSi, and
\fref{Fe2VAlexp} for Fe$_2$VAl. 
								}
      \label{overviewrho}
      \end{center}
\end{figure*}

Correlated narrow-gap semiconductors display a number of intriguing properties. 
In particular the temperature-dependence of transport, spectroscopic and magnetic quantities is highly unusual.

\paragraph{Charge degrees of freedom.}

Basically all materials considered here exhibit one or more temperature regimes in which the resistivity follows activated behaviour, as expected for conventional semiconductors,
see \fref{overviewrho}. In systems with comparatively large gaps ($\Delta\gtrsim200$meV), e.g., FeAs$_2$, RuSb$_2$, or FeGa$_3$, there can be several regions
that are amendable to an activation-law fit with gaps of different sizes. The largest of the latter will give a description of the activation regime that is highest
in temperature and is identified as the fundamental gap, oftentimes in congruence with
the direct gap extracted from optical experiments. The smaller gaps are then interpreted as arising from impurity or defect states inside the intrinsic gap.
The different activated regimes may be connected by regions in which the slope of the resistivity is positive, albeit 
in absence of any metallic characteristics in optical or photoemission spectra.

The situation is different for the materials on our list that have the narrowest gaps ($\Delta\lesssim100$meV), e.g., FeSi or FeSb$_2$.
These systems behave as renormalized but coherent semiconductors at low temperatures. However, activated behaviour 
persists only up to a temperature $T^*_\rho$ that is still significantly smaller than the respective gap $\Delta$: $k_BT^*_\rho\ll\Delta$.
Here, we arbitrarily define $T^*_\rho$ as the temperature where the relative deviation from activated behaviour, $(\rho-\rho_{\hbox{\tiny act}})/\rho_{\hbox{\tiny act}}$,
exceeds $3\%$, see \fref{FeSiCe3trans}(a) for the case of FeSi.
There, the resistivity clearly surpasses the exponential decay above $\sim T^*_\rho=160$K
and even develops a positive slope, while $k_BT^*_\rho=14$meV$\ll\Delta\approx50$meV. Concomitantly, 
optical\cite{PhysRevLett.71.1748,PhysRevB.55.R4863,PhysRevB.56.1366,PhysRevB.79.165111,perucci_optics,PhysRevB.82.245205}, 
Raman\cite{PhysRevB.51.15626},
photoemission\cite{PhysRevB.52.R16981,SAITOH1995307,1367-2630-11-2-023026,PhysRevB.72.233202,PhysRevB.77.205117} and 
tunnelling\cite{PhysRevB.58.15483} spectroscopy witness in FeSi and FeSb$_2$ a clear crossover to a metallic phase, see \fref{FeSiCe3spec}(a) \& (b).
These results are  in clear defiance of a mere thermal activation across a fundamental gap. 
Moreover, with changing temperature, transfers of spectral-weight in the optical conductivity occur over energy-scales much larger than the size of the gap\cite{PhysRevLett.71.1748,PhysRevB.79.165111,perucci_optics,PhysRevB.82.245205}, suggestive
of many-body effects\cite{PhysRevB.54.8452} that are controlled by large interactions $U$ rather than the size of the gap, $U\gg\Delta$.
This will be discussed in more detail in \sref{KI}.

Besides the intriguing temperature dependence, also the energies of one-particle excitations, such as the gap, are beyond expectations from band-theory:
In the case of FeSi, the low temperature gap as well as band-dispersions are largely overestimated by density functional theory (DFT\cite{RevModPhys.71.1253,RevModPhys.61.689}) (cf.\ \sref{band}), indicative of notably
enhanced masses $m^*/m_{\hbox{\tiny band}}$ caused by dynamical correlations.

The mentioned effects are most pronounced for the materials that have the smallest gaps. Within a given class, these are the systems build from
a $3d$ transition metal and a group 13-15 element with large atomic radius. 
Indeed, correlation effects, and thus band-width and gap renormalizations, are expected to be less pronounced for transition metals with more extended $4d$-orbitals.
For comparisons of $3d$-transition metal systems to their $4d$ homologues, see \sref{bandFeSi} and \sref{marcasites} for the cases of silicides and antimonides, respectively.

Also at very low temperature deviations to activated behaviour occur in the resistivities,  see \fref{overviewrho}.
Indeed, they show pronounced tendencies towards saturation for basically all materials considered here.
Commonly, this behaviour is ascribed to the presence of impurity or defect states inside the gap that pin the chemical potential and supply residual conduction.%
\footnote{See \sref{pdrag} for a modelling of such impurity contributions to the conductivity in FeSb$_2$.}
This scenario is supported by the presence of different activation regimes in some compounds (see above), as well as a
strong sample dependence of low-temperature transport observables.
However, finite lifetimes of {\it intrinsic} valence and conduction states---ubiquitous in correlated materials---provide an alternative mechanism for resistivity saturation, as will be discussed in \sref{KubBoltz}.

The importance of electronic correlation effects in narrow-gap intermetallic semiconductors with open $3d$-shells is further strengthened by the following: Doping the discussed stoichiometric compounds yields metallic phases in which mobile charge carriers exhibit largely enhanced effective masses.
Some examples are:

\begin{itemize}
	\item FeSi$_{1-x}$Al$_x$\cite{ANDP:ANDP2065080206,PhysRevLett.78.2831,PhysRevB.58.10288,PhysRevB.87.184304} yields masses 14 times larger than the
	free electron value, as extracted from, both, the specific heat and the susceptibility for $x=1\%-5\%$\cite{PhysRevLett.78.2831}.
	\item Fe$_{1-x}$Co$_x$Si has masses up to $m^*/m_e=30$\cite{PhysRevB.56.1366}, as extracted from the specific heat.
	\item FeSb$_{2-x}$Te$_x$ yields $m^*/m_e=10-30$ as extracted from the thermopower\cite{10.1063/1.3556645}.
	\item FeSb$_{2-x}$Sn$_x$ exhibits for $x>0$ masses $m^*/m_e=10-15$ as extracted from the specific heat\cite{bentien:205105}.
	\item CoSb$_3$'s Hall coefficient \cite{doi:10.1063/1.363405}, as well as the thermopowers of
  Co$_{1-x}$Ni$_x$Sb$_3$\cite{doi:10.1063/1.371287,PhysRevB.65.115204} and CoSb$_3$ with Yb-fillings \cite{Tang2015} display enhanced masses: $m^*\sim 2-5m_e$.
	\item Fe$_2$VAl: the residual in-gap density in typical (possibly off-stoichiometric) samples causes a metallic specific heat from which
	effective masses enhanced by a factor of 5 have been extracted\cite{PhysRevB.60.R13941}.
\end{itemize}

\paragraph{Magnetic properties.}

\begin{figure*}[th]
  \begin{center}
		{\includegraphics[angle=0,width=.45\textwidth]{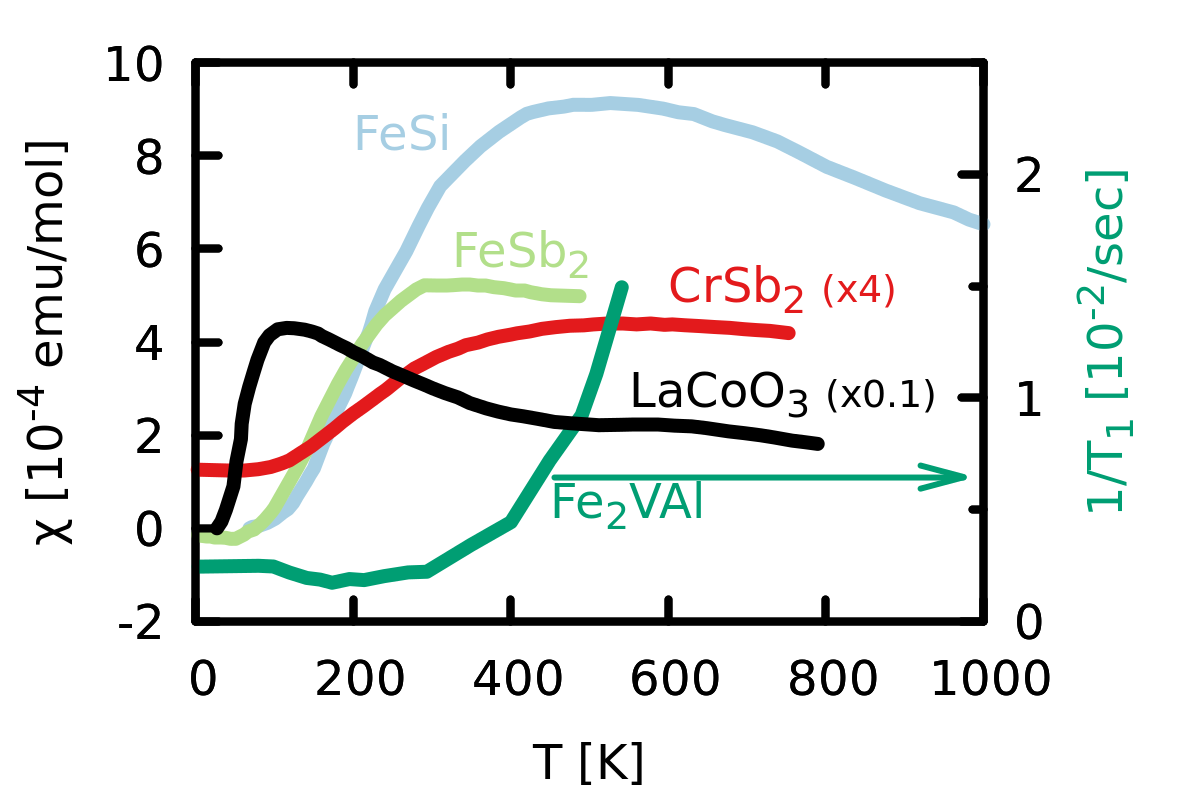}}
		{\includegraphics[angle=0,width=.45\textwidth]{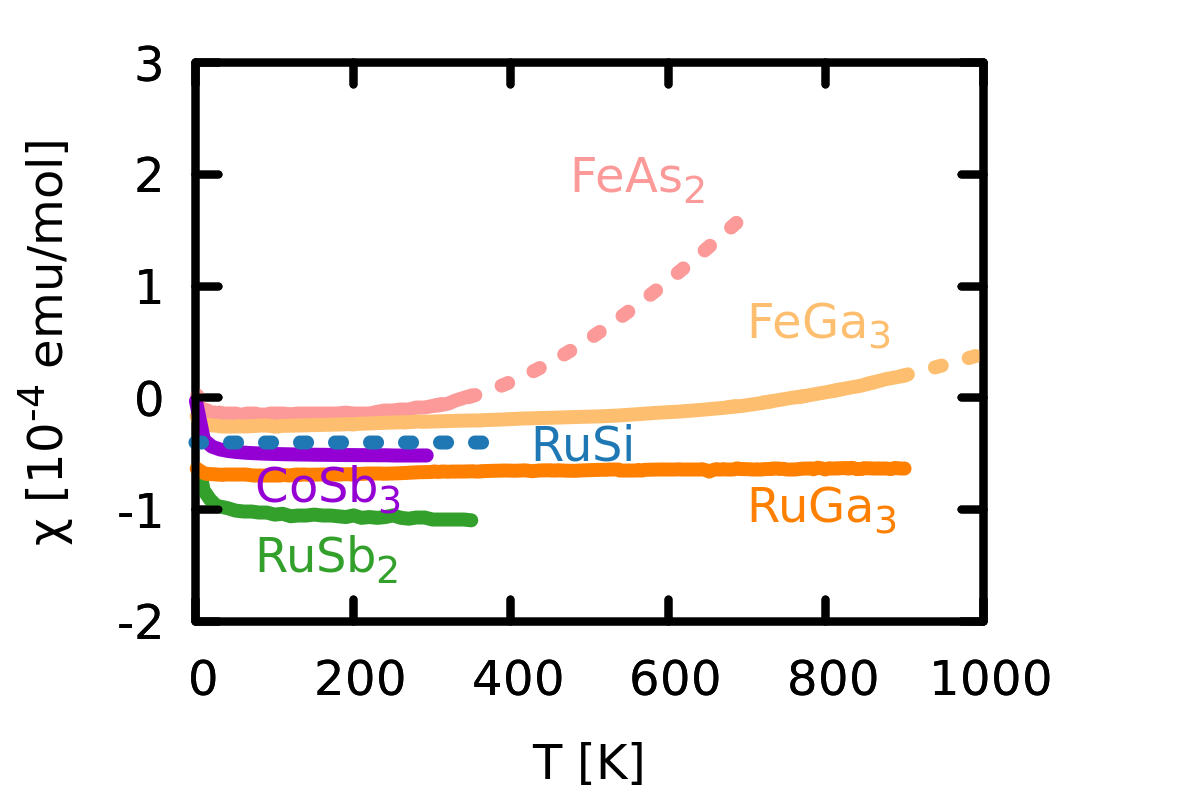}}
	      \caption{{\bf Overview: magnetic susceptibilities.}  Shown is a selection of susceptibilities
				for some materials from \tref{table1}.
				FeSi: Jaccarino \etal\ \cite{PhysRev.160.476}, Takagi \etal\ \cite{JPSJ.50.2539},
				RuSi: Buschinger \etal\ \cite{BUSCHINGER199757} and Hohl \etal \cite{Hohl199839} stated the intrinsic $\chi$ to be basically temperature-independent up to 320K,
				FeSb$_2$ Koyama \etal\ \cite{koyama:073203},                    
				RuSb$_2$, FeAs$_2$: Sun \etal\ \cite{APEX.2.091102},
				CrSb$_2$: Sales \etal\ \cite{PhysRevB.86.235136},     
				FeGa$_3$, RuGa$_3$: Gamza \etal\ \cite{monika_fega3} (field $H\parallel ab$),
				CoSb$_3$: Morelli \etal\ \cite{PhysRevB.56.7376},
				LaCoO$_3$: Yamaguchi \etal\ \cite{PhysRevB.53.R2926}.
				For
				Fe$_2$VAl the $^{27}$Al NMR relaxation rate of Lue and Ross\ \cite{PhysRevB.58.9763} is shown.
				Dotted lines for FeAs$_2$ and FeGa$_3$ are activation-law fit extrapolations.
				CrSb$_2$ is antiferromagnetic below 273K;  a complex magnetic order has also been suggested for FeGa$_3$\cite{monika_fega3,0953-8984-30-4-045601}.
								}
      \label{overviewchi}
      \end{center}
\end{figure*}

The intriguing insulator-to-metal crossover in the small-gap materials is accompanied by large changes in the magnetic response, see \fref{overviewchi}:
Starting from low temperatures, the uniform spin susceptibility of some systems rises to quite large absolute values.
Following activated behaviour, the increase in the response is largest for the materials with the smallest fundamental gaps.
Beyond that, one can surmise that electronic correlations---greater in $3d$ than $4d$ transition metal-based compounds---play a role in
the prefactor of the exponential rise: Indeed while FeAs$_2$ and RuSi, as well as FeGa$_3$ and RuGa$_3$, have pairwise comparable gaps (see \tref{table1}), only the $3d$ compounds
show an activated susceptibility (within the available experimental temperature window).
In the theory \sref{theo}, see in particular \sref{PAM}, we suggest that this distinction is mainly driven by the difference in the Hund's rule coupling.

Still more intriguing is that the magnetic susceptibilities of FeSi, FeSb$_2$, and CrSb$_2$ reach a maximum---at a temperature that we shall call
$T^{max}_\chi$---beyond which they realize a Curie-Weiss-like decay.%
\footnote{The onset of a peculiar antiferromagnetic order in CrSb$_2$ below 273K\cite{Holseth1970,PhysRevLett.108.167202}  has no visible signature in the susceptibility.} 
Together with the insulator-to-metal crossover, this signature in the spin-response is the most salient and unexpected property of correlated semiconductors.

While all materials considered here%
\footnote{with the exception of CrSb$_2$ below its N{\'e}el temperature $T_N=273$K, and possibly FeGa$_3$\cite{monika_fega3,0953-8984-30-4-045601}.}
are paramagnetic, many of them are situated in direct proximity to 
a spin-ordered phase. Magnetism can be induced by isoelectronic substitutions that expand the lattice, off-stoichiometry, as well as doping.
Examples include:

\begin{itemize}
	\item FeSi$_{1-x}$Ge$_x$: ferromagnetic metal above $x=0.25$\cite{BAUER19981401,PhysRevLett.91.046401}.
	\item Fe$_{1-x}$Ru$_x$Si: ferromagnetic metal for $x=0.1$ below 15K \cite{Paschen1999864}.
	\item Fe$_{1−x}$Co$_x$Si: ferromagnetic metal for $0.05\le x\le 0.8$ \cite{Beille1983399,Paschen1999864,doi:10.1143/JPSJ.59.305,Manyala2000,PhysRevB.72.224431}.
	\item Fe$_{1−x}$Co$_x$Sb$_2$: weak metallic ferromagnetism for $0.2\le x\le 0.4$ \cite{hu:224422,PhysRevB.74.195130}.
	\item Fe$_{1−x}$Cr$_x$Sb$_2$: antiferromagnetic insulator for $0.2\le x\le 1$ \cite{PhysRevB.76.115105}.
	\item Fe$_2$VAl orders ferromagnetically via a multitude of small changes with respect to stoichiometry:
	Fe$_{2+x}$V$_{1-x}$Al\cite{PhysRevB.85.085130,KANOMATA2001390,doi:10.1143/JPSJ.69.1004,0953-8984-12-8-318},
   Fe$_2$V$_{1-x}$Cr$_x$Al\cite{PhysRevB.79.174423},		Fe$_2$VAl$_{1-\delta}$\cite{PhysRevB.82.104408},
		Fe$_{2−x}$V$_{1+x}$Al\cite{SODA2004338}. 
\end{itemize}
Details can be found in \sref{KI} (FeSi), \sref{marcasites} (FeSb$_2$), and \sref{heuslers} (Fe$_2$VAl).

\subsubsection{Thermoelectricity.}

\begin{figure}[!h]
  \begin{center}
	{\includegraphics[angle=0,width=.45\textwidth]{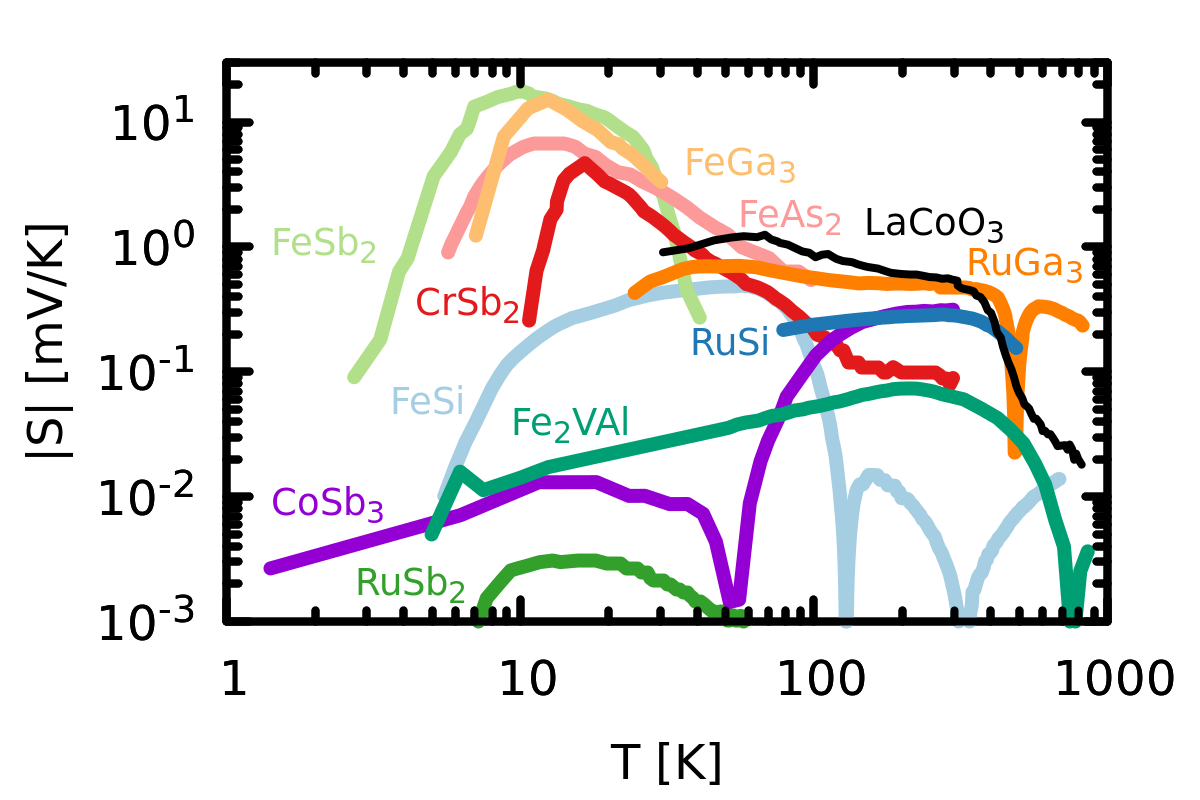}}
		      \caption{{\bf Overview: thermopowers.}
					Shown is a selection of thermopowers
				for some materials from \tref{table1}.
				FeSi: Sales \etal\ \cite{PhysRevB.50.8207},
				RuSi: Hohl \etal\ \cite{Hohl199839},
				FeSb$_2$, RuSb$_2$, FeAs$_2$: Sun \etal\ \cite{APEX.2.091102},
				CrSb$_2$: Sales \etal\ \cite{PhysRevB.86.235136},     
				FeGa$_3$, RuGa$_3$: Wagner-Reetz \etal\ \cite{PhysRevB.90.195206},
        CoSb$_3$: Dyck \etal\ \cite{PhysRevB.65.115204},
				Fe$_2$VAl: Knapp \etal\ \cite{PhysRevB.96.045204},
				LaCoO$_3$:  Se{\~n}arı{\'i}s-Rodr{\'i}guez and Goodenough \cite{SENARISRODRIGUEZ1995224}.
				For more experimental data see \fref{FeSiCe3trans}(c), \fref{FeSiS} for FeSi, 
\fref{Takahashi} for FeSb$_2$, \fref{CoSb3} for CoSb$_3$, \fref{Fe2VAlexp} for Fe$_2$VAl.     
								}
      \label{overviewS}
      \end{center}
\end{figure}

The observation that warrants the title of this review is that correlated narrow-gap semiconductors
exhibit large thermopowers. 
For stoichiometric compositions, the largest response is usually achieved at 
rather low temperatures when these systems are in their insulating regime, see \fref{overviewS}.
Indeed, the typical temperature dependence of the thermopower is a $U$-like shaped curve.
The thermopower of a generic, coherent insulator behaves roughly as $\propto 1/T$ (cf.\ \sref{limit}).
In our systems of interest, this decay with rising temperature is accelerated by the insulator-to-metal crossover
(cf.\ \fref{FeSicoh} in \sref{silicides} for the case of FeSi).
On the low-temperature side, the thermopowers vanish towards absolute zero, as expected from the third law of thermodynamics
(cf.\ \fref{BoKu} in \sref{KubBoltz} for the influence of finite lifetimes on this decay).

Some of the considered materials have competitive powerfactors $S^2\sigma$ in a temperature-range that is 
useful for Peltier-refrigeration, or even waste-heat recovery (cf.\ \sref{skutts} and \sref{heuslers} for skutterudite and Heusler compounds, respectively).
However, most of the correlated narrow-gap semiconductors discussed here (notably those which exhibit the most pronounced
correlation effects), have large powerfactors only at very low temperatures (cf.\ \sref{silicides} for FeSi and \sref{dianti} for FeSb$_2$).
An application of these stoichiometric compounds in thermoelectric devices is thus only envisagable 
for sensor or cooling technologies at cryogenic temperatures\cite{Heremans2016}. 
However, owing to large lattice thermal conductivities, no viable alternative for liquid-He-cooling is on the horizon.
The impact of correlation effects onto thermoelectric properties, as well as individual materials will be discussed in \sref{thermo}.

\subsubsection{Other applications.}

Besides thermoelectricity, there are a number of proposals 
for the use and relevance of the unconventional physics of correlated narrow-gap semiconductors, both, in
fundamental science as well as for technological applications. Examples include:
\begin{itemize}
  \item Spin transport electronics: the very large anomalous Hall effect in Fe$_{1-x}$Co$_x$Si\cite{Manyala2004}
is a harbinger for spintronic applications.
	\item Astrophysics: from infrared absorption measurements, 
FeSi has been hinted to occur in circumstellar dust shells\cite{Ferrarotti2000},
as previously suggested by calculations\cite{Lodders1999}.
Ferrarotti \etal\ \cite{Ferrarotti2000} proposed to exploit the 
strongly temperature-dependent properties of FeSi  as a ``thermometer'' to probe 
environmental temperatures near distant stars.
\item Earth and planetary science: FeSi (B20 or B2 structure) is possibly a relevant composition in the core of the Earth and other terrestrial (rocky) planets
\cite{GRL:GRL8102,GRL:GRL18651,JGRB:JGRB13523}. Its presence near the core-mantle boundary might explain the anomalously high electrical conductivity of 
this region\cite{Dubrovinsky2003}.
\item CrSb$_2$ and FeSb$_2$ have been considered as high-capacity anode materials in lithium-based batteries\cite{FERNANDEZMADRIGAL2001205,PARK20104987}.
\item Heusler materials have applications in, e.g., spintronics, solar cells, or magneto-calorics\cite{Graf20111,Galanakis2016}.
\end{itemize}

\subsubsection{Perspective.}

From the above overview, we can extract some general tendencies for the signatures of correlation effects in our materials of interest.
Both, the insulator-to-metal crossover and the Curie-Weiss-like susceptibility are more evident
the smaller the semiconducting gap is. Typically, narrow gaps occur in the $3d$-transition metal compounds rather than their $4d$ analogues.
Indeed, dynamical correlations as well as spin-fluctuations are naturally expected to be of greater importance the more localized
the orbitals that host them are:
Larger effective masses in the $3d$ materials cause more strongly renormalized gaps, as well as magnitudes of fluctuating moments that approach
the values of the respective isolated ion.
In all, the physics behind the correlation signatures in the charge and the spin sector, as well as their characteristic
temperatures, $T^*_\rho$ and $T^{max}_\chi$, are likely to be microscopically linked (see \sref{micro}).
Pertinent energy scales are the charge and the spin gap, the strength of electron-electron interactions,
in particular (see sections \sref{real}) the Hund's rule coupling.
Depending on the interplay of these energies, manifestations of many-body effects will vary notably, as evidenced in
the figures \ref{overviewrho}, \ref{overviewchi}, and \ref{overviewS}.

\begin{framed}
\noindent
	{\bf Signatures of correlation effects in narrow-gap semiconductors}
	\begin{itemize}
		\item small upper bound for activation laws in the resistivity $T^*_\rho\ll\Delta/k_B$,
		 and the magnetic susceptibility $T^{max}_\chi\lesssim\Delta/k_B$.
		\item insulator-to-metal crossovers for $k_BT\ll\Delta$.
		\item enhanced paramagnetism at low $T$, Curie-Weiss-like behaviour for $T>T^{max}_\chi$\\
		(if experimentally accessible).
		\item transfer of optical spectral weight over energies much larger than the gap $\Delta$.
		\item proximity to (spin) ordered phases.
		\item large effective masses under doping.
		\item large thermopowers at low temperatures.
	\end{itemize}
\end{framed}

\begin{figure*}[p]
  \begin{center}
	FeSi $\qquad\qquad\qquad\qquad\qquad\qquad\qquad\qquad\qquad\qquad$ Ce$_3$Bi$_4$Pt$_3$
	\subfloat[{\bf resistivity.} FeSi (left): Bocelli \etal\ \cite{Bocelli_FeSi}; Ce$_3$Bi$_4$Pt$_3$ (right): Katoh \etal\ \cite{Katoh199822}.
	Dashed (blue) lines are activation law fits with gap $\Delta$; $T^{max}_\chi$ indicates the peak temperature of the susceptibility (see \fref{FeSiCe3trans}(b)); 
	at $T^{*}_\rho=160$K$/120$K (FeSi/\cbp) the relative deviation of the resistivity to the shown activation law exceeds $3\%$.
	]{%
	{\includegraphics[angle=0,width=.45\textwidth]{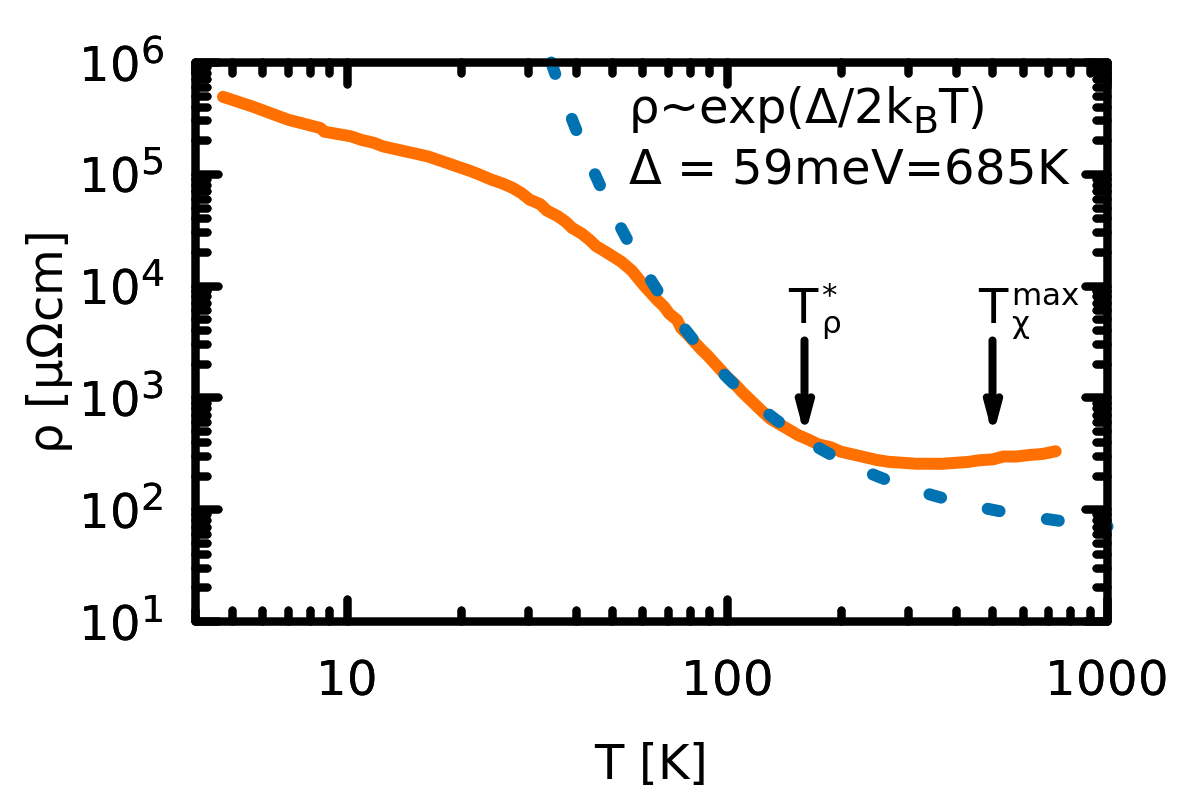}}
	{\includegraphics[angle=0,width=.45\textwidth]{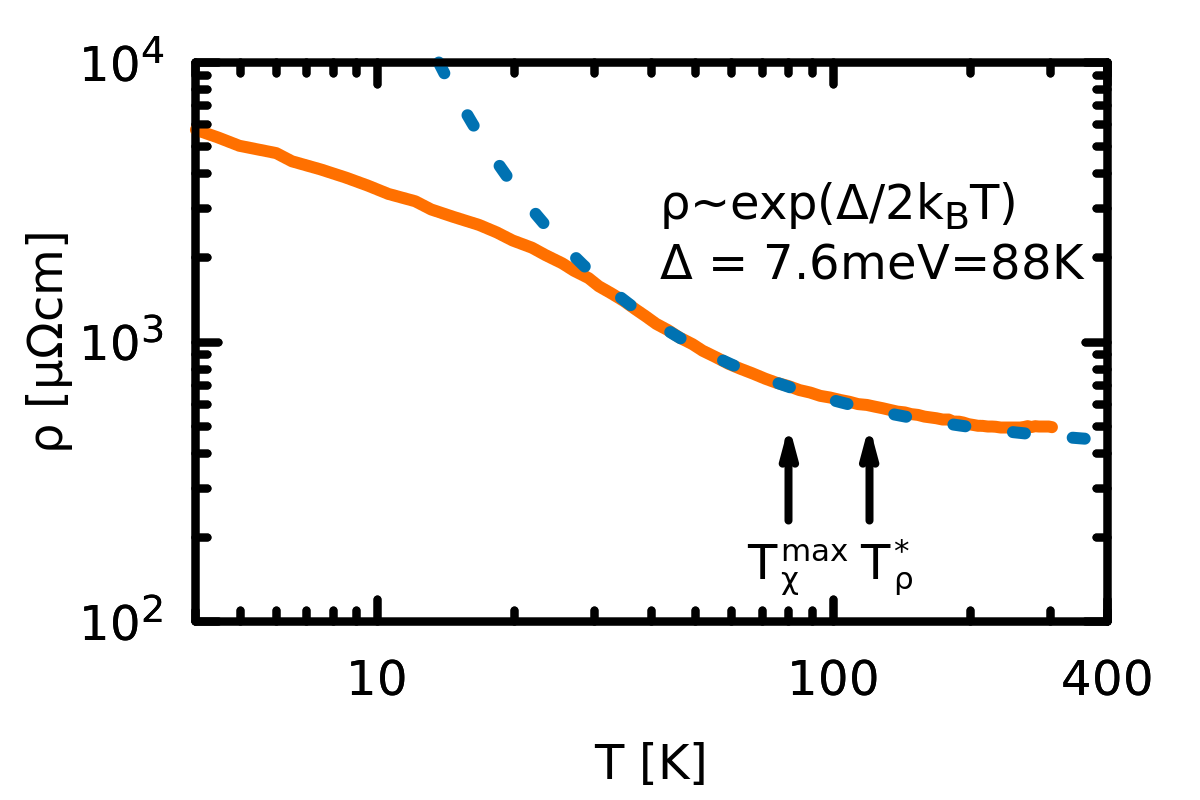}}
	}
	
\subfloat[{\bf magnetic susceptibility.} FeSi (left): data combined from Jaccarino \etal\ \cite{PhysRev.160.476} and Takagi \etal\ \cite{JPSJ.50.2539};
Ce$_3$Bi$_4$Pt$_3$ (right): Hundley \etal\ \cite{PhysRevB.42.6842}; dashed (blue) lines are activation law fits $\sim \exp(-\Delta_s/k_BT)/T$ in the local moment picture; 
dashed (red) lines indicate Curie-Weiss-like behaviour, $\chi=N_A\mu_{eff}^2/ \left( 3k_B(T-\theta)\right) $. In between, $\chi$ peaks at $T^{max}_\chi=500$K (75K) for FeSi (\cbp ).]
{{\includegraphics[angle=0,width=.45\textwidth]{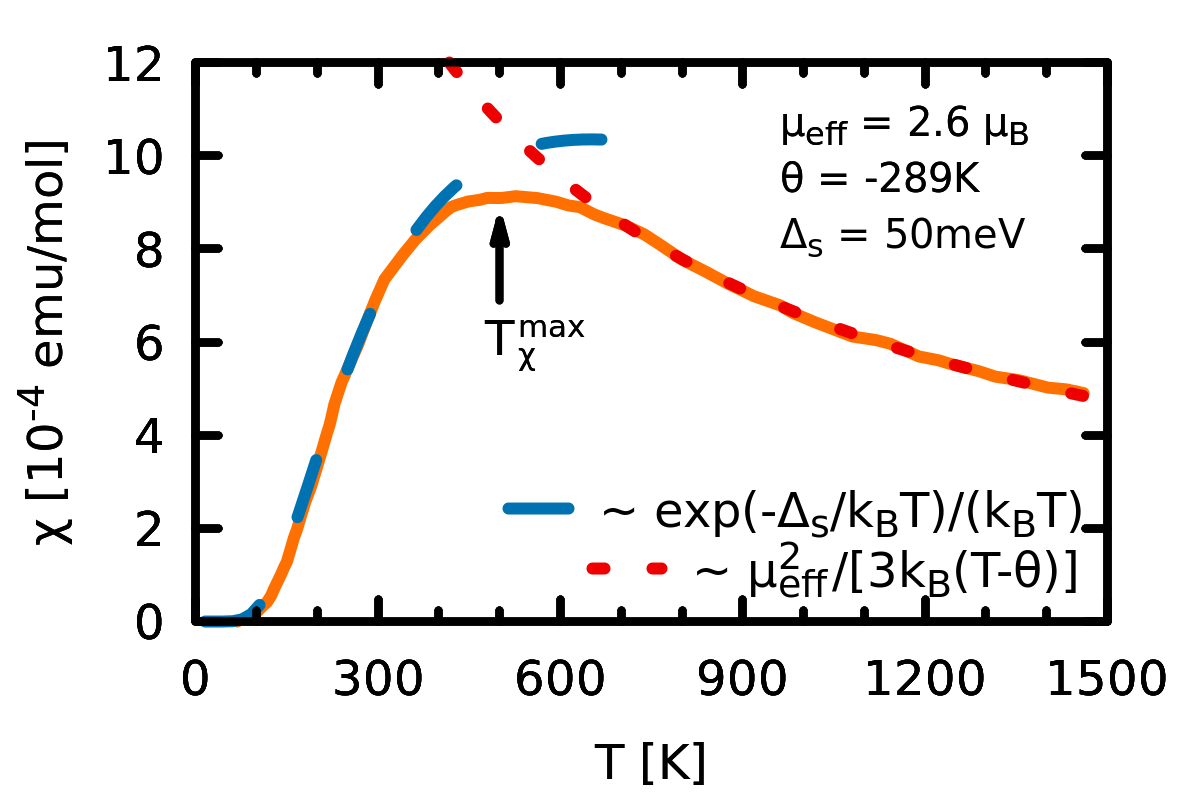}}
{\includegraphics[angle=0,width=.45\textwidth]{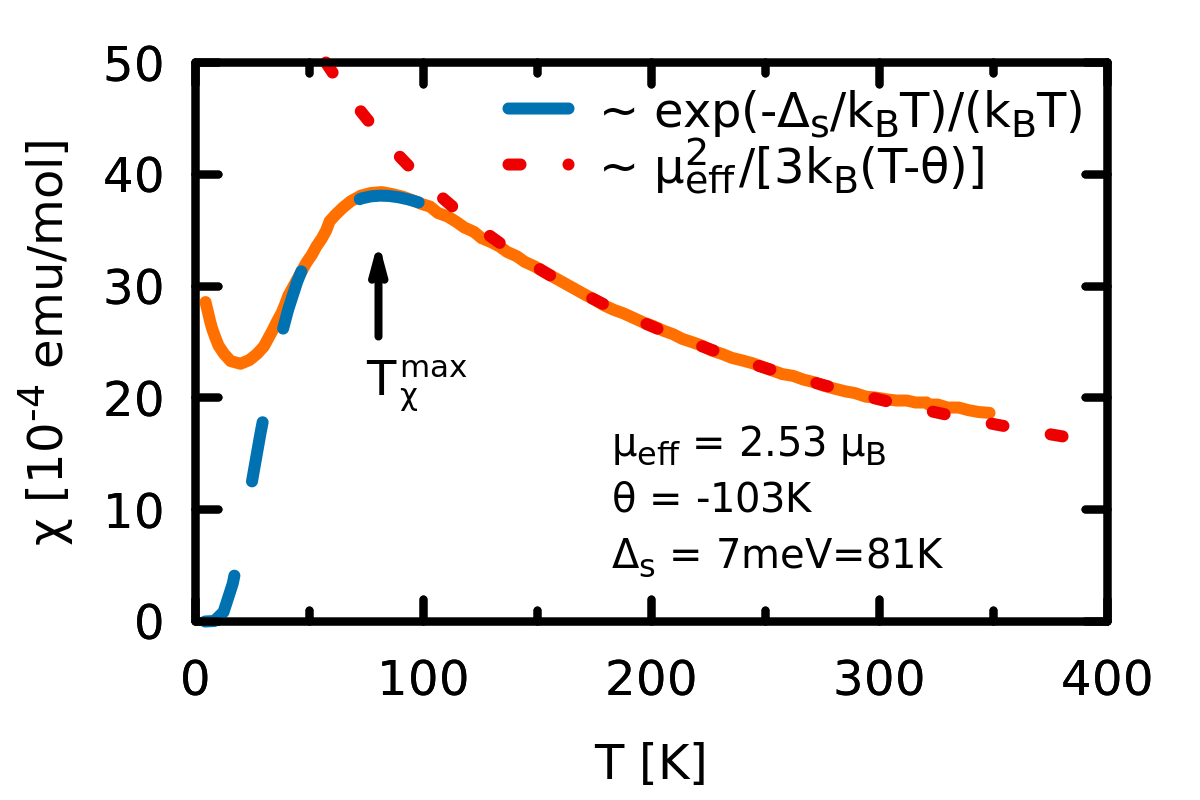}}}

	\subfloat[{\bf thermopower.} FeSi (left):  
	Wolfe \etal\ \cite{Wolfe1965449},
	Sun \etal\ \cite{PhysRevB.90.245146},
	Sales \etal\ \cite{PhysRevB.50.8207,PhysRevB.83.125209};
	Ce$_3$Bi$_4$Pt$_3$ (right): Hundley \etal\ \cite{PhysRevB.50.18142} and Katoh \etal\ \cite{Katoh199822}.]
	{ {\includegraphics[angle=0,width=.45\textwidth]{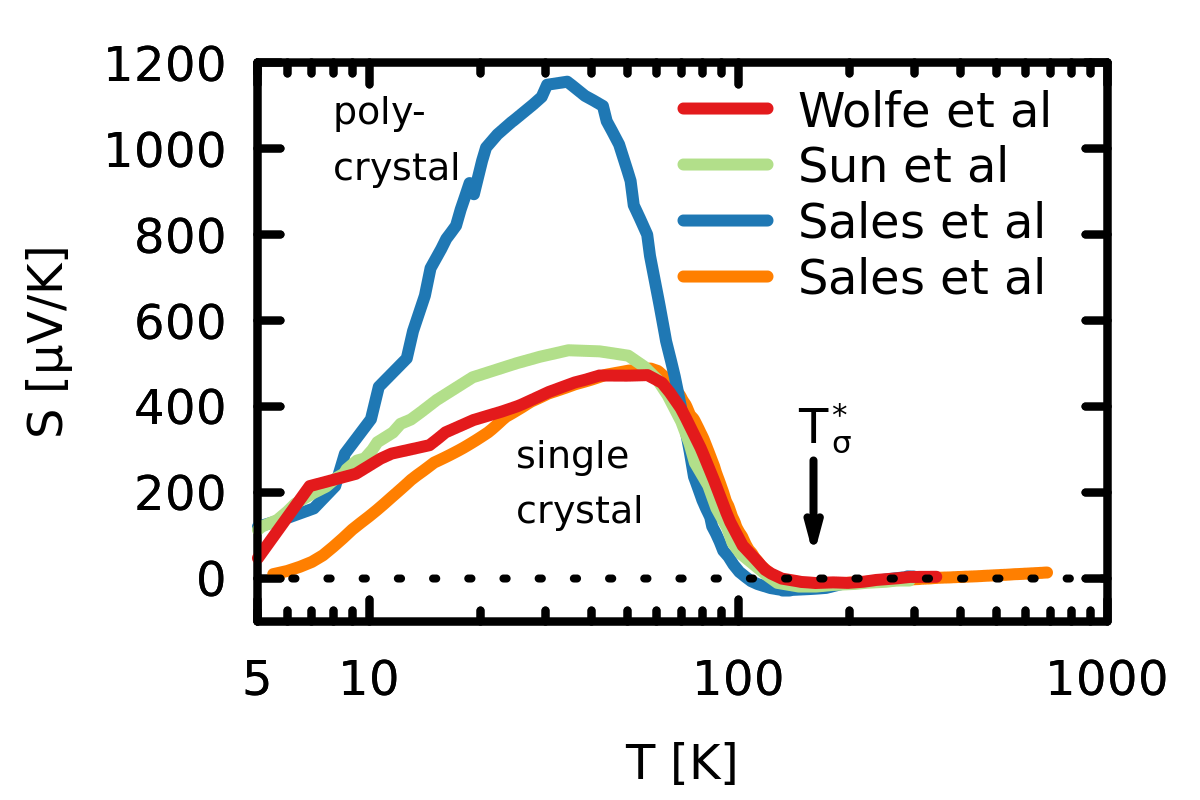}}
	{\includegraphics[angle=0,width=.45\textwidth]{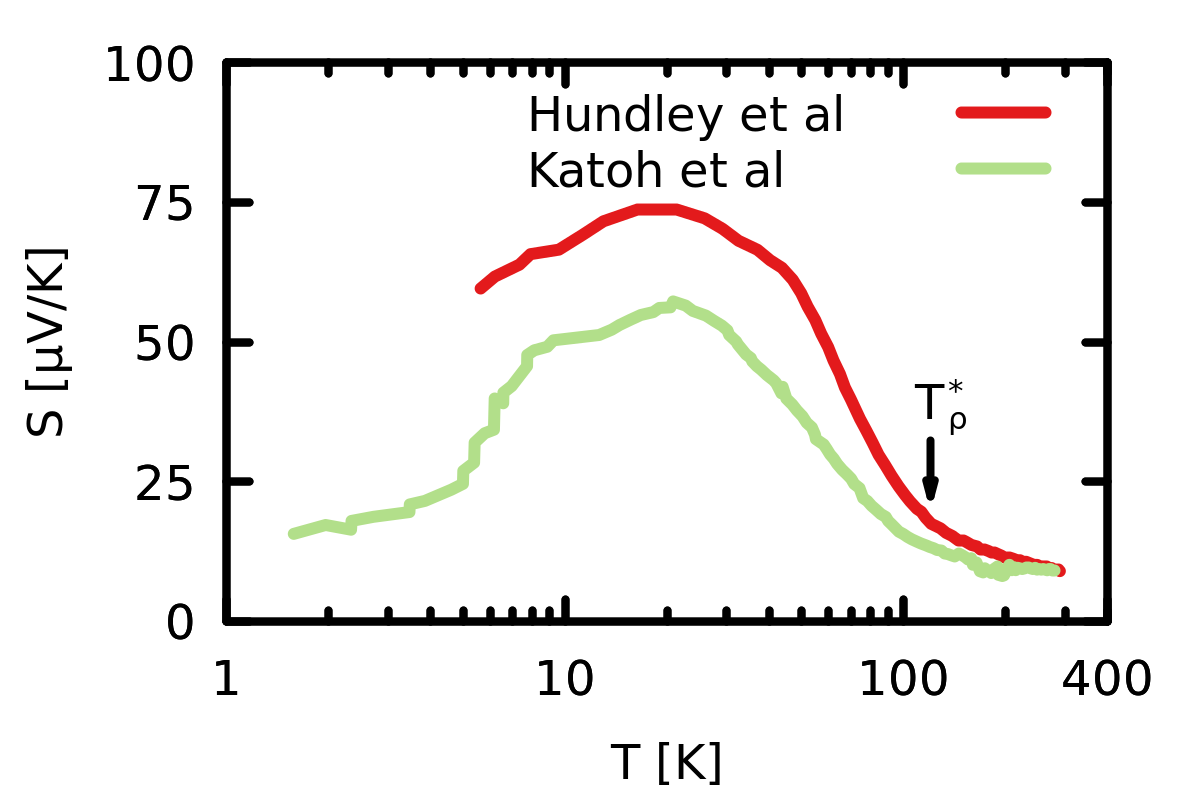}}}
	      \caption{{\bf Transport and magnetic properties:} The correlated narrow-gap semiconductor FeSi (left) vs.\ the Kondo insulator Ce$_3$Bi$_4$Pt$_3$ (right).
				}
      \label{FeSiCe3trans}
      \end{center}
\end{figure*}

\subsection{Correlated narrow-gap semiconductors vs.\ Kondo insulators}
\label{KI}

Early on it has been noted \cite{ki,FISK1995798,PhysRevB.51.4763,PhysRevB.67.155205,PhysRevB.72.045103,PhysRevB.58.10288,LACERDA19931043,Schlesinger1997460,PhysRevB.50.14933,1742-6596-273-1-012055,Colemann2007}
 that the anomalies observed in the most preeminent of $d$-electron narrow-gap semiconductors 
are strikingly reminiscent of the physics of heavy-fermion semiconductors. 
The latter, also known as Kondo insulators\cite{ki}, are systems with partially filled $f$-shells 
that show insulating behaviour at low temperatures in the absence of any magnetic long-range order.
In the standard (Doniach) picture \cite{DONIACH1977231}, the local moment of the $f$-electrons fluctuates freely at high-temperatures, giving rise to a Curie-Weiss susceptibility.
Concomitantly, light ($s$, $p$) conduction states are decoupled from the localized ($f$) states and their density is finite at the Fermi level, causing metallic behaviour.
Upon lowering the temperature the local moments get screened by the conduction electrons via the Kondo effect, and the resulting hybridization between $f$- and conduction states
leads to the liberation of mobile charges with large effective masses. 
In Kondo insulators, however, these heavy charges lead to a completely filled band, i.e., the chemical potential happens to fall into the hybridization gap,
quenching spin and charge excitations at low temperatures.

In this section, we will detail empirical similarities and differences between such $f$-electron heavy-fermion Kondo insulators
and the $d$-electron based correlated narrow-gap insulators that are the main subject of this review.
We will focus the juxtaposition mainly on a direct comparison of 
two compounds prototypical for their class: FeSi and Ce$_3$Bi$_4$Pt$_3$.%
\footnote{For a wider discussion of the family of Kondo insulating materials, including complications arising from the multiplet and crystal-field structure, as well as the interplay of electronic correlations
and spin-orbit coupling, see the reviews in Refs.~\cite{FISK1995798,RevModPhys.71.687,Riseborough2000,Dzero2016}.
}
Our survey of experimental and phenomenological observations will find a continuation in subsequent chapters: 
In \sref{band}, we discuss the electronic structures of FeSi and \cbp\ within standard band theory.
In \sref{model}, we present a survey of results for the  minimal many-body setups that contain the salient
features of correlated semiconductors and Kondo insulators---the two-covalent-band Hubbard model and the periodic Anderson model, respectively.
This comparison in the context of reductionist models is extended in \sref{PAM}.
Finally, we turn in \sref{real} to realistic many-body electronic structure calculations: Results for FeSi are reviewed 
in \sref{pnas} and \sref{micro}, while \sref{Ce3DMFT} presents new findings for \cbp.
The comparison on these two systems culminates in \sref{sec:covalent}.

\subsubsection{Charge degrees of freedom.}

We begin our one-to-one comparison of FeSi and \cbp\ with properties related to the charge degrees of freedom: the resistivity, the optical conductivity and
photoemission spectra.

\begin{figure*}[!t!h]
  \begin{center}
	FeSi $\qquad\qquad\qquad\qquad\qquad\qquad\qquad\qquad\qquad\qquad$ Ce$_3$Bi$_4$Pt$_3$
	\subfloat[{\bf photoemission spectra.} FeSi (left): Klein \etal\ \cite{1367-2630-11-2-023026}; Ce$_3$Bi$_4$Pt$_3$ (right): Takeda \etal\ \cite{Takeda1999721}. In both cases the photoemission intensity was (i) divided by the Fermi function, and, (ii) a fit of the resulting spectra at the highest temperature (120K for FeSi, 300K for Ce$_3$Bi$_4$Pt$_3$) was subtracted to emphasize relative changes.]
	{{\includegraphics[angle=0,width=.45\textwidth]{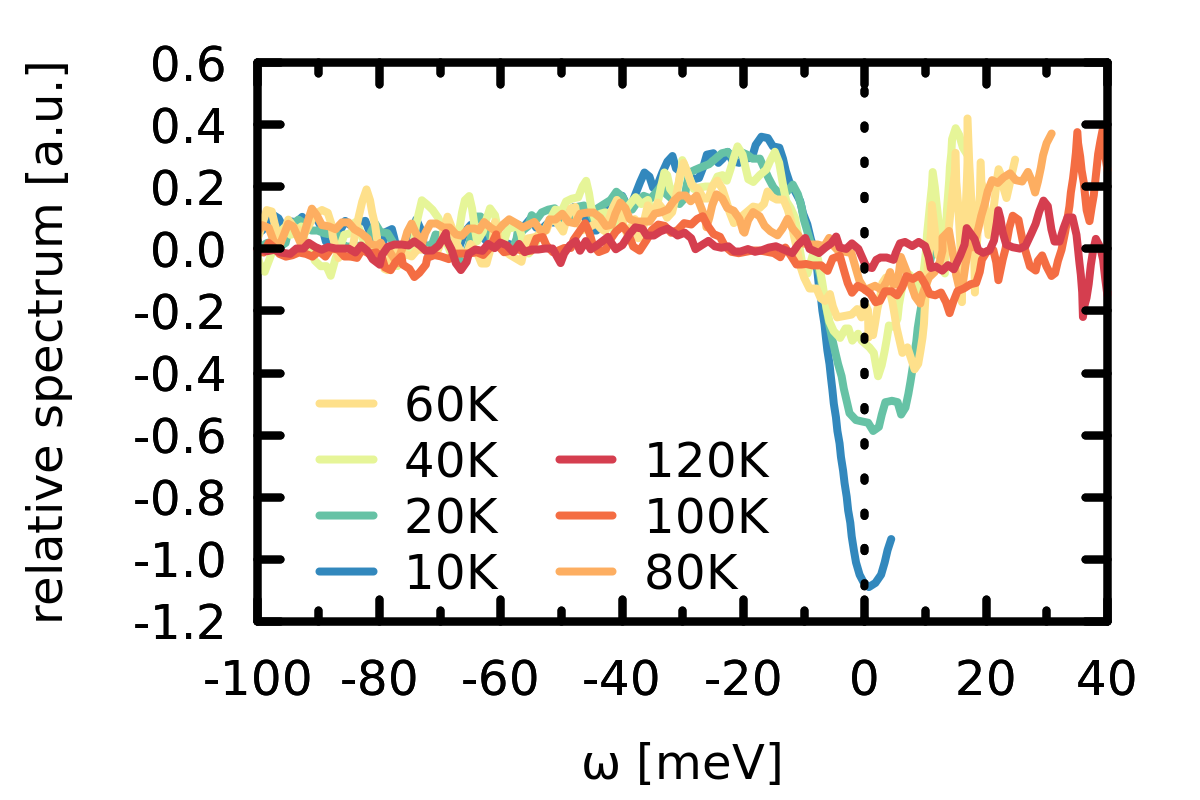}}
	{\includegraphics[angle=0,width=.45\textwidth]{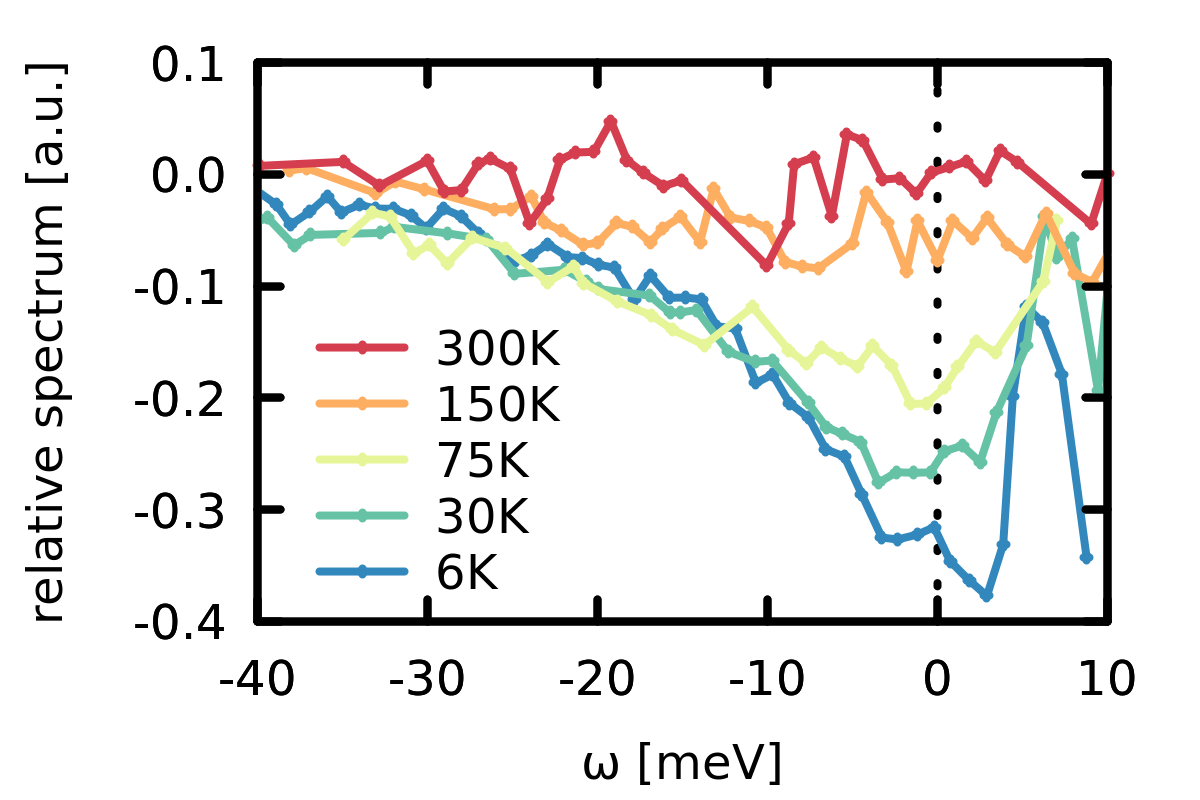}}
	}

	\subfloat[{\bf optical conductivity.} FeSi (left): ellipsometry data from Menzel \etal\ \cite{PhysRevB.79.165111} and normal-incidence infrared spectroscopy from Damascelli \etal\ \cite{PhysRevB.55.R4863}; Ce$_3$Bi$_4$Pt$_3$ (right): data from Bucher \etal\ \cite{PhysRevLett.72.522}. FeSi (\cbp) surpasses $2\cdot 10^3/(\Omega \cm)$ 
	at around $160$K$\approx T^*_\rho$ ($60$K$\lesssim T^*_\rho=120$K).
		]{%
	{\includegraphics[angle=0,width=.45\textwidth]{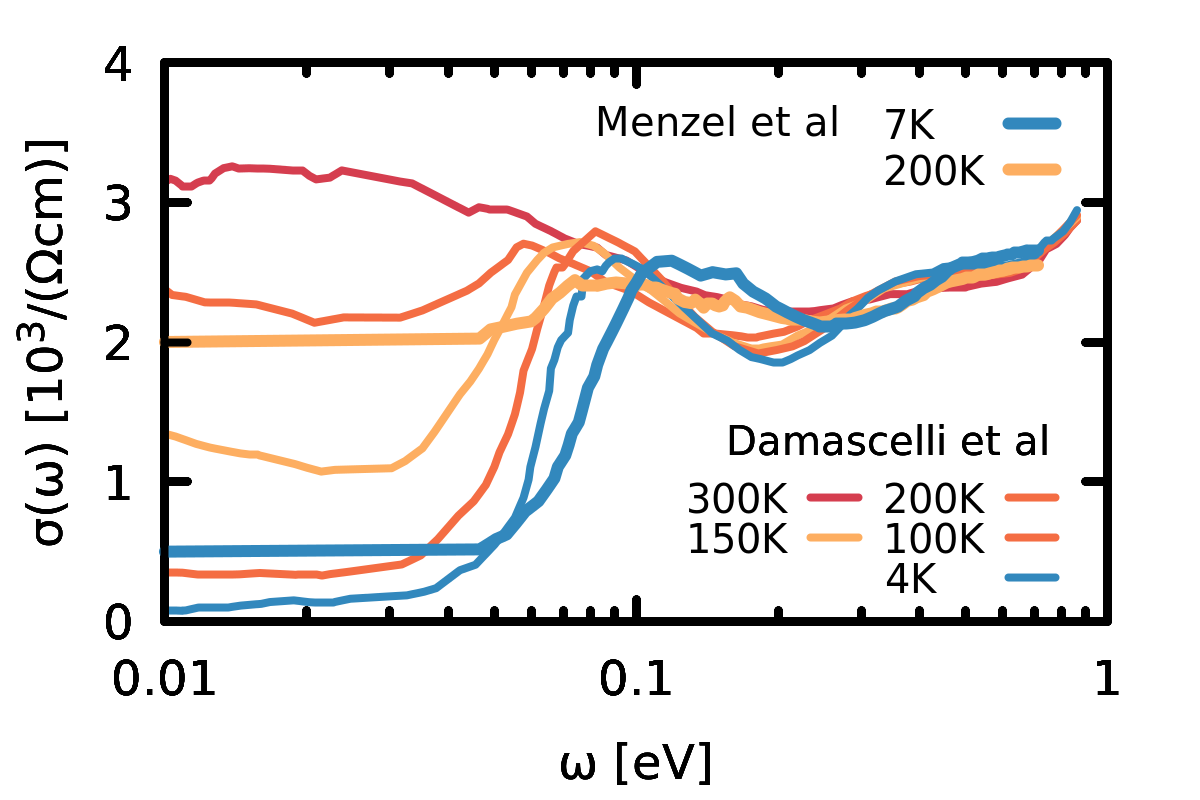}}
	{\includegraphics[angle=0,width=.45\textwidth]{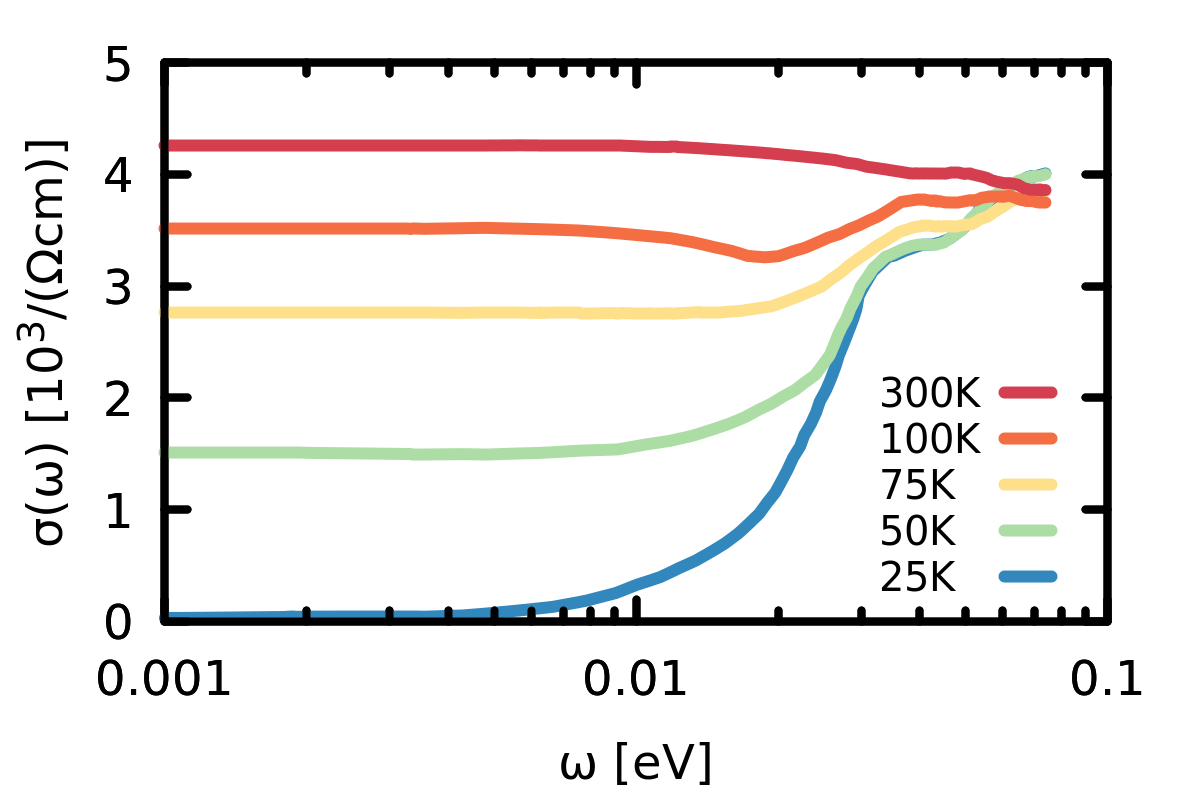}}
	}
	      \caption{{\bf Spectroscopic observables:}   The correlated narrow-gap semiconductor FeSi vs.\ the Kondo insulator Ce$_3$Bi$_4$Pt$_3$. 
								}
      \label{FeSiCe3spec}
      \end{center}
\end{figure*}

\paragraph{Insulator-to-metal crossover.}

As shown in \fref{FeSiCe3trans}(a) the resistivity of FeSi and \cbp\ both follow activation laws
at intermediate temperatures---60-120K for FeSi and 30-150K for \cbp---with charge gaps 
of about 59meV and 7.6meV, respectively.
At lower temperatures, the resistivity of both compounds
increases more slowly than expected from thermal activation. This tendency towards resistivity saturation is commonly interpreted as
caused by impurities or other defects, but it can also be a sign for the presence of other residual bulk scattering mechanisms
 (cf.\ the discussion in \sref{KubBoltz}). 
More important in our context are the deviations at high temperatures, where the resistivity surpasses the activation law.
In the case of FeSi, the slope in temperature, $d\rho/dT$, even becomes positive, clearly indicative of metallic behaviour.
This has led some authors to propose that the value of the charge gap in these systems is temperature-dependent.
Indeed, it is common practice to collect deviations from activated behaviour into a non-constant $\Delta(T)$.
Doing this, e.g., Hundley \etal\ extracted for \cbp\ a gap from the resistivity and the thermopower that halves in size when going from 50K up to 200K\cite{Hundley1994443},
in congruence with slave-boson calculations for the periodic Anderson model\cite{PhysRevB.47.6879}.
Without complementary observables, it is, however, difficult to ascertain whether this scenario is realized in \cbp\ or FeSi.

The metallization process is more pronounced in spectral and optical properties:
\Fref{FeSiCe3spec}(a)  displays the change in the FeSi photoemission intensity of Klein \etal\ \cite{1367-2630-11-2-023026} relative to a high-temperature reference spectrum (see the figure caption for details, and Refs.~\cite{SAITOH1995307,PhysRevB.72.233202,PhysRevB.52.R16981,PhysRevB.77.205117} for other photoemission works).
Panel (b) shows similarly processed data for \cbp\ from Takeda \etal\ \cite{Takeda1999721}.
At low temperature, both materials exhibit a pronounced depletion at the Fermi level.
In the case of FeSi (left), this gap is flanked by notably sharp features in the valence band. These are particularly evident in angle-resolved spectra\cite{PhysRevB.77.205117,PhysRevB.52.R16981}. 
For \cbp, changes are much less pronounced. Indeed, as seen in \fref{DMFTCe3}, also the valence part of the theoretical many-body spectrum of this $f^1$-compound is rather broad at all temperatures.%
\footnote{Spectra obtained more recently for Kondo insulators with more $f$-electrons feature much narrower valence-state peaks,
see, e.g., the results for SmB$_{6}$\cite{1367-2630-15-4-043042,PhysRevLett.112.226402,doi:10.7566/JPSCP.3.017038,0953-8984-28-36-363001,Min2017}.}

The charge gap at low temperatures is equally visible in the optical conductivity of FeSi, shown in \fref{FeSiCe3spec}(b,left).
In the case of FeSi, the direct (optical) gap probed by dipolar transitions is comparable to the indirect gap extracted from the resistivity.
In \cbp, \fref{FeSiCe3spec}(b,right), the optical gap is notable bigger ($\Delta_{indir}\approx 37$meV$=429$K$\times k_B$ \cite{PhysRevLett.72.522}) than the indirect gap ($\Delta_{dir}\approx 7.6$meV),
as expected in a hybridization gap scenario in the spirit of the periodic Anderson model (cf.\ \sref{PAM}).

With growing temperature the gap in the photoemission and the optical spectrum gets increasingly filled. Indeed, while the gap edges soften in FeSi,
there is no discernible trend in the photoemission spectrum towards a gap closure by virtue of the valence peak moving towards the Fermi edge.
In the optical conductivity, the absorption feature around 100meV does slightly move towards lower energies\cite{PhysRevB.79.165111,PhysRevB.55.R4863}%
\footnote{see also Fig. 2.5 in Ref.~\cite{damascelli}, and results for electronic Raman scattering in Ref.~\cite{PhysRevB.51.15626}.}%
, but it is clearly not responsible for the rising conductivity in the dc limit. 
In the case of \cbp, the first shoulder in the optical spectrum even moves to slightly higher energies upon increasing temperature.
In all, this suggests that in both materials, {\it the gap does not close by the displacement of quasi-particle states, but instead
it is filled by incoherent spectral weight as temperature rises}. 
This picture is further supported by differential-conductance measurements from point-contact spectroscopy for \cbp\cite{LAUBE1999303}
and tunnelling spectroscopy for FeSi\cite{PhysRevB.58.15483}, that also see practically temperature-independent peak positions.

At $T^*_\rho=160$K FeSi reaches $\sigma(\omega=0)\approx 2\cdot 10^3/(\Omega \cm)$, the conductivity of a bad metal:
FeSi metallizes at a temperature
that is smaller  than the low-temperature gap $\Delta/k_B=685$K by a factor of more than four.
\cbp\ surpasses the same value of the conductivity already around 60K, which is slightly smaller than the coherence temperature $T^*_\rho=100$K that we defined above.
These temperatures are comparable to the indirect gap $\Delta_{indir}/k_B\approx 88$K, but they are still much smaller than the direct gap $\Delta_{dir}/k_B\approx 429$K. 
Consequently, the crossover to a metallic state is, in both systems,
qualitatively far beyond thermal excitations across a fundamental gap.

\paragraph{Spectral weight transfers.} 

One of the powers of optical spectroscopy is the existence of sum-rules that allow for a quantitative analysis of spectral weight transfers.
Indeed, the integral over the optical conductivity, 
\begin{equation}
F(\Omega,T)=\int_0^\Omega d\omega\Re\sigma(\omega,T) 
\label{fsum}
\end{equation}
yields, for $\Omega\rightarrow\infty$, a value, $F(\infty,T)=\frac{\hbar\pi ne^2}{2m}$, that is
solely determined by the total density $n$ of carriers participating in optical absorption and the bare electron mass $m$\cite{millis_review}.
Therewith, the integral of spectral weight is independent of external parameters, such as temperature, or pressure.
In other words, the spectral weight that is lost in the low-energy optical conductivity, as FeSi and \cbp\ become insulating
upon cooling, must be shifted to higher energies.
The frequency $\Omega=\Omega_c$ above which the integral $F(\Omega_c,T)$ becomes independent of the external parameter 
sets the energy scale over which the reshuffling of spectral weight occurs.
This scale can give clues about the mechanisms that underlie the evolution of the system
for changing conditions:
In a conventional semiconductor, thermal activation leads to a finite electron(hole) population in the conduction(valence) band, that is however
restricted to the direct vicinity of the band edges. Changes in inter-band transitions in the optical conductivity  are thus limited to a few $k_BT$ above the
charge gap $\Delta$,  implying $\Omega_c\gtrsim\Delta$.
The same applies, e.g., for gap-opening phase transitions owing to density-waves (examples are Cr\cite{PhysRevLett.20.384} or NdNiO$_3$\cite{PhysRevB.51.4830}).

In materials with strong electronic correlations, the energy scale $\Omega_c$ required to 
reach a T-independent $F(\Omega_c,T)$ can be much larger. 
In Mott-Hubbard systems, for example, spectral weight transfers occur over a frequency range set by the
Hubbard $U$ interaction\cite{PhysRevLett.75.105,PhysRevB.54.8452}. The latter can be of the order of, or larger than, the bandwidth.
In case it triggers a metal-insulator transition in a multi-orbital system, the fundamental gap $\Delta$ can be much smaller than $U$.
Examples with $\Omega_c\gg\Delta$ are, e.g., the transition metal oxides VO$_2$ and V$_2$O$_3$\cite{PhysRevB.73.165116,PhysRevB.77.115121,optic_prb}.

In the case of FeSi, the energy needed to recover the transferred spectral weight with changing temperature was controversial\cite{PhysRevLett.71.1748,PhysRevB.55.R4863,0295-5075-28-5-008,PhysRevB.56.1366}, until accurate ellipsometry measurements\cite{PhysRevB.79.165111}%
\footnote{that do not require Kramers-Kronig transforms to be applied to reflectivity data obtained for a restricted frequency range.}
 ruled in favour of a $\Omega_c$ much larger
than the charge gap.
The situation is similar for the Kondo insulator \cbp: there, the reshuffling of spectral weight occurs over an energy range of $\sim250$meV \cite{FISK1995798} 
  corresponding to more than six times the size of the optical gap\cite{PhysRevLett.72.522}.


\subsubsection{Spin degrees of freedom.}

\paragraph{Magnetic susceptibility and magnetic neutron scattering.}
\label{sus}

\Fref{FeSiCe3trans}(b) displays the uniform magnetic susceptibilities for FeSi \cite{PhysRev.160.476,JPSJ.50.2539} (left) and \cbp\ \cite{PhysRevB.42.6842} (right).
At low temperatures spin excitations in both compounds are gapped 
and the responses have an activation-type behaviour.%
\footnote{The low-T upturn in the susceptibility of \cbp\ is not a property intrinsic to the bulk, as shown by neutron scattering measurement\cite{PhysRevB.44.6832}. Also in the case of FeSb$_2$ it has been clearly demonstrated that a residual susceptibility at low temperatures derives from defects\cite{doi:10.1143/JPSJ.80.054708}. 
}
Without any further microscopic insight, it is however not obvious to deduce the origin of enhanced or temperature-induced paramagnetism from the experimental data.
In \fref{FeSiCe3trans}(b) we have deliberately fitted both low temperature susceptibilities with the same expression, $\chi(T)\propto\exp(-\frac{\Delta_s}{k_BT})/T$,
which corresponds to the behaviour of local moments. The applicability of this form and further details will be discussed in \sref{pheno}.
Here, we only note that in this picture the extracted spin gap---$\Delta_s=50$meV for FeSi and $\Delta_s=7$meV for \cbp---is of similar magnitude than the indirect 
charge gap $\Delta_{indir}$ obtained from fitting the resistivity,
yet much smaller than the direct, optical gap $\Delta_{dir}$.%
\footnote{Inelastic neutron spectra suggest a slightly larger spin gap of $12$meV\cite{PhysRevB.44.6832} that is, however, also still much smaller than the optical gap.} 
Using instead a simplified band picture to extract the spin-gap from the susceptibility yields, via $\chi(T)\propto\exp(-\Delta_s/(2k_BT))$ a larger value $\Delta_s=96$meV for FeSi,
advocating, in all, the hierarchy $\Delta_{indir}<\Delta_s<\Delta_{dir}$.

This finding suggest that a non-trivial electronic structure is realized in these systems. Indeed in a band insulator, activation of spin and charge excitation are governed by the same energy scale $\Delta$. In a Mott insulator, on the other hand, a finite charge gap is accompanied by gap-less spin fluctuations ($\Delta_s=0$).
More insight will be discussed in the context of many-body models, see \sref{model}.

At larger temperatures, after going through a maximum at $T^{max}_\chi$, see \fref{FeSiCe3trans}(b), the activated behaviour in $\chi(T)$ gives way to a crossover to Curie-Weiss-like decay
\begin{equation}
\chi(T)=N_A\mu_{eff}^2/\left(3k_B(T-\theta)\right).
\label{localmoment}
\end{equation}
This crossover in the magnetic response is seen for \cbp\ also in Knight shift \cite{PhysRevB.49.16321}.
For FeSi we find $\mu_{eff}=2.6\mu_B$, which is close to the behaviour expected for local moments with $S=1$: $\mu_{eff}/\mu_B=g_s\sqrt{S(S+1)}=2\sqrt{2}\approx 2.8$ ($g_s=2$). In the case of \cbp, we extract $\mu_{eff}=2.53\mu_B$. This value is very close to the fluctuating moment of isolated Ce$^{3+}$ ions:
 $\mu_{eff}/\mu_B=g_J\sqrt{J(J+1)}=2.54$ ($J=5/2$, $g_J=0.857$).
The Curie-Weiss temperatures are negative in both cases, $\theta=-289$K ($\theta=-103$K) for FeSi (\cbp), suggestive of antiferromagnetic interactions.
In the Kondo lattice picture, this is in line with the Ruderman-Kittel-Kasuya-Yosida (RKKY) coupling that competes with the Kondo screening\cite{DONIACH1977231}. \cbp\ (and other Kondo insulators) are believed to be in the strong $J$-coupling regime, where the RKKY coupling fails to impose long range magnetic order because of a dominating Kondo effect.%
\footnote{A mechanism as to why the RKKY interaction is most often of antiferromagnetic character has recently been proposed in Ref.~\cite{Ahamed2017}.} 
In the case of FeSi, the negative Curie-Weiss temperature is contrasted with results from magnetic neutron spectroscopy
results of Shirane \etal\ \cite{PhysRevLett.59.351} and Tajima \etal\ \cite{PhysRevB.38.6954} (see \fref{FeSiN} and the discussion in \sref{pnas}).
These measurements revealed a strongly peaked magnetic cross section for $\vek{q}=0$ at $T^{max}_\chi=500$K\cite{PhysRevLett.59.351} and above\cite{PhysRevB.38.6954}, indicative of {\it ferromagnetic} fluctuations.%
\footnote{Indeed, a ferromagnetic state can also be reached by doping, Fe$_{1-x}$Co$_x$Si\cite{Beille1983399,doi:10.1143/JPSJ.59.305,Manyala2000,PhysRevB.72.224431}, expansion of the lattice by chemical pressure, FeSi$_{1-x}$Ge$_x$\cite{BAUER19981401,PhysRevLett.91.046401}, as well as ultra-high magnetic fields\cite{Kudasov1998,Kudasov1999} (cf.\ the discussion below).}

\paragraph{Magnetoresistance and high-field experiments.}
\label{MR}

In a Kondo insulator, the formation of Kondo singlets below the coherence temperature leads to the gapping of spin- and charge excitations.
These singlets, formed by a strongly renormalized hybridization between conduction electrons and $f$-levels, can be broken up by
a large enough magnetic field $H$. To do so, the Zeeman energy needs to be comparable to the spin gap,
i.e., $g_JJ\mu_BH\approx\Delta_s$, where $g_J$ is the gyromagnetic ratio, and $J$ the electron's total angular momentum.
In the simplest picture, such a field $H$ will induce a phase transition from the Kondo phase to a light (conduction) metal with fully polarized $f$-electrons.%
\footnote{Note however, that numerical calculations for the Kondo lattice model\cite{PhysRevLett.92.026401} and the periodic Anderson model\cite{PhysRevB.70.245104} suggest that with an applied magnetic field the spin gap collapses and a transverse AF order develops, while the charge gap remains
finite up to higher fields.}

In \cbp\ a partial polarization of the $f$-states is signalled by a large negative magnetoresistance (MR), $(\rho(H)-\rho(0))/\rho(0)$, measured below 50K in fields $H$ up to 10T\cite{Hundley1993425}.
With a spin gap of about 12meV\cite{PhysRevB.44.6832}, the critical field $H_c$ needed to metallize \cbp\ (Ce$^{3+}$: $J=5/2$) is however much larger. The estimate from the above formula is 40T.
Reaching such fields, the insulator-to-metal transition has indeed been observed in measurements of the resistivity\cite{BOEBINGER1995227}($H_c\sim50$T), as well as 
the specific heat\cite{Jaime2000} ($H_c\sim40$T). These findings support the Kondo picture for \cbp.

In $d$-electron derived narrow-gap semiconductors, the much larger size of the spin gap prohibits a clear picture in magnetic fields of conventional size.
In fields up to 9T experiments find a large positive MR for FeSb$_2$\cite{0295-5075-80-1-17008,PhysRevB.67.155205,PhysRevB.88.245203}.
For FeSi, even the sign of the MR strongly depends on the experiment.
The samples of highest quality---according to the  residual-resistance ratio criterion---are those of Paschen \etal\ \cite{PhysRevB.56.12916} that yield a negative magnetoresistance  below 50K in a field of 7T.%
\footnote{
The MR reported by Otha \etal~\cite{Ohta1997463} for FeSi is positive up to 16T for all temperatures, but
samples with lower RRR show a less positive MR; Lisunov \etal~\cite{LISUNOV199637} found a positive signal up to 35T;
more recently, Sun \etal~\cite{PhysRevB.90.245146} reported for 8T a sign change from MR$>0$ at low temperatures to MR$<0$ above 70K that interestingly correlates with a maximum (minimum) in the Nernst (Hall) coefficient, while the absolute value of the magnetoresistance is much smaller than in older experiments.
}
The notable sample dependence, as well as a positive contribution to the magnetoresistance could be linked to the presence of impurities.
Alternatively, a positive MR could be due to the proximity 
to a ferromagnetic instability.%
\footnote{While the size of the charge gap is insensitive to, e.g., dopings with up to 3\% cobalt\cite{PhysRevB.56.1366},
dopings (T=Co, Rh) and also isoelectronic substitutions (T=Ru) of 10\%, Fe$_{0.9}$T$_{0.1}$Si, induce 
itinerant ferromagnetism\cite{Paschen1999864}, concomitant with a large positive MR\cite{Paschen1999864,Manyala2000,PSSB:PSSB201248237}.}

Using ultra-high magnetic fields of up to spectacular 450T as produced by an explosively pumped flux compression generator, Kudasov \etal\ \cite{Kudasov1998,Kudasov1999}
found the resistivity of FeSi to continuously decrease by several orders of magnitude.
While at $T=77$K the semi-conductor to metal crossover was continuous also in the magnetic signal,
a sample cooled to $T=4.2$K displayed a jump to a  magnetic moment of $0.95\mu_B$ at a field of 355T\cite{Kudasov1999},
suggestive of a ferromagnetic metallic state.
Moreover, the field-dependent resistivity was found to be strongly non-exponential, indicative of a Zeeman splitting beyond
the rigid-band picture. Indeed, a conventional collapse of the experimental spin-gap of $\Delta_s\approx50$meV (see \tref{table1}),
is naively expected at 435T (using $S=1$ and $g=2$), i.e., for fields larger than found in experiment.

\begin{figure*}[!t]
  \begin{center}
	\subfloat[FeSi]{
	{\includegraphics[angle=0,width=.45\textwidth]{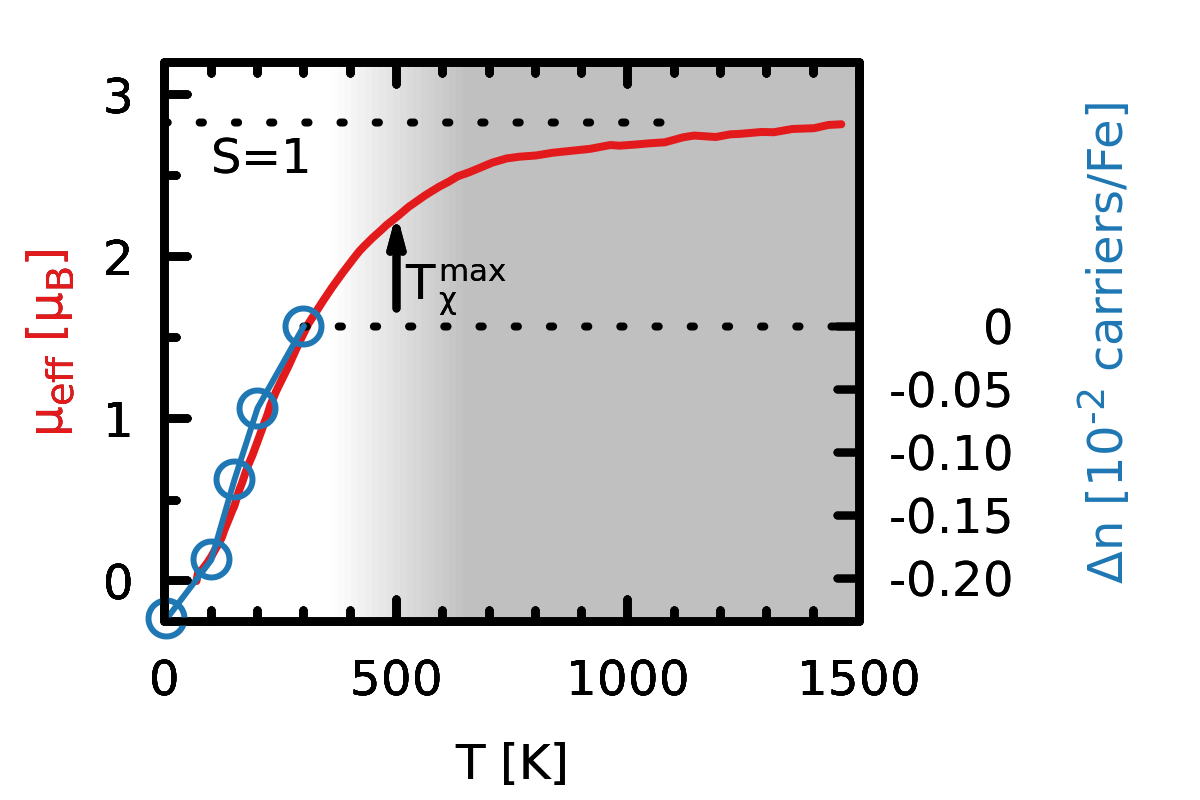}}}
		$\qquad$
	\subfloat[\cbp]{
	{\includegraphics[angle=0,width=.45\textwidth]{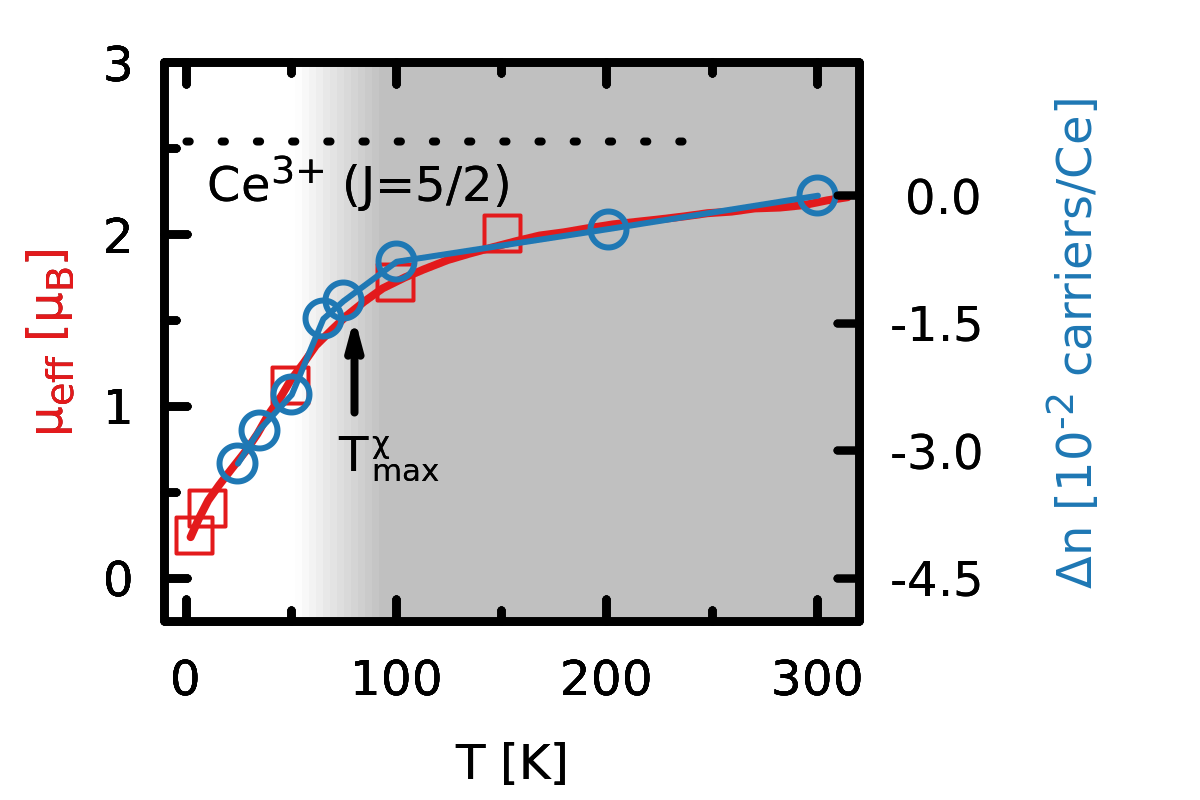}}
	}
	      \caption{{\bf interplay of charge and spin degrees of freedom.} 
				Comparison of the temperature dependence of (i) the effective fluctuating moment $\mu_{eff}(T)$
				(red, extracted from the local moment expression \eref{localmoment}) with (ii)
				the density $\Delta n$ of low-energy charge carriers that are transferred to higher energies upon cooling below $T_0=300$K
				(blue, determined from the optical spectral weight integral \eref{effn}).
				For FeSi (left) we use the optical conductivity data of Damascelli \etal\ \cite{PhysRevB.55.R4863} with a cutoff frequency $\Omega=100$meV.
				The susceptibility data are combined from from Jaccarino \etal\ \cite{PhysRev.160.476} and Takagi \etal\ \cite{JPSJ.50.2539}.
				Also indicated is the value of $\mu_{eff}=g_s \sqrt{S(S+1)}\mu_B=2\sqrt{2}\mu_B$ expected for local moments with $S=1$ ($g_s=2$).				
				For Ce$_3$Bi$_4$Pt$_3$ (right) data is reproduced from Bucher \etal\ \cite{PhysRevLett.72.522}, using a cutoff $\Omega=37$meV in \eref{effn}.
				The effective moment $\mu_{eff}(T)$ is obtained from the susceptibility of Hundley \etal\ \cite{PhysRevB.42.6842}. Square symbols are 
				results from neutron measurements of Severing \etal\ \cite{PhysRevB.44.6832}. The dashed line indicates the moment of isolated Ce$^{3+}$
				($J=5/2$, $g_J=0.857$) ions.
								}
      \label{interplay}
      \end{center}
\end{figure*}

\subsubsection{Interplay of charge and spin degrees of freedom}
\label{KIinter}

\paragraph{Correlation in the temperature dependence of charge and spin observables.}

Despite different values for the charge and spin gap, electrical and magnetic properties are intimately linked in the Kondo picture:
When Kondo coherence is destroyed as temperature rises, the $f$-charge decouples from the conduction electrons and its moment begins to fluctuate freely. 
Concomitantly, conduction electron density transfers to lower energies and fills the hybridization gap.
That both effects occur on the same temperature scale in \cbp\ has been beautifully illustrated by Bucher \etal\ \cite{PhysRevLett.72.522}. 
In \fref{interplay} we reproduce their analysis and extend it to FeSi.

Fitting the local moment expression, $\chi(T)=\frac{N_A}{3k_BT}\mu_{eff}(T)\exp(-\frac{\Delta_s}{k_BT})$,
to the experimental susceptibility, a temperature-dependent effective moment $\mu_{eff}(T)$ is extracted. At high temperature
we recover, as mentioned before, for both compounds, values that are close to the respective single ion limit (cf.\ \sref{sus}). For \cbp: 
$\mu_{eff}/\mu_B=\sqrt{J(J+1)}g_J=5.92$ as expected for Ce$^{3+}$ ($J=5/2$, $g_J=0.857$); for FeSi
$\mu_{eff}/\mu_B=\sqrt{S(S+1)}g_s=2.8$ supposing $S=1$ ($g_s=2$).
Upon lowering temperature, however, the magnitude of these effective moments decreases.
In \cbp\ this observation is associated with the Kondo effect: conduction electrons begin to screen the $f$-moment.
The concomitant gapping of the density of states is heralded by the loss of optical weight at low energy.
The amount of charge participating in this transfer can be obtained from the spectral weight integral of \eref{fsum}.
By tracking the integral for a cut-off frequency $\Omega$ that roughly delimits the regime in which weight is lost,
with respect to a base temperature $T_0$, we can compute the amount of the charge transferred away from the Fermi level to energies {\it above} $\Omega$:
\begin{equation}
\Delta n(T)=\Delta n(\Omega,T,T_0)= \frac{2m}{\hbar\pi e^2} \left[ F(\Omega,T)-F(\Omega,T_0)\right]
\label{effn}
\end{equation}
Using $T_0=300$K and $\Omega=37$meV for \cbp\cite{PhysRevLett.72.522}, and $\Omega=100$meV for FeSi (cf.\ \fref{FeSiCe3spec}), we see that
the amount of charge $\Delta n(T)$ that participates in conduction upon heating indeed correlates with the emergence of the effective local moment $\mu_{eff}(T)$.
Unfortunately, there are no results for the optical conductivity at high enough temperatures to reach the Curie-Weiss-like regime for FeSi.
It would be interesting to see whether the ferromagnetic-like fluctuations evidenced in magnetic neutron scattering {above} 300K\cite{PhysRevLett.59.351,PhysRevB.38.6954} induce, for FeSi, a deviation from the current analysis which is based on the local moment picture.

In fact, from our
(admittedly somewhat arbitrary) definition of characteristic temperatures for the charge and spin response, $T^*_\rho$ and 
$T^{max}_\chi$, we note a quantitative difference between FeSi and \cbp:
In the latter, the temperature below which Kondo screening sets in, $T^{max}_\chi$, is slightly smaller than the crossover temperature $T^*_\rho$
of the conductivity (cf.\ \fref{FeSiCe3trans}(a-b), \fref{FeSiCe3spec}(b)): $T^{max}_\chi/T^*_\rho=75/120=0.625$.
In FeSi, on the other hand, this ratio evaluates to five times as much: $T^{max}_\chi/T^*_\rho=3.125$.

\paragraph{Nature of the low-temperature charge and spin gap and proximity to ordered states.} 

Information about the nature of the charge gap
can be deduced from experimental measurements in which the electronic structure is perturbed.

\begin{figure}[t!h!]
{\includegraphics[angle=0,width=.45\textwidth]{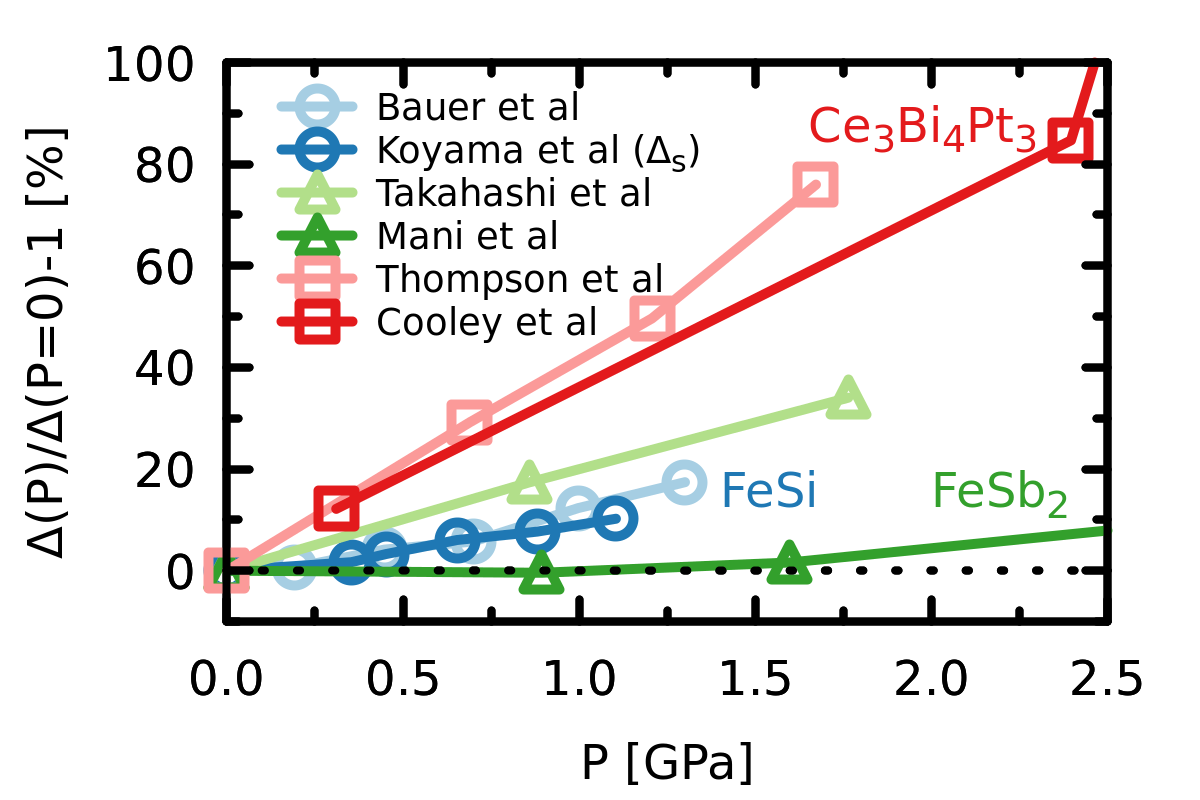}}
\caption{{\bf Pressure dependence of the gap.} Shown is the relative evolution 
of the charge gap of FeSi from Bauer \etal\ \cite{Bauer1997794}, the spin-gap of FeSi by Koyama \etal\ \cite{doi:10.1143/JPSJ.68.1693}
			(see also Refs.~\cite{REICHL1999866,PhysRevB.63.115103}), the charge gap of
			FeSb$_2$  from Mani \etal\ \cite{0953-8984-24-7-075601} and Takahashi \etal\ \cite{Takahashi2013}, as well as 
			the charge gap of \cbp\ from Thompson \etal\ \cite{Thompson1993} and Cooley \etal\ \cite{PhysRevB.55.7533}.}
\label{deltaP}%
\end{figure}

\subparagraph{Pressure.}
One such perturbation is external pressure. Commonly, compression increases orbital overlaps, causes a greater delocalization of charges
and thus larger transfer integrals. In a typical band-insulator, this increase in bandwidths leads to  a reduction of the gap.
Also in most correlated systems, hydrostatic pressure is a control parameter to enhance or even induce metallicity.
A famous example is the bandwidth-controlled Mott metal-insulator transition in V$_2$O$_3$\cite{Limelette89,Imada}.

If, however, the band-gap originates from hybridization, i.e., 
at least one side of the gap consists of bonding/anti-bonding states,
the opposite tendency can be found.
In fact, since also the transfer integrals at the origin of the bonding/anti-bonding splitting are enhanced by pressure, the gap typically increases.
This behaviour is indeed found for, e.g., FeSi\cite{GRECHNEV1994835,PhysRevB.63.115103,BAUER19981401,doi:10.1143/JPSJ.68.1693,PELZER2001227}, FeSb$_2$\cite{0953-8984-24-7-075601,Takahashi2013}  and \cbp\cite{PhysRevB.55.7533,Thompson1993}
as extracted from the activated behaviour of the resistivity and the magnetic susceptibility. 
As depicted in \fref{deltaP}, the increase of the gap scales roughly linearly with the applied pressure and is notably larger (on a relative scale) for the Kondo
insulator \cbp.%
\footnote{
In view of other Kondo insulating systems, a word of caution is in order.
While for Ce-based compounds, the above rationale holds, the gap in hole-like Kondo insulators
is actually suppressed under pressure, as shown for SmB$_6$\cite{PhysRevB.28.7397} or YbB$_{12}$\cite{IGA1993419}.
Interestingly, a critical pressure of 6GPa suppresses the gap in SmB$_6$ and magnetic long-range order with a saturated moment of $0.5\mu_B$ develops below 12K\cite{PhysRevLett.94.166401}.
It was suggested that the opposing trends in, e.g., \cbp\ and SmB$_6$ are rooted in the different signs for the change in unit-cell volume upon
adding/removing an electron to the system (ion-size difference of the $f^n$, $f^{n\pm 1}$ configurations)\cite{PhysRevB.54.12993}.
}%
\footnote{For a study of doped FeSi under pressure, see Refs.~\cite{PhysRevB.72.224431,PhysRevB.83.085101}.}

It would be interesting to extend the above analysis of charge and spin gap---$\Delta_{indir}$, $\Delta_s$---to the coherence scale $T^{max}_\chi$:
Is the dependence of $T^{max}_\chi$ on the value of the gap the same in both compounds? Or is the crossover to Curie-Weiss-like decay
in the $d$-electron compounds different, as it is proposed to be driven by Hund's rather than Hubbard physics
(cf.\ \sref{PAM})? Answering these questions
requires, however, measurements of FeSi under pressure up to temperatures above $T^{max}_\chi$, currently unavailable.

\subparagraph{Isoelectronic substitutions.}

Another perturbation of the gap is achieved by isoelectronic substitutions.
One route is the isoelectronic replacement of ligands.%
\footnote{Another route of isoelectronic substitution is the replacment of the 3$d$-transition metal(lanthanoid)
in FeSi(\cbp) by their homologues one row down in the periodic table.
The interesting case of  Fe$_{1-x}$Ru$_x$Si will be discussed in \sref{FeRuSi}.
Interchanging 4$f$ orbitals with 5$f$ ones is unfortunately not possible in \cbp, as the actinoid thorium prefers
the 5$f^0$6$d^2$7$s^2$ rather than the 5$f^1$6$d^1$7$s^2$ configuration.
}
Here, the most prominent example  is the alloy FeSi$_{1-x}$Ge$_x$.
While the end-member compounds of this series are isoelectronic as well as isostructural,
their electronic and magnetic properties are very different. Contrary to FeSi, which is insulating at low temperature
and does not show any signs of magnetic order, FeGe is metallic and shows spin helical long-range order\cite{LUNDGREN1968175,0953-8984-1-35-010}.
The latter roots in the noncentrosymmetric B20 crystal structure that allows for
a chiral spin-spin Dzyaloshinskii-Moriya interaction, which is particularly strong in FeGe.
In fact, FeGe\cite{0953-8984-1-35-010} has an ordering temperature ($T_C\approx 279$K)  and a helix period length ($\lambda\approx700$\AA) that are, both, much larger than  in the  prototypical helimagnet MnSi\cite{ISHIKAWA1976525} which also crystallizes in the B20 structure.
Strong correlation effects manifest themselves in FeGe by notable effective masses $m^*/m_e=3-5$ extracted from both specific heat measurements
($\gamma\approx10$mJ/(mol$\cdot$K$^2$)) \cite{1402-4896-9-1-007,PhysRevB.94.144424,PhysRevLett.91.046401} and optical spectroscopy\cite{PhysRevB.75.155114}.
Also, transfers of optical spectral weight occur over an energy range of the order of 1eV\cite{PhysRevB.75.155114}.

\begin{figure}[th]
{\includegraphics[angle=0,width=.425\textwidth]{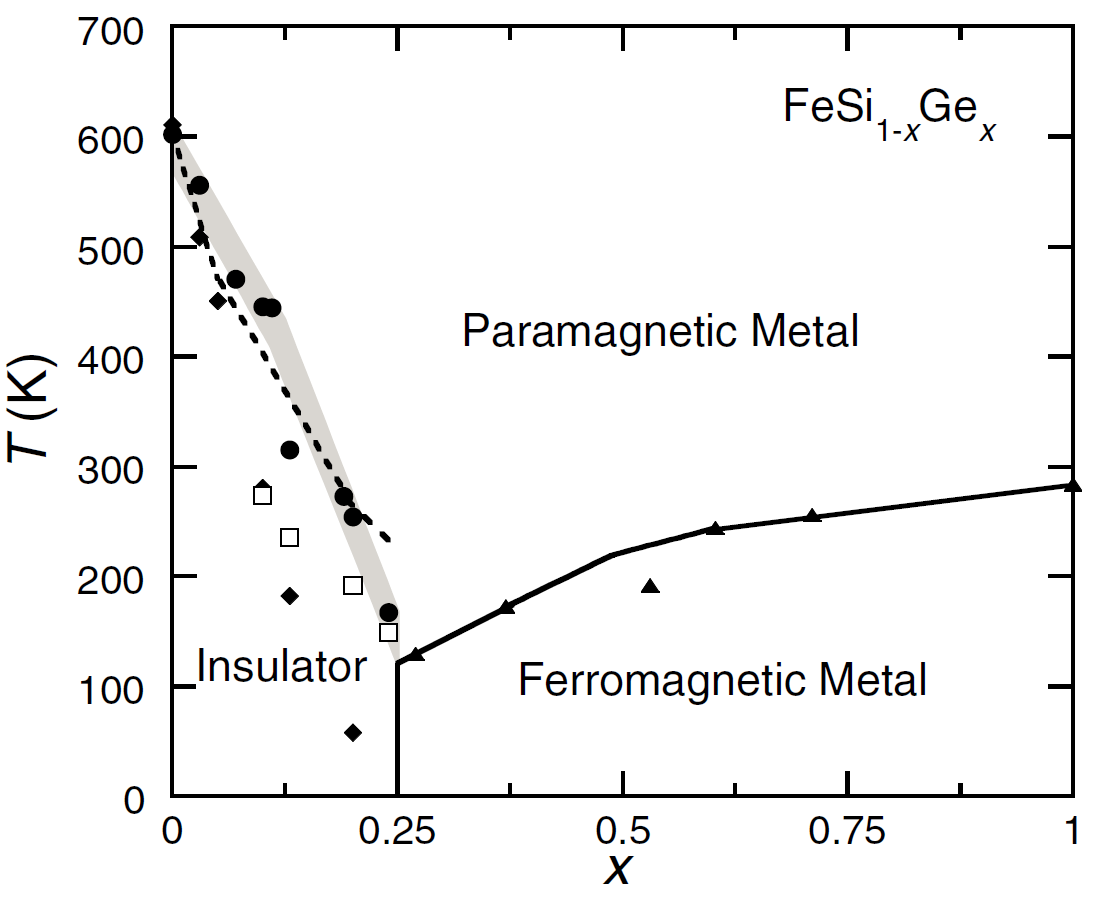}}
\caption{
{\bf Composition vs.\ temperature phase diagram of FeSi$_{1-x}$Ge$_x$.}
Phases of the isostructural, isoelectronic substitution series FeSi$_{1-x}$Ge$_x$ according to Yeo \etal\ \cite{PhysRevLett.91.046401}
obtained from activation-law fits to the resistivity and magnetic susceptibility (solid diamonds, solid circles, respectively),
the minima in the resistivity (open squares), the ferromagnetic ordering temperature (solid triangles).
The dashed line follows the calculation of Schlottmann and  Hellberg \cite{doi:10.1063/1.362014},
other lines are guides to the eye. From Yeo \etal\ \cite{PhysRevLett.91.046401}.}
\label{FeSiGeYeo}
\end{figure}

Starting at FeSi, already small concentrations of Ge in FeSi$_{1-x}$Ge$_x$ have a large impact on electronic and magnetic properties.
Activation law fits to the resistivity find a linear decrease of the gap with concentration $x$, with the gap vanishing at $x=0.5$\cite{BAUER19981401}.
More recently, Yeo \etal\ \cite{PhysRevLett.91.046401} established the phase diagram shown in \fref{FeSiGeYeo} based on resistivity (charge gap) and magnetic susceptibility (spin gap) measurements:
Both charge and spin gap decrease with the Ge-concentration and the maximum in the magnetic susceptibility $T^{max}_\chi$ moves towards lower temperature. Interestingly, spin excitations are moving notably quicker towards gapless behaviour with $x$ than charge excitations, reminiscent of the above discussion on the non-equivalence of charge and spin gaps.
Concomitantly, a Pauli-paramagnetic contribution appears in the magnetic susceptibility\cite{1742-6596-868-1-012009}.
At low temperatures, the introduction of Ge drives a first order transition from a non-magnetic semiconductor to a ferromagnetic metal at $x=0.25$. The size of the charge gap in FeSi$_{1-x}$Ge$_x$ correlates with the lattice constant\cite{BHARATHI19971}, suggesting that
the principal origin of the transition is chemical pressure, i.e., the volume expansion. 
Indeed, the unit-cell volume of FeGe 12.5\% larger than that of
FeSi\cite{doi:10.1016/j.stam.2007.04.004}.
Therefore, the (spin spiral) long-range order can be counteracted by external pressure: Magnetism in the end compound FeGe is suppressed by 19GPa, although metallicity is found to persist\cite{PhysRevLett.98.047204}. 

Again, it would be quite valuable to measure the magnetic susceptibility
of FeSi$_{1-x}$Ge$_x$ for $x\le0.25$ up to higher temperatures so as to extract $T^{max}_\chi$ as well as the magnitude of the effective fluctuating moment
in the Curie-Weiss-like regime.

\medskip

According to Doniach's phase diagram\cite{DONIACH1977231}, external parameters that reduce the Kondo coupling
can drive also Kondo insulators through a phase transition to a magnetic metal.
In fact, the quantum criticality associated with the crossover of dominant Kondo to preponderant RKKY interaction is a very active field of research\cite{Schroder2000,Gegenwart2008,Wirth2016,PSSB:PSSB201300005}.  Some examples:
Antiferromagnetism in the dense Kondo system CePdSn is suppressed above 6GPa while the resistivity develops an insulating slope\cite{IGA1993419}.
A critical pressure $P_c\approx 6$GPa drives the Kondo insulator SmB$_6$ through a transition (likely to be of first order) into a magnetically ordered metallic state
\cite{PhysRevLett.94.166401}.
Electron-doping CeRhSb via substitution of Pd for Rh drives a crossover from a paramagnetic insulator to an antiferromagnetic metal\cite{PhysRevB.55.14100}.
Further, several experiments have found a crossover from an antiferromagnetic metal to a non-magnetic ``insulator''
 for diverse
alloyings in the CeNiSn system, e.g., CeNi$_{1−x}$Pt$_x$Sn \cite{ADROJA1996275,SAKURAI19921415,NISHIGORI1993406,KALVIUS2000256}, CeNi$_{1-x}$Pd$_x$Sn\cite{doi:10.1143/JPSJ.60.2542}
and CeNi$_{1−x}$Cu$_x$Sn\cite{KALVIUS2000256}.%
\footnote{Yet, CeNiSn is often referred to as a ``failed Kondo insulator''. In fact, the system is thought to be rather a Kondo semimetal\cite{TAKABATAKE1996413,NAKAMOTO1995840}
owing to an anisotropic hybridization amplitude\cite{JPSJ.65.1769,PhysRevLett.117.216401}.}

Hence, while $d$-electron-based narrow-gap semiconductors and Kondo insulators may both exist in proximity to magnetically ordered phases, the nature of
that ordering is typically different in both cases (FM vs.\ AF).

\subparagraph{Doping.}
\label{Ce3dope}

Finally, charge and spin excitations can be altered by doping.
We first discuss non-isoelectronic substitutions in FeSi.
Substituting up to 3\% of Fe with Co
causes only a minute reduction of, both, the direct and indirect charge gap\cite{PhysRevB.56.1366}.%
\footnote{Other transition metal substitutions that support the B20 crystal-structure (but will not be discussed here) include electron doping with Ni\cite{Panfilov1975,ACKERBAUER2009414,0953-8984-29-33-335701}, and hole doping by replacing Fe with Mn\cite{Manyala2008}, or Cr\cite{Panfilov1975,YADAM2016311}.}
Moreover no anomalies in thermodynamic and transport quantities have been observed, pointing towards
the persistence of paramagnetism. 
The specific heat, however, was found to increase roughly linearly with doping, reaching 7.6mJ/(mol K$^2$) at $x=3$\%.
Assuming as origin an increased density of states at the Fermi level, Chernikov \etal\ \cite{PhysRevB.56.1366} extracted a surprisingly
large effective mass ratio $m^*/m_e\approx 30$.

Co-concentrations beyond 5\% lead to a transition to a ferromagnetic metallic state\cite{Beille1983399,doi:10.1143/JPSJ.59.305,Manyala2000,PhysRevB.72.224431}.
This is quite notable, since the end compound, CoSi, is a diamagnetic semimetal. 
Indeed for $x>0.8$, Fe$_{1-x}$Co$_{x}$Si becomes again paramagnetic.%
\footnote{Additional electrons also induce metallic ferromagnetism in Fe$_2$VAl (see \fref{Fe2VAlprox}) and in FeSb$_2$\cite{hu:224422,PhysRevB.74.195130}.}
Lacerda \etal\ \cite{LACERDA19931043} showed that the magnetic susceptibility of electron doped Fe$_{1-x}$Co$_x$Si and FeSi$_{1-x}$P$_x$  
are qualitatively akin, prompting the authors to conclude that magnetism in the Co-substituted samples is not due to the moment of Co.
By comparing Co-doping with Fe$_{1-x}$Rh$_x$Si and Fe$_{1-x}$Ru$_x$Si---which all become ferromagnetic above a critical doping---Paschen \etal\ \cite{Paschen1999864}
substantiated that magnetism in Fe$_{1-x}$Co$_x$Si does not arise from local moments on Co. Indeed the 4$d$-orbitals orbitals of Rh have too large
a radial extend to allow for local magnetism. 
That also isoelectronically substituted Fe$_{1-x}$Ru$_x$Si orders ferrromagnetically points 
to an influence of the unit-cell volume onto magnetism (as in FeSi$_{1-x}$Ge$_x$, see above); that the ordered moment for insulating Ru-substituted samples is the smallest among the considered alloys indicats that (electron doping-induced) metallicity boosts magnetism in FeSi\cite{Paschen1999864}, further strengthening the itinerant picture.
The effect of the unit-cell volume onto magnetism is corroborated by pressure experiments that see the disappearance of long-range order
in doped samples above a critical pressure\cite{PhysRevB.83.085101}; also, epitaxially strained samples have a largely different critical doping level than the doped bulk
\cite{PhysRevB.89.134426}\footnote{Note, however, a further complication: these thin films exhibit a rhombohedral distortion.}.

It is crucial to note the influence of cobalt-doping onto the spin-gap $\Delta_s$ and the crossover temperature $T^{max}_\chi$ at which the magnetic susceptibility
peaks before assuming a Curie-Weiss-like decay:
The introduction of Co may affect the impurity-derived low-temperature upturn in $\chi$. Otherwise, the magnetic susceptibility of Fe$_{1-x}$Co$_x$Si
for concentrations up to $10$\% looks (above $T_C$)  remarkably similar to that of FeSi, albeit with a constant upwards shift\cite{LACERDA19931043}.
This means that both $\Delta_s$ (as extracted from an activation-law fit) and, apparently, also $T^{max}_\chi$ are basically unaltered by doping. The added carriers only introduce a $T$-independent Pauli-like contribution, congruent with the above findings from specific heat measurements.
As will be discussed below, this is a marked difference to the behaviour of doped \cbp.

Alternatively also the ligand can be substituted. Hole doping
can be achieved, e.g., in FeSi$_{1-x}$Al$_x$\cite{ANDP:ANDP2065080206,PhysRevLett.78.2831,PhysRevB.58.10288,PhysRevB.87.184304}.
DiTusa \etal\ \cite{PhysRevLett.78.2831} interpreted their results such that FeSi is a renormalized realization of silicon,
in the same sense as proposed for Kondo insulators: 
Doping the system leads to a (possibly disordered) Fermi liquid ground state, albeit with largely enhanced
effective masses.
Indeed, Al-doping causes the charge gap---as extracted from fitting the resistivity to an activation law above 100K---to rapidly decrease, while a metallic slope appears at lower temperatures already above about $x=1$\%\cite{ANDP:ANDP2065080206,PhysRevB.58.10288}.
From the specific heat coefficient, a considerable effective mass $m^*/m=14$ was extracted\cite{PhysRevLett.78.2831}.
Concomitantly, as in the case of Co-doping, a Pauli-like contribution appears in the magnetic susceptibility. The electrons introduced by Al affect
neither $T^{max}_\chi$ nor the (impurity-derived) Curie-Weiss upturn at very low temperatures.
As a consequence, contrary to the charge gap, the spin gap remains roughly constant up to $10\%$ Al\cite{PhysRevB.58.10288}, 
indicating a non-trivial interplay of charge and spin degrees of freedom.

\medskip

Hole-doping the prototypical Kondo insulator \cbp\ by substituting La for Ce notably suppresses the electrical resistivity\cite{PhysRevB.42.6842}
and leads to a finite specific heat at low-temperature\cite{PhysRevB.42.6842,CANFIELD1992217}. From the latter, much enhanced  effective masses of up to $m^*/m\approx100$ at $x=30$\% have been extracted\cite{PhysRevB.77.115134}.
However, doping also largely modifies the spin degrees of freedom: In fact the peak temperature, $T^{max}_\chi$, consistently moves to lower temperatures with increasing doping
\cite{PhysRevB.42.6842,PhysRevB.77.115134}. 
Indeed, as expected from, e.g.,\ the Coqblin-Schrieffer model\cite{PhysRevLett.51.308}, the linear coefficient of the specific heat $\gamma$ scales with $1/T^{max}_\chi$.
This dependency is realized in (Ce$_{1-x}$La$_x$)$_3$Bi$_4$Pt$_3$ above $25$\% La \cite{CANFIELD1992217}: Ce ions then behave as impurities in a metallic host.
In polycrystalline samples, additionally, the overall magnitude of $\chi$ decreases \cite{PhysRevB.77.115134}, contrary to the Pauli-like increase seen in doped FeSi (see above). Moreover,  the spin-gap decreases notably with $x$\cite{Severing1994480}.
As discussed by Kwei \etal\ \cite{PhysRevB.46.8067},
there is a direct correlation between the magnetic response in the local moment regime and the change in lattice constant
along the (Ce$_{1-x}$La$_x$)$_3$Bi$_4$Pt$_3$ series: Indeed the square of the effective moment $\mu_{eff}^2\propto T\cdot \chi$ increases linearly with the shrinking
lattice constant towards the  Ce ($x=1$) end-member. This suggests that the change in lattice constant tracks the number of $f$-electrons.
Interestingly, also the
metallic $f^2$-compound Pr$_3$Bi$_4$Pt$_3$ has a larger volume than its Ce cousin\cite{FISK1995798}.
Also in other heavy fermion compounds, e.g., rare-earth chalcogenides, (mixed) valency and lattice constants are found to correlate\cite{RevModPhys.48.219}.
In \cbp\ the valency that realizes Kondo insulating volume coincides with the lowest equilibrium volume.%
\footnote{Also in some $d$-electron narrow-gap semiconductors, the smallest lattice constant is found for insulating phases, 
see, e.g., \fref{Fe2VAlprox} (top): the lattice constant of Fe$_2$VAl is minimal for perfect stoichiometry. Deviations cause the volume to expand,
while the system metallizes (see the specific heat in panel (c)) and develops long-range magnetic order (see panel (b)).
In FeSi, off-stoichiometry Fe$_{1-x}$Si$_{1-x}$ does not show any significant change in the lattice parameter for small $x$ \cite{doi:10.1080/08957950211339},
while electron doping via Fe$_{1-x}$Co$_x$Si increases the lattice\cite{Manyala2004}. However, in the latter potential effects of the valency onto the unit-cell volume are masked by the contraction induced by the smaller atomic radius of Co.}

\subsubsection{Spin-orbit effects.}
\label{SO}

Lately, it was advocated that Kondo insulators might harbour topologically non-trivial band-structures\cite{PhysRevLett.104.106408,Dzero2016}.
Crucially, Kondo insulators might be systems in which topological effects---derived from a one-particle description---are not severely encumbered by strong correlation physics. This is has to be contrasted to Mott-Hubbard physics, that is able to completely destroy any topological protection, see Di Sante \etal\ \cite{DiSante_cubio}.
Indeed, the spin-orbit coupling has even been suggested to play a role in the stabilization of Kondo insulating behaviour.
Recently, this question has been investigated for the prototypical Kondo insulator \cbp\ by Dzsaber \etal\ \cite{PhysRevLett.118.246601}:
It was shown that---contrary to all the above cases---the unit-cell volume in the isoelectronic substitution series Ce$_3$Bi$_4$(Pt$_{1-x}$Pd$_x$)$_3$ is virtually constant for all compositions $0\le x\le 1$. Consequently, there are only {two} fundamental changes along the series: (i) a reduction of the spin-orbit coupling $\lambda_{SOC}$ because Pd is a much lighter element than Pt, and (ii) a change in the Ce-4$f$-to-noble-metal-$d$ hybridization caused by the different radial character of the 4$d$ and 5$d$ orbitals of Pd and Pt, respectively---an effect no considered in Ref.\ \cite{PhysRevLett.118.246601}.

Dzsaber \etal\ \cite{PhysRevLett.118.246601} found a collapse of the charge gap for growing $x$, with the end member Ce$_3$Bi$_4$Pd$_3$ exhibiting a nearly temperature-independent resistivity.
Concomitantly, $T^{max}_\chi$, that marks the onset of Kondo screening (cf.\ \fref{FeSiCe3trans}(b)), vanishes.
Excluding antiferromagnetic ordering, it was suggested that Ce$_3$Bi$_4$Pd$_3$ is a Weyl-Kondo semimetal\cite{PhysRevLett.118.246601,Lai2016}.

If a Kondo insulator can be destabilized by decreasing the spin-orbit coupling, this would be a large difference to systems like FeSi,
that only involve elements light enough to safely neglect relativistic effects for all practical purposes.%
\footnote{While spin-orbit coupling effects are certainly smaller in $3d$ transition-metal-based narrow-gap semiconductors, let us mention that 
also FeSi, RuSi, FeGe, and other compounds crystallizing in the non-symmorphic B20 structure (cf.\ the discussion of their band-structures in \sref{bandFeSi})
allow for non-trivial topological effects, such as non-vanishing electronic Berry phases\cite{0295-5075-104-3-30001}.
Also, see the review by Martins \etal\ \cite{0953-8984-29-26-263001} for the interplay of correlation effects and the spin-orbit coupling in $4d$ and $5d$ systems.
}
In \sref{Ce3dft} we will discuss the electronic structure of \cbp\ from the band-theory perspective
and shed light onto the modifications caused by the substitution of Pt with Pd.
In \sref{Ce3DMFT} we further present new realistic many-body calculations for \cbp.

\subsubsection{Thermoelectricity.}
\label{KIthermo}

Finally, FeSi and \cbp\ also exhibit empirical similarities in their thermoelectric response.
As evident in \fref{FeSiCe3trans}(c) both compounds develop large thermopowers at low temperatures.
Combined with the resistivity data, \fref{FeSiCe3trans}(a), these yield quite notable powerfactors, see \tref{table1}.
This is expected for insulating systems which possess a significant particle-hole asymmetry.
In an asymmetric, coherent semiconductor the thermopower roughly behaves as $1/T$, cf.\ \sref{limit}.
Effects of incoherence or other mechanisms towards metallization will quench the thermopower.
In the photoemission spectra of FeSi discussed above (cf.\ \fref{FeSiCe3spec}(a))  notable spectral weight appears at the Fermi level starting at 100K.
In this range, the thermopower is already below $100\mu$V/K. At $T^*_{\sigma}$ (see \sref{KI} and, there, \fref{FeSiCe3spec}(b)), the thermopower has already become negligible in FeSi.
Using the same measure, the decay in \cbp\ is slower: At its $T^*_{\sigma}$, the thermopower is still about a third of its maximum value.

While the insulator-to-metal crossover with rising temperature is antagonistic to a large thermoelectric response, it is
difficult to assess, on the basis of the experimental data alone, whether the metallization {\it causes} the quenching of the thermopower.
As will be discussed in \sref{silicides} (see, in particular, \fref{FeSicoh}), theoretical calculations strongly suggest an affirmative answer.

\paragraph{Response of doped samples.}
Besides the difference between a metallic and an insulating thermoelectric response, the thermopower---being
 a measure for particle-hole asymmetry---is extremely sensitive to doping.
In the case of FeSi, \cbp, and other narrow-gap semiconductors both effects conspire:
Introduced electrons or holes may not only change the character of preponderant charge carriers but they also
suppress the insulating state.
Indeed, the thermopower of the Kondo insulator Ce$_3$Sb$_4$Pt$_3$ is extremely sensitive
to doping both via ligand alloying (Ce$_3$Cu$_x$Pt$_{3-x}$Sb$_4$\cite{PhysRevB.60.5282}), as well as
lanthanoid substitution (${\mathrm{Nd}}_{x}{\mathrm{Ce}}_{3\ensuremath{-}x}{\mathrm{Pt}}_{3}$\cite{PhysRevB.58.16057}).
Both kinds of doping quickly suppress the thermopower.
The same applies to FeSi: As seen in \fref{Sakai} (taken from Sakai \etal\ \cite{JPSJ.76.093601,Sakai4569473})
stoichiometric FeSi sticks out as an island of large low-temperature thermopower among transition metal mono-silicides TMSi (TM=Cr, Mn, Fe, Co) and their alloys.
Also doping FeSi on the ligand site, e.g., hole-doping via  FeSi$_{1-x}$Al$_x$\cite{PhysRevB.58.10288}, suppresses the thermopower.
As will be explained in \sref{limit} electron-doping FeSi is expected to drive the thermopower $S$ through a sign-change and realize a
negative response of a magnitude comparable to that of stoichiometric FeSi.
Indeed, this is seen when substituting some Co for Fe, \fref{Sakai}. 
Realizing electron-doping on the ligand site might have the advantage of less perturbing the low-energy electronic structure in terms of disorder.
Since FeSi$_{1-x}$P$_{x}$ can be synthesized in the B20 structure\cite{LACERDA19931043}\footnote{FeP has, at ambient conditions, the regular MnP-type structure (space group Pnma) \cite{GU2011154}.},
it would be quite interesting to perform thermoelectric measurements on it.

\subsubsection{Summary.}

We end this section with a list of empirical similarities and differences of FeSi and \cbp, as well as some proposals for future experiments that 
could extend the presented comparison.

\paragraph{Empirical similarities of FeSi and \cbp.}
\begin{itemize}
	\item absence of long-range order at all $T$ in stoichiometric samples.
	\item charge and spin excitations are gapped at low-$T$ with activated behaviour in resistivity and magnetic susceptibility.
	\item spin gap tends to be smaller than the direct charge gap: $\Delta_s<\Delta_{dir}$.
	\item crossover to a metallic state for $T\ll\Delta_{dir}/k_B$;
	      towards the metallization, the gap does not shrink in size, but it gets filled by incoherent weight.
	\item spectral weight transfers in the optical conductivity over scales much larger than the direct charge gap $\Delta_{dir}$.
	\item crossover to a Curie-Weiss-like magnetic susceptibility above $T^{max}_\chi\lesssim\Delta_{indir}$, with
	a magnitude $\mu_{eff}$ of fluctuating moments close to the single ion limit.
	\item pressure increases spin and charge gaps, indicative of a gap caused by hybridization.
	\item large thermopowers at low-$T$ when the systems are coherent and insulating.
	\item possibility to drive a paramagnetic-insulator to ferromagnetic-metal transition with magnetic fields
	that induce Zeeman-splittings of the order of the spin gap.
	\item doping induces metallization; emerging quasi-particles have substantially enhanced effective masses $m^*$.
	\item proximity to ordered phases that can be reached by doping, isoelectronic substitutions (volume expansion and/or change of
	radial component of orbitals), or applied fields. 
\end{itemize}

\paragraph{Empirical differences between FeSi and \cbp.}
\begin{itemize}
	\item proximity to different spin-ordered phases: non-magnetic $d$-electron narrow-gap semiconductors have a propensity for nearby {\it ferromagnetic} instabilities.%
	\footnote{One exception is Fe$_{1-x}$Cr$_x$Sb$_2$ which is an antiferromagnetic insulator for $0.2\le x\le 1$\cite{PhysRevB.76.115105}.}
	Long range order is likely of {\it itinerant} origin, with enhancements from the Hund's-rule coupling (cf.\ Refs.~\cite{doi:10.1143/JPSJ.65.1056,PhysRevB.57.6896,Vollhardt1999,PhysRevLett.99.216402}, and the discussion in \sref{PAM}),
  and can be reached under volume expansion or doping. 
	Some Kondo insulators can instead be driven to an {\it antiferromagnetic} (RKKY-mediated) instability by (inverse) physical or chemical pressure, 
	changes in the Kondo coupling by isoelectronic substitutions with different radial quantum number, or doping. Moment-carrying charges tend to be {\it localized}.
	\item interplay of charge and spin degrees of freedom under doping: While doping largely changes the activated behaviour in the conductivity and induces a conducting state at low-temperatures in both FeSi and \cbp, the effect on the magnetic response is markedly different. Indeed doping only mildly affects the spin-gap $\Delta_s$ and crossover temperature $T^{max}_\chi$
	in FeSi, while in \cbp\ both are quite sensitive to the introduction of holes.
	\end{itemize}

\paragraph{Experiments to further the current understanding.}
\begin{itemize}
	\item high temperature optical conductivity measurements of FeSi to access spectral weight transfers: 
	Does the potentially different microscopic origin of Curie-Weiss-like decay in the susceptibility (local vs.\ itinerant fluctuations) result in 
	a disparate interplay of charge and spin degrees of freedom above $T^{max}_\chi$? 
	\item dependence of $T^{max}_\chi$ in FeSi and \cbp\ under pressure (or FeSi$_{1-x}$Ge$_x$): does quasi-particle coherence follow the same scaling with $\Delta$ in both materials?	
	\item measurements of the magnetic susceptibility of Fe$_{1-x}$Co$_x$Si and FeSi$_{1-x}$Al$_x$ up to $T^{max}_\chi$ and beyond:
	We stated that $T^{max}_\chi$ is seemingly unaffected by doping. However, this was a statement made by an educated extrapolation. Actual measurements of $\chi$ only extend to 400K,
	which is still slightly below $T^{max}_\chi$.
	\item electron-doped FeSi$_{1-x}$P$_{x}$: 
	It would be interesting to extend the work of Lacerda \etal\ \cite{LACERDA19931043} to magnetoresistance and thermopower measurements.
	The former could substantiate the claim of an itinerant nature of magnetism in Fe$_{1-x}$Co$_x$Si. The latter is motivated by the
	finding that Co-doping yields a large $n$-type thermopower\cite{JPSJ.76.093601} in Fe$_{1-x}$Co$_x$Si.
  Does FeSi$_{1-x}$P$_{x}$ perform better (because the transition-metal site is less perturbed)?
	\item dependence of $T^{max}_\chi$ in Fe$_{1-x}$Ru$_x$Si: Is coherence controlled by the fundamental gap or the Hund's rule coupling? (cf.\ the discussion of the resistivity of the substitution series in \sref{silicides}, and the model \sref{PAM} on Hund's physics)
\end{itemize}

\section{Theories of correlated narrow-gap semiconductors}
\label{theo}

This chapter deals with theoretical efforts to understand the peculiar behaviour of correlated narrow-gap semiconductors.
First (\sref{pheno}) we will discuss early phenomenological scenarios that invoke electronic correlation effects of some sort to
explain the experimental observations in the prototypical compounds FeSi and FeSb$_2$.
Next (\sref{band}) we review the merits and failures of band-theory applied to several classes of relevant materials,
and include new calculations for the Kondo insulator \cbp.
Based on the insights of band-theory, solutions of many-body models have shed light onto the qualitative impact of correlation
effects in these systems. These are recapitulated in \sref{model}.
Building on this knowledge, realistic many-body calculations (\sref{real}) were able to quantitatively reproduce, for the prototypical FeSi,  the signatures of correlation effects
in diverse experimental observables. Importantly, the congruence with experiments allowed for the deducing
of microscopic insights (\sref{micro}) beyond what was included in the reductionist models studied before. 
Indeed, the emerging physical picture refutes some earlier scenarios, while reconciling others. Crucially, the Hund's rule coupling was
identified as a new key ingredient to the  physics observed in FeSi. 
Further, new realistic many-body calculations for \cbp\ (\sref{Ce3DMFT}) are used to discuss potential microscopic differences between $d$-based narrow-gap semiconductors
and Kondo insulators.
With this insight, we performed new model calculations (\sref{PAM}) 
that will put the essence of correlated narrow-gap semiconductors into a wider context.
Finally (\sref{lattice}), is devoted to the role played by lattice degrees of freedom and
effects of the electron-phonon coupling in systems such as FeSi.


\subsection{Early phenomenological scenarios} 
\label{pheno}

The endeavour to deduce a phenomenological scenario from experimental findings for FeSi was pioneered by Jaccarino \etal\ in their seminal work~\cite{PhysRev.160.476}.
There, they modelled in particular the intriguing magnetic susceptibility (see \fref{FeSiCe3trans} (b)) with (i) a density-of-states model for itinerant
spin paramagnetism, and (ii) a spin-model of localized moments.
See also the extensive discussion of Tsujii \etal\ for FeGa$_3$\cite{JPSJ.77.024705}, as well as Petrovic \etal\ \cite{PhysRevB.67.155205,PhysRevB.72.045103} and Hu \etal\ \cite{PhysRevB.74.195130} for FeSb$_2$.

\begin{figure}[th]
{\includegraphics[angle=0,width=.4\textwidth]{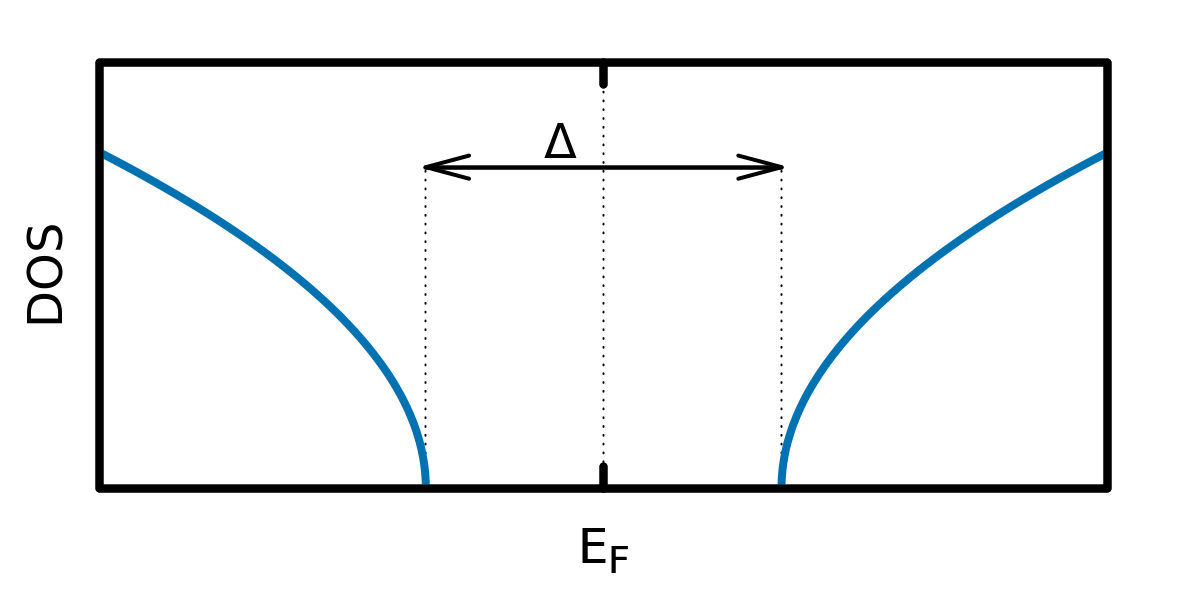}%

\includegraphics[angle=0,width=.4\textwidth]{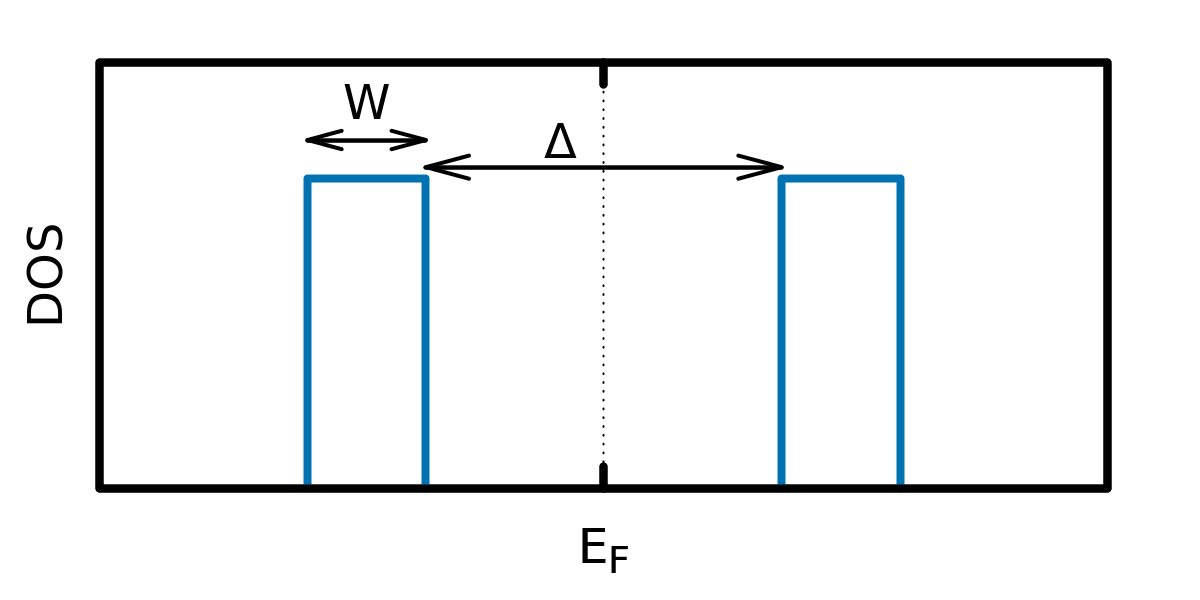}%

\includegraphics[angle=0,width=.4\textwidth]{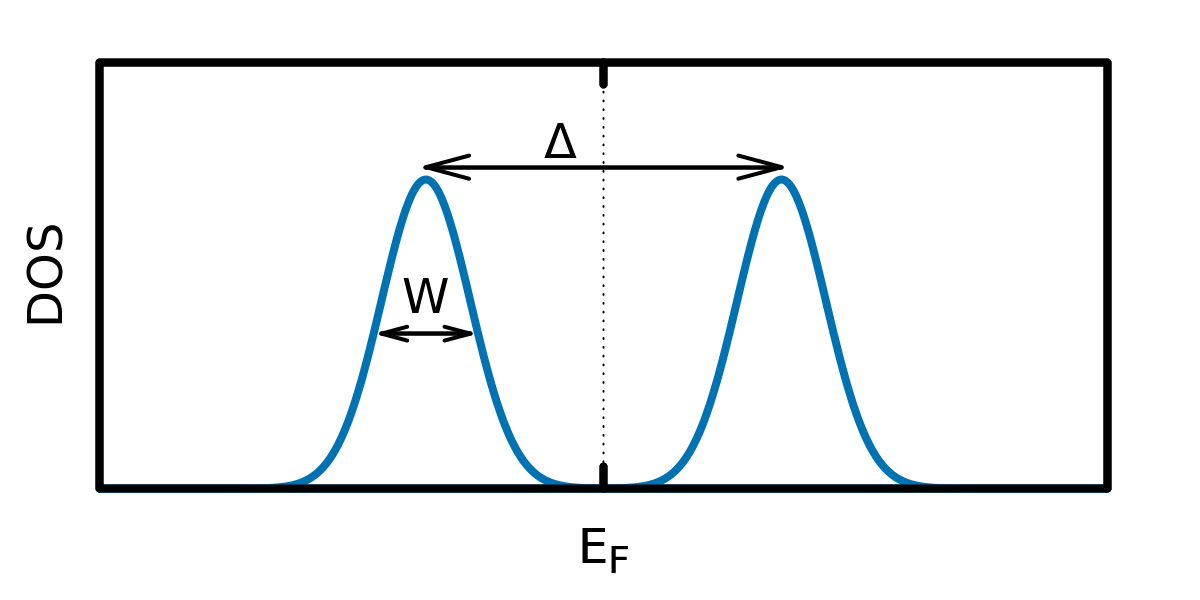}%
}\caption{ 
{\bf Schematics of density-of-states (DOS) models for FeSi.} Regular DOS with dispersive valence and conduction bands, separated by a gap $\Delta$ (top);
narrow-band DOS ($W<\Delta$) as proposed by Jaccarino \etal\ \cite{PhysRev.160.476} (middle); Gaussian DOS model as proposed by Mani \cite{MANI2004551} (bottom).}
\label{FeSiDOSscheme}%
\end{figure}

\paragraph{Density-of-states models.}
A regular semiconductor has dispersive valence and conduction states, with a gapped density of states (DOS) $D(E)$ as is schematically illustrated
in \fref{FeSiDOSscheme} (top). In the one-particle picture, such a DOS will produce a Pauli spin-susceptibility ($f(E)$ is the Fermi function and $\mu_B$ is the Bohr magneton)
\begin{equation}
\chi(T)=2\mu_B^2\int dE \left( -\frac{\partial f(E)}{\partial E}\right) D(E)
\label{PauliS}
\end{equation}
that results in activated behaviour of a form $\chi(T)=2\mu_B^2 A(T) \exp(-\frac{\Delta}{2k_BT})$, where $\Delta$ is the band-gap and $A(T)$ a function that depends on the details of the electronic structure.%
\footnote{For example, $A(T)=4\rho_0$ for a constant density of states 
$\rho(\epsilon)=\rho_0[\theta(\epsilon-\Delta/2)+\theta(-\epsilon-\Delta/2)]$, and $A(T)=2\sqrt{\pi k_B T} A_0$ for parabolic dispersions in three dimensions (as shown in \fref{FeSiDOSscheme} (top)), $\rho(\epsilon)=A_0[\theta(\epsilon-\Delta/2)\sqrt{\epsilon-\Delta/2}+\theta(-\epsilon-\Delta/2)\sqrt{-\epsilon-\Delta/2}$\cite{JPSJ.77.024705}.}
Spin excitations are thus governed by the same energy scale, $\Delta$, as charge excitations, stating that in band-insulators the
charge and the spin gap are equal. Using a constant density of states for the low-temperature susceptibility of FeSi, we find $\Delta=96$meV.
However, the fit is of worse quality than the one shown in \fref{FeSiCe3trans}(b,left)  that uses a spin model (see paragraph below).
In order to account for a maximum in $\chi(T)$ at $T^{max}_\chi$ within the one-particle framework, Jaccarino \etal\ \cite{PhysRev.160.476} and later Mandrus \etal\ \cite{PhysRevB.51.4763}
proposed a strong renormalization of the bandwidth, as schematically shown in \fref{FeSiDOSscheme}(middle). Indeed the finite bandwidth $W$ can lead to anomalies
in thermodynamic and transport properties. 
Via \eref{PauliS} a good fit of the experimental susceptibility is achieved for $\Delta\approx80$meV and $W=65$meV\cite{PhysRevB.51.4763}. Also the electronic contributions to the specific heat
and the thermal expansion coefficient were faithfully modelled with parameters of similar magnitude\cite{PhysRevB.51.4763}.
In this picture, the DOS has to exhibit features that are narrower than the gap itself, $W<\Delta$, as realized, e.g., for the $f$-electron density in Kondo insulators.
It was however concluded that the renormalization effects necessary to reach such an electronic structure for FeSi are largely beyond conventional band-theory \cite{PhysRevB.51.4763} and deemed rather unrealistic\cite{PhysRev.160.476}.
Finally, Mani \cite{MANI2004551} proposed a DOS-model with a Gaussian density, see \fref{FeSiDOSscheme} (bottom). The Gaussian tails of the excitations
are supposed to describe a finite in-gap density arising from disorder-derived localized states, but could also, qualitatively, account for an incoherence-induced (yet temperature independent) broadening.
This model was shown to yield a reliable fit to the conductivity under pressure for pristine FeSi, as well as for the FeSi$_{1-x}$Ge$_x$ data of Ref.~\cite{PhysRevB.63.115103},
foreshadowing the influence of quasi-particle incoherence seen in more sophisticated approaches discussed later in this chapter.

\paragraph{Spin models.}
\label{spinmodel}

A scenario opposite to the extended Bloch band picture is that of localized moments. For an excited state of spin $S$ separated from the ground state by a spin gap $\Delta_s$,
one expects in the limit $k_BT\ll\Delta_s$ a magnetic susceptibility $\chi=\frac{N_Ag_s^2\mu_B}{3k_BT}S(S+1)(2S+1)\exp(-\frac{\Delta_s}{k_BT})$, where $N_A$ is Avogadro's constant and $g_s$ the  gyromagnetic ratio. Using the standard value $g_s=2$, Jaccarino \etal\ \cite{PhysRev.160.476} found a spin gap $\Delta_S=69$meV$=795$K$\times k_B$ for a spin state $S=1$, while the best fit was achieved for $S=1/2$, $\Delta=65$meV$=750$K$\times k_B$, albeit with an enhanced $g_s\approx 4$.
From the data in \fref{FeSiCe3trans}(b)  we find $\Delta_s=50$meV and $S(S+1)(2S+1)g_s^2=14.6$, which yields for $g_s=2$ a value $S=0.8$,
to give a satisfactory description. 
Above 700K, as discussed in \sref{sus} (see again \fref{FeSiCe3trans}(b,left)), a Curie-Weiss law, $\chi=N_A\mu_{eff}^2/\left(3k_B(T-\theta)\right)$, gives a faithful representation of the experimental data, with a fluctuating moment $\mu_{eff}=2.6\mu_B$ that is actually close to the single-ion limit of iron for $S=1$, where $\mu_{eff}/\mu_B=g_s\sqrt{S(S+1)}=2\sqrt{2}\approx 2.8$ ($g_s=2$).
This good phenomenological description of the different temperature regimes in the susceptibility notwithstanding, further insights are needed to establish  a conclusive microscopic picture.
In fact the presented local-moment scenario is at odds with magnetism under doping or lattice expansion, established to be of itinerant origin (see the discussion in \sref{MR}).
Also in the case of FeSb$_2$, inelastic neutron spectra point towards an itinerant origin of magnetic fluctuations\cite{PhysRevB.83.184414}.

\paragraph{Other scenarios.}
Varma \cite{PhysRevB.50.9952} stressed the importance of going beyond a description in terms of the Kondo-lattice model,
to include effects of itinerancy and charge fluctuations for both FeSi and heavy-fermion materials.
He proposed that a temperature-induced mixed valence could be at the origin of the anomalous behaviour of FeSi.

Misawa and Tate \cite{MISAWA1996617} reproduced the experimental susceptibility of FeSi
in an itinerant model when going beyond the Pauli expression, \eref{PauliS}, 
by including a scattering amplitude in the Fermi-liquid picture. The inclusion of scattering effects
allowed to employ the model density-of-states of Refs.~\cite{PhysRev.160.476,PhysRevB.51.4763} and shown in \fref{FeSiDOSscheme}(middle), albeit
in a more realistic parameter regime, namely $W>\Delta$.

Finally, different variants of spin-fluctuation theory were invoked to explain magnetic and thermal properties of FeSi:
Takahashi and Moriya \cite{JPSJ.46.1451} discussed FeSi as a nearly itinerant ferromagnetic semiconductor with a temperature-induced local moment;
Evangelou and Edwards \cite{0022-3719-16-11-015} moreover included important effects of a temperature-dependent density-of-states self-consistently
and found qualitative agreement with experiment.

\subsection{Band-structures and many-body perturbation theory} 
\label{band}

{
\begin{figure*}[!t]%
\begin{tabular}{cccc}
{ 0\% ``NaCl''} & 50\% & 75\% & 100\% FeSi (B20) \\
{ \includegraphics[angle=0,width=0.225\textwidth]{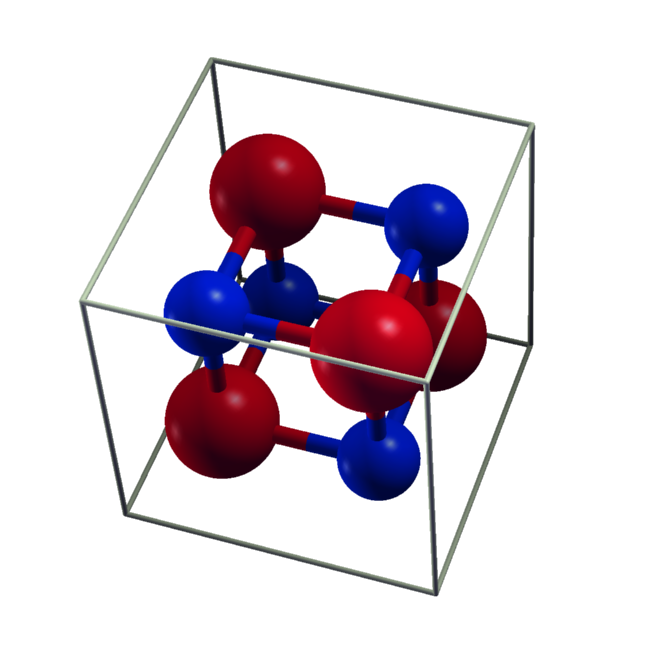}} &
{ \includegraphics[angle=0,width=0.225\textwidth]{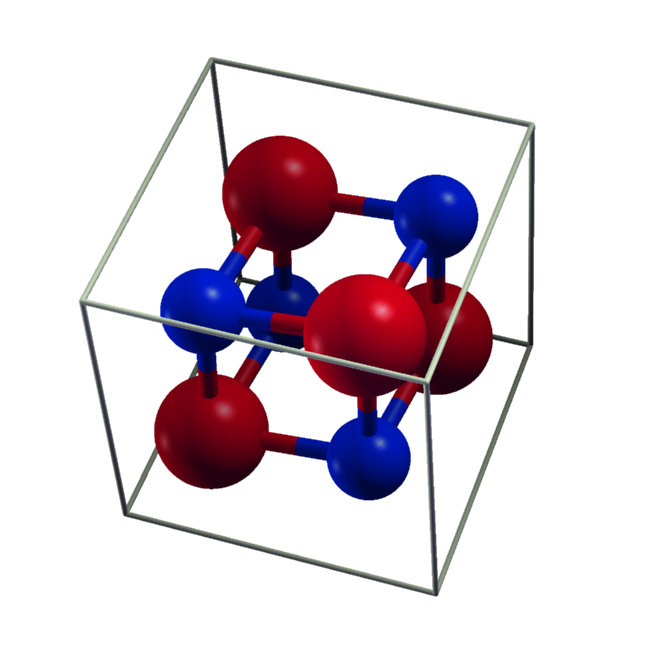}} &
{ \includegraphics[angle=0,width=0.225\textwidth]{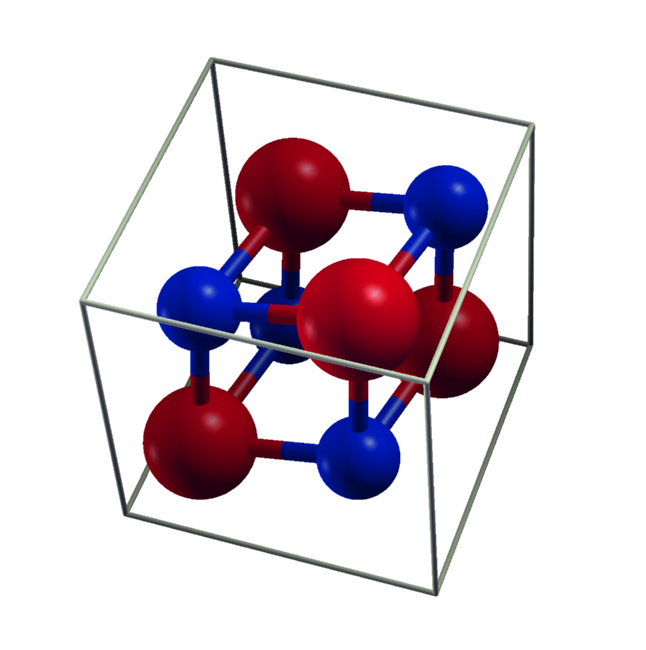}} &
{ \includegraphics[angle=0,width=0.225\textwidth]{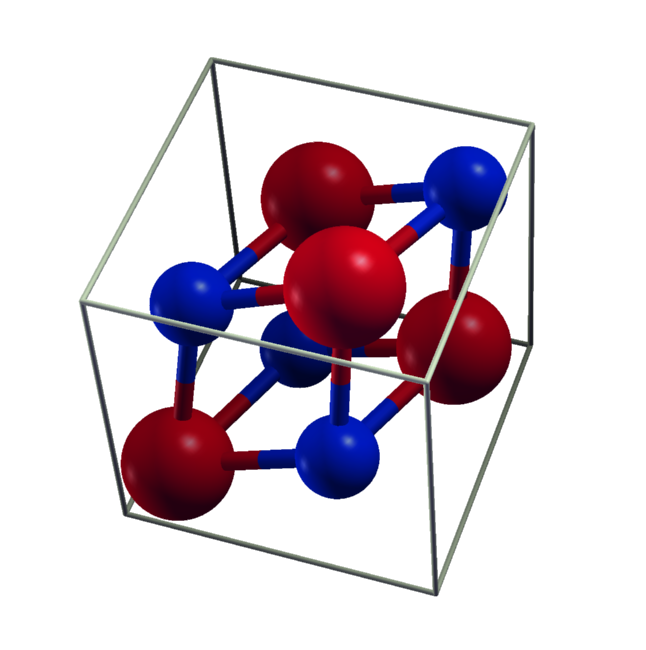}} \\
{ \includegraphics[clip=true,trim=0 50 20 20, angle=0,width=0.235\textwidth]{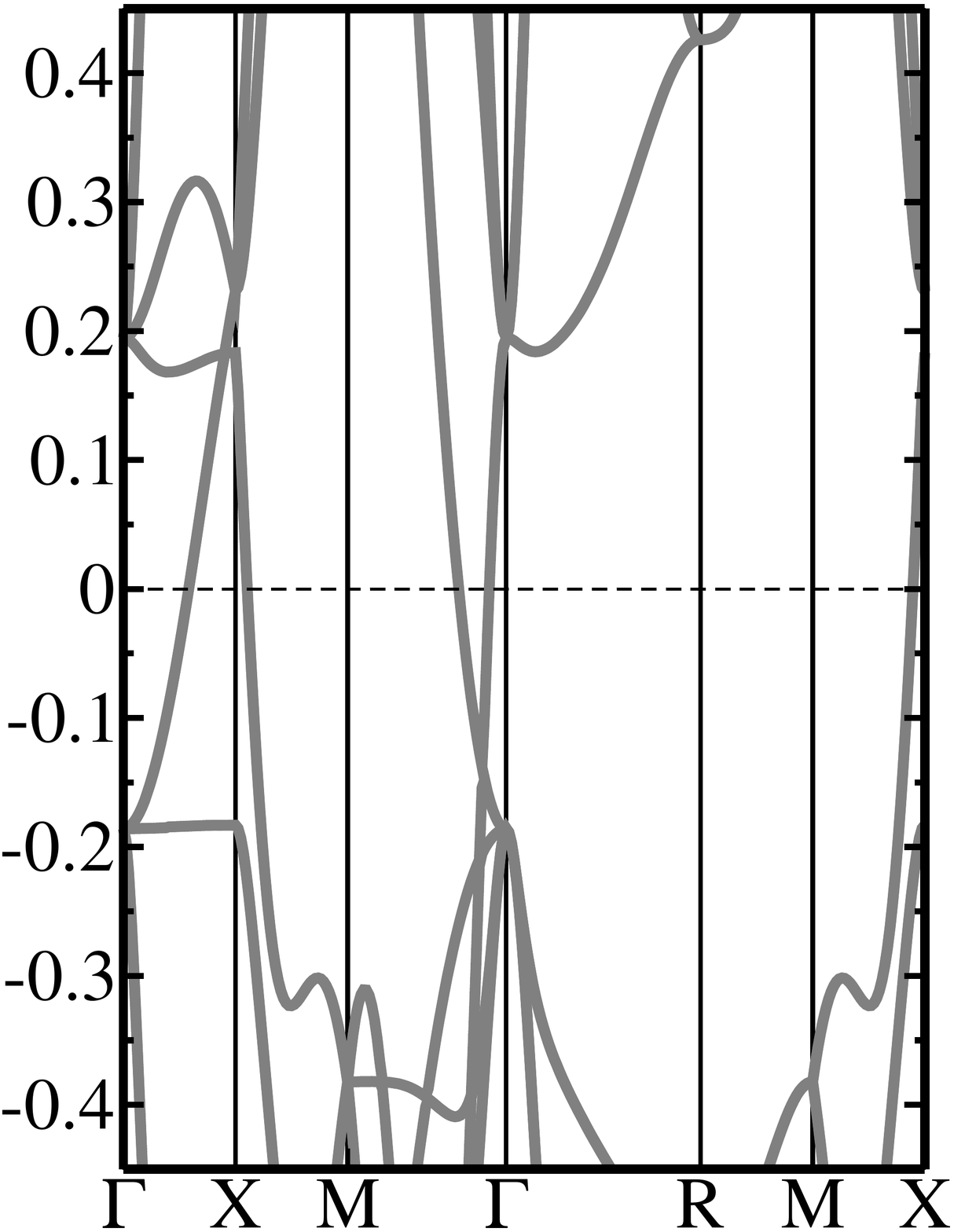}} &
{ \includegraphics[clip=true,trim=0 50 20 20, angle=0,width=0.235\textwidth]{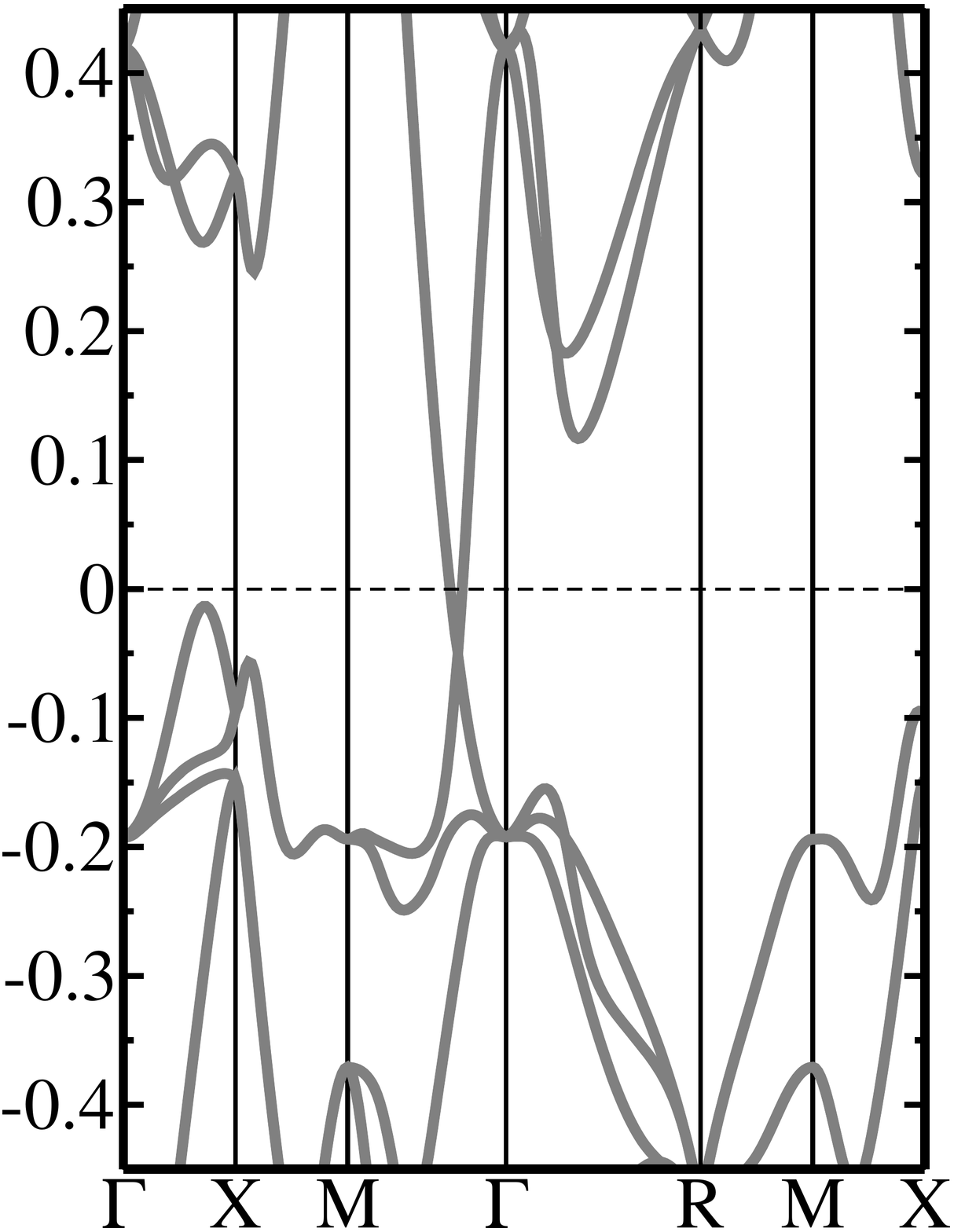}} &
{ \includegraphics[clip=true,trim=0 50 20 20, angle=0,width=0.235\textwidth]{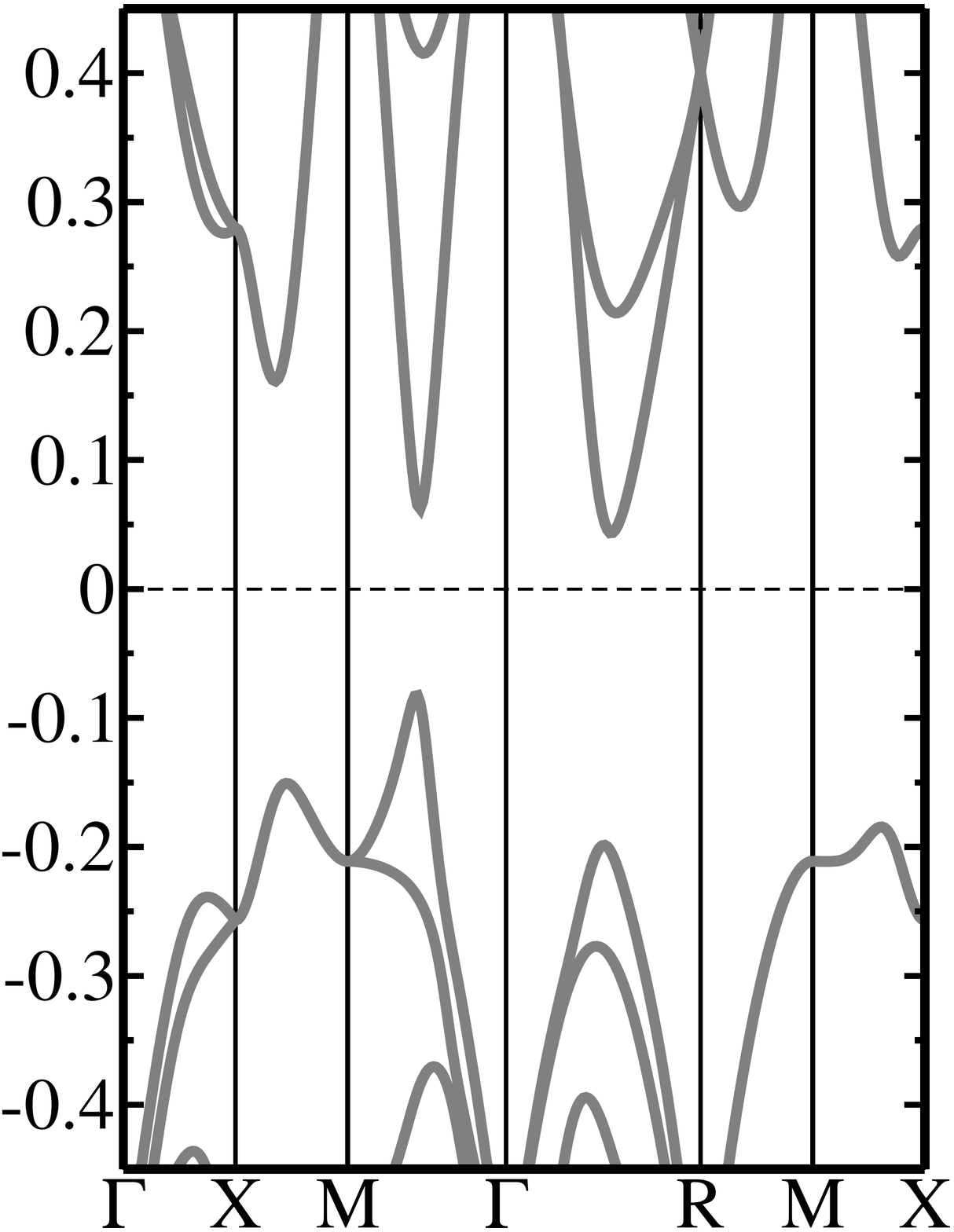}} &
{ \includegraphics[clip=true,trim=0 50 20 20, angle=0,width=0.235\textwidth]{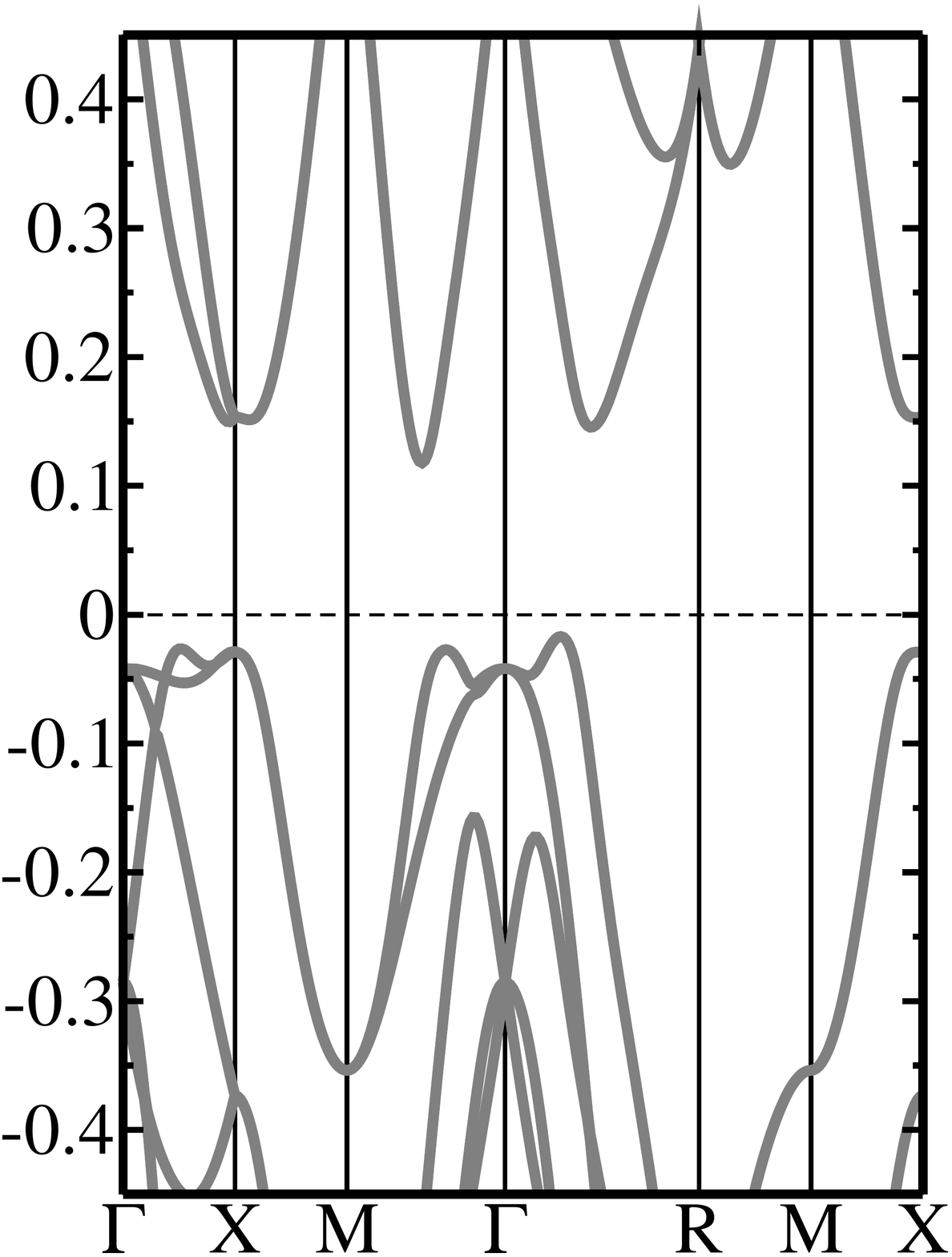}} 
\end{tabular}
\caption{ 
{\bf Band structure of FeSi.} The B20 crystal structure of FeSi can be viewed as a distortion of a simple cubic rock-salt structure:
The positions of the iron (large red spheres) and silicon (smaller blue spheres) atoms are ($u$,$u$,$u$), ($\oo{2}+u$,$\oo{2}-u$,$1-u$), ($1-u$,$\oo{2}+u$,$\oo{2}+u$), and ($\oo{2}-u$,$1-u$,$\oo{2}+u$)
with $u$(Fe)=1/4 and $u$(Si)=3/4 for the ``NaCl'' structure, and $u$(Fe)=0.136 and $u$(Si)=0.844 for FeSi\cite{PhysRevB.47.13114}.
The top panel shows (from left to right) the crystal structures (rendered with Ref.~\cite{Anton1999176}) for 0\%, 50\%, 75\% and 100\% of the distortion
as measured by a linear interpolation of the $u$ parameters at constant unit-cell volume ($a=4.488$\AA\cite{PhysRevB.56.12916}).
The lower panel shows the corresponding GGA band-structures (energies in eV with respect to the Fermi level).
Figure adapted from Ref.~\cite{jmt_hvar}. See also the discussion in Refs.~\cite{PhysRevB.81.125131,PhysRevB.75.024116} and Ref.~\cite{0295-5075-104-3-30001} for the relation of the $u$ parameters to the chirality of the crystal. 
}
\label{fig_bnd_fesi}
\end{figure*}
}

In this section, we will review applications of {\it ab initio} electronic structure methods to correlated narrow-gap semiconductors,
as well as to the prototypical Kondo insulator \cbp.
We will discuss results from effective one-particle theories, such as density functional theory (DFT\cite{RevModPhys.71.1253,RevModPhys.61.689}), hybrid functionals, and static mean-field DFT+$U$\cite{PhysRevB.44.943}.
These methods provide very valuable insights into the complexities and chemistry of real materials, such as 
the nature of bonding (ionic vs.\ covalent), crystal-field splittings, multi-orbital effects,
as well as qualitative information on instabilities towards (itinerant) magnetism.
Such results are instrumental for constructing effective low-energy many-body models, that will be discussed in \sref{model}.
Going beyond density functional based methods, we also include here some results from Hedin's {\it GW} method\cite{hedin,ferdi_gw,RevModPhys.74.601}, a many-body perturbation theory,
as well as constrained random phase approximation (cRPA) \cite{PhysRevB.70.195104} estimates of local (Hubbard $U$, Hund's $J$) interactions for use in low-energy many-body models.

\subsubsection{FeSi, other B20 compounds, and a comparison to oxides.}
\label{bandFeSi}

FeSi and FeGe as well as  their 4$d$ homologues RuSi and RuGe crystallize in the so-called B20 structure. Despite being cubic (space group P2$_1$3)
the compounds' crystal-structure is quite complex: There are four formula units per unit-cell, each Ru (Fe) has seven Si (Ge) neighbours, and the absence of inversion symmetry
causes a Dzyaloshinskii-Moriya-derived spin spiral structure in the magnetically ordered phase of FeGe\cite{0953-8984-1-35-010}. Indeed these compounds have the same atomic arrangement
as the prototypical itinerant helimagnet MnSi\cite{doi:10.1063/1.1708422,ISHIKAWA1976525}.
As suggested by Mattheiss and Hamann\cite{PhysRevB.47.13114}, the B20 structure can be seen as a simple cubic rock-salt structure (space group Fm3m) with a considerable distortion along the [111] direction. 
Nonetheless FeSi shows no signs of any long-range order\cite{doi:10.1143/JPSJ.18.995} and is paramagnetic down to 40mK\cite{PhysRevB.50.14933},
while RuSi is diamagnetic with a temperature-independent susceptibility\cite{BUSCHINGER199757,Hohl199839}.

\begin{figure*}[!t]
  \begin{center}
		\subfloat[FeSi: hybridization gap]
		{{\includegraphics[angle=0,width=.48\textwidth]{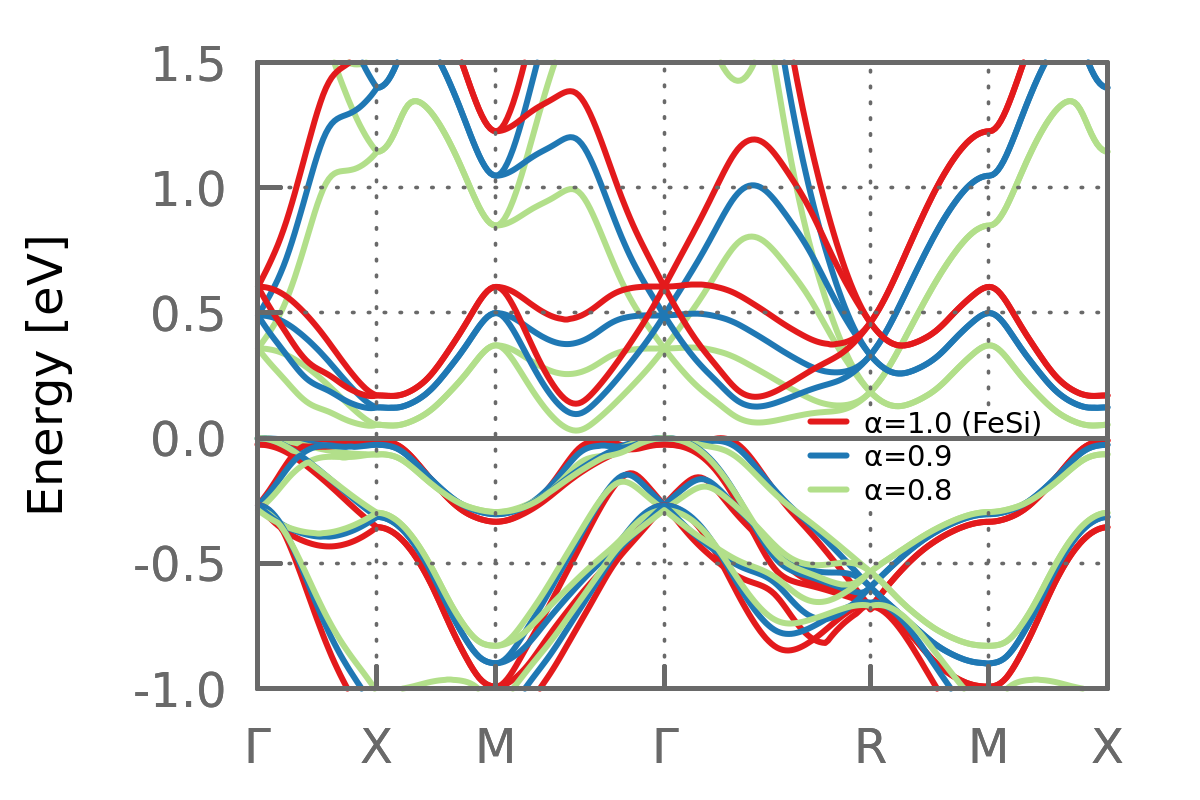}}}
	$\quad$
	\subfloat[LaCoO$_3$: crystal-field gap]
	{{\includegraphics[angle=0,width=.48\textwidth]{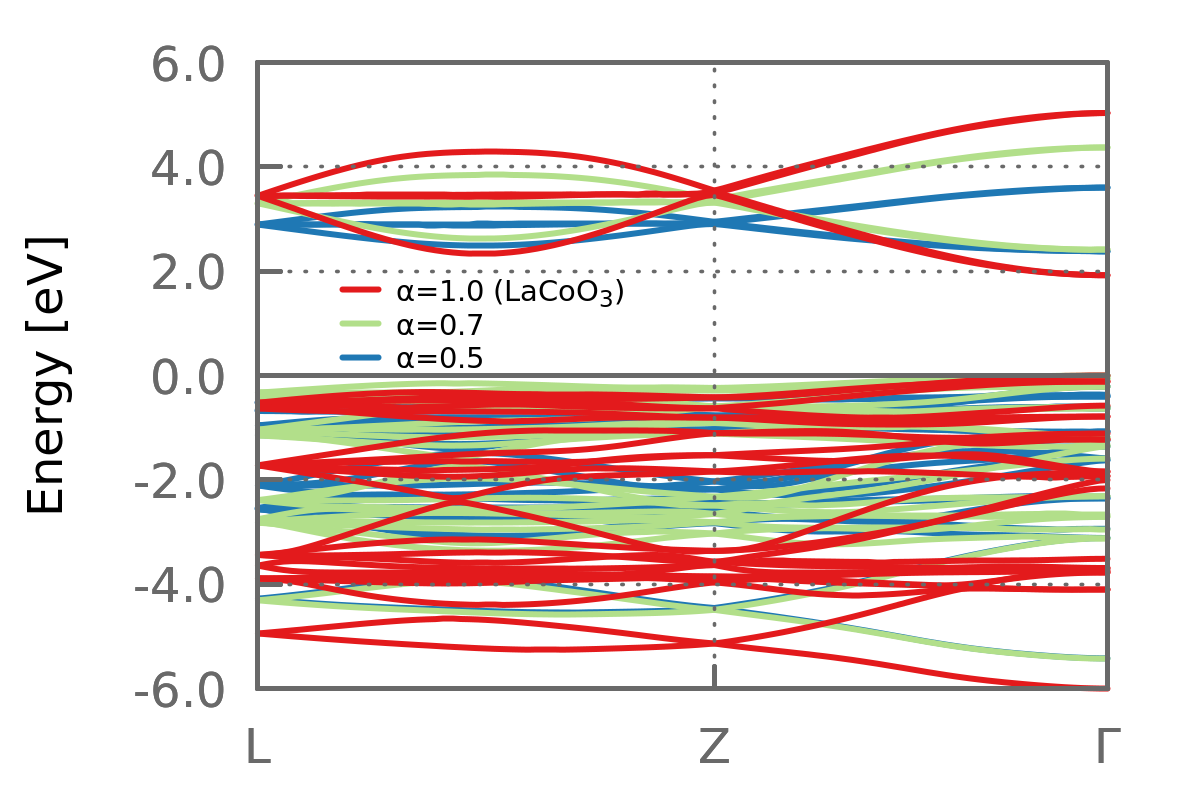}}}
				\caption{{\bf Hybridization gap in FeSi vs.\ crystal-fields in LaCoO$_3$.}
					On the left (right) is shown, in red, the GGA (mBJ) band-structure of FeSi (LaCoO$_3$) . The band-structures for $\alpha<1$ have been obtained
					by scaling the hybridization of each transition-metal atom (Fe or Co) to all other atoms with a factor $\alpha$ in a maximally localized Wannier function setup.
          The local basis includes the Fe-3$d$, Si-3$p$, and Si-3$s$ orbitals for FeSi and the Co-3$d$ and O-2$p$ orbitals for LaCoO$_3$.
					In the case of FeSi, the gap is found to {\it decrease} linearly with $\alpha$, suggesting a dominant hybridization-origin of the band-gap---a signature of covalent bonding. 
					In LaCoO$_3$, reducing the hybridization {\it increases} the charge gap, as band-widths shrink. See text for details.					
					}				
      \label{FeSiscalehyb}
      \end{center}
\end{figure*}


\paragraph{FeSi and FeGe: insights from band theory.}
The described distortion is crucial for the insulating ground-state in the silicides: Indeed, as shown in \fref{fig_bnd_fesi} (left), rock-salt FeSi is metallic
within DFT.
Nonetheless the density of states (DOS) at low energies is quite small, as the bands that cross the Fermi level are quite dispersive. 
With increasing distortion the Fermi level crossings are avoided and an indirect gap of 100-130meV is opened\cite{PhysRevB.47.13114,PhysRevB.49.2219,PhysRevB.81.125131,jmt_hvar}.
The atomic displacements of the distortion conserve the cubic Bravais lattice but reduce the point group to tetrahedral.
As a consequence the transition metal 3$d$-orbitals split into a low-lying $z^2$ and two doublets $x^2$-$y^2$, $xy$ and $xz$, $yz$.
The latter two are sufficiently separated in energy so that---for a 3$d^6$ configuration for Fe/Ru---the three lowest orbitals are completely filled and the paramagnetic DFT solutions become insulating.

For the silicides, DFT correctly finds this paramagnetic insulator to be the  ground-state. 
As discussed by Mazurenko \etal\ \cite{PhysRevB.81.125131} for FeSi, the band-gap does not primarily originate 
from crystal-fields, but from distortion-induced changes in the hybridization between the iron atoms and their environment.
To illustrate this, we have constructed maximally localized Wannier functions for a subspace consisting of the Fe-3$d$ and Si-3$s$ and 3$p$ orbitals of a GGA band-structure calculation.
The use of this atom-centred local basis allows to empirically study the influence of inter-atomic hybridizations onto the electronic structure.
Indeed, we can scale all hybridizations between each iron atom and all other atoms in the unit-cell with a factor $\alpha\leq 1$.
\Fref{FeSiscalehyb}(a) displays the results:
Akin to the (direct gap of the) periodic Anderson model, the gap in FeSi changes linearly with the strength $\alpha$ of the hybridization.
In fact, below $\alpha\lesssim 0.75$ the system becomes metallic.
Hence FeSi can be categorized as a covalent hybridization-gap semiconductor\cite{PhysRevB.47.13114,PhysRevB.59.15002,PhysRevB.78.033109,PhysRevB.81.125131,jmt_fesi}.
However, there are qualitative differences to the case of canonical Kondo insulators: 
An individual scaling of inter-atomic hybridizations in FeSi (results not shown) divulges that the gap is mainly driven by inter-iron hybridization---as opposed
to hybridizations between localized orbitals and conduction-electron bands as in, e.g.,\ \cbp\ (see \sref{Ce3dft}).
In fact, Mazurenko \etal\ \cite{PhysRevB.81.125131} investigated the spread of their Wannier functions and found that the ones centred on the iron atoms 
have substantial weight at neighbouring iron sites---in stark contrast to localized $f$-orbitals in Kondo insulators.
As a consequence the highest valence and lowest conduction bands have different orbital characters. 
Instead, we will see in \sref{Ce3dft} that in DFT calculations for \cbp, the dominant character is the same on both sides of the gap. 
This distinction between these two systems will be further discussed in \sref{PAM} for prototypical many-body models, and in \sref{abinitiohyb} in the context
of realistic many-body calculations.

At this point, we find it instructive to compare the covalently bonded hybridization-gap insulator FeSi to the oxide LaCoO$_3$.
Both compounds share, at first glance, several empirical similarities, such as the insulator-to-metal transitions with rising temperature and 
a strongly non-monotonous magnetic susceptibility (see \fref{overviewchi}). 
Here, we limit the comparison to the ground-state as described within band-structure methods.
To obtain an insulating Kohn-Sham spectrum for LaCoO$_3$, the inclusion of exchange effects beyond the LDA or GGA functionals is required.
\Fref{FeSiscalehyb}(b) displays the band-structure obtained using the mBJ functional\cite{mBJ}. There, the gap, $\Delta^{mBJ}=1.9$eV vastly overestimates
the experimental finding of 100meV%
\footnote{QS{\it GW} results (unpublished) exhibit an even larger gap of 2.4eV.} (see \tref{table1}).
In the shown DFT solution, the nominal six $d$-electrons of cobalt fully populate the $t_{2g}$ orbitals that are well separated
from the thus empty $e_g$ states, resulting in a low-spin configuration.
In order to elucidate the dominant origin of the $t_{2g}$-$e_g$ splitting, we have performed the same analysis as for FeSi, i.e.,
we have constructed a local Wannier basis and scaled down the hybridizations of the cobalt atoms with their environment.
Contrary to FeSi, this procedure causes the gap to in LaCoO$_3$ to {\it increase}. Indeed the visible shrinking of the
$e_g$-dispersion trumps the slight downshift of its centre-of-mass. Therefore, the gap-formation in this setup is dominated by local
crystal-fields, as previously observed by K\v{r}{\'a}pek \etal\ \cite{PhysRevB.86.195104}.
For a discussion of how the covalent vs.\ ionic gap-formation impacts finite-temperature properties, including
electronic correlation effect, see \sref{micro} for a sequel of the FeSi-LaCoO$_3$ comparison and \sref{sec:covalent} for a discussion of the bigger picture.

We now return to band-theory results for FeSi:
Incorporating Hubbard-$U$-like interactions via static mean-field LDA+$U$ calculations, Anisimov \etal\ \cite{PhysRevLett.76.1735}
proposed that FeSi could be driven through a first-order transition to a ferromagnetic metal by applying a magnetic field of $B_c\approx 170$T.
The field would induce a low-spin (paramagnetic insulator) to high-spin (ferromagnetic metal with a moment of $1\mu_B$/Fe) transition.
They further argued that the anomalous behaviour of FeSi in the absence of any field could be propelled by the proximity (in total energy) to the  critical end point, $T_c=280$K, of that transition. This interpretation gives {\it ab initio} support to the early theory that FeSi is a nearly ferromagnetic system (see Refs.~\cite{JPSJ.46.1451,0022-3719-16-11-015} and \sref{pheno} above).
Subsequently, measurements in ultra-high fields (cf.\ the discussion in \sref{MR}) confirmed the existence of a discontinuous metamagnetic semiconductor-to-metal transition with $B_c\approx355$T and $T_c<77$K\cite{Kudasov1998,Kudasov1999}.

Allowing for spin-polarization, iron germanide, FeGe, is predicted to be a ferromagnetic metal 
with an ordered moment of $\sim 1\mu_B$ within DFT\cite{YAMADA20031131,PhysRevB.80.035122,0953-8984-24-9-096003,PhysRevLett.98.047204}, DFT+U\cite{PhysRevLett.89.257203}, or hybrid functional methods\cite{PhysRevB.80.035122}, in congruence with experiment\cite{WAPPLING1968173,LUNDGREN1968175,0953-8984-1-35-010}.
Also some of the observations of the substitutional series FeSi$_{1-x}$Ge$_x$ (see \sref{KIinter} for details of experimental findings) are qualitatively captured within band-theory:
Yamada \etal\ \cite{YAMADA20031131} found a transition from a non-magnetic-insulator to ferromagnetic-metal for $x\approx 0.5$ in DFT.
Using DFT supercell calculations, Jarlborg\cite{PhysRevLett.98.047204} found ferromagnetism above $x=0.3$, suggested to arise from a combination of an
increased volume and substitutional disorder.
Within LDA+$U$, Anisimov \etal\ \cite{PhysRevLett.89.257203} 
found a first order transition to a ferromagnetic metal at $x=0.4$ for a reasonable value of $U$.
These findings are in rough agreement with the experimental critical substitution $x\approx0.25$\cite{PhysRevLett.91.046401}.

If the dominant control parameter in the FeSi$_{1-x}$Ge$_x$ series was the unit-cell volume, applying pressure
to FeGe should suppress magnetism and drive the system insulating. 
Indeed, FeSi has a 12.5\% smaller unit-cell than FeGe, which can roughly be reached at 25GPa\cite{doi:10.1016/j.stam.2007.04.004}.
In experiment\cite{PhysRevLett.98.047204} the Curie temperature is indeed suppressed at a critical pressure of $P_c\approx 19$GPa. However, even above this pressure a residual conductivity persists, so that it was suggested to arise from zero-point motion\cite{PhysRevLett.98.047204}.
Neglecting the latter effect,
band theory finds a pressure-driven phase transition from a ferromagnetic-metal to a non-magnetic-insulator.
The critical pressure varies however by a factor of at least four, depending on the
functional used in DFT (LDA: $P_c\le 10$GPa, PW91: $P_c=40$GPa) \cite{PhysRevB.80.035122}.
Best agreement with experiment is found when optimising the crystal structure with GGA and use LDA for the electronic structure ($P_c=18$GPa)\cite{0953-8984-24-9-096003}.

Also some aspects of the non-isoelectronic substitution series Fe$_{1-x}$Co$_x$Si\cite{Beille1983399,doi:10.1143/JPSJ.59.305} are qualitatively captured within band-theory.
Using the virtual crystal approximation in DFT, Morozumi and Yamada\cite{MOROZUMI20071048} 
found a transition to a half-metallic ferromagnet for finite dopings up to 50\%.
These results are consistent with the experimentally evidenced linear increase of the magnetic moment 
with $x$ up to 30\%. 
Indeed, as was proven experimentally by Paschen \etal\ \cite{Paschen1999864}, the spontaneous moment in Fe$_{1-x}$Co$_x$Si
does not originate from moments localized on the Co-dopants, but is of itinerant origin, and thus, in principle, amenable to band-theory
via Stoner physics.

At higher dopings, the theoretical moment\cite{MOROZUMI20071048}  in Fe$_{1-x}$Co$_x$Si decreases less quickly than the experimental one\cite{doi:10.1143/JPSJ.59.305}, probably for lack of treating the randomness of the alloy properly. Indeed, the theoretical magnetic structure is very sensitive to disorder\cite{PhysRevB.73.024426},
before the diamagnetic metallic state of the end member CoSi is reached. 

\begin{figure}[!t]
{\includegraphics[angle=0,width=.45\textwidth]{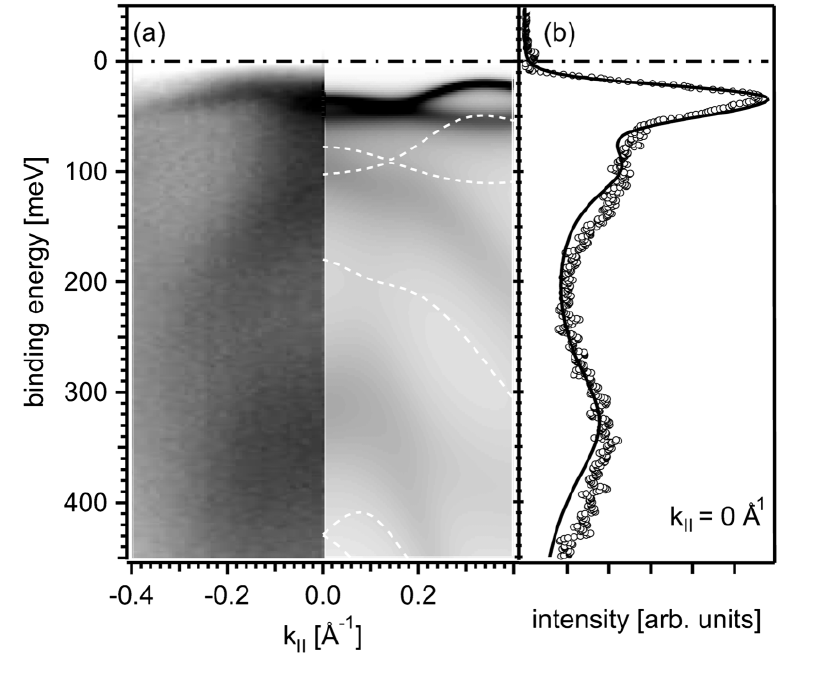}}
\caption{{\bf Angle-resolved photoemission spectrum of FeSi.} 
(a) left: ARPES intensity obtained from a He-I source ($21.23$eV) from the (100) surface at $T=10$K;
right: DFT bands (white dashed lines) superimposed on an intensity plot obtained
from fitting a phenomenological self-energy correction on top of DFT to the experimental data:
$\Sigma(\omega)=g_h\frac{\omega}{(\omega+i\gamma_h)^2}-\frac{g_l}{\omega+i\gamma_l}-i\frac{g_l}{\gamma_l}$
with $g_h=10.5$eV$^2$, $\gamma_h=6.5$eV, $g_l=0.0093$eV$^2$, $\gamma_l=0.1103$eV.
(b) energy-distribution curve at $k_\parallel=0$\AA$^{-1}$. 
 From Klein \etal\ \cite{PhysRevLett.101.046406}.}
\label{FeSiKlein}%
\end{figure}

Finally, other band-structure intricacies of B20-compounds include the recent findings of K{\"u}bler \etal\ \cite{0295-5075-104-3-30001} of a non-vanishing electronic Berry phase in
the insulating silicides FeSi, RuSi, and OsSi, corresponding to a crystalline chirality that heralds the possibility of realizing a macroscopic electric polarization
in these systems.

\medskip
In all, for the discussed properties, band-structure methods give very valuable information, as well as an often semi-quantitative description.
We now turn to physical properties where this is not the case.

\paragraph{FeSi: failures of band-theory.}
\label{fail}

\begin{figure}[h]
{\includegraphics[angle=0,width=.51\textwidth]{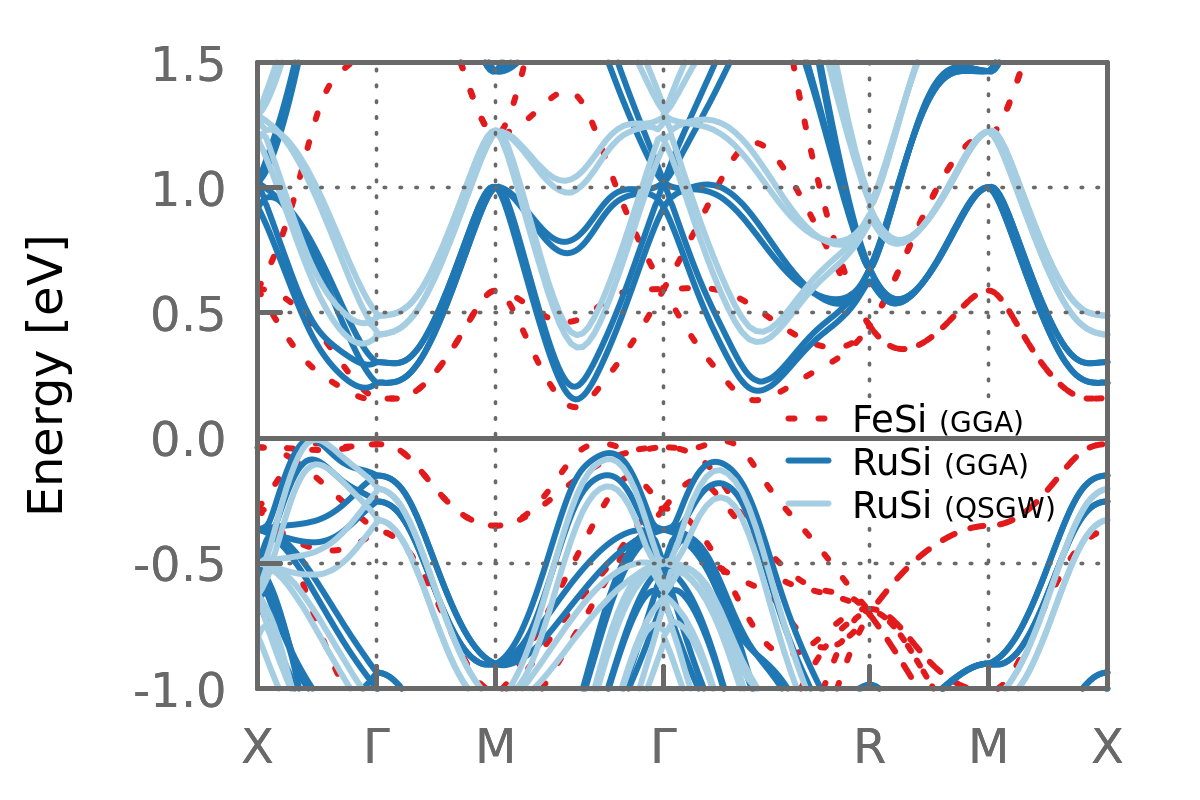}}
\caption{{\bf Comparison of FeSi and RuSi.} Shown is the band-structure of the isostructural and isoelectronic
compounds FeSi (red, dashed) and RuSi (dark blue) in the generalized gradient (GGA) approximation. The indirect charge gaps are $\Delta_{indir}=127$meV and $\Delta_{indir}=172$meV
for FeSi and RuSi, respectively. In RuSi the inclusion of the spin-orbit coupling leads to a notable splitting of the bands.
Many-body corrections within QS{\it GW} (light blue) boost the gap to $\Delta_{indir}=370$meV, slightly above the experimental $\Delta_{indir}=260-310$meV.}
\label{RuSi}%
\end{figure}

Standard band-structure methods work at zero temperature. Therefore, they cannot access the unusual spectral and magnetic behaviour of FeSi at finite temperatures.
Yet, we can compare band-theory results to the experimental properties at low-temperatures.

\subparagraph{Spectral properties.}
While DFT-related methods in principle only produce an auxiliary Kohn-Sham spectrum, it has become common practice to compare the latter to the excitation spectrum
of the solid.
In the case of FeSi, the Kohn-Sham band-gap is notably {\it over}estimated, which is in stark contrast to the usual
{\it under}estimation of band-gaps in conventional semiconductors.
Indeed while activation law fits e.g.,\ to the resistivity extract  $\Delta_{indir}=59$meV (see \fref{FeSiCe3trans}(a,left)),
DFT finds a value about twice as large: $\Delta_{indir}^{GGA}=130$meV\cite{PhysRevB.47.13114,PhysRevB.49.2219,PhysRevB.59.15002,PhysRevB.81.125131,jmt_hvar}.%
\footnote{Nonetheless, DFT calculations give good estimates for the {\it change} of the gap with the unit-cell volume, $d\ln\Delta/d\ln V\approx -6$, see Refs.~\cite{GRECHNEV1994835}. Also note, that there is a non-negligible dependence of the charge gap on the exchange-correlation potential used, e.g., $\Delta_{indir}^{LDA}=85$meV, congruent with the gap originating from hybridization effects, cf.\ below.}

As will be discussed in the theory \sref{pnas}, the discrepancy between band-theory and experiment is rooted in dynamical electronic correlation effects that renormalize the excitation spectrum. 
On an empirical level, this physics can be particularly well illustrated on the basis of photoemission\cite{SAITOH1995307}
and angle-resolved photoemission spectroscopy (ARPES)\cite{PhysRevB.52.R16981,PhysRevB.77.205117,PhysRevLett.101.046406,1367-2630-11-2-023026} experiments.
Comparing the photoemission intensity with band-structure results, see \fref{FeSiKlein}, one witnesses (i) a notable shrinking of the charge gap, and (ii)
a narrowing of the bands near the Fermi level. 
This renormalization of bands has been analysed in terms of phenomenological self-energies $\Sigma(\omega)$\cite{SAITOH1995307,PhysRevB.52.R16981,PhysRevB.77.205117,PhysRevLett.101.046406,1367-2630-11-2-023026} that allow for a Brinkman-Rice-like bandwidth-narrowing but neglect orbital- and momentum dependencies. 
Performing a low-energy expansion, $\Sigma(\omega)=(1-1/Z)\omega-i\gamma \omega^2+\mathcal{O}(\omega^3)$, of the self-energy form used by Klein \etal\ \cite{PhysRevLett.101.046406}
yields at $T=10$K a substantially reduced quasi-particle weight $Z=0.5$ and a scattering coefficient $\gamma=7$/eV.
Despite the simplicity of the Ansatz, the resulting  effective mass $m^*/m^{DFT}=1/Z=2$ is quite close to the value found in realistic many-body approaches (discussed in \sref{pnas}).

\begin{figure*}[!th]
  \begin{center}
	\subfloat[{\fesb.}]
	{\includegraphics[angle=0,width=.45\textwidth]{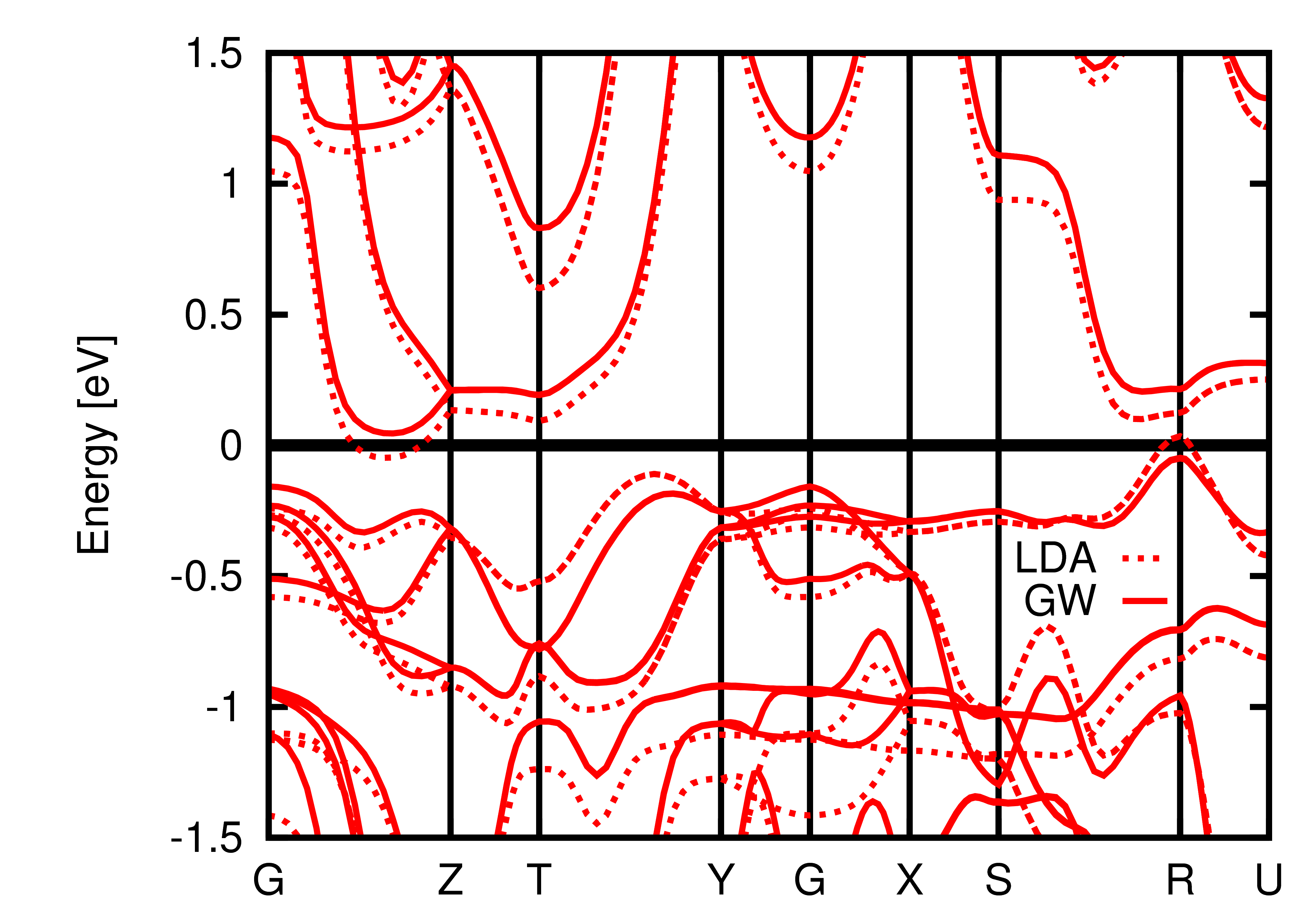}}
	$\qquad$
	\subfloat[{\feas.}]
	{\includegraphics[angle=0,width=.45\textwidth]{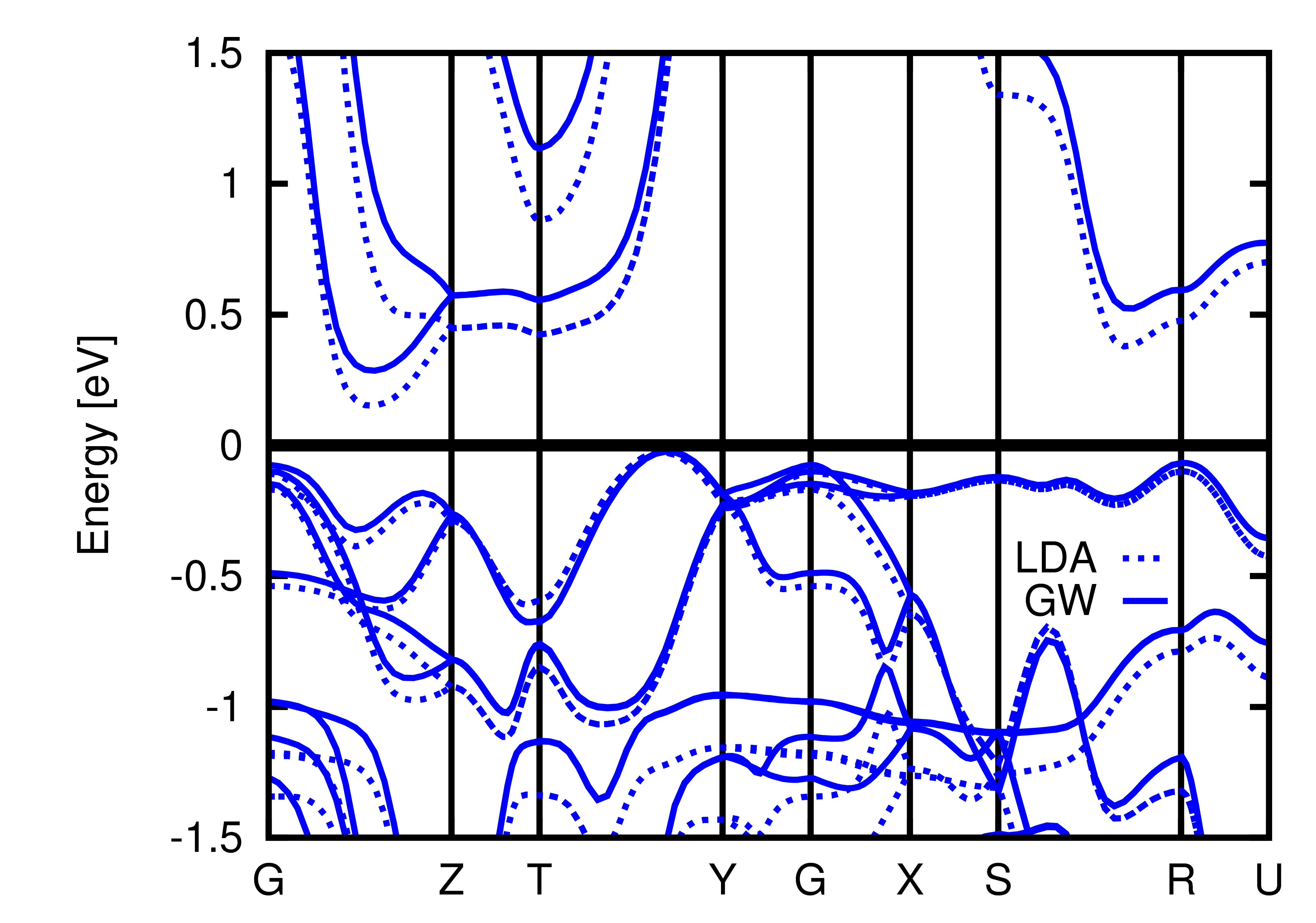}}
		      \caption{{\bf Density functional and many-body perturbation theory.}  Band-structures of \fesb\ and \feas\ as obtain from LDA (full lines) and
					Hedin's {\it GW} approximation (dashed lines). Note that, contrary to LDA, {\it GW} correctly finds \fesb\ to be an insulator. From Ref.~\cite{jmt_fesb2}.
								}
      \label{FeSb2GW}
      \end{center}
\end{figure*}

\subparagraph{Ground-state properties.}
Also in the true realm of DFT, ground-state properties, there are findings that are at odds with experiment,
or are at least quite unusual:
The equilibrium unit-cell volume of FeSi (experimentally: $V_0\approx 89.6$\AA$^3$\cite{PSSA:PSSA2210230244,Vocadlo2002}) comes out slightly too small in DFT.
While overbinding is expected for the LDA functional ($V_0=84$\AA$^3$\cite{PhysRevB.59.12860}), surprisingly also the typically underbinding GGA
underestimates the volume ($V_0=88.96$\AA$^3$\cite{PhysRevB.59.12860}, 
$V_0=88.1$\AA$^3$\cite{PhysRevB.94.144304}).
More dramatically are the discrepancies in the bulk modulus: While DFT yields values above 200GPa (LDA: $B_0=260$GPa, GGA: $B_0=220$GPa\cite{PhysRevB.59.12860}),
the experimental values scatter in the range $B_0=115-185$GPa\cite{0953-8984-7-36-001,PSSA:PSSA2210230244,Wood:gl0431,Guyot277,JGRB:JGRB13523} with, however, a single outlier of 209GPa\cite{GRL:GRL8102}. 
These findings may be interpreted as arising from electronic correlation effects that decrease participation of charges in the bonding.
More electron-lattice properties will be discussed in \sref{lattice}.

\paragraph{The 4$d$ homologue: RuSi.}

The 4$d$-analogue of FeSi, RuSi, is a semiconductor with a medium-sized gap of $260-310$meV as inferred from resistivity measurements\cite{Buschinger1997238,Hohl199839} and $\sim400$meV from optical spectroscopy\cite{Buschinger1997238,Vescoli1998367}.
As mentioned in the introduction, there are no anomalies in the temperature dependence of spectroscopic or magnetic observables of RuSi.

It is plausible that the extended 4$d$-orbitals of Ru make for a less correlated electronic structure that is, then, more amenable to band-theory.
In fact, density functional theory yields a band-structure similar in shape to FeSi, see \fref{RuSi} for a comparison within GGA. Yet the dispersions are enhanced and a larger indirect 
gap is present: 230meV\cite{Imai2006173,0295-5075-85-4-47005,springerlink:10.1134/S1063782609020031,jmt_hvar} without, 170meV with the inclusion of spin-orbit coupling.  Thus, contrary to FeSi, the gap is smaller within band theory than in experiment, as is usually the case for uncorrelated semiconductors.
The origin of the gap-underestimation is typically attributed to an insufficient treatment of the exchange part\cite{doi:10.1063/1.464913} of the DFT exchange-correlation functional.
An {\it ab initio} many-body approach that excels, among others, at describing band-gaps is Hedin's {\it GW} method\cite{hedin,ferdi_gw,RevModPhys.74.601}.
Here, we employ the so-called quasi-particle self-consistent variant QS{\it GW}\cite{PhysRevLett.93.126406} to RuSi%
\footnote{In QS{\it GW} a hermitianized self-energy is fed back into the band-structure code. Here we use a full-potential linearized muffin-tin orbital (FPLMTO) method\cite{fplmto} so as to eliminate any dependence
on the DFT starting point.}.
QS{\it GW} was shown to yield improved band-gaps\cite{schilfgaarde:226402}, particularly for $sp$-semiconductors, while it overestimates gaps in some $d$-electron systems\cite{schilfgaarde:226402} (see also the 
discussion of FeSb$_2$ and related materials below).
As can be seen in \fref{RuSi}, the screened exchange self-energy from the {\it GW} indeed widens the band-gap; we find a value of about 370meV, slightly above experimental values.
We will discuss the thermopower of RuSi and the substitution series Fe$_{1-x}$Ru$_{x}$Si in \sref{silicides}.
Model calculations in \sref{PAM} will be used to rationalize the different behaviours of FeSi and RuSi.


\subsubsection{The marcasites: FeSb$_2$, FeAs$_2$ \& Co.}
\label{marcasites}

The family of FeSb$_2$\cite{fesb2_struct,Holseth1970,Fan1972136}, FeAs$_2$\cite{Fan1972136}, FeP$_2$\cite{Hulliger1959,1402-4896-4-3-010} , CrSb$_2$\cite{Holseth1970}, their 4$d$-analogues RuSb$_2$\cite{Holseth1968}, RuAs$_2$\cite{HULLIGER1963}, as well as their 5$d$ osmium-based members\cite{Kjekshus1977}  
crystallize in the regular FeS$_2$ marcasite structure (orthorhombic space group Pnnm) with two formula units per unit cell.

In this structure, the transition metal ions are surrounded by distorted pnictogen octahedra, that
share corners along the $c$-axis.
According to ligand field theory, the transition metal 3$d$-orbitals split into the lower $t_{2g}$ and higher-lying $e_g$ orbitals.
The existence of two different transition metal--pnictogen distances causes the $t_{2g}$ to further split
into a lower doublet $\Lambda$ and a higher single orbital $\Xi$.
In the ionic picture, the transition metal would be in a $3d^4$ low-spin configuration, with an empty $\Xi$ orbital\cite{Goodenough1972144,PhysRevB.72.045103}.
From this point of view,  a (thermal) population of the $\Xi$ orbital may cause
the metallization\cite{Goodenough1972144}, as well as---via a spin-state transition\cite{PhysRevB.67.155205,PhysRevB.72.045103}---the enhanced paramagnetism in \fesb.
As will be briefly discussed in \sref{micro}, such a scenario was shown to be realized in LaCoO$_3$.

\paragraph{Density functional theory.}
Subsequent band-structure calculations by Madsen \etal\ \cite{Madsen_fesb2}, however, suggested a more covalent rather than ionic picture of bonding in the marcasites.
Indeed
 a stabilization of those $d$-orbitals pointing towards the ligands was evidenced, which in particular causes a lowering of the $e_g$ bands. 
As a consequence the transition metal configuration of all iron-, ruthenium-, and osmium-based marcasites from above is close to $d^6$ rather than $d^4$
(see also the discussion surrounding \fref{valence}).

In \fesb\ and RuSb$_2$ the stabilization of the $e_g$ orbitals with respect to ligand field theory occurs to the extent that density functional theory in fact yields a metallic ground-state~\cite{Madsen_fesb2,luko_fesb2,bentien:205105,jmt_fesb2},
with small electron pockets halfway between the $\Gamma$ and $Z$ high symmetry
points, and corresponding hole pockets at all corners, $R$, of
the orthorhombic Brillouin zone,  see \fref{FeSb2GW}(a) for FeSb$_2$.
 
It is important to notice that these pockets are
of different orbital characters\cite{luko_fesb2,Madsen_fesb2} (cf.\ also the discussion for FeSi):
In the global coordinate system, the electron pocket is dominantly of $z^2$ and the hole of $x^2-y^2$ character\cite{luko_fesb2}%
\footnote{In a local coordinate system in which the local projection of the $d$-block of the Hamiltonian is as diagonal as possible (see Ref.~\cite{jmt_fesb2} for details),
the electron pocket is mainly of $d_{xy}$ character, and the hole pocket is formed by the now degenerate $d_{xz}$ and $d_{yz}$ orbitals.}.

{
\begin{table}[b]%
\begin{tabular}{|l|c|c|}
\hline
\rowcolor{blue}
$\Gamma$ [eV$^{-1}$]&\fesb &\feas\\
\hline
\cellcolor{lblue}$\omega<0$ &0.15 & 0.08 \\
\hline
\cellcolor{lblue}$\omega>0$ &0.02-0.05 & 0.02-0.03\\
\hline
\end{tabular}
\caption{{\bf {\it GW} scattering rates.} Particle-hole-asymmetry of the scattering amplitude $\Gamma$ within the {\it GW} approximation, as extracted by fitting the average $d$-orbital self-energy (in the Kohn-Sham basis) to $\Im\Sigma(\left|\omega\right|<5~\hbox{eV}) = -\Gamma\omega^2$. From Ref.~\cite{jmt_fesb2}.}
\label{gamma}
\end{table}
}

With FeSb$_2$ and RuSb$_2$ being insulators at low temperatures, band-structure methods fail qualitatively to give the correct ground-state spectrum
(contrary to the case of FeSi).
This shortcoming notwithstanding, conventional band-theory does reproduce some experimental findings for iron antimonide
even on a quantitative level: The unit-cell volume and bulk-modulus~\cite{PhysRevB.72.045103} are indeed well
captured\cite{0953-8984-21-18-185403}.

It is instructive to further compare FeSb$_2$ to its isostructural, isoelectronic relatives FeAs$_2$ and FeP$_2$. The
LDA band-structure of the former is shown in \fref{FeSb2GW}(b), for the latter compound, see Refs.~\cite{0953-8984-21-18-185403,BRAHMIA20131249}.
With respect to \fesb, the chemical pressure of the larger As atoms is almost isotropic, and the $c/a$ ratio remains virtually constant%
\footnote{We use
$a=5.3$\AA, $b=5.98$\AA, $c=2.88$\AA \cite{Fan1972136}; as a function of external pressure, the $c/a$ ratio slightly decreases\cite{PhysRevB.72.045103}.}.
Consequently, the bands of \feas\ are similar to those of \fesb, albeit a finite gap opens at the Fermi level.
We find $\Delta_{indir}=225$meV within LDA and $\Delta_{indir}=275$meV in GGA\cite{jmt_fesb2}, only slightly larger than the experimental 200-220meV\cite{Fan1972136,APEX.2.091102}. 
Below, we will discuss why---in moderately correlated materials---such good agreement for band-gaps is often fortuitous. 
First, however, we will review works in which methodologies beyond DFT were employed to yield an insulating ground-state for FeSb$_2$.

\begin{figure*}[!t]
  \begin{center}
		\subfloat[density of states]
		{{\includegraphics[angle=0,width=.45\textwidth]{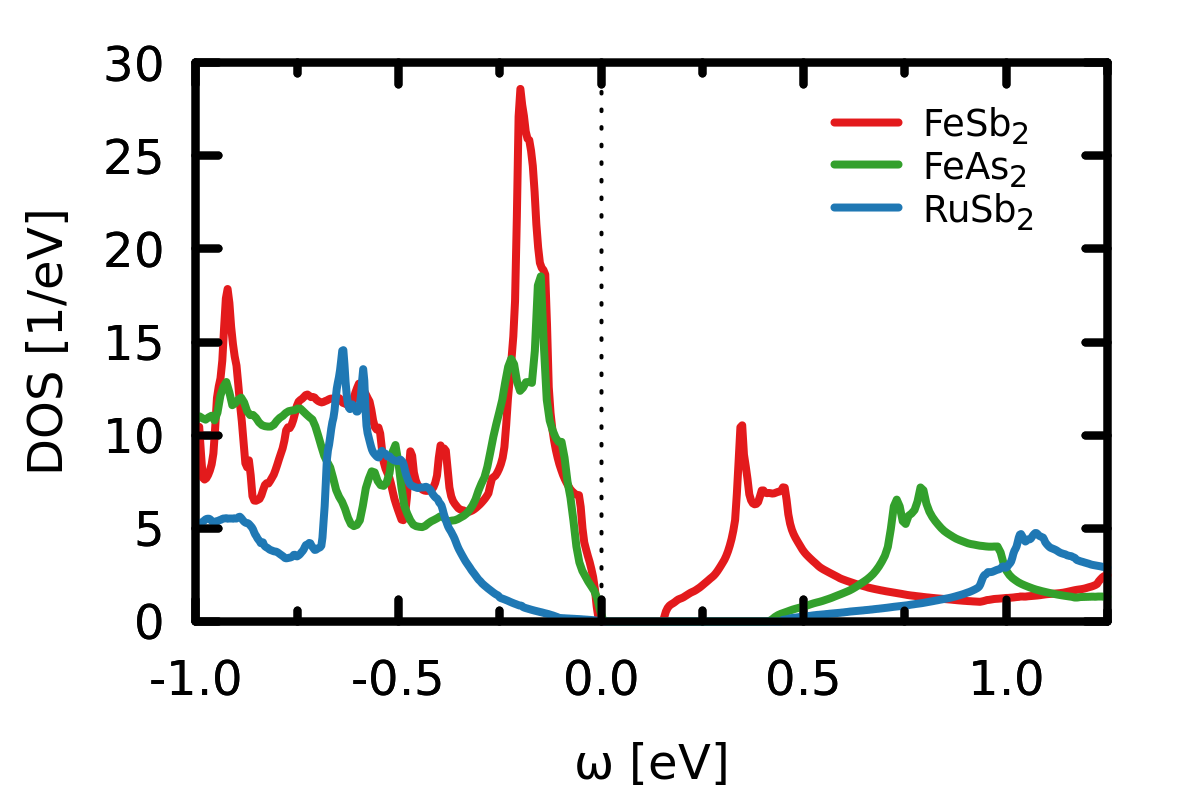}}}
		$\quad$
		\subfloat[band-structure]
		{{\includegraphics[angle=0,width=.45\textwidth]{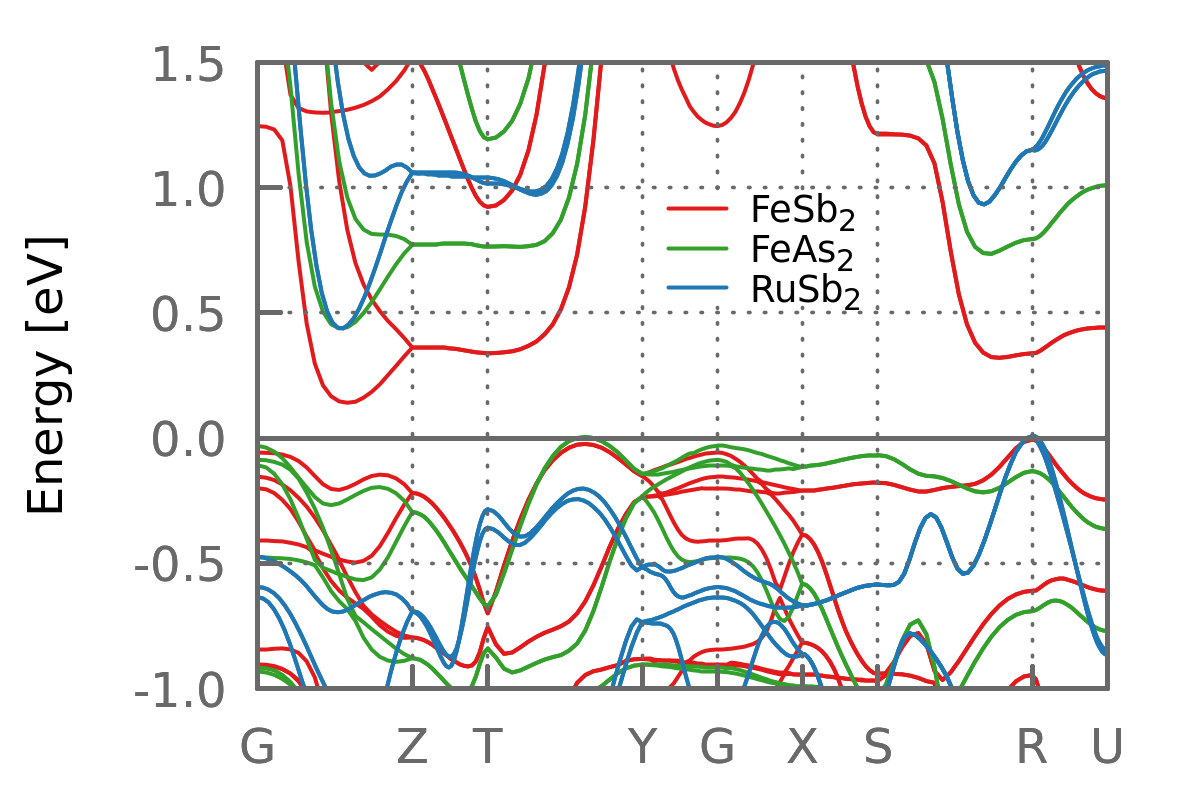}}
	}
	      \caption{{\bf  Comparison of FeSb$_2$, FeAs$_2$, and RuSb$_2$ within QS\textbfit{GW}.} 
Panel (a) shows the QS{\it GW} DOS for FeSb$_2$ (red), FeAs$_2$ (green), and RuSb$_2$ (blue).
Panel (b) displays the corresponding band-structures.
								}
      \label{marcasitesqsgw}
      \end{center}
\end{figure*}

\paragraph{DFT+U.}
Lukoyanov \etal\ applied the static mean-field LDA+U scheme to FeSb$_2$ and found strong similarities to FeSi. In particular, they
found the paramagnetic state to be stable up to a critical $U=2.6$eV above which ferromagnetism with an ordered moment of 1$\mu_B$ becomes
energetically favourable. In analogy to claims on FeSi\cite{PhysRevLett.76.1735}, these findings may suggest that FeSb$_2$ is a nearly ferromagnetic semiconductor whose properties are
strongly influenced by the proximity to a critical end point of a magnetic transition.
Weak ferromagnetism was indeed subsequently found experimentally under doping in Fe$_{1-x}$Co$_x$Sb$_2$\cite{hu:224422,PhysRevB.74.195130}.

\paragraph{Hybrid functional methods.}
That band-gaps are underestimated in DFT methods is owing to an insufficient treatment of exchange effects.
One way to improve upon this, is to include fractions of exact exchange via so-called hybrid functionals\cite{becke:1372}.
Using the B3PW91 functional\cite{doi:10.1063/1.464913} for the $d$-orbitals of the iron atoms, a paramagnetic solution with a much too large gap of
600meV (900meV) was found for FeSb$_2$ (FeAs$_2$)\cite{jmt_fesb2}. See also \fref{excessFe} that shows the electronic structure of FeSb$_2$ 
in the presence of an antisite defect, using the modified Becke-Johnson exchange potential\cite{mBJ}.

\paragraph{GW approximation.}
To put the latter findings as well as dynamical effects of electronic correlations into context, Hedin's (non-self-consistent) {\it GW} approximation~\cite{hedin} was applied to FeSb$_2$ and FeAs$_2$\cite{jmt_fesb2}.
The resulting one-particle dispersions---computed from the {\it GW} self-energy $\Sigma$ via
$\epsilon_{GW}\approx Z\biggl[\epsilon_{LDA}+\Re\Sigma(\epsilon_{LDA})\biggr]$ with $Z^{-1}=1-\partial_\omega\left.\Re\Sigma\right|_{\omega=\epsilon_{LDA}}$---are shown
in \fref{FeSb2GW} in comparison to LDA results:
In FeSb$_2$ a small gap opens, in agreement with experiment, while {\it GW} mildly increases the gap in FeAs$_2$, slightly deteriorating the congruence with experiment.

To analyse the {\it GW} band-structures, we note that, quite generally, there are two competing effects\cite{jmt_pnict} in electronic structures beyond DFT: (i) the inclusion of exchange contributions to the self-energy
 {\it widens} gaps and bandwidths\cite{hedin,doi:10.1063/1.464913}, and (ii) dynamical renormalizations, that (to linear order) give rise to the quasi-particle weight $Z$, {\it reduce}
gaps and band-widths\cite{PhysRevB.78.033109,sentef:155116}.

When applying {\it GW} to \fesb\ the competition of both effects is noticeable: While occupied bands are visibly narrowed and renormalized towards the Fermi level,
unoccupied states are moved up in energy, resulting in the opening of a gap.
Turning off dynamical renormalizations, i.e., setting the quasi-particle weight to unity $Z=1$, yields a larger gap of about 200meV.
From this observation, one can learn two things: (i) the {\it ab initio} screened exchange (SEX) included in the {\it GW} approach
yields a smaller gap-{\it enhancement} than the {\it ad hoc} admixtures of bare exchange in hybrid functional approaches;
(ii) effects of dynamical renormalizations substantially {\it shrink}
the band-gap\cite{PhysRevB.78.033109,sentef:155116} with respect to a (screened) Hartree-Fock-like reference. 
Within {\it GW}, values of $Z\approx0.5$ ($Z\approx0.6$) are found
for FeSb$_2$ (FeAs$2$)\cite{jmt_fesb2}.
For the case of FeAs$_2$ one could thus claim that, LDA finds a good value for the indirect gap merely because of an error cancellation, namely the joint neglecting
of both exchange and dynamical correlations.

Concomitant with the linear slope of the real-part giving rise to $Z$, also the imaginary part of the self-energy $\Sigma$ is Fermi-liquid-like, i.e., quadratic in frequency. 
Interestingly, however, the scattering rate is notably asymmetric with respect to the Fermi level, see \tref{gamma}.

{
\begin{table}[b]%
\begin{tabular}{|l|l|l|l|}
\hline
\rowcolor{blue}
 & FeSb$_2$ & FeAs$_2$ & RuSb$_2$\\
\hline
\cellcolor{lblue}$\Delta^{exp}_{indir}$ [meV]& 30 & 200-220 & 290-330\\
\hline
\cellcolor{lblue}$\Delta^{LDA}_{indir}$ [meV]& --   & 220  & --   \\
\cellcolor{lblue}$\Delta^{QSGW}_{indir}$  [meV]& 150 & 430 & 430 \\
\hline
\cellcolor{lblue}$U^{cRPA}$ [eV] & 3.9 & 3.4 & 2.6\\ 
\cellcolor{lblue}$J^{cRPA}$ [eV] & 0.63 & 0.63 & 0.44 \\ 
\hline
\end{tabular}
\caption{{\bf Gaps and interaction of the marcasites.} Tabulated are the indirect charge gaps from experiment (see \tref{table1} for references), LDA and QS{\it GW}.
Moreover the Hubbard $U$ and Hund's $J$ for the transition-metal $d$ orbitals are given.
These have been obtained by applying cRPA on top of the QS{\it GW} electronic structure in a maximally-localized Wannier setup\cite{miyake:085122} that includes transition metal $d$ and pnictogen $p$-orbitals. Screening has been eliminated in a window [-3,+3]eV around the Fermi level. Shown are $d$-orbital averages.}
\label{tmarcs}
\end{table}
}

\paragraph{QS{\it GW} and cRPA.}

For a more in-depth comparison of FeSb$_2$, FeAs$_2$, and RuSb$_2$ we performed new QS{\it GW}\cite{PhysRevLett.93.126406} and cRPA\cite{PhysRevB.70.195104} calculations.
The QS{\it GW} DOS and band-structures are shown in \fref{marcasitesqsgw}(a) and (b), respectively, while
\Tref{tmarcs} summarizes key results. There, we also include estimates for the values of the Hubbard $U$ and Hund's $J$ interactions (see caption for computational details).

In all marcasites studied here, the self-consistency increases the gap with respect to the above one-shot {\it GW} calculations.
In \fesb\ $\Delta_{indir}$ reaches 150meV---a value five-times as large as in experiment. 
Also in the case of FeAs$_2$, the gap is significantly overestimated, while QS{\it GW} yields a more reasonably sized gap for RuSb$_2$.

The tendency to overestimate band-gaps (see also the case of RuSi above) is rooted in two deficiencies: 
(i) insufficient screening of the Coulomb interaction within RPA\cite{schilfgaarde:226402,PhysRevLett.99.246403} (e.g., neglecting particle-hole correlations),
(ii) an underestimation of dynamical self-energy effects in the perturbative (first order) {\it GW} approximation.
Both problems can be addressed by combining {\it GW} with dynamical-mean field theory (DMFT)\cite{bible,vollkot} in so-called {\it GW}+DMFT\cite{PhysRevLett.90.086402} methods (see Refs.~\cite{0953-8984-26-17-173202,0953-8984-28-38-383001,Tomczak2017review} for recent reviews,
and Ref.~\cite{jmt_sces14,Choi2016} for the QS{\it GW}+DMFT variant).

These sources of error notwithstanding, we can still analyse the {\it trends} in our results and speculate about the reasons why these three materials behave very differently in experiment (see \fref{overviewrho} and \fref{overviewchi} for the resistivities and magnetic susceptibilities, respectively).
To do so, we further compute interaction matrix elements within cRPA, see \tref{tmarcs}. Our estimates for the Hubbard $U$ and Hund's $J$ are quite similar for FeSb$_2$ and FeAs$_2$, while significantly smaller for RuSb$_2$.%
\footnote{The bare (i.e., unscreened) interaction is slightly larger in the arsenide than the antimonide\cite{jmt_fesb2}, in line with trends of the Coulomb interaction under external pressure\cite{jmt_wannier,jmt_mno}.
Indeed the ionic radius of As is larger than Sb and causes chemical pressure.
The slightly smaller value of $U$ for FeAs$_2$, on the other hand, is mostly owing to using the same energy window for which screening is eliminated. Since the band-width of FeAs$_2$ is larger
than that of \fesb\ there are more excitations outside that window and contribute to the screening. In RuSb$_2$ the dominant change with respect to the iron compounds is the larger extend of the 4$d$-orbitals, resulting in smaller interaction matrix elements.}
From the point of view of the charge gap, however, FeAs$_2$ and RuSb$_2$ can be loosely grouped together, while FeSb$_2$ is the odd one out.

Therefore it can be speculated that the difference in physics between FeSb$_2$ and FeAs$_2$ is mainly controlled by the different sizes of the gaps.
This is congruent with the observation that observables, such as the resistivity and the magnetic susceptibility, are akin in shape,
albeit their characteristic temperature scales differ by the ratio of the respective gaps, $\Delta(\hbox{FeAs}_2)/\Delta(\hbox{FeSb}_2)$.
In this scenario one would expect that the activated behaviour of the magnetic susceptibility of FeAs$_2$\cite{APEX.2.091102} crosses over to a Curie-Weiss-like decay at
$T^{max}_\chi\approx 2200$K---which is, however above the melting point of the material.

The susceptibility of RuSb$_2$, on the other hand, does not exhibit any signs of activated behaviour\cite{APEX.2.091102} (see \fref{overviewchi}), despite having a gap comparable in size to that of FeAs$_2$.
Here the difference lies in the vastly disparate interactions. Given the propensity for ferromagnetic order in these systems, one can surmise
that the dominant control parameter of spin-fluctuations is the Hund's rule coupling $J$. Since the latter is smaller by one third in ruthenium antimonide,
the prefactor of the activated behaviour is expected to be significantly smaller.
Model calculations presented in \sref{PAM} substantiate these claims (see also \sref{micro} for the influence of the $J$ onto the mass enhancement of FeSi).

\subsubsection{The Kondo insulator \cbp.}
\label{Ce3dft}

The treatment of 4$f$-states in effective one-particle theories such as density functional theory is very poor:
strongly localized states contribute too much to the bonding as their hybridization with valence and conduction electrons is largely overestimated. Moreover, interaction-driven multiplet effects are completely neglected.
Nonetheless, even for \cbp, we can gain some interesting qualitative insights from band theory.

\begin{figure*}[!t]
  \begin{center}
	\subfloat[\cbp: an insulator]
	{{\includegraphics[angle=0,width=.45\textwidth]{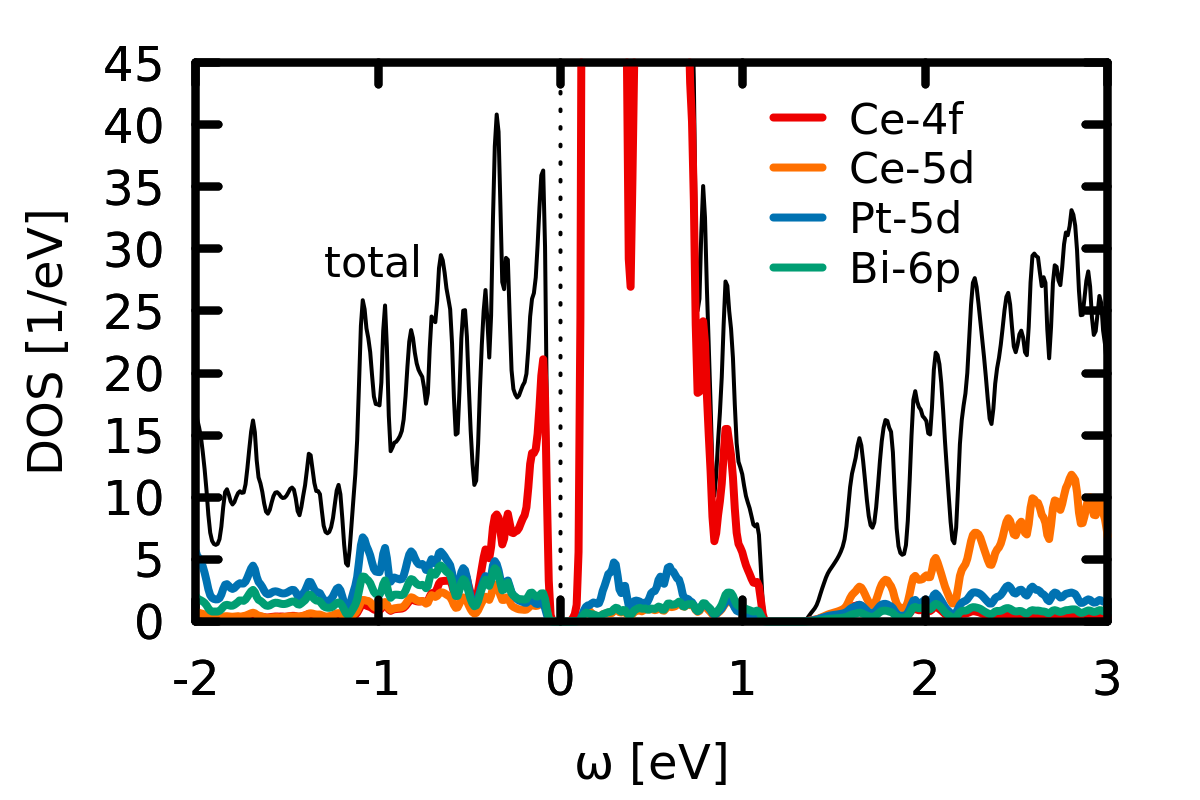}}}
	\subfloat[4$f^1$ removed from valence: a metal]
	{{\includegraphics[angle=0,width=.45\textwidth]{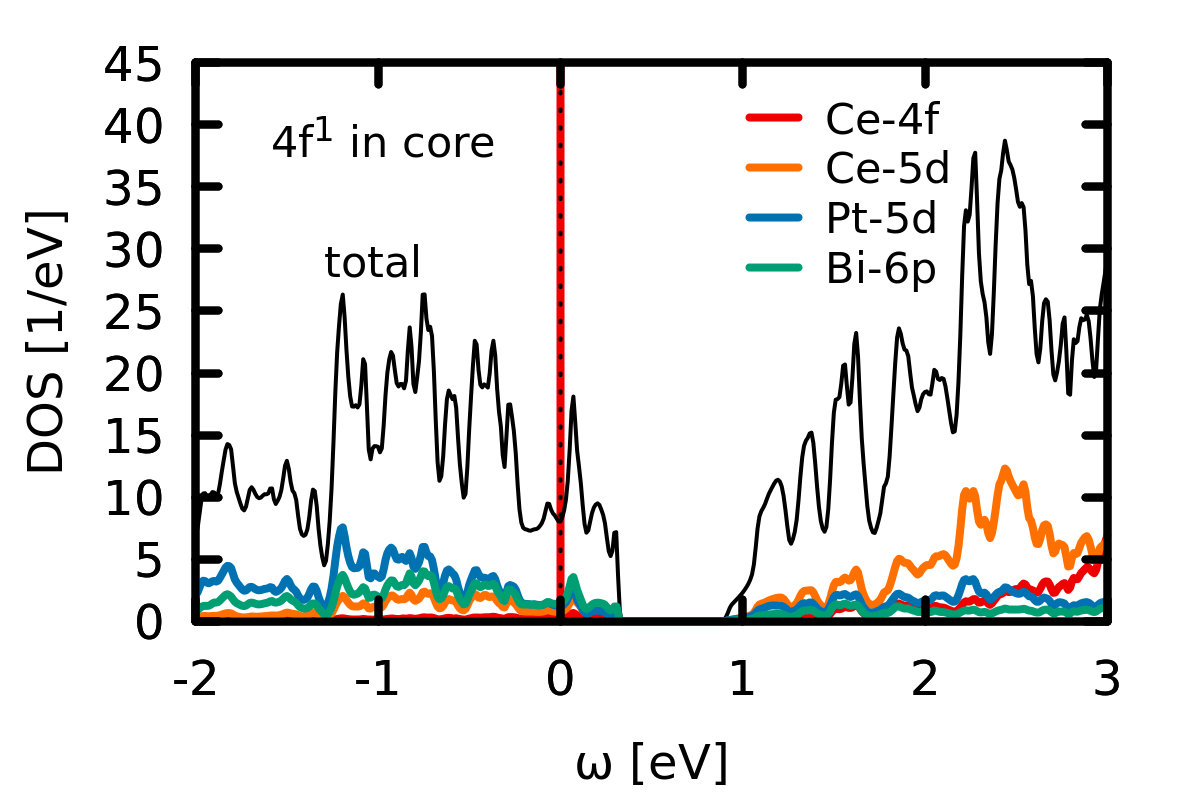}}
	}
	      \caption{{\bf Density of states (DOS) of \cbp.} 
				Both panels show the DOS of \cbp\ within DFT using the GGA-PBE functional and including spin-orbit coupling
				as implemented in wien2k. (a) shows the results of the full calculation which is insulating with a gap 
				$\Delta=140$meV. 
				In (b) the occupied 4$f^1$ electron was removed from the valence and transferred into the core, i.e., it is treated as purely atomic and does no
				longer participate in the bonding.  The unhybridized $f$-level appears as a delta peak at the Fermi level.
				The resulting DOS is metallic, as the number of valence electrons is odd.
				Since no refined atomic positions have been reported, we follow Ref.~\cite{doi:10.1143/JPSJ.62.2103} and use for Bi the ideal
				position $(u,u,u)$ with $u=1/12$. Orbital characters have been obtained from a angular decomposition within atomic spheres of radius 2.5\AA. Other weight
				is mainly accounted for by interstitial regions. 
								}
      \label{Ce3dos}
      \end{center}
\end{figure*}

\cbp\ is a cubic intermetallic that	
crystallizes in the body-centred Y$_3$Sb$_4$Au$_3$
structure\cite{Dwight:a14686} with non-symmorphic space-group I-43d, 
two formula units per unit-cell of lattice constant
$a=10.051$\AA \cite{PhysRevB.42.6842,PhysRevB.46.8067}. 
Ce atoms sit in the Wyckoff position 12a ($3/8$, 0, $1/4$), Pt occupies site 12b
($7/8$, 0, $1/4$), and Bi is located at ($u$,$u$,$u$). Since no value for $u$ has been reported in the literature, we follow Ref.~\cite{doi:10.1143/JPSJ.62.2103} and use the ideal position, $u=1/12$.

Density functional calculations for this compound were pioneered by Takegahara \etal\ \cite{doi:10.1143/JPSJ.62.2103}
within the local density approximation. This work established that \cbp\ has an insulating ground-state within band theory.
In the data shown here, we employed the generalized gradient approximation (GGA) with the PBE functional, and include 
(unless otherwise specified) spin-orbit coupling for all atoms within the wien2k\cite{wien2k} implementation.

\begin{figure}[!th]
{\includegraphics[angle=0,width=.45\textwidth]{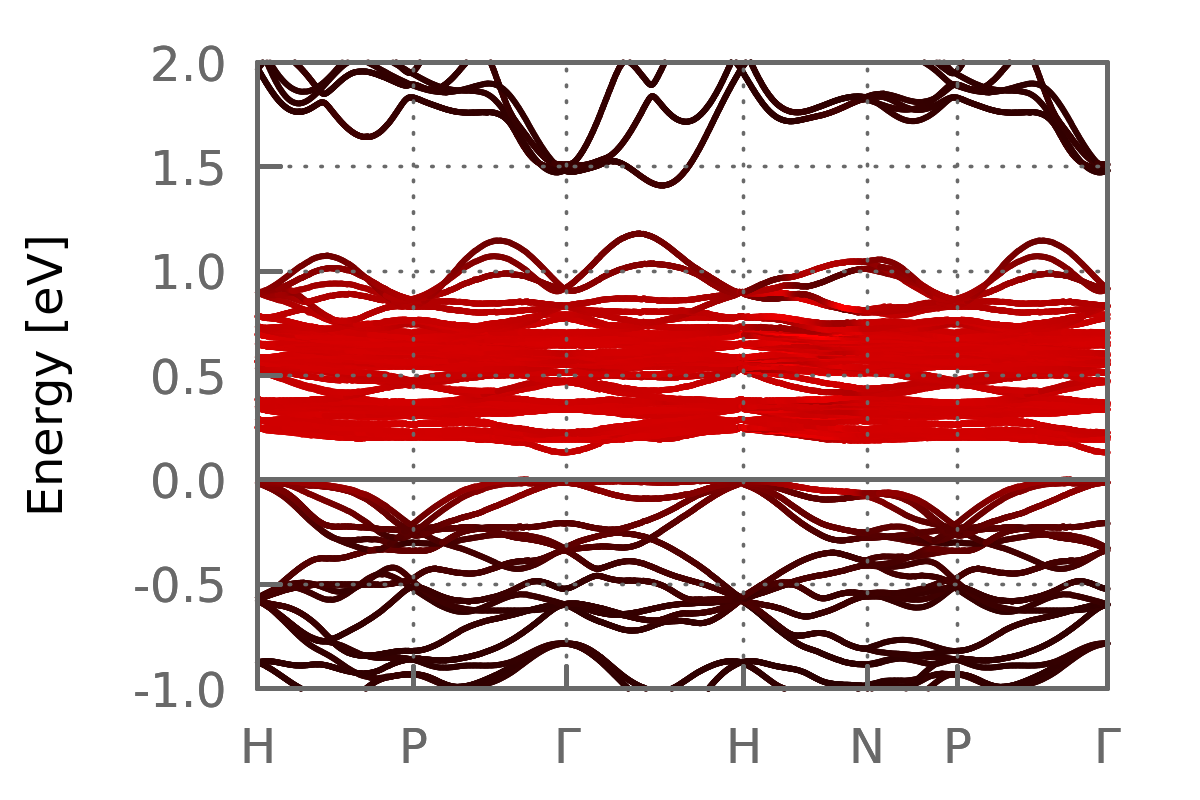}}
\caption{{\bf  Band structure of \cbp.} 
Shown is the GGA band-structure corresponding to the DOS of \fref{Ce3dos}(a). The charge gap is weakly indirect: $\Delta_{dir}=141$meV, $\Delta_{indir}=131$meV.
The shade of red indicates the degree
of Ce-4$f$ orbital character; the bands below the Fermi level contain about one Ce-4$f$-electron.}
\label{Ce3bnds}%
\end{figure}

\begin{figure*}[!t]
  \begin{center}
	\subfloat[density of states]
	{{\includegraphics[angle=0,width=.45\textwidth]{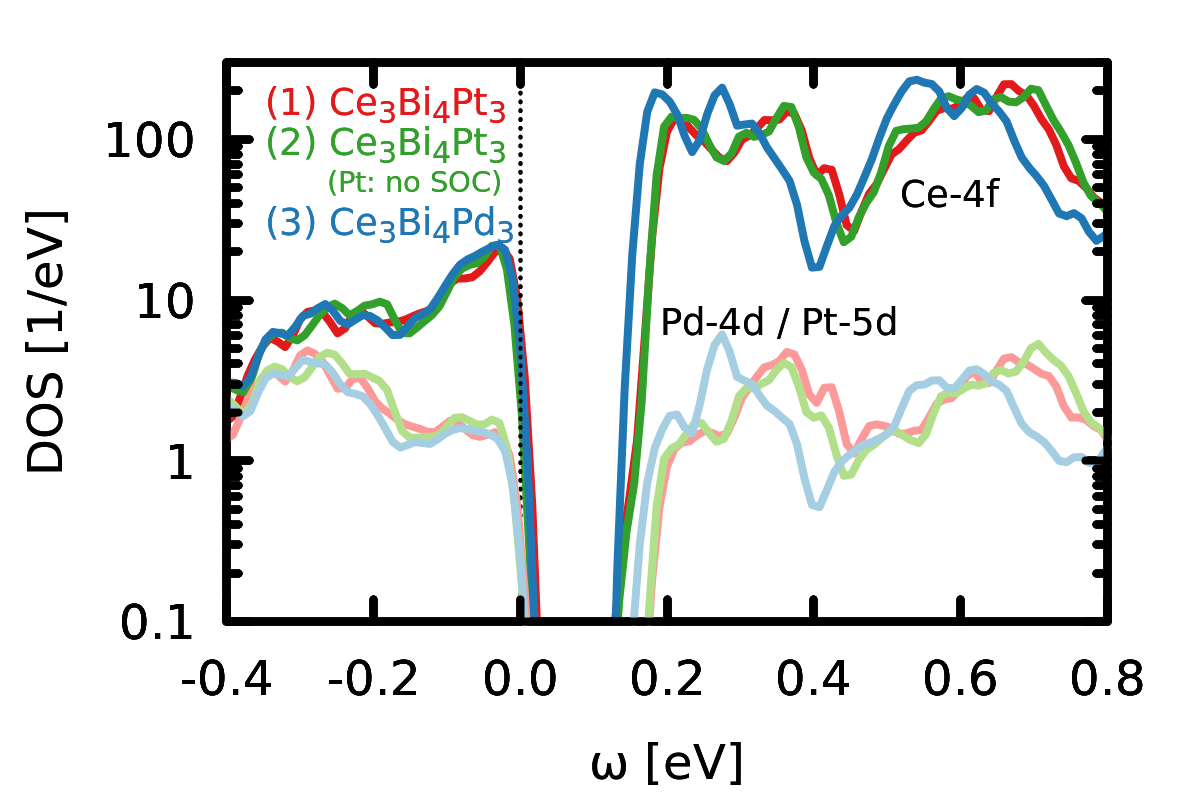}}}
	\subfloat[band-structure]
	{{\includegraphics[angle=0,width=.45\textwidth]{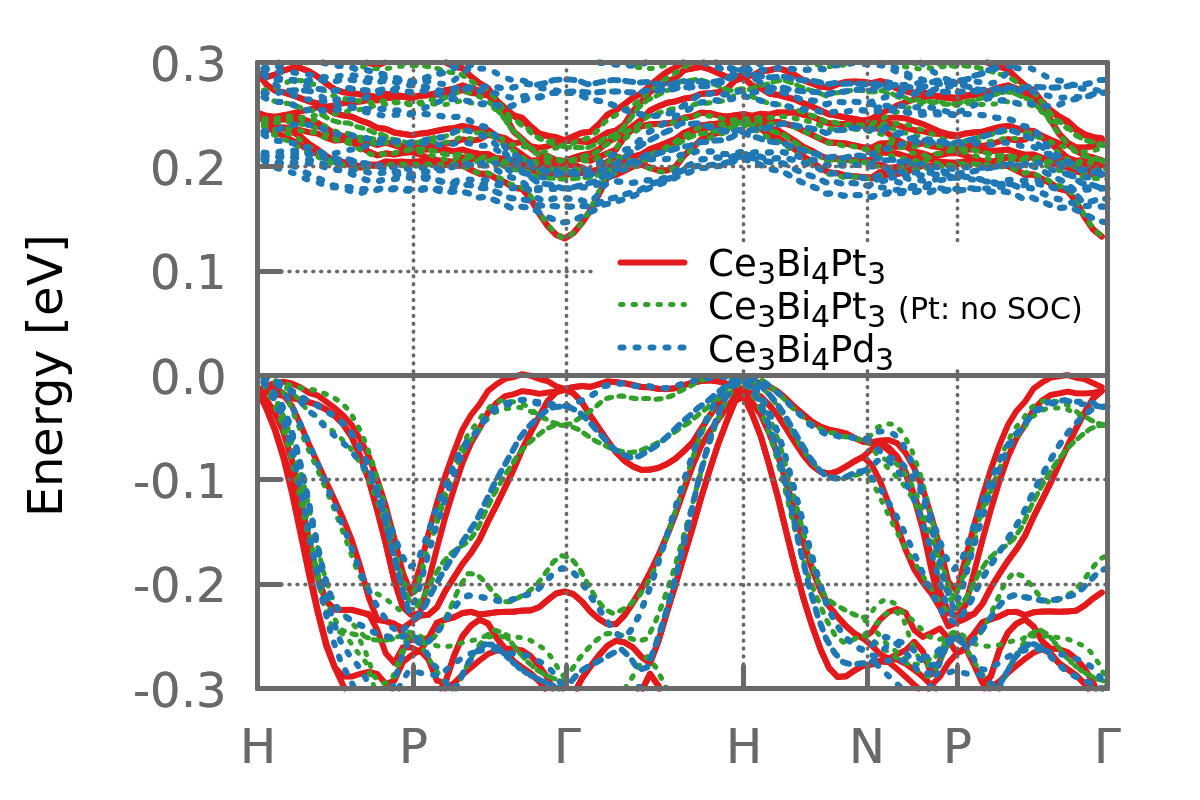}}
	}
	      \caption{{\bf  Comparison of \cbp\ and Ce$_3$Bi$_4$Pd$_3$: importance of radial character and spin-orbit coupling.} 
Panel (a) shows the DOS for (1) \cbp\ (red), (2) fictitious \cbp\ with the spin-orbit coupling turned off for Pt (green), and (3) Ce$_3$Bi$_4$Pd$_3$ (blue).
Panel (b) displays the band-structure of \cbp\ (red) and Ce$_3$Bi$_4$Pd$_3$ (blue).
These calculations illustrate that---on the DFT-level---the change in the spin-orbit coupling along the substitution series mostly affects the valence states,
while important modifications on the conduction side of the hybridization gap results from the 
change in the radial character (Pt-$5d$ to Pb-$4d$). We use the same lattice constant for \cbp\ and Ce$_3$Bi$_4$Pd$_3$ as motivated in Ref.~\cite{PhysRevLett.118.246601}.
								}
      \label{Ce3PtPd}
      \end{center}
\end{figure*}

As seen from the density of states in \fref{Ce3dos} and the band-structure in \fref{Ce3bnds}, \cbp\ is insulating within DFT,
with a weakly indirect gap: We find $\Delta_{dir}=141$meV, $\Delta_{indir}=131$meV.%
\footnote{the gap within LDA is slightly smaller, we find $\Delta_{indir}=110$meV, while legacy calculations\cite{doi:10.1143/JPSJ.62.2103} gave only 30meV.}
On the conduction state side the gap is delimited by a single, rather dispersive band that dips down through the bulk of unoccupied $f$-states around the $\Gamma$-point
(see also \fref{Ce3PtPd}(b)).
Among the valence states, there is roughly one charge carrier of $4f$-character per Ce-atom. As will be illustrated below, in a purely atomic scenario the 4$f^1$ configuration would lead to a  half-filled level, and thus
a metallic band-structure. Hybridizations with other orbitals---most notably the Pt-$5d$ and Bi-$6p$ (see \fref{Ce3dos}(a))---lead to the formation of dispersive
bonding/anti-bonding states, separated by a hybridization gap. 
The total number of valence electrons in \cbp\ is even and thus allows for
an insulating solution. Indeed, the valence $f$-electrons of Ce then reside in the bonding bands filling them completely, causing the chemical potential
to fall inside the gap.
As an illustration, we performed a calculation in which the 4$f^1$ configuration of each Ce-atom was moved from the valence into the core sector, 
i.e., it was treated as completely localized. As a consequence, the total number of electrons treated as valence states becomes
odd, and the system is necessarily metallic, as shown in \fref{Ce3dos}(b).
The situation is thus qualitatively akin to the generic case of the periodic Anderson model (PAM, see \sref{PAM}) with the $c$-$f$ hybridization
being finite (\cbp\ \fref{Ce3dos}(a)) and zero ($f^1$ in core, \fref{Ce3dos}(b)), respectively.
Our and previous\cite{doi:10.1143/JPSJ.62.2103} DFT results can be interpreted as providing the bare, i.e., non-interacting, dispersions of the materials.
Including many-body renormalizations, e.g., via dynamical mean-field theory in DFT+DMFT\cite{RevModPhys.78.865}, strongly decreases the hybridization and thus the gap by virtue of dynamical self-energy effects (see \fref{DMFTCe3} in \sref{Ce3DMFT} for the DFT+DMFT spectral function of \cbp).

Remaining on the level of DFT, we can comment on the recent study of Dzsaber \etal\ \cite{PhysRevLett.118.246601} who advocate the importance
of the spin-orbit coupling (SOC) for the stabilization of Kondo insulating behaviour in \cbp\ (see also \sref{SO}).
Tracking the physical properties along the isoelectronic, isostructural, and isovolume substitution series, Ce$_3$Bi$_4$(Pt$_{1-x}$Pd$_x$)$_3$,
a Kondo insulator to Kondo semi-metal transition was evidenced\cite{PhysRevLett.118.246601}.
It was argued that the consequence of replacing Pt with Pd is a change in the strength of the noble-metal spin-orbit coupling.
Hence the latter was identified as a control parameter to destabilize Kondo insulating behaviour.

To investigate this from the band-structure perspective, we show in \fref{Ce3PtPd}(a) the DOS of (1) \cbp, (2) fictitious \cbp\ where the spin-orbit
coupling of the Pt atoms is turned off, and (3) the fully substituted compound ($x=1$): Ce$_3$Bi$_4$Pd$_3$. 
These calculations allow to disentangle the two major changes when substituting Pd for Pt:
(i) the change in mass of the noble metal ion (which controls the strength of the spin-orbit coupling), and (ii) the change
in the radial character in the isoelectronic substitution (4$d$ vs.\ 5$d$ electrons of Pd and Pt, respectively).

We find that (i) spin-orbit coupling of platinum affects the valence bands: 
When turning off the SOC for Pt, the gap becomes strongly indirect, as the valence band maximum
moves from near the $\Gamma$-point to the $H$-point.
As far as the valence states are concerned the electronic structure of this fictitious \cbp\ indeed strongly resembles that of Ce$_3$Bi$_4$Pd$_3$.
These considerations identify the spin-orbit coupling as the major control parameter for valence states along the Ce$_3$Bi$_4$(Pt$_{1-x}$Pd$_x$)$_3$ series.
Moreover, we find  the (bare) hybridization gap---that is the one-particle ingredient to the Kondo coupling---to be insensitive to the spin-orbit coupling of Pt.

For conduction electrons the spin-orbit coupling of Pt plays, however,  no appreciable role: The dispersions in \fref{Ce3PtPd}(b) are virtually 
the same with or without SOC on Pt.
The conduction bands are however significantly modified by (ii) the change in the radial character of the noble-metal $d$-orbitals ($5d$ for Pt, $4d$ for Pd):
In fact, the charge gap of Ce$_3$Bi$_4$Pd$_3$, $\Delta_{indir}=147$meV, is {\it larger} than for \cbp\ (with or without SOC, cf.\ \fref{Ce3PtPd}(b)). 
Yet, the hybridization gap as seen in the DOS, \fref{Ce3PtPd}(a), is dominated
by the flat 4$f$-derived bands at around 200meV. These states move down for growing $x$ along the substitution series, engulfing the dispersive band that forms the
conduction band minimum in \cbp. As a consequence, the edge feature in the DOS---that is dominated by the Ce-4$f$ and Pt-$5d$/Pd-$4d$---is displaced towards the Fermi level; the gap as apparent in the DOS shrinks, while the actual gap value increases.
The relative stabilization of the noble-metal $d$ states is mirrored also in their filling: Within atomic spheres
of the same radius (2.5\AA), Pt in \cbp\ harbours 7.6 $d$-electrons, while in Ce$_3$Bi$_4$Pd$_3$ there are in fact 8.2.
Thus, for conduction states the relevant change along the substitution series is the radial character of the noble metal orbitals.

In all, these band-structure considerations suggest that, 
the bare (i.e., unrenormalized) electronic structure in the substitution series Ce$_3$Bi$_4$(Pt$_{1-x}$Pd$_x$)$_3$
is not only controlled by the spin-orbit coupling but also by the radial character of ligand orbitals.
As far as the $d$-$f$ hybridization is concerned, the effect of the latter is in fact dominating and, as a consequence, the Kondo coupling is not constant along the series.
Future many-body studies for this system therefore need to go beyond model calculations and include the interplay of both these realistic ingredients.
For realistic many-body calculations for stoichiometric  \cbp, see \sref{Ce3DMFT}.

\subsection{Many-body models} 
\label{model}

The above band-structures account for the structural and multi-orbital complexity of the materials under consideration.
From them, one can construct\cite{PhysRevB.70.195104,ferdi_down,miyake:155134}---or at least take inspiration for---reductionist many-body models that are to be solved
with more sophisticated techniques, allowing for a more proper treatment of electronic correlation effects and finite temperatures.

The models that have hitherto been used to describe correlation effects in FeSi can be grouped into two classes: Those based
on the periodic Anderson model (PAM), and those related to  Hubbard models.%
\footnote{We shall not discuss the Kondo lattice model that is often invoked for heavy-fermion intermetallics, since effects
of itinerancy are certainly of importance in FeSi.}
We certainly do not aim for completeness in this section and strictly focus on studies and parameter regimes 
that are directly linked to either FeSi or \cbp.
Beyond the literature reviewed here, a contrasting juxtaposition of the two quintessential models will be presented in \sref{PAM}.

\subsubsection{Periodic Anderson model (PAM).}

The PAM is believed to contain the essence of the physics relevant in heavy-fermion materials. 
It describes a periodic array of atomic levels ($f$-states) subjected to a local Coulomb interaction $U$ and a hybridization to
a dispersive band of non-interacting conduction ($c$) electrons:
\begin{widetext}
\begin{equation}
H=\sum_{\svek{k}\sigma}\epsilon_\svek{k}\cc_{\svek{k}\sigma}\ca_{\svek{k}\sigma}+
\epsilon_f\sum_{i,\sigma}n^f_{i\sigma}+\sum_{\svek{k}\sigma}V_{\svek{k}}(\cc_{\svek{k}\sigma}\fa_{\svek{k}\sigma}+\fc_{\svek{k}\sigma}\ca_{\svek{k}\sigma})
+U\sum_{i}n^f_{i\sigma}n^f_{i-\sigma}
\label{HPAM}
\end{equation}
\end{widetext}
Here $\cc$, \ca (\fc, $\fa$) are the creation and annihilation operators of the conduction (atomic) states, and we have used a mixed representation
where $\svek{k}$ labels the momentum in the Brillouin zone, while $i$ indexes a lattice site; the spin is denoted by $\sigma$; $\epsilon_\svek{k}$ is the dispersion of the decoupled $c$-electrons, $\epsilon_f$ the location of the $f$-level, $V_\svek{k}$ the hybridization between the two electron species.
Finally, $n^f_{i\sigma}=\fc_{i\sigma}\fa_{i\sigma}$ is the local number operator of the $f$-electrons, and $U$ their on-site Coulomb repulsion.
In the non-interacting limit, $U=0$, the symmetric PAM ($\epsilon_f=0$, $\epsilon_{\svek{k}}$ particle-hole symmetric, e.g., hyper-cubic lattice with only nearest neighbour hopping, $\mu=U/2$) with local hybridization $V_\svek{k}=V$, is gapped for finite $V$. The direct gap is proportional to the hybridization, $\Delta_{dir}=2V$, while the indirect gap can be much smaller, one finds $\Delta_{indir}=\sqrt{D^2+4V^2}-D$, where $D$ is half the bandwidth of the (unhybridized) $c$-electron (see \fref{hubpam} for a specific case).

In the context of \cbp,
Riseborough \cite{PhysRevB.45.13984} studied the PAM within mean-field theory and found a renormalization
 of the hybridization $V$, that leads---in the symmetric, half-filled case---to a shrinking of the low-temperature gap.
Further he obtained spin susceptibilities in good qualitative agreement with experimental findings for \cbp.
Sanchez-Castro \etal\ \cite{PhysRevB.47.6879} used a slave boson technique in the strong coupling limit, and applied it to the PAM
with a parabolic conduction band. Owing to the particle-hole asymmetry of their model, the authors could in particular study
the temperature dependence of the thermopower and the Hall coefficient. It was found that these quantities had opposite signs 
at low temperatures, congruent with experimental findings for \cbp\
 (thermopower hole-like\cite{PhysRevB.50.18142}, Hall coefficient electron-like\cite{FISK1995798}%
\footnote{See however Ref.~\cite{Katoh199822} where a positive Hall signal was found.})
as commonly found also in other heavy-fermion materials, e.g.,\ CeCu$_2$Si$_2$\cite{PhysRevLett.110.216408}.

Dynamical mean-field theory (DMFT)\cite{bible,vollkot} was first applied to the symmetric PAM by Jarrell \etal\ \cite{PhysRevLett.70.1670,PhysRevB.51.7429}.
They evidenced that spectra, susceptibilities and thermodynamic quantities behave like heavy-fermion metals at high temperatures (large masses, Curie-Weiss law),
and like insulators at low temperatures (activated behaviours). 
The DMFT approach was in particular shown to capture the competition between the Kondo effect and the magnetic RKKY interaction\cite{PhysRevB.51.7429}.
Indeed, the DMFT phase-diagram of the half-filled symmetric PAM on the infinite Bethe lattice (see
Sun \etal\ \cite{PhysRevB.48.16127} and Rozenberg \cite{PhysRevB.52.7369}) exhibits antiferromagnetic long-range order
above a second order critical line  given by  $U_c\propto V_c^2$ for large $U$. These findings extend Doniach's ideas \cite{DONIACH1977231}
for the Kondo lattice to the PAM.

Subsequently, also Rozenberg \etal\ \cite{PhysRevB.54.8452} followed the Kondo insulator route  and interpreted experimental findings for FeSi and \cbp\
in terms of DMFT calculations for the PAM.
They showed that the PAM yields, for $U>0$,  the same hierarchy of energy scales, $\Delta_{indir}<\Delta_s<\Delta_{dir}$, that was evidenced for FeSi and \cbp\ (see \sref{KI}).
Rozenberg \etal\ in particular modelled the optical conductivity: Congruent with experiments for both compounds [see \sref{FeSiCe3spec}(b)],
the direct gap $\Delta_{dir}$ is virtually independent of temperature, yet, with rising temperature, it is filled with incoherent spectral weight
as the local moment of the $f$-states begin to fluctuate freely.%
\footnote{See also Ref.~\cite{Franco2009} for optical conductivities, and Ref.~\cite{PhysRevB.60.11361} for susceptibilities in the strong coupling limit.}
A similar observation was made by Mutou and Hirashima \cite{doi:10.1143/JPSJ.64.4799} for the ($f$-contribution to the) spin-excitation spectrum.

Having in mind FeSi, Figueira \etal\ \cite{Figueira2012} studied the PAM with a slave-boson approach and computed transport observables. 
The authors found good qualitative agreement with experimental measurements at low temperatures, when using an {\it ad hoc} 
scattering rate. This finding advocates the presence of  sizable many-body renormalizations for the dispersion near the Fermi level.

That dynamical mean-field theory provides an accurate picture of the symmetric PAM  was motivated by Tanaskovi{\'c} \etal\cite{PhysRevB.84.115105}: 
Employing cellular DMFT, they demonstrated the absence of non-local (yet short-range) correlations in regimes sufficiently above
the model's N{\'e}el temperature. This dominance of local correlation physics also serves to motivate the accurateness of, e.g., our realistic calculations
for the Kondo insulator \cbp\ in \sref{Ce3DMFT}.

Having in mind anisotropic Kondo insulators, Yamada and Ono\cite{PhysRevB.85.165114}
studied the PAM with non-local hybridizations. They evidenced, among others, that the uniform and the local
magnetic susceptibility virtually coincide down to $T^{max}_{\chi}$. Further, they studied how
interaction effects suppress fluctuations of charge, while enhancing spin fluctuations with respect to
the spin and charge susceptibilities that coincide in the non-interacting $U=0$ (limit).

A model that goes beyond the PAM in that it allows for the correlated band to have finite dispersions was 
introduced by Continentino, Japiassu and Troper \cite{PhysRevB.49.4432}. Indeed the model was designed to encompass
both, systems like \cbp\ (in which the correlated states are close to atomic-like) and compounds such as FeSi
(in which $d$-orbital-derived bands have a finite dispersion), through a scalable hopping amplitude.
The model was solved using an equation-of-motion technique in the strong coupling limit.
The authors found that, contrary to the symmetric PAM, a critical hybridization was needed to yield an insulating
spectrum. It was speculated that this is the reason why covalent insulators are not that common
among $d$-electron systems. 

While the PAM exhibits notable similarities to the one-band Hubbard model\cite{PhysRevLett.85.373,Held2000},
the band-structure results from the preceding section motivate that narrow-gap semiconductors, such as FeSi and FeSb$_2$,
require a description in terms of a multi-band Hubbard model. These are the subject of the next section.

\subsubsection{Hubbard models.}

The first multi-band Hubbard model
\begin{widetext}
\begin{equation}
H=-\sum_{\svek{k},L,\sigma}\epsilon^L_{\svek{k}}\cc_{\svek{k}L\sigma}\ca_{\svek{k}L\sigma}+\sum^{L\ne L\pr}_{\svek{k}LL\pr\sigma}V^{LL\pr}_{\svek{k}}(\cc_{\svek{k}L\sigma}\ca_{\svek{k}L\pr\sigma}+\cc_{\svek{k}L\pr\sigma}\ca_{\svek{k}L\sigma})
+\sum_{iLL\pr\sigma\sigma\pr}U^{LL\pr}_{\sigma\sigma\pr} n_{iL\sigma}n_{iL\pr\sigma\pr}
\label{HDon}
\end{equation}
\end{widetext}
 devised for studying FeSi is owing to Fu and Doniach \cite{PhysRevB.51.17439}. They considered
two bands ($\epsilon_{\svek{k}}^1=-\epsilon_{\svek{k}}^2$) on a hyper-cubic lattice that hybridize locally ($V_\svek{k}=V$). This setup yields an insulating density of states 
in the non-interacting limit with a gap that is direct, $\Delta_{dir}=2V$.
Turning on the Hubbard $U$ interaction (i) renormalizes downwards the bare gap and (ii) introduces a scattering rate.
The latter rises with temperature and causes one-particle excitations to broaden, to the extend that incoherent spectral weight
appears at the Fermi level. Using an approximate expression for the magnetic susceptibility, good agreement was found with experiment.
The model of Fu and Doniach \cite{PhysRevB.51.17439} contains the minimal ingredients to describe the essence of correlated narrow-gap semiconductors.

Using a density-of-states inspired from {\it ab initio} calculations,
Kune\v{s} and Anisimov \cite{PhysRevB.78.033109} solved a one-band Hubbard model with DMFT and stressed the importance of the covalent
nature of the hybridization gap in systems like FeSi and FeSb$_2$. Further, they computed
optical conductivities and obtained a resistivity that accounts for the non-monotonous temperature dependence shown in \fref{FeSiCe3trans}(a,left).
Finally, the authors also computed within DMFT the local and uniform magnetic susceptibilities of their model with and without doping.
It was found that susceptibilities were much enhanced with respect to the non-interacting system, with a Curie-Weiss-like tail emerging at high temperatures,
or, in the doped case, throughout the considered temperature range.
Interestingly, the effective fluctuating moment was shown to be notably momentum-dependent---indicating that a picture of spatially localized moments (which is at the origin
of Curie-Weiss behaviour) is not appropriate in this setting. Notably, the local susceptibility was found to be much larger than the uniform one
(we will come back to this point in \sref{pnas}).

Sentef \etal\ \cite{sentef:155116} studied within DMFT a half-filled two-band model similar to that of Doniach and Fu \cite{PhysRevB.51.17439} 
and investigated the interaction-temperature phase diagram as well as more formal aspects of correlation effects in covalent insulators.
The Hubbard $U$ was found to drive the system through a first order transition from a correlated covalent insulator to a Mott insulator.
The coexistence region terminates in a critical end point at finite temperatures, above which the hybridization insulator and the Mott insulator
are continuously connected via a crossover through a bad-metal phase. While in the non-interacting limit, spin and charge gap coincide,
their degeneracy was found to be lifted at finite $U$, with the spin gap decreasing faster than the charge gap upon approaching the insulator-to-insulator transition,
in congruence with the experimental gaps of FeSi and \cbp\ (see \sref{KI}).

The effect of correlation effects on the optical conductivity of FeSi was also studied by Urasaki and Saso \cite{JPSJ.68.3477}
using perturbation theory for a two-band Hubbard model, finding qualitative agreement with the experimental temperature-dependence.
They also found remarkable agreement for the thermopower when using a realistic density-of-states\cite{Saso20021475}.

Inspired from their band-structure results (see \sref{band}), Mazurenko \etal\ \cite{PhysRevB.81.125131} constructed an effective three-band model
to mimic FeSi, and performed one-band DMFT calculations for its density-of-states.
In agreement with previous works and experimental observations, they found an
effective mass enhancement of about two. In particular, however, Mazurenko \etal\ studied the doped system and rationalized the appearance
of ferromagnetism in Fe$_{1-x}$Co$_x$Si in terms of the Stoner criterion, arriving at the conclusion that magnetism is of itinerant origin.
Later, Yang \etal\ \cite{0295-5075-95-4-47007} solved the full three-band model of Mazurenko \etal, and made the important
observation that the Hund's rule coupling $J$ plays a significant role in stabilizing ferromagnetism.%
\footnote{For the importance of the Hund's rule coupling for metallic ferromagnetism, see, e.g., Refs.~\cite{doi:10.1143/JPSJ.65.1056,PhysRevB.57.6896,Vollhardt1999,PhysRevLett.99.216402}.}

\medskip

Thus, as for experimental observables, also the minimal models for FeSi (two-band Hubbard model) and \cbp\ (PAM) are exhibiting a quite similar behaviour.
This begs the questions whether the physics underlying both the models and the actual compounds is in fact akin, or whether the
microscopic origin can be distinguished. 
We will try to answer this question in the following sections using available and new realistic many-body calculations.

\subsection{Many-body electronic-structure theory} 
\label{real}

While many-body {\it models} have enlightened our understanding about many physical phenomena in correlated systems,
they cannot account for all the complexity encountered in real materials. This concerns in particular multi-band effects, such
as crystal-fields, multiplet structures, ligand hybridizations, spin-orbit coupling, etc.
Given the enormous sensitivity of even fundamental properties of correlated materials, such details do matter.
Hence, at times, it is advantageous to turn around the usual {\it modus operandi} in which reductionist models are
solved before using state-of-the-art methods for realistic many-body calculations. In fact, realistic calculations
could serve instead as a guide to which control parameters should be included in pertinent many-body models.
Indeed, most model setups for FeSi and others have used rather unspecific dispersions, at times even particle-hole symmetric ones
(that result in a vanishing thermopower). Also for Kondo insulators,
the importance of realistic electronic structures was noted already early on\cite{Doniach1994450}.
In particular, the influence of the Hund's rule coupling---a genuine multi-orbital effect---was mentioned only in very few works. 

In this section we will discuss recent {\it ab initio} DFT+DMFT calculations for FeSi that build on realistic band-structures and Coulomb interactions.
We will first show (\sref{pnas}) that for a considerable panoply of physical observables
which probe different (yet interlinked) degrees of freedom, excellent agreement is found with respect to experiment.
With this validation of the theoretical setup, the microscopic ingredients of the theory can be analysed (\sref{micro}), allowing
for a better fundamental understanding of the physics relevant to correlated narrow-gap semiconductors.
In \sref{Ce3DMFT}, we will complement these results for FeSi with new calculations for \cbp\ to highlight microscopic similarities and differences.
In \sref{PAM} the thus gained insight will be used to go back to model-setups and scan relevant parameter regimes so as to 
put FeSi and \cbp\ into the wider context of related compounds.

\subsubsection{Theoretical description of experimental observables of FeSi.}
\label{pnas}

In this section, we will review results from realistic many-body (DFT+DMFT) calculations for FeSi for spectral and optical observables,
as well as magnetic susceptibilities\cite{jmt_fesi}. These theoretical findings compare favourably  with experimental results from photoemission and optical spectroscopies, as well
as neutron spectroscopy and other magnetic probes. 
A discussion of transport observables (resistivity, thermopower, etc.) can be found in the thermoelectricity chapter in \sref{silicides}.
The methodology employed for the following results is the realistic extension of DMFT\cite{bible,vollkot}, dubbed DFT+DMFT\cite{RevModPhys.78.865},
as implemented by Haule \etal\ \cite{PhysRevB.81.195107}. 
We refer to the original works\cite{jmt_fesi,jmt_hvar} for technical details.

\paragraph{Spectral properties.}

\begin{figure*}[!t]
{\includegraphics[angle=0,width=.45\textwidth]{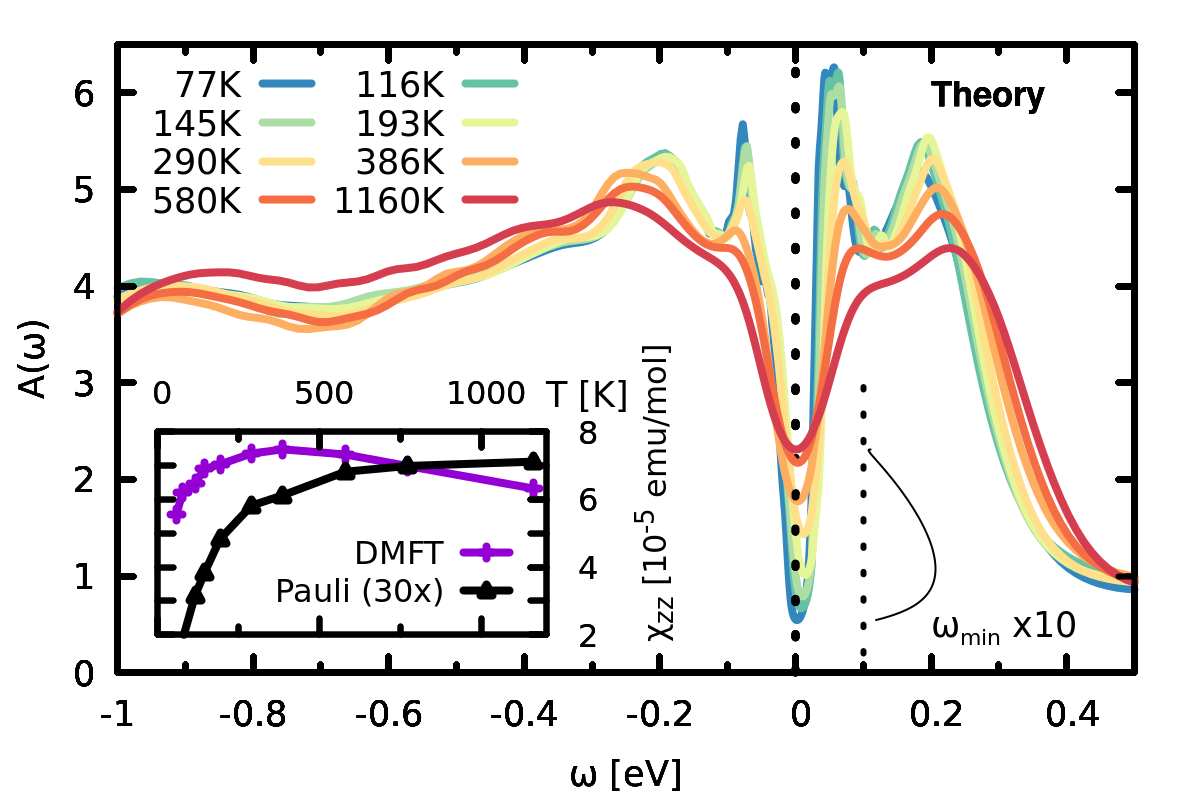}}
$\quad$
{\includegraphics[angle=0,width=.45\textwidth]{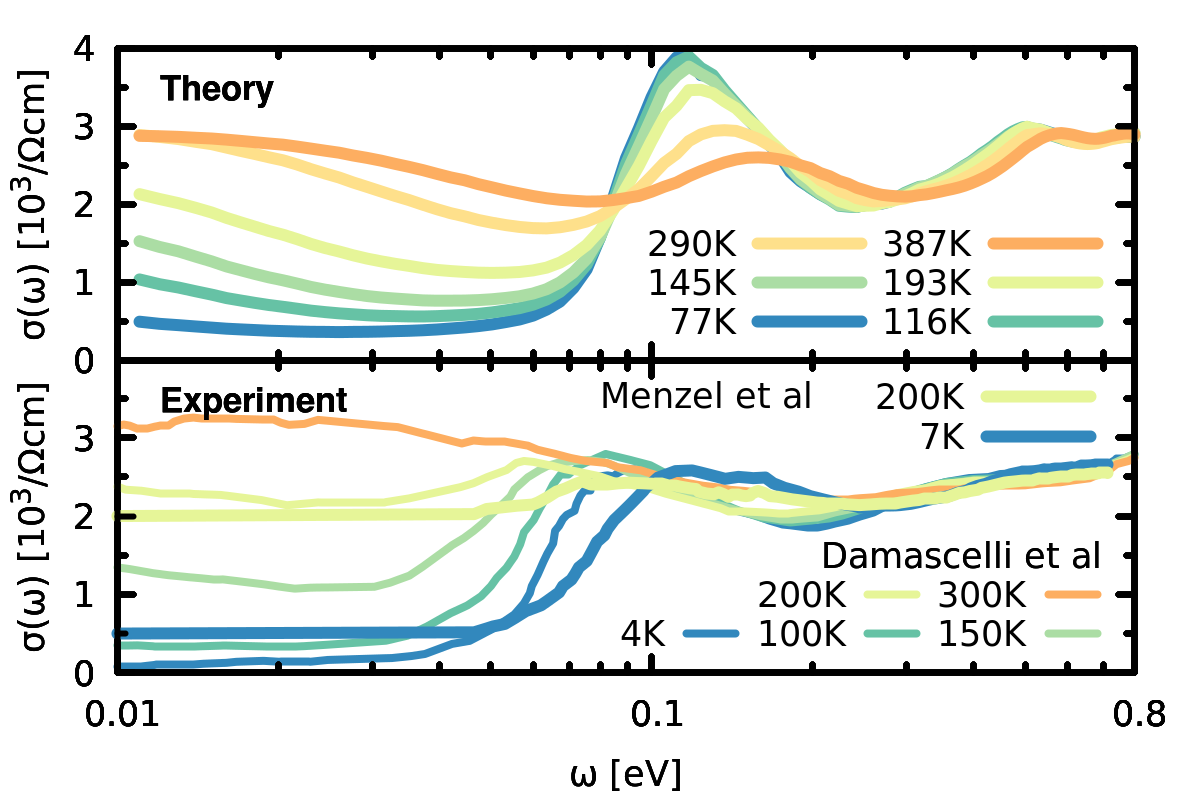}}%

{\includegraphics[angle=0,width=.45\textwidth]{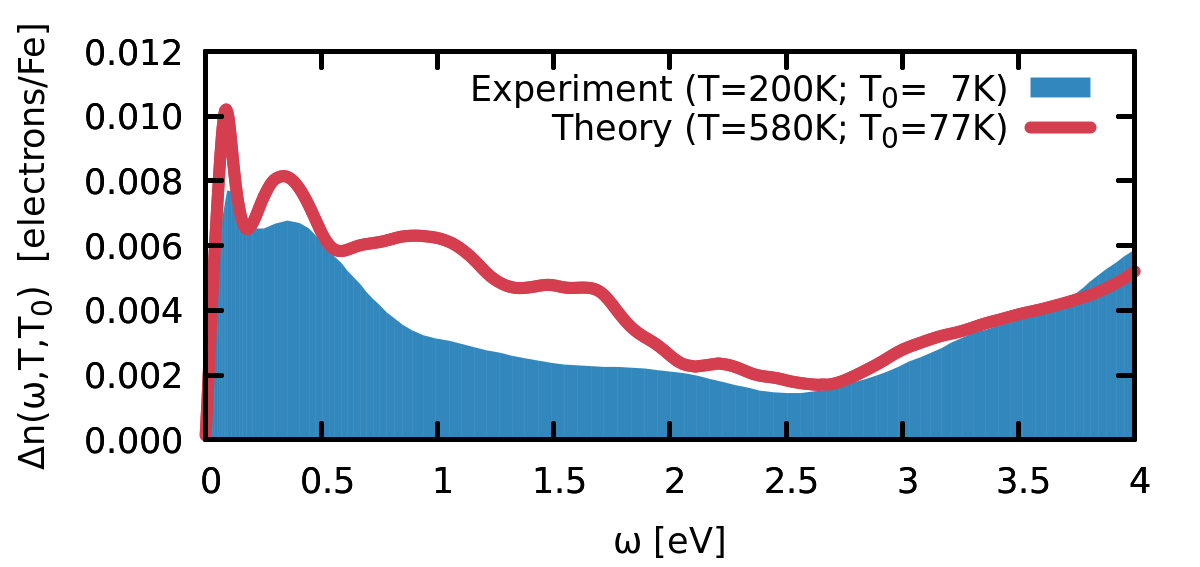}}
\caption{{\bf FeSi within DFT+DMFT: spectral, optical, and magnetic properties.} 
Local spectral function for different temperatures (top left). 
There, the minimum in the spectral function with respect to the Fermi level is depicted (10x magnified) by $\omega_{min}$.
The inset shows the local DMFT spin susceptibility and (30x magnified) the Pauli susceptibility according to \eref{PauliS}.
Optical spectra (top right) are shown in comparison to the experimental data of Refs.~\cite{PhysRevB.55.R4863,PhysRevB.79.165111}.
Optical spectral weight transfers (bottom left) according to \eref{effn} and compared to experimental data of Ref.~\cite{PhysRevB.79.165111}.
Adapted from Ref.~\cite{jmt_fesi}.}
\label{FeSiDMFT1}%
\end{figure*}

\Fref{FeSiDMFT1} (top left) displays the DFT+DMFT spectral function for various temperatures.
At low temperatures (blue) FeSi is gapped, with a charge gap of $\sim 50$meV that is narrowed by about a factor of two as compared to band-theory (cf.\ \fref{RuSi}),
in congruence with photoemission spectroscopy (cf.\ \fref{FeSiKlein} and \fref{FeSiCe3spec}(a)).
As temperature rises, peaks in the spectral function broaden and incoherent spectral weight spills into the gap and the system increasingly metallizes.
Similar spectra have subsequently been obtained by Yanagi and Ueda\cite{PhysRevB.93.045125} using a perturbative approach.

\paragraph{Optical properties.}
The crossover to a (badly) metallic state is also quantitatively reproduced in the optical conductivity, as shown in \fref{FeSiDMFT1} (right).
As noted already in \sref{KI}, the metallic response is not caused by the gap edges shifting to the Fermi level. 
Indeed, the first maximum in the absorption 
does not move substantially with temperature.%
\footnote{We note, however, that the trend in experiment and theory is opposite: The peak moves up with $T$ in theory, while it moves
down in experiment. We attribute this the expansion of the lattice with increasing temperature, which tends to reduce the hybridization gap.
The calculations were done at constant volume.}
Instead the intensity of that peak diminishes as spectral weight is transferred to lower energies.
This transfer of spectral weight is analysed in more detail in \fref{FeSiDMFT1} (bottom).
According to \eref{effn} $\Delta n(\omega,T,T_0)$ describes
the  redistribution of spectral weight up to the energy $\omega$ with respect to a base temperature $T_0<T$.
A positive value of $\Delta n(\omega,T,T_0)$ thus indicates that, for $T>T_0$, there is a net transfer of spectral weight to energies lower than $\omega$.
An intersection of $\Delta n$ with the $x$-axis would correspond to a full recovery of spectral weight as imposed by the $f$-sum rule, \eref{fsum}.
First, we note that the theoretical results closely trace the experimental temperature dependence.
Second, there are several isosbectic points\cite{PhysRevB.87.195140} in the optical conductivity---$\partial/\partial_T\sigma(\omega)=0$---which translate into extrema in $\Delta n$.
The first of these peaks is at 80meV, the scale of the semiconducting gap, above which spectral weight aggregates at low temperatures.
The first minimum in $\Delta n(\omega)$ occurs at the second isosbectic point of $\sigma(\omega)$ at $\sim 180$meV. There, the theoretical curve reaches a compensation of only $\sim 35\%$  of the excess carriers from the first feature. In fact, $\Delta n(\omega)$ does not vanish up to 4eV, indicating that a total compensation according to the sum-rule is not achieved below the scale of the screened Coulomb repulsion of $U=5$eV used here.

\begin{figure*}[!t]
  \begin{center}
	\subfloat{
	{\includegraphics[clip=true,trim=0 0 0 0,angle=0,width=.45\textwidth]{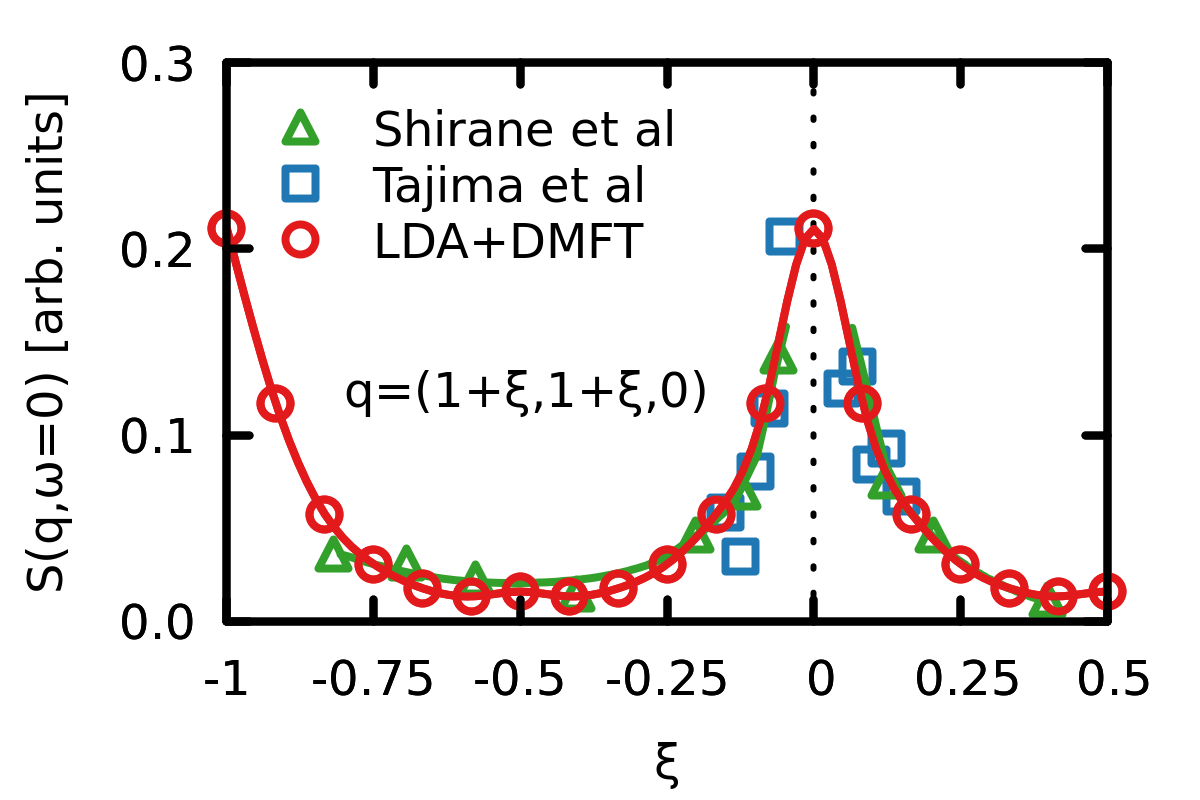}}
	$\quad$
	{\includegraphics[clip=true,trim=0 0 0 0,angle=0,width=.45\textwidth]{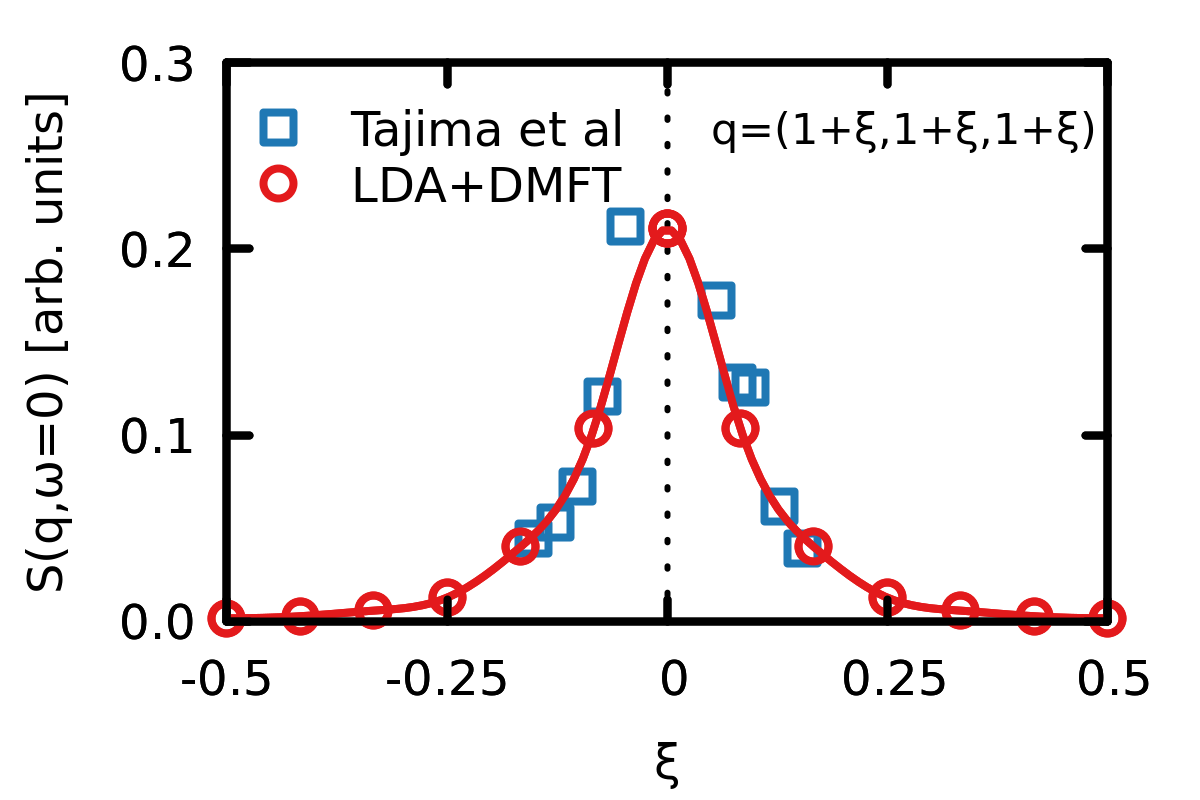}}
		}
	      \caption{{\bf FeSi: Neutron spectroscopy.}  Magnetic (quasi) elastic structure factor displaying scattering near the $\vek{q}$-points $(1+\xi,1+\xi,0)$ and $(1+\xi,1+\xi,1+\xi)$. Experimental data from Shirane \etal\ \cite{PhysRevLett.59.351} and Tajima \etal\ \cite{PhysRevB.38.6954} obtained at 300K with 41meV neutrons at $\Delta\omega=0$. Theory~: LDA+DMFT $\Delta\omega=2$meV at $290$K. Note that the intensities have been arbitrarily scaled.
	}
      \label{FeSiN}
      \end{center}
\end{figure*}

\paragraph{Magnetic properties.}
The inset of \fref{FeSiDMFT1} (top left) shows a comparison of the local spin-susceptibility as obtained from DFT+DMFT and the Pauli susceptibility
according to \eref{PauliS}.
The local susceptibility has the expected characteristic temperature dependence as seen in experiment (growing with $T$ at low-$T$, Curie-Weiss-like decay above $500$K, cf.\ \fref{FeSiCe3trans}(b)). The Pauli expression, as computed from the DFT+DMFT spectral function is very low in magnitude%
\footnote{note that the latter has been multiplied by a factor of 30.}
 and is, moreover, a monotonously increasing function of temperature,
and thus does not account for the observed physics.
We further note, that the local susceptibility has a magnitude smaller by a factor of $\sim 10$
as compared to the measured uniform magnetic susceptibility (cf.\ \fref{FeSiCe3trans}(b)).
To make things worse, previous (one-band) model calculations\cite{PhysRevB.78.033109} found that the local susceptibility
is actually {\it larger} than the uniform susceptibility, potentially
aggravating congruency with experiment. In the periodic Anderson model, the local and uniform susceptibility coincide above $T^{max}_\chi$\cite{PhysRevB.85.165114}.
This impasse is solved as follows:
(i) experimentally: As mentioned before, magnetic neutron spectroscopy experiments\cite{PhysRevB.38.6954,PhysRevLett.59.351} evidenced a ferromagnetic-like susceptibility,
i.e., a $\chi(\vek{q},\omega\rightarrow 0)$ that is peaked at $\vek{q}=0$ and equivalent points.%
\footnote{Interestingly, extracted fluctuating moments and linewidths of FeSi are reminiscent of results for the metallic ferromagnetic Heusler compound Pd$_2$MnSn well above its Curie
temperature\cite{PhysRevB.34.1762}.
}
This finding strongly suggests that the uniform susceptibility should instead be
much larger than the $\vek{q}$-averaged (=local) susceptibility.
(ii) theoretically: Using the formalism of Park \etal\ \cite{Park_prl11}, we compute the structure factor $S(\vek{q},\omega\rightarrow 0)$ from our DFT+DMFT data.%
\footnote{Details will be published elsewhere. Note that the same formalism
has recently also been successfully applied to another intermediate valence compound, namely CePd$_3$\cite{Goremychkin186}.}
The agreement with the experimental neutron results as shown in \fref{FeSiN} is excellent and clearly evidences an enhancement of the susceptibility at ferromagnetic wavevectors.
Krannich \etal\ \cite{Krannich2015} analysed the shown elastic structure factor of Tajima \etal\ \cite{PhysRevB.38.6954} with a phenomenological paramagnetic scattering function,
$S(\svek{q},\omega=0)=\frac{k_BT}{\hbar}\frac{\chi_0/\gamma}{\left(1+(\svek{q}\xi)^2 \right)^2}$, with the static susceptibility $\chi_0$, a damping factor $\gamma$,
and a correlation length $\xi$. Our theoretical simulation reproduces their finding of $\xi\approx3.6$\AA. This is a quite remarkable result, since the correlation length is therewith
larger than the distance 2.75\AA\ between neighbouring iron atoms.
This ferromagnetic enhancement 
in particular implies a uniform susceptibility that is significantly larger than the local one.
This qualitative change as compared to previous single-band calculations (see \sref{model}) advocates the importance of genuine multi-band effects in FeSi.
It is in particular the influence of the Hund's rule coupling that favours tendencies towards ferromagnetism.

Indeed, using realistic many-body setups solved with a perturbative technique, Yanagi and Ueda\cite{PhysRevB.93.045125}
discussed the different magnetic ground-states of the isostructural and isoelectronic Fesi and FeGe (cf.\ \sref{bandFeSi}).
In particular they evidenced a strong dependence of the magnetic ordering temperature and moment in FeGe on the strength of the Hund's rule coupling $J$.
The important influence of $J$ for spectral properties is discussed in the next paragraph.

\subsubsection{A microscopic understanding: spin fluctuations and Hund's physics}
\label{micro}

Having validated the theoretical setup through quantitative congruence with experiments, one can take a step back, and analyze the microscopic origin
of the evidenced behaviour.

\begin{figure}[!b]
{\includegraphics[angle=0,width=.45\textwidth]{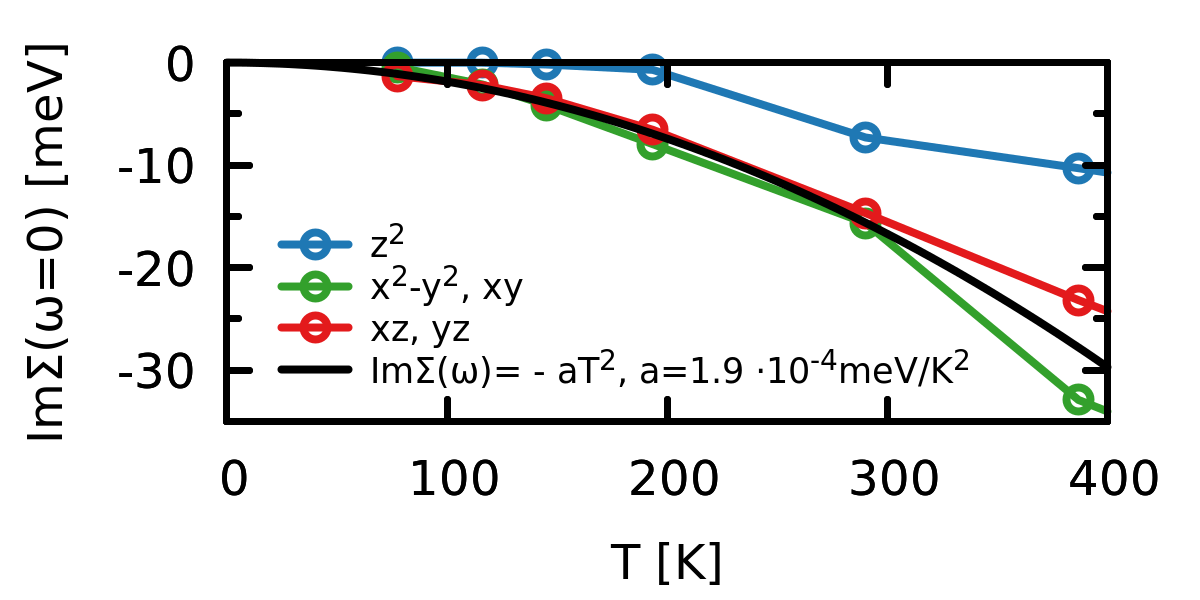}}
\caption{{\bf FeSi: Coherence of excitations.} Shown is the imaginary part of the DFT+DMFT self-energy at zero frequency, $\Im\Sigma_L(\omega=0)$,
as a function of temperature. For the orbital components $L=x^2$-$y^2$, $xy$, $xz$, $yz$, the inverse lifetime scales with $T^2$. Adapted from Ref.~\cite{jmt_fesi}.}
\label{FeSiImS}%
\end{figure}

\paragraph{The self-energy and the Hund's rule coupling.}
As evident from the spectral function and the optical conductivity, the metallization crossover in FeSi within DFT+DMFT is not caused by a temperature-induced
narrowing of the charge gap. Instead the latter is filled with incoherent weight as spectral features broaden substantially with increasing temperature.
Information on the coherence of one-particle excitations is encoded in the imaginary parts of the self-energy.
As displayed in \fref{FeSiImS}, $\Im\Sigma(\omega=0)$ follows---for orbital components that account for the majority of spectral weight near the Fermi level---a
Fermi liquid-like $T^2$ behaviour. While the overall magnitude of the scattering rate remains small in comparison e.g., to some metallic transition metal oxides\cite{poter_v2o3,byczuk-2007-3}, it reaches values of $\sim 30$meV that are comparable to half the size of the charge gap at around 400K. Indeed at this temperature only a pseudo gap remains in  the spectral function (\fref{FeSiDMFT1}(a)) and
spectral weight at the Fermi level is finite. The fact that such moderate correlation effects play a significant role in FeSi is thus rooted in the proximity of energy scales:
$-\Im\Sigma\sim\Delta\sim k_BT$.

\begin{figure*}[!t]
  \begin{center}
	\subfloat{
	{\includegraphics[clip=true,trim=0 0 300 0,angle=0,width=.32\textwidth]{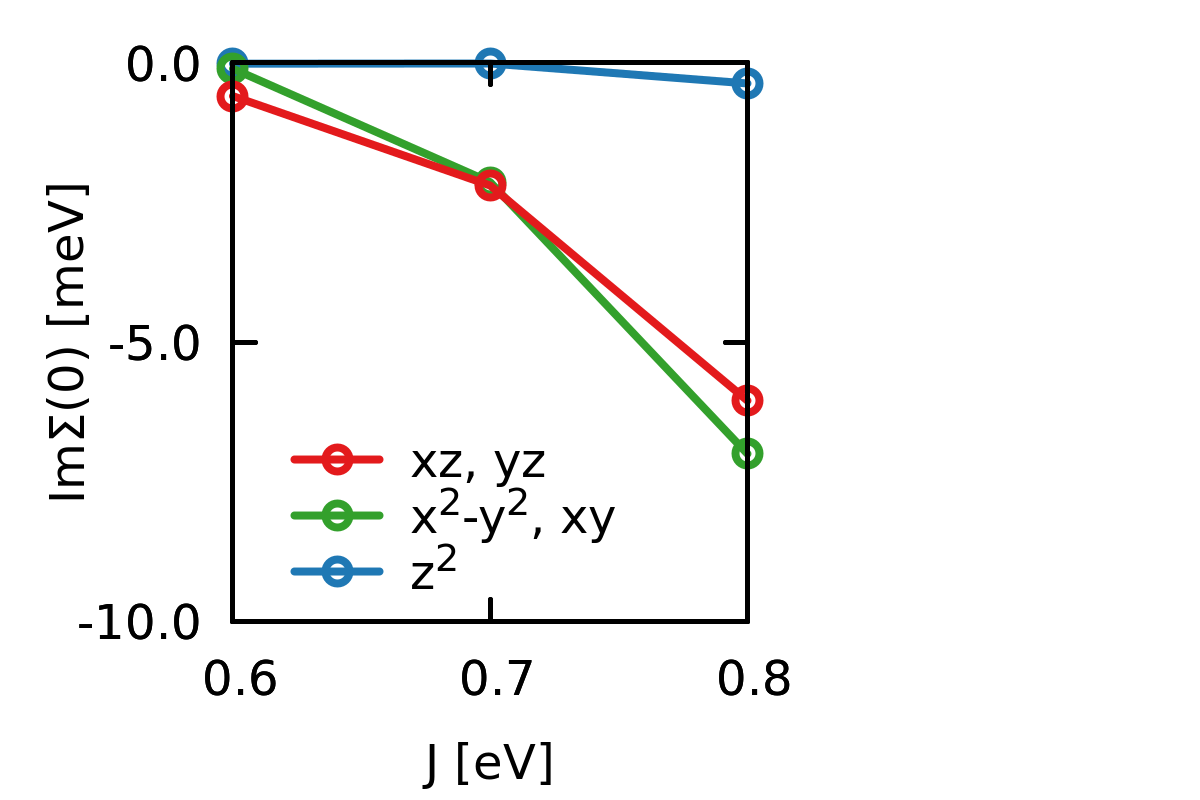}}
	{\includegraphics[clip=true,trim=0 0 300 0,angle=0,width=.32\textwidth]{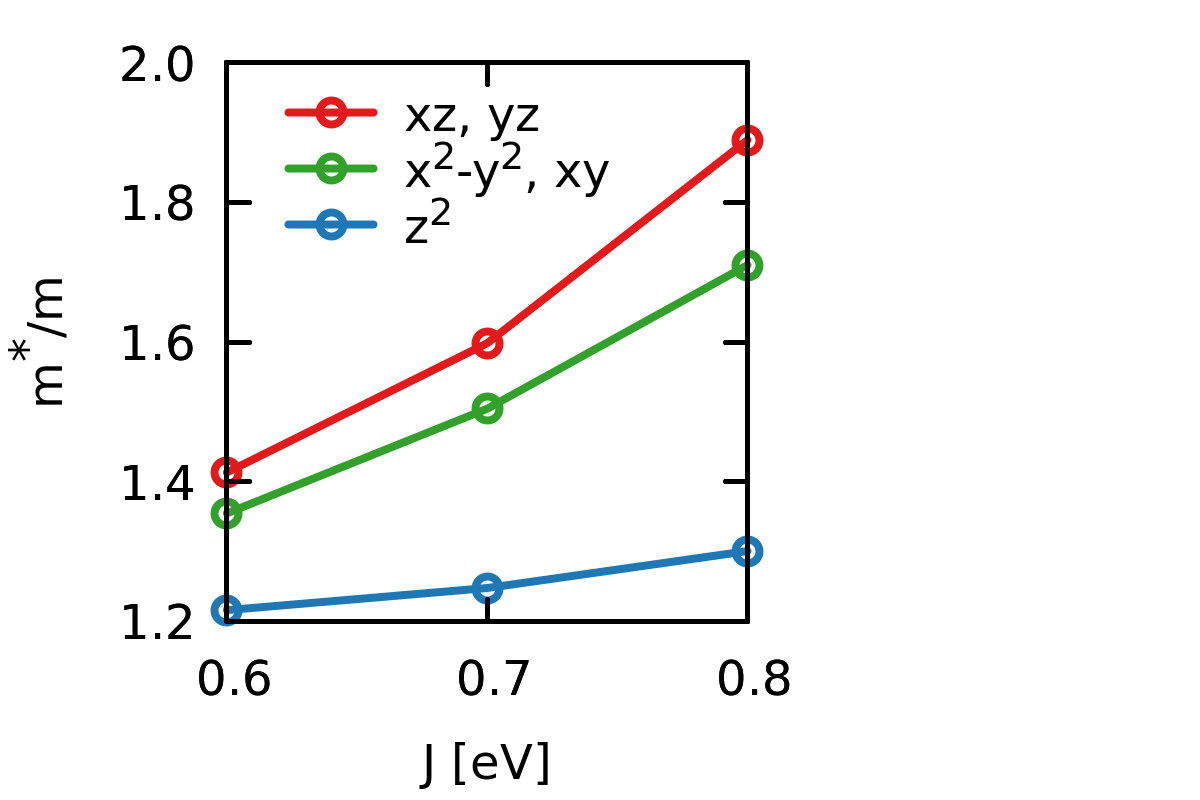}}
	{\includegraphics[clip=true,trim=0 0 300 0,angle=0,width=.32\textwidth]{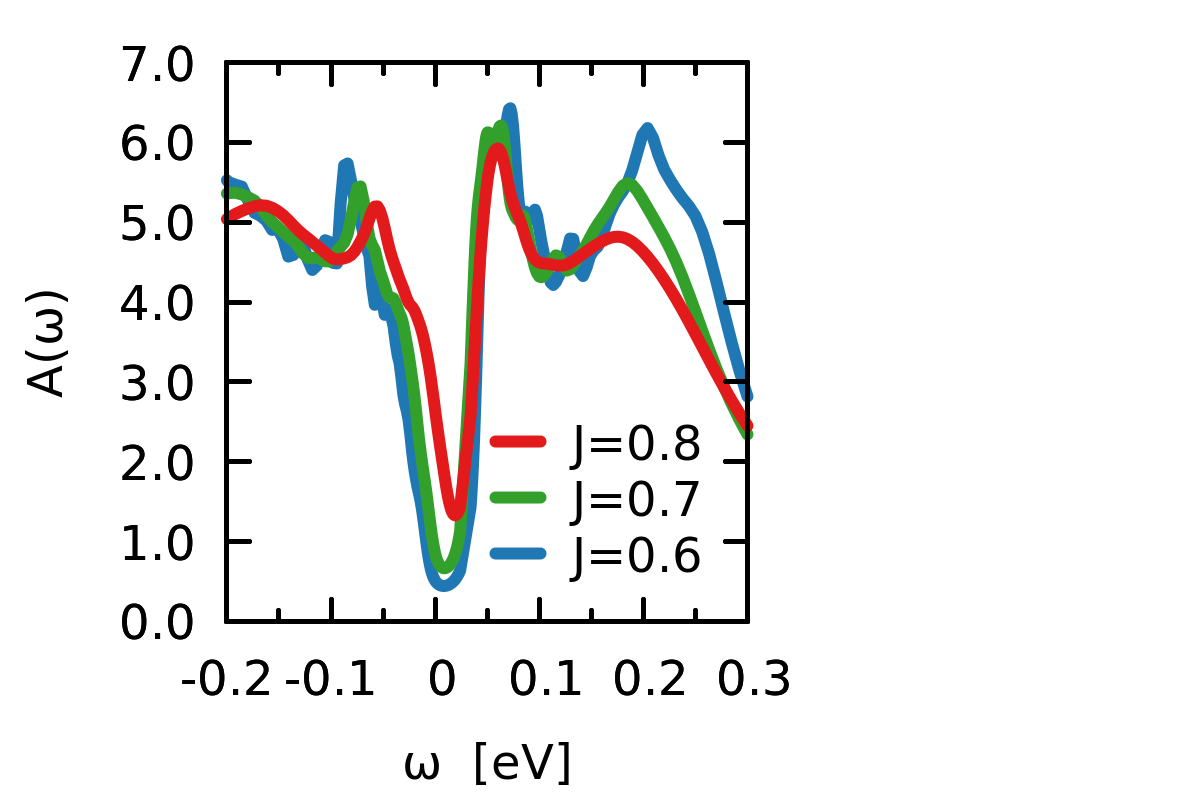}}
	}
	      \caption{{\bf FeSi: Influence of the Hund's rule coupling $J$.} (a) scattering rate $\Im\Sigma(\omega=0)$, (b) effective mass $m^*/m$, (c) spectral function
				for different Hund's $J$ at constant Hubbard $U=5.0$eV at $T=116$K, using only density-density interactions. 
								Adapted from Ref.~\cite{jmt_fesi}.}
      \label{FeSiJ}
      \end{center}
\end{figure*}

What controls the strength of quasi-particle incoherence in FeSi?
In the many-body model of Fu and Doniach\cite{PhysRevB.51.17439} and others, see \sref{model}, the local Coulomb interaction---the Hubbard $U$---was identified as
the driving force behind the metallization crossover. 
In the realistic calculations (that use $U=5$eV and $J=0.7$eV%
\footnote{in rough accordance with constrained DFT results\cite{2018arXiv180103496D}}.), however, changes
 in $U$ do not yield substantial modifications of the spectral function. Indeed, the large bandwidth ($\gtrsim 10$eV) of the Fe-3$d$ and Si-3$s$,3$p$ conglomerate dominates the Hubbard $U$ for reasonable values of the latter.
However, a significant dependence of many-body renormalizations on the Hund's coupling $J$ was evidenced\cite{jmt_fesi}:
As seen in \fref{FeSiJ}(a) the scattering rate at $T=116$K dramatically changes with $J$. In fact, when going from the realistic value $J=0.7$eV of iron compounds\cite{PhysRevB.82.045105} up to only $J=0.8$eV, $\Im\Sigma(\omega=0)$ more than doubles. Decreasing $J$ to 0.6eV, on the other hand, causes $\Im\Sigma(\omega=0)$ to almost vanish.

Concomitantly, also renormalizations encoded in the real-parts of the self-energy strongly depend on $J$: The mass enhancement---in DMFT: $m^*/m=1/Z$---increases
notably with $J$, see \fref{FeSiJ}(b), while remaining at a moderate magnitude ($<2$).%
\footnote{
The mass enhancement is albeit larger than in other Fe-Si binary compounds, e.g., metallic $\alpha$-FeSi$_2$\cite{miiller_fesi2}.
}
Combined, the effects of $J$ in the self-energy cause largely different spectra: As seen in panel (c) of \fref{FeSiJ}
both the width of the pseudogap at $T=116$K as well as the amount of spectral weight at the Fermi level is drastically dependent on the Hund's rule coupling.

What is the origin of this sensitivity?
First, in FeSi there is no direct competition between the Hund's coupling $J\sim 0.7$eV and crystal-field splittings $\delta=\mathcal{O}(0.3\hbox{eV})$.
That  $J>\delta$ indicates that a high-spin state is always favoured.
This clear hierarchy also rules out that a spin-state transition is at the origin of FeSi's anomalous behaviour (see also below).
Second, such a strong impact of $J$ was previously found  also for other iron-based materials\cite{1367-2630-11-2-025021,Yin_pnictide,PhysRevLett.104.197002}, adatoms\cite{Khajetoorians2015}, as well as for some 4$d$-systems\cite{PhysRevLett.106.096401} (see also the recent Refs.~\cite{Hausoel2017,Hariki2017}). 
The general mechanism is investigated in Refs.\cite{PTP.49.1483,PhysRevLett.103.147205,PhysRevLett.106.096401,Yin_pnictide,PhysRevLett.106.096401,PhysRevB.83.205112,PhysRevLett.107.256401,annurev-conmatphys-020911-125045}:
A large Hund's $J$ constrains electrons in different orbitals towards having the same spin, therewith ``orbitally blocking''\cite{Yin_pnictide} the Kondo interaction.
As a consequence the coherence temperature is shown to be suppressed exponentially\cite{PTP.49.1483}, resulting in a larger $\Im\Sigma$ at a given temperature.
For these systems the name Hund's metal has been coined\cite{Yin_pnictide}. 
Besides the effective mass enhancement, also spin- and charge fluctuations in these Hund's systems have genuine characteristics, as discussed for FeSi in the next section.

\paragraph{The spin and charge state.}

Above, we have discussed the momentum-structure of the magnetic susceptibility, and explained the reasons why the local component is smaller than
the measured uniform susceptibility. Besides the differentiation in momentum-space, the spin response also has an interesting structure in the time domain:
The static local spin susceptibility, $\chi_{loc}(\omega=0)\sim\frac{1}{\beta}\int d\tau \langle S_z(\tau)S_z(0)\rangle$,
displays, as discussed above, a strong and non-monotonous temperature dependence (see inset of \fref{FeSiDMFT1} top left).
$\chi_{loc}(\omega=0)$ describes the time-averaged response. Complementarity, we can compute 
the instantaneous response $\lim_{\tau\rightarrow 0}\langle S(\tau)S(0)\rangle$.
Interestingly, the latter is found to be virtually independent of temperature:
It yields, {\it for all temperatures}, an effective moment $\mu_{eff}=\sqrt{S(S+1)}g_s\approx3$ ($g_s=2$), which is consistent with the Curie-Weiss-like decay
manifest in the experimental uniform susceptibility at {\it high temperatures}. Indeed we find good agreement with $\chi=N_A{\mu_{eff}^2}/[{3k_B(T-\theta)}]$ for $\mu_{eff}=2.6\mu_B$ which is close to the single-ion limit in case of a $S=1$ configuration (see discussion in \sref{KI}).

We can analyse this in more detail: \Fref{FeSiHISTO} shows the quantum Monte Carlo 
histogram\cite{PhysRevB.75.155113,PhysRevLett.99.126405}
 for the DFT+DMFT calculation at low (left) and high (right) temperature. The bars indicate the probability distribution of the many-body wave-function for being in an eigenstate of the effective iron atom, decomposed into the number of particles $N$ and the spin state $S$.
We observe that:

\begin{figure*}[!t]
  \begin{center}
	\subfloat[$T=116$K]
	{\includegraphics[clip=true,trim=0 0 0 0,angle=0,width=.45\textwidth]{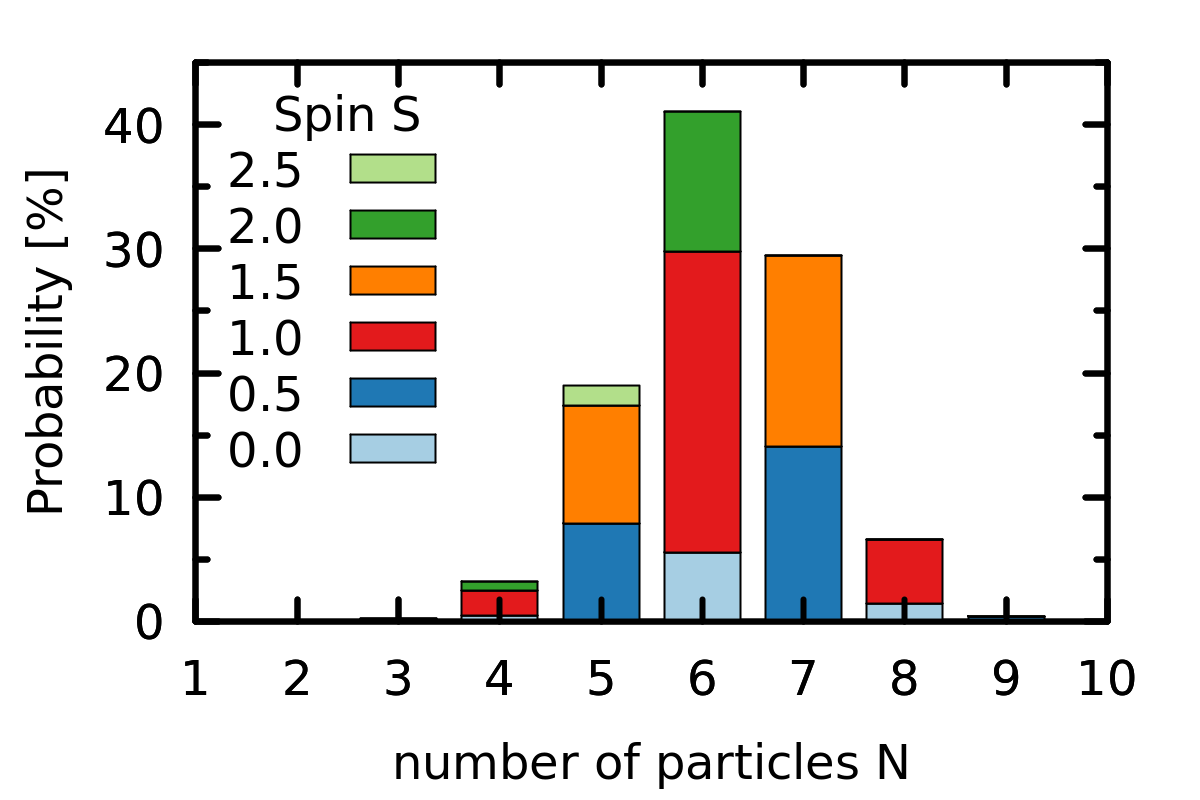}}
	\subfloat[$T=1161$K]
	{\includegraphics[clip=true,trim=0 0 0 0,angle=0,width=.45\textwidth]{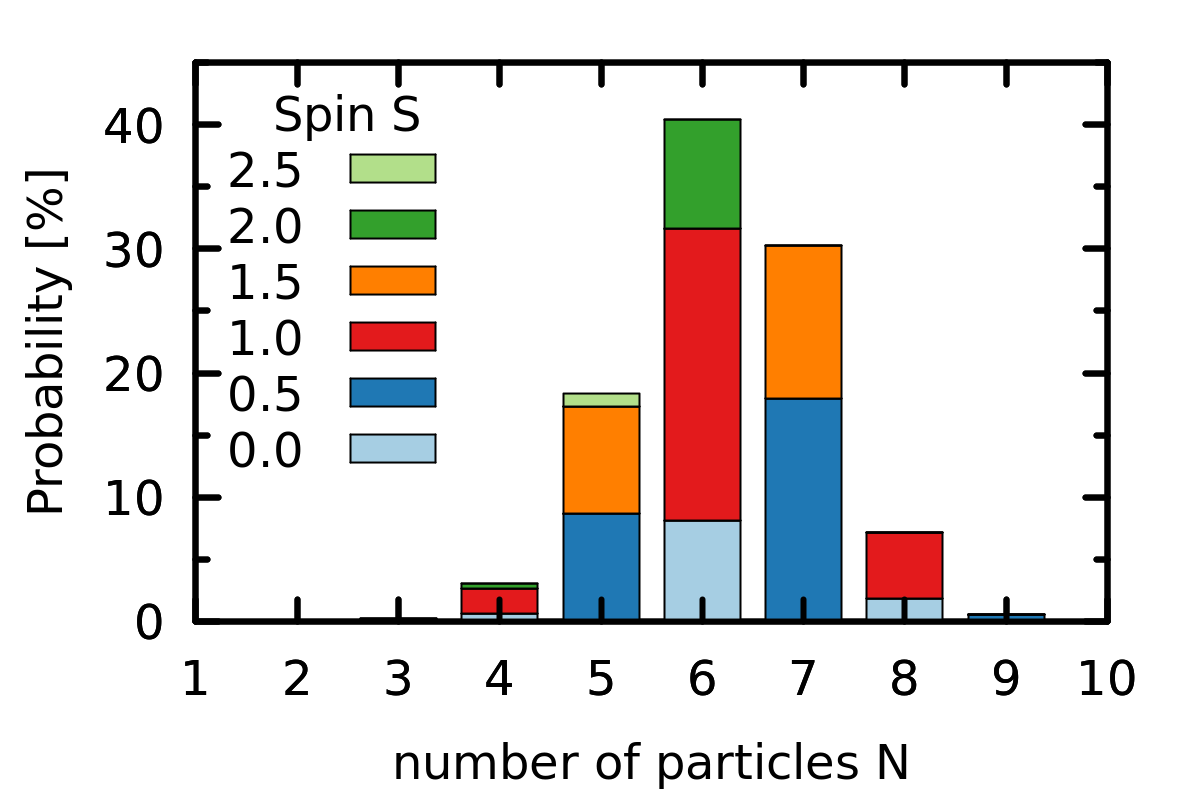}}
	      \caption{{\bf FeSi: DMFT histograms.} The bar-charts display the probability distribution of the system being in an eigenstate
				of the effective iron atom, resolved into states of occupation $N$ and spin-state $S$ for (a) $T=116$K and (b) $T=1161$K.
								Adapted from Ref.~\cite{jmt_hvar}.}
      \label{FeSiHISTO}
      \end{center}
\end{figure*}

\begin{enumerate} 
	\item the decomposition onto $N$ and $S$ is almost independent of temperature.
The scenario of a temperature-induced moment\cite{PhysRevLett.59.351} 
is hence not realized in our theory.
Also, as anticipated above, a spin-state-transition does not occur.
This situation has to be contrasted with systems in which the crystal-field splitting and the Hund's rule coupling are of comparable size\cite{PhysRevLett.99.126405}:
	For example, spin-state crossovers have been identified to underlie the pressure-induced (low-spin)insulator-to-(high-spin)metal transition
	in (the B8 phase of) MnO\cite{kunes_mno}, or the anomalous behaviour in the magnetic susceptibility of LaCoO$_3$\cite{PhysRevB.79.014430,PhysRevB.81.035101,PhysRevB.86.195104,PhysRevLett.106.256401,PhysRevLett.110.267204} (see \fref{overviewchi} or Refs.\ \cite{SENARISRODRIGUEZ1995224,PhysRevB.53.R2926}).
	Indeed, in LaCoO$_3$ X-ray absorption spectroscopy estimates a crystal-field splitting between (at low-$T$ fully occupied) $t_{2g}$ and (empty) $e_g$
orbitals of $\delta\approx 0.7$eV. This value corresponds to
	typical estimates of the Hund's $J$ in transition-metal based compounds.
	In LaCoO$_3$, the competition between $\delta$ and $J$ is tipped by a	
	thermal activation of the high-spin state (in cooperation with the expansion of the lattice).
	As shown by K\v{r}{\'a}pek \etal\ \cite{PhysRevB.86.195104} this leads to a pronounced temperature-dependence
	of the DMFT histogram for both charge- and spin-resolved probabilities---in qualitative difference to the case of FeSi.
	An important factor that distinguishes FeSi from LaCoO$_3$---the degree of ionic vs.\ covalent bonding---will be discussed in a broader context in \sref{sec:covalent}.
\item the iron valence of $\left\langle N \right\rangle\approx6.2$ is consistent with the covalent band-picture of \sref{bandFeSi} 
(see also \fref{valence}).
\item the spin state $S=1$ has the overall largest probability at all temperatures, consistent with the above discussed effective moment.
\item the probability distribution is not dominated by only few configurations of $N$ and $S$. This indicates strong fluctuations at short time-scales.
	In fact, the system has a strongly intermediate valence, as the charge variance is of order 1~: $\delta N=\langle\left( N-\langle N\rangle\right)^2\rangle\approx 0.93$.
However, as the variance is  insensitive to temperature, a thermally induced mixed valence \cite{PhysRevB.50.9952} can be excluded.
Also spin fluctuations at short time-scales, $\delta S=\langle\left( S-\langle S\rangle\right)^2\rangle\approx 0.33$, are large.
\end{enumerate}

The above analysis shows that the physics of correlations in FeSi is very different from the Mott-Hubbard physics in, say, transition metal oxides\cite{Imada}.
There, the Hubbard $U$ quenches charge fluctuations and reduces the time-scale associated with them, while the time averaged local spin susceptibility is strongly enhanced.
In fact, in a Mott insulator, charge fluctuations are basically absent, while the local moment of charge carriers fluctuates freely, giving
rise to a Curie-Weiss susceptibility.
In FeSi, on the contrary, spin fluctuations are short-lived:
The (time-averaged) local spin susceptibility is only moderate enhanced, 
while the short-time (energy-averaged) fluctuating magnetic moment is strongly enhanced.
Contrary to Mott insulators, fluctuating magnetic moments in systems like FeSi do not require the quenching of charge fluctuations.
This is indeed a hallmark of Hund's metals (see the preceding section). Still, FeSi is an example where the Hund's metal concept exists for an only poorly metallic state, 
in which the hybridization of the iron atoms with their surrounding is weak (see also the discussion of the DMFT hybridization function in \sref{Ce3DMFT}).

\paragraph{The physical picture.}
With the above details of the realistic many-body simulations, one can address the fundamental picture of FeSi:
(a) What is the microscopic origin of the crossover in the magnetic susceptibility? (b) How is the latter linked to the emergence of incoherent spectral weight. In other words,
what is the relation between the crossovers for the spin and charge degrees of freedom?
(c) What is the difference, if any, to the physics of Kondo insulators?

At low temperatures, FeSi is a band-like semiconductor:
The band gap in the one-particle spectrum inhibits, both, spin and charge fluctuations at finite time-scales.
This results in (i) a small, activated (uniform) spin susceptibility, and (ii) a lack of phase-space available for electronic scattering, resulting in only a small
imaginary part of the self-energy.
As such, the system can be approximately described by an effective one-particle Hamiltonian that is
diagonalizable in momentum space, stating that $\svek{k}$ is a good quantum number.

With increasing temperature (beyond $T^{max}_\chi$) a fluctuating moment develops at the iron sites, which is a manifestation of the single-ion physics that dominates
at large temperatures when spatial correlations become short-ranged\cite{ANDP:ANDP201100042}.  In this limit, the index conjugate to the momentum, the lattice site $i$, becomes a good quantum number.
The fluctuating moment thus provides a link to the real-space lattice, and the description in terms of Bloch-bands breaks down as $\mathbf{k}$ is no longer a good quantum number.
As a consequence, the one-particle spectrum decomposes over many momenta, or, equivalently acquires a finite lifetime for a given momentum. This lifetime is encoded
in the imaginary parts of the self-energy and leads to the appearance of incoherent spectral weight that eventually fills the gap.
We stress that despite the notion of locality of the moments on the iron sites, magnetic fluctuations are ferromagnetic (see \fref{FeSiN}) due to the strong on-site Hund's rule coupling. 

This scenario qualitatively reconciles the behaviour seen for spectral properties in the pioneering model of Fu and Doniach\cite{PhysRevB.51.17439} 
with the seminal spin-fluctuation theory of Takahashi and Moriya\cite{JPSJ.46.1451}: In the DFT+DMFT picture, 
the crossovers in the spin and charge response in FeSi are intimately linked to each other.
In the microscopic details, however, the realistic many-body calculations\cite{jmt_fesi,jmt_hvar} revealed notable differences to previous model calculations\cite{PhysRevB.51.17439,PhysRevB.81.125131,PhysRevB.78.033109}.
It is likely that the scenario for FeSi reviewed here is also relevant for other narrow-gap intermetallics.
For example, it can be expected that---once issues related to exchange contributions to the self-energy (see \sref{marcasites}) are overcome---the 
microscopic behaviour of FeSb$_2$ is qualitatively akin to that of FeSi.

The above discussions answers the first two questions, (a) and (b), raised at the beginning of this paragraph. 
In order to answer---in \sref{sec:covalent}---the third question, (c), we performed realistic many-body calculations for \cbp\ that are discussed below.

\subsubsection{Realistic many-body calculations for \cbp.}
\label{Ce3DMFT}

\begin{figure}[!t]%
  \begin{center}
\includegraphics[width=0.49\textwidth]{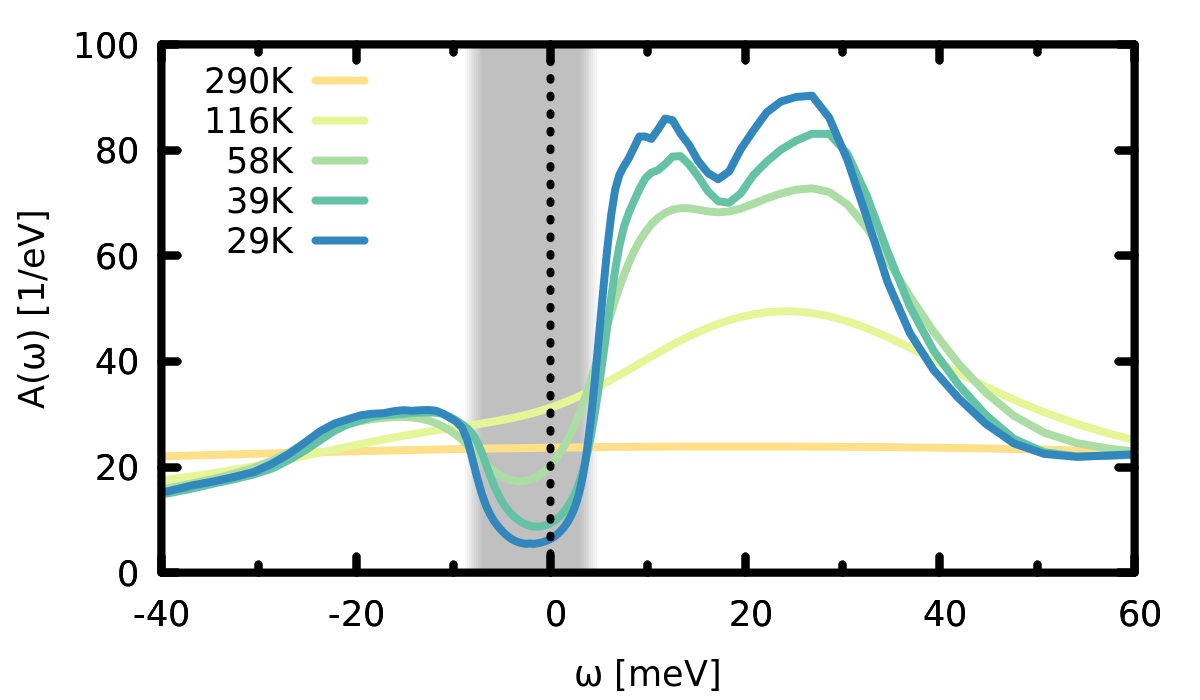}%
\caption{{\bf \cbp: spectral functions.}  Shown is the DFT+DMFT local spectral function of \cbp\ for different temperatures. Clearly visible is the opening of a gap
of about 10meV (indicated by the grey-shaded area) when cooling below $\sim 100$K.
}%
\label{DMFTCe3}
\end{center}
\end{figure}

For a direct comparison of FeSi to heavy-fermion-based Kondo insulators, 
we performed DFT+DMFT calculations for the prototypical compound \cbp.%
\footnote{We employed DFT+DMFT with a CTQMC solver as implemented in Ref.~\cite{PhysRevB.81.195107} with rotationally invariant interactions parametrized by $U=5.5$eV and $J=0.68$eV (similar to what was used previously for elemental Ce\cite{PhysRevLett.87.276403,PhysRevB.81.195107,PhysRevB.89.125113,PhysRevB.89.195132} and other Ce-based compounds\cite{PhysRevB.81.195107,jmt_cesf,Goremychkin186}), 
spin-orbit coupling, charge self-consistency, and nominal double-counting. Details will be published elsewhere.
}
\Fref{DMFTCe3} displays the spectral function obtained for different temperatures.
As was the case in FeSi, lowering temperature results in a depletion of spectral weight at the Fermi level. 
The width of the emerging gap, $\Delta\lesssim 10$meV, is consistent with the experimental indirect charge gap of 8-12meV (see \fref{FeSiCe3trans}(a), \fref{FeSiCe3spec}(a), and \tref{table1}). With respect to band-structure results (see \sref{Ce3dft}) the renormalization of the gap corresponds to 
an effective mass-enhancement of about a factor of 10. 
Also the temperature scale, $\lesssim 100$K, around which the metallization occurs in the spectral function is congruent with both photoemission \cite{Takeda1999721}
and optical spectra \cite{PhysRevLett.72.522} (see \fref{FeSiCe3spec}).
This coherence-incoherence crossover is encoded in the imaginary parts of the self-energy (not shown).
As was the case for FeSi (see \fref{FeSiImS}), $\Im\Sigma(\omega=0)\propto T^2$  for orbitals relevant for spectral weight
near the Fermi level.
These changes in spectral properties are accompanied by a rearrangement of the spin degrees of freedom. Indeed, an analysis of the DMFT
magnetic susceptibility (not shown) is compatible with the standard picture of the unlocking of local moments above a coherence temperature $T^{max}_\chi$.

The next section will put these results into a broader context by discussing differences between FeSi, \cbp, and LaCoO$_3$.\\

%

\subsection{Kondo insulators, transition-metal intermetallics and oxides: A categorization}
\label{sec:covalent}

\subsubsection{The DMFT perspective on insulators}
\label{dmftins}

In order to illustrate qualitative differences in narrow-gap semiconductors with partially-filled correlated orbitals,
it is convenient to take a look under hood of the DMFT method\cite{bible,vollkot}.
There, the periodic solid is mapped onto an effective (e.g., iron or cerium) atom (called an ``impurity'') which is coupled to a reservoir (``bath'') with which it can exchange particles.
Translational invariance is restored through a self-consistency condition which identifies the impurity Green's function with the local Green's function of the lattice for the subset ``$d$'' of correlated orbitals,
$G_{dd}^{loc}(\omega)\stackrel{!}{=}G^{imp}(\omega)$, i.e.,
\begin{widetext}
\begin{equation}
\left[\sum_{\svek{k}}\left[\omega+\mu-H(\svek{k})-\Sigma(\omega) \right]^{-1}\right]_{dd}=\left[\omega+\mu-H_{dd}^{loc}-\Delta(\omega)-\Sigma_{imp}(\omega)  \right]^{-1}
\label{SCC}
\end{equation}
\end{widetext}
Here, $\mu$ the chemical potential, $H(\svek{k})$ the one-particle Hamiltonian, and $H_{dd}^{loc}$ its local ($\svek{k}$-summed) component in the correlated $d$-subspace
that encodes {\it local crystal-fields}.
$\Sigma$ is the lattice self-energy which is approximated as the impurity self-energy $\Sigma_{imp}$ in the $d$-subspace, $\Sigma_{dd}=\Sigma_{imp}$, and is zero otherwise.
Finally, $\Delta(\omega)$ is the so-called hybridization function that describes the coupling of the effective correlated atom to the bath.

Using \eref{SCC}, we can characterize the origin of an insulating state, i.e., a vanishing  of $G^{loc}(\omega=0)$, through its manifestations in  $H(\svek{k})$, $\Sigma(\omega)$, $H_{dd}^{loc}$, and $\Delta(\omega)$.
In this DMFT perspective, we can distinguish:%
\footnote{Note that, here, we limit the discussion to paramagnetic insulators without spin, charge, or structural orders.}

\begin{framed}
\noindent
	{\bf Three prototypes of paramagnetic insulators} (see also \tref{insulators})
\begin{enumerate}
  \item {\bf Mott insulators:}
	$\Sigma$ diverges at low energies (at least for some orbitals), while $H(\svek{k})$ is metallic. This is the Brinkmann-Rice\cite{PhysRevB.2.4302} scenario of a Mott insulator: spectral weight is destroyed ($Z\rightarrow 0$) by diverging effective masses, as realized e.g., in YTiO$_3$\cite{pavarini:176403}, V$_2$O$_3$\cite{PhysRevLett.86.5345,poter_v2o3}, or thin films of SrVO$_3$\cite{PhysRevLett.114.246401}.
	\item {\bf (ionic) band insulators:}
	Here, both $\Im\Sigma$ and $\Im\Delta$ vanish at the Fermi level at low $T$, while $H^{loc}_{dd}+\Re\Delta(0)+\Re\Sigma(0)-\mu$ is finite for all orbitals.
	In other words, it is local crystal-fields encoded in $H_{dd}^{loc}$ (possibly enhanced by the static renormalizations $\Re\Sigma(0)$, $\Re\Delta(0)$) that are responsible for the gapping.
	This is the case of LaCoO$_3$ where the chemical potential falls between the (at low temperatures) filled $t_{2g}$ and the empty $e_g$ orbitals (see \sref{bandFeSi} and below).
	Also the dimerized M1 phase of VO$_2$, a correlation-enhanced Peierls insulator\cite{me_vo2}, will fall into this category.
	\item {\bf Kondo insulators (covalent band insulators):}
	In these systems $\Delta$ diverges at low energies to suppress spectral weight, while $\Sigma$ is benign and crystal-fields in $H^{loc}_{dd}$ are mostly irrelevant for realizing the insulator.
	This is the case of \cbp, see below.
\end{enumerate}
\end{framed}

Since at least the last two scenarios do not exclude each-other, it may be difficult to provide a clean-cut categorization for a given material.
Before looking at {\it ab initio} results for FeSi and \cbp, let us thus illustrate some characteristics of Kondo insulators, as well as the grey area between purely covalent and ionic insulators first for a toy model, then from the DFT perspective.

\begin{table*}[!t]%
\begin{tabular}{|l|c|c|c|c|l|}
\hline
\rowcolor{blue}
type / property                  & $H(\svek{k})$ &$-\Im\Sigma$ & $\Delta H^{loc}$&$-\Im\Delta$ & examples\\
\hline
\cellcolor{lblue}band insulator & gapped          & 0          &  large      & 0 &LaCoO$_3$, FeSi ($d_{z^2}$,$d_{xz}$,$d_{yz}$)\\
\hline
\cellcolor{lblue}Kondo insulator & gapped & 0            &  small    & $\infty$ & \cbp, FeSi ($d_{x^2-y^2}$,$d_{xy}$)\\
\hline
\cellcolor{lblue}Mott insulator & metallic & $\infty$           & small      & 0 & YTiO$_3$, thin films SrVO$_3$\\
\hline
\end{tabular}
\caption{{\bf Different types of insulators.} Typical properties of the one-particle Hamiltonian $H(\svek{k})$, and the low-temperature and low-energy behaviour of the imaginary parts of the self-energy $\Im\Sigma$ and the hybridization function $\Im\Delta$, and local crystal-field splittings $\Delta H^{loc}$ encoded in $H^{loc}_{dd}$ 
(e.g., $\Delta H^{loc}=H^{loc}_{e_g}-H^{loc}_{t_{2g}}$) for different kinds of paramagnetic insulators with partially filled $d$- or $f$-shells.
}
\label{insulators}
\end{table*}

\subsubsection{A toy-model.} 
\label{toy}
Let us first focus on the left side of \eref{SCC}:
From the lattice point of view the difference between ionic and covalent insulators is hidden in $H(\svek{k})$. 
This is seen in a simple toy model in the quasi-particle picture. Consider:
\begin{eqnarray}
H(\svek{k})=\begin{bmatrix}
\epsilon_{\svek{k}}^d & V \\
V^* & \epsilon_{\svek{k}}^p
\end{bmatrix},
\qquad
\Sigma(\omega)=\begin{bmatrix}
\Sigma(\omega) & 0 \\
0 & 0
\end{bmatrix}.
\label{model1}
\end{eqnarray}
i.e., one correlated band $\epsilon^d_\svek{k}$ coupled to a ligand state $\epsilon^p_\svek{k}$ via a hybridization $V$, that for simplicity we have assumed to be local.
Many-body renormalizations are described by a self-energy, for which we take a Fermi-liquid-like form $\Sigma(\omega)=(1-1/Z)\omega-i0^+$ that neglects lifetime effects.

The first goal is to illustrate generic behaviours of the hybridization function for Kondo insulators.
To this end, we consider a setup in the spirit of the periodic Anderson model to mimic \cbp: $\epsilon^d_{\svek{k}}=0$ is an atomic-like $f$-level, renormalized by $Z$, and $\epsilon^p_{\svek{k}}=\epsilon_{\svek{k}}$ are the non-interacting conduction electrons of atoms other than Ce. Assuming particle-hole symmetry yields $\mu=0$ and $\sum_{\svek{k}}\epsilon_{\svek{k}}=0$, implying the absence of any crystal-fields.
The bare one-particle spectrum is then gapped, with the direct gap determined by $2V$.
The interacting dispersions are then given by $\omega_{\svek{k}}=\epsilon_{\svek{k}}/2\pm\sqrt{\epsilon_{\svek{k}}^2/4+Z|V|^2}$,
stating that only the {\it inter-atomic} hybridization (hence also the direct gap) is renormalized by electron-electron interactions. 
Equivalently, this can be seen in the hybridization function $\Delta(\omega)=\omega-(G_{loc, dd}(\omega))^{-1}-\Sigma(\omega)$:
For the left side of \eref{SCC} we have
$G_{loc, dd}=\sum_\svek{k}Z/(\omega-Z\Delta_\svek{k}(\omega)+i0^+)$
with a momentum-dependent quantity $\Delta_\svek{k}(\omega)=\frac{|V|^2}{\omega-\epsilon_{\svek{k}}+i0^+}$. $\Delta_\svek{k}(\omega)$ in particular diverges at the conduction-electron positions.
In order to match the right-hand side of \eref{SCC}, $G^{imp}(\omega)=Z/(\omega-Z\Delta(\omega)+i0^+)$, the DMFT hybridization function $\Delta(\omega)$
has to diverge, here---because of particle-hole symmetry---at the Fermi level.
Consequently, the divergence of $\Delta$ is a pure one-particle effect. Instead of being a signature of the Kondo effect, $\Delta$ here merely encodes
inter-atomic hybridizations that couple the effective correlated site to the DMFT bath.
From this band-structure perspective, the influence of many-body effects is to renormalize down the strength of hybridization $\Delta\rightarrow Z\Delta$.
Further, effects of incoherence (neglected here) will, when iterating to self-consistency, mitigate the divergence above a coherence temperature
(see the {\it ab initio} results in \sref{abinitiohyb}).

\begin{figure*}[!t]%
  \begin{center}
\includegraphics[width=0.65\textwidth]{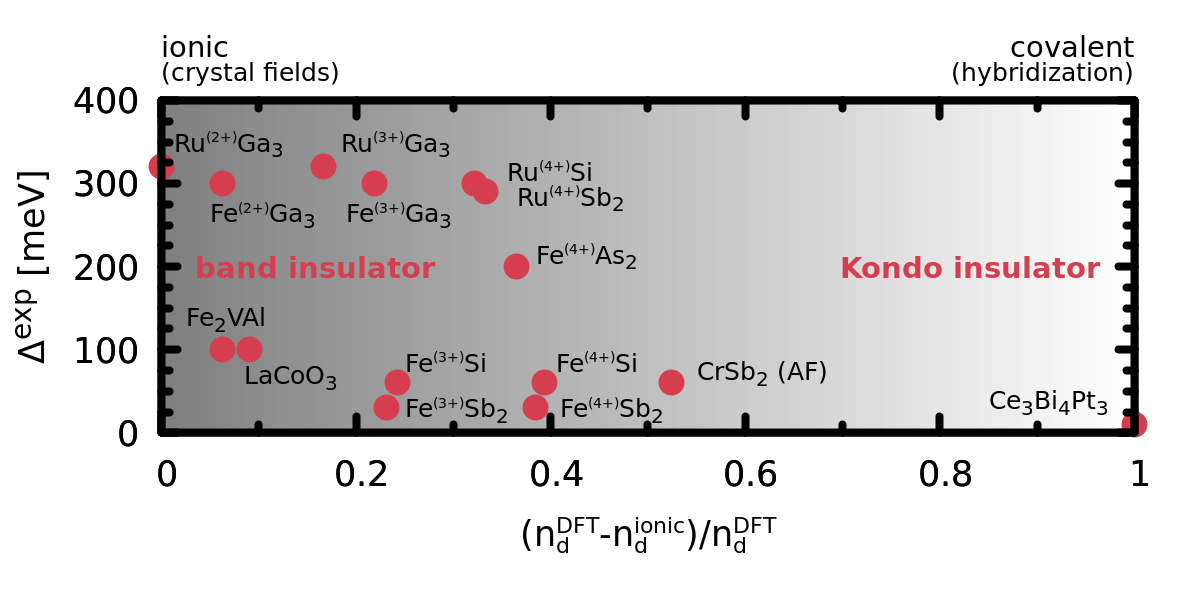}%
\caption{{\bf Ionic vs.\ covalent bonding.} 
Shown is a collection of materials distinguished by 
the degree of covalent bonding for their partially-filled correlated orbitals ($3/4d$ or $4f$). As a rough measure, we use
the relative deviation of the DFT-occupation of these orbitals, $n^{DFT}_d$, from the expectation in the purely ionic picture, $n^{ionic}_d$. 
For some materials different ionic states are indicated.
The gapping of correlated electrons in materials on the right (left) are dominated by hybridizations $\Delta(\omega)$  (local crystal-fields $H^{loc}_{dd}$), respectively.
Using this proxy, Kondo-insulating characteristics prevail in compounds on the right and become less pronounced towards the left.
DFT occupations were determined from a character projection inside the muffin-tin spheres (of typical radii) using wien2k. The dependence on the exchange-correlation potential (LDA, PBE, mBJ) is negligible for all practical purposes. The $y$-axis shows the experimental (indirect) charge gap (see \tref{table1}).
}
\label{valence}
\end{center}
\end{figure*}

The model can also shed some insights onto the differences of FeSi and LaCoO$_3$. 
We model both compounds by two decoupled copies of the above model: the first copy describes a high-lying valence $d$-band $\epsilon^{d}_{\svek{k}}=+(\epsilon_{\svek{k}}+\delta/2)$
and a ligand band $\epsilon^{p}_{\svek{k}}=-(\epsilon_{\svek{k}}+\delta/2)$ lowered by a crystal-field $\delta$, both of which are
coupled via a local hybridization $V$. In the second copy, $\epsilon^{d}_{\svek{k}}$ and $\epsilon^{p}_{\svek{k}}$ switch places, e.g.,  $\epsilon^{d}_{\svek{k}}=-(\epsilon_{\svek{k}}+\delta/2)$ describes a correlated conduction band. Then, not the individual but the sum of both systems is particle-hole symmetric for $\mu=0$.
For the first copy, we then obtain for the left side of \eref{SCC}
$G_{loc, dd}(\omega)=\sum_\svek{k} Z/(\omega-Z(\epsilon_{\svek{k}}+\delta/2)-Z\Delta_\svek{k}(\omega)+i0^+)$ with $\Delta_\svek{k}(\omega)=\frac{|V|^2}{\omega+\epsilon_{\svek{k}}+\delta/2+i0^+}$.
Now, there are two degrees of freedom that can cause insulating behaviour:
For small crystal-fields, $\delta$, insulating behaviour requires the same mechanism as in the above model for \cbp: a pole-like structure in $\Delta_{\svek{k}}$ (and, correspondingly,
in the DMFT hybridization function $\Delta$). Indeed, for $\delta=0$ the band-gap is proportional to $V$ (at least for $Z=1$).
Empirically this situation corresponds to FeSi within DFT, as illustrated by \fref{FeSiscalehyb}(a).
In the opposite limit, i.e., a vanishing hybridization $V=0$, insulating behaviour requires that the crystal-field $\delta/2$ exceeds the half-bandwidth of $\epsilon_{\svek{k}}$.
This is the prevailing mechanism in LaCoO$_3$, as shown within DFT in \fref{FeSiscalehyb}(b).
From this point of view, one can call LaCoO$_3$ a mainly {\it ionic} insulator, and FeSi a dominantly {\it covalent}, i.e., Kondo-like insulator.
Naturally, there exists a continuous spectrum between the pure prototypes of Kondo-like and (ionic) band insulators.
Also, the hybridization function of an ionic system can still be strongly peaked outside the charge-gap, encoding
charge-fluctuations at higher energies.

\subsubsection{a DFT-based proxy to distinguish ionic from covalent insulators.}
\label{proxy}
As motivated by the preceding model as well as \sref{band}, interesting insights can be gained from one-particle band-structures even in the case of Kondo systems.
Indeed, DFT-based hybridization functions $\Delta(\omega)$ can divulge {\it trends} in gap-sizes or Kondo temperatures among families of compounds.
For a recent discussion of band-structure-derived hybridization functions for Ce-compounds, see Ref.~\cite{PhysRevMaterials.1.033802}.

Here, we are interested in determining---without
performing costly many-body calculations---whether an insulating material is rather a Kondo or a (ionic) band insulator.
For paramagnetic insulators with partially filled $d$/$f$-shells we
propose---as a crude
indicator for Kondo insulating behaviour---the relative deviation of DFT-occupancies with respect to nominal expectations in a purely ionic picture: $(n_d^{DFT}-n_d^{ionic})/n^{DFT}_d$. This proxy can hence be seen as a measure for charges of correlated orbitals to participate in covalent bonding.
In \fref{valence} we have collected a number of relevant compounds. The Kondo insulator \cbp\ appears on the right-most side, as it does not donate all of its Ce-$4f^1$
electron to more electronegative atoms. 
The oxide LaCoO$_3$, on the other hand, is found much towards the left.
Indeed the Co-valence within DFT is rather close to that of a Co$^{3+}$ ion. 
The intermetallic semiconductors FeSi and FeSb$_2$ rank in the middle of the diagram, pointing to the influence of, both, covalent and ionic
characteristics.

\begin{figure*}[!t]
  \begin{center}
		\subfloat[\cbp: spectral function (29K)]
		{{\includegraphics[angle=0,width=.48\textwidth]{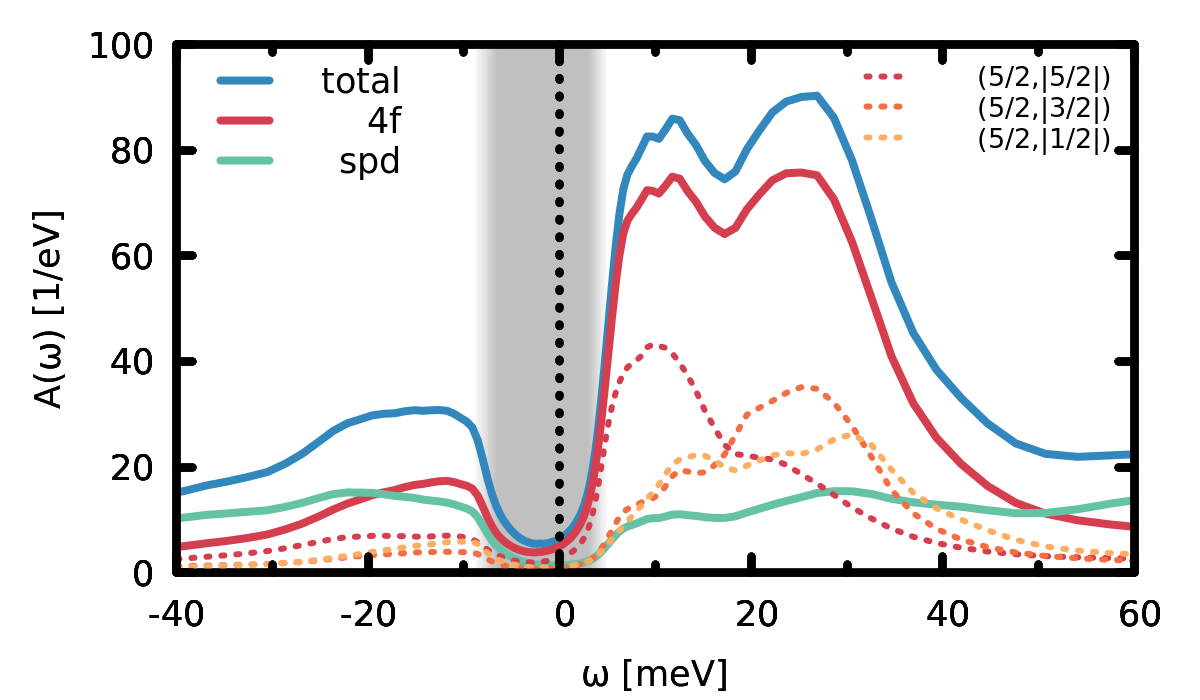}}}
	$\quad$
		\subfloat[FeSi: spectral function (116K)]
		{{\includegraphics[angle=0,width=.48\textwidth]{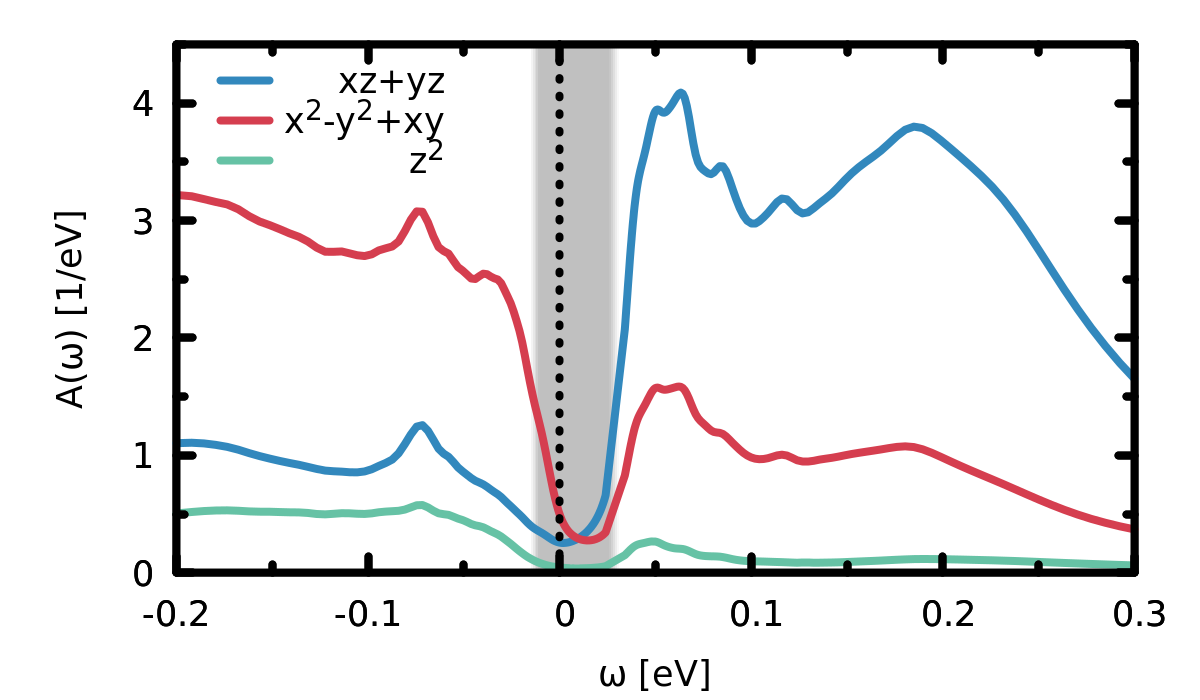}}}

\subfloat[\cbp: hybridization function] 
	{{\includegraphics[angle=0,width=.48\textwidth]{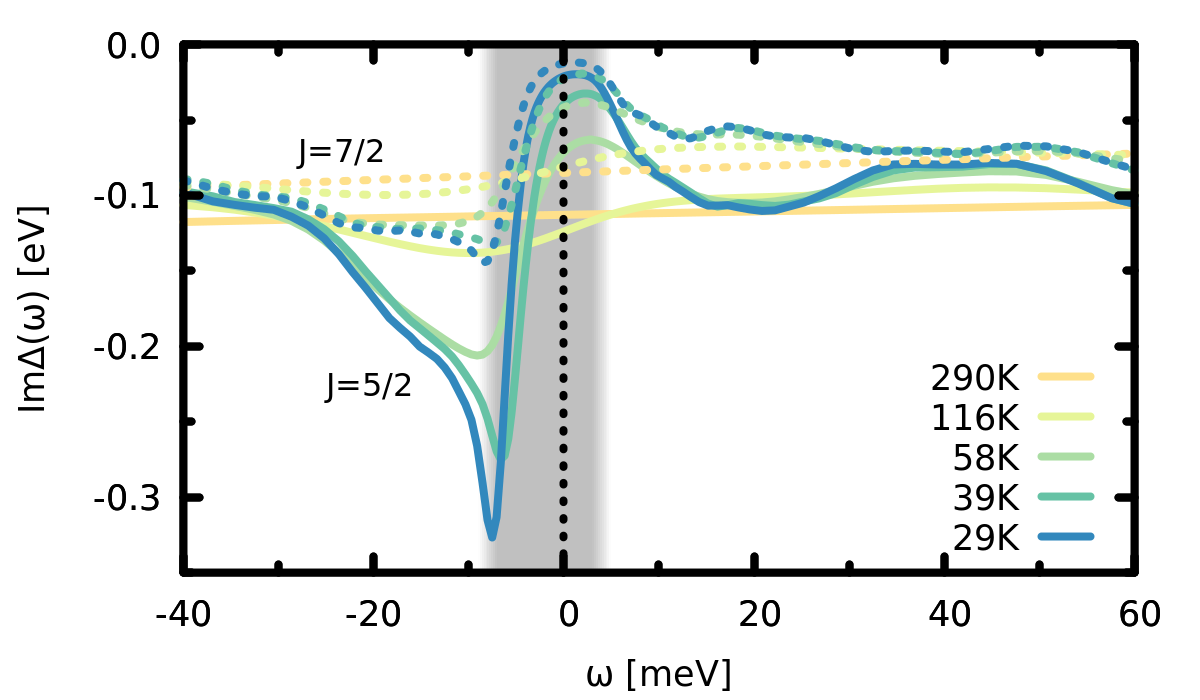}}}
$\quad$
	\subfloat[FeSi: hybridization function] 
	{{\includegraphics[angle=0,width=.48\textwidth]{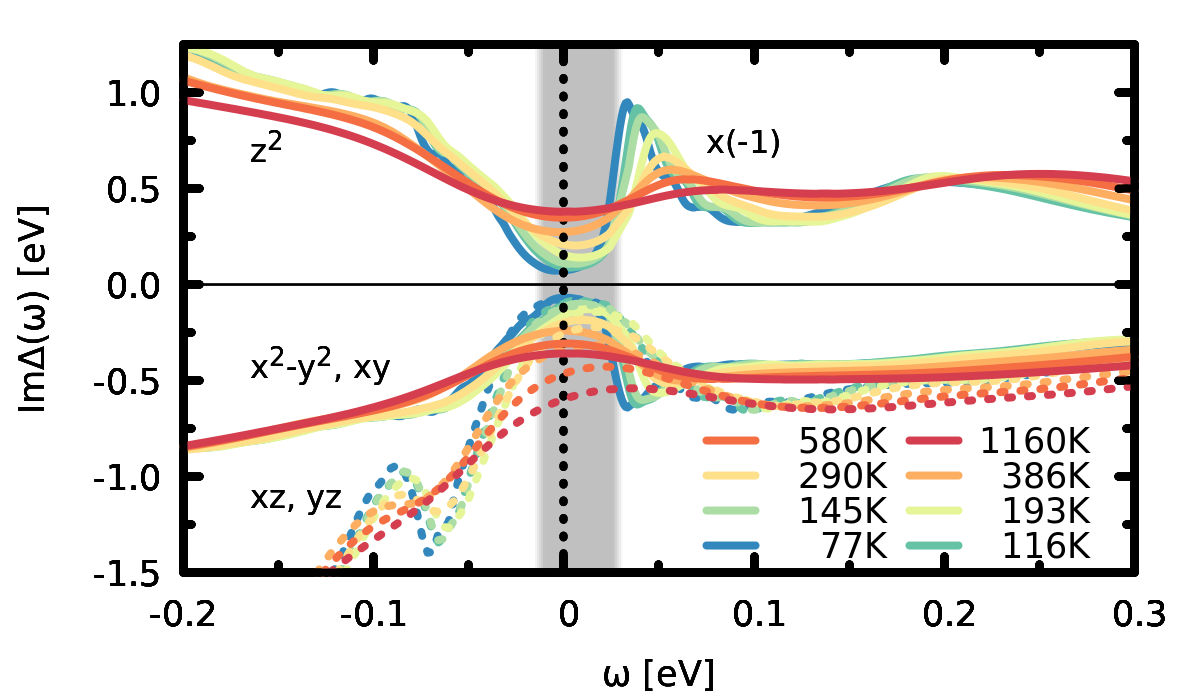}}}
	      \caption{{\bf  \cbp\ (left) and FeSi (right) within DFT+DMFT.} (a) \& (b) spectral functions:
				For \cbp\ we show the total spectral function at 29K, resolved into the contributions of the Ce-$4f$ (and its 5/2-multiplets) and the conduction electron ($spd$) states.
				For FeSi we show the spectrum resolved into its Fe-$3d$ components. In both panels, we have marked in red the spectral contributions that are gapped via the Kondo insulator
				mechanism.
				(c) \& (d) temperature dependence of the hybridization functions: 
				Shown are the imaginary parts $\Im\Delta(\omega)$ for relevant orbital components and different temperatures.
				In panel (a) components with the same $\left|J_z\right|$ have been averaged; in panel (c) components have been averaged over $J_z$.
				For clarity, the sign of the $d_{z^2}$ hybridization function has been inverted in panel (d).
}
      \label{DMFTCe32}
      \end{center}
\end{figure*}

\subsubsection{Comparison of the {\it ab initio} results of FeSi and \cbp.}
\label{abinitiohyb}
We are now in the position to analyse and compare the realistic many-body simulations for FeSi and \cbp:
\Fref{DMFTCe32} (a) \& (b) display the respective DFT+DMFT spectral functions, $A(\omega)$, at low temperatures, resolved into relevant orbital contributions.
Panels (c) \& (d) show how the imaginary parts of the corresponding hybridization functions, $\Delta(\omega)$, evolve with temperature.

In the local spectral function, we see a pronounced difference between \cbp\ (left) and FeSi (right):
In the heavy-fermion material, there is no orbital polarization. Indeed, both, the valence and the conduction side
of the low-temperature gap are dominated by the correlated $4f$-orbitals. 
Furthermore, the $J=5/2$ multiplets are basically degenerate in what concerns the gap-formation. In particular, all of them
have their centre of mass above the Fermi level.  The gapping thus occurs similarly for {\it all} $J=5/2$ multiplets.

From the DMFT perspective, this situation is described by the development of a strong peak in the hybridization function for the $J=5/2$ components.
In the many-body language, this is the building up of the Kondo effect. It restores band-like coherence below a characteristic temperature through
the formation of singlet states between $4f$-electrons of the effective Ce-atom and conduction electrons from its surrounding. 
Seen from the low temperature (band-structure) perspective,
 the large $\Delta(\omega)$ is merely a way to encode---within the single-site impurity construction---the effect of {\it inter-atomic} (i.e., out-of-impurity) hybridizations,
that, with growing temperature, are rendered ineffective through the emergence of local spin-fluctuations.
Note that, unlike for the simple model in \sref{toy}, the enhancement of $\Delta(\omega)$ occurs at the lower edge of the charge gap
since \cbp\ is strongly particle-hole asymmetric.

Contrary to \cbp, the orbital character of the valence and conduction states are not the same in FeSi. Indeed, the occupied spectrum is
dominated by the Fe-$3d_{x^2-y^2}+3d_{xy}$ orbitals,  while unoccupied states derive mainly from Fe-$3d_{xz}+3d_{yz}$ weight.
Still, all orbital components are individually gapped.
We can analyse the origin of for the different orbitals from the DMFT perspective:
We see that insulating behaviour of the $3d_{xz}+3d_{yz}$ components is realized
by a hybridization function that increasingly vanishes inside the gap:
There is (averaged over time) no spontaneous exchange of particles of these orbitals with the surrounding of the effective iron atom.
Feature in $\Delta(\omega)$ that mark charge fluctuations only appear for energies well above gap.
In the categorization of \sref{dmftins}, these orbitals are hence trivially band-insulating.

The situation for the $3d_{x^2-y^2}+3d_{xy}$ components that account for most of the valence electrons, as well as the $d_{z^2}$ components
is more intricate.
In fact, upon lowering temperature, 
a feature emerges in $\Im\Delta(\omega)$ just outside the upper edge of the charge gap for both these kind of orbitals.
It can be shown, that the $d_{z^2}$ orbitals are gapped out,
meaning that the denominator of the $d_{z^2}$ impurity Greens function, $\omega+\mu-H^{loc}_{dd}-\Re\Delta(\omega)-\Re\Sigma(\omega)$,
is finite for energies inside the gap (see Figure S1B in the supplementary material of Ref.~\cite{jmt_fesi}). 
While hybridizations gain coherence upon cooling, they are not responsible for the gap formation.
According to our classification, \sref{dmftins}, also the $d_{z^2}$-orbitals are hence band-insulating.
The $3d_{x^2-y^2}$, $3d_{xy}$ components, however, bear all the marks of a Kondo-insulator, as seen in \cbp:
These orbitals' spectral weight accumulates to the most part of one side of the Fermi-level, while on the other side a 
sharp low-energy feature develops in $\Delta(\omega)$ that suppresses states and opens a gap.
In conclusion, we can call FeSi an {\it orbital-selective Kondo insulator}.

In the light of this discussion, there are three main differences between FeSi and \cbp:
(1) the low-temperature gap in \cbp\ is purely of Kondo-insulating nature, whereas in FeSi the origin of the gapping depends on the orbital character.
(2) for orbitals in FeSi that suppress quantum fluctuations through the Kondo effect, the relevant non-local hybridization dominantly links these orbitals
to other iron atoms. In \cbp, the gap-inducing hybridization connects the ``impurity'' Ce-$4f$-$5/2$ orbitals to uncorrelated, dispersive states of Bi and Pt. 
(3) in the case of FeSi, the origin of the destruction of asymptotic freedom with rising temperature has multi-orbital (Hund's rule) characteristics.


\subsection{Covalent insulators vs.\ Kondo insulators: Hubbard vs.\ Anderson}
\label{PAM}

\begin{figure*}[!t]
  \begin{center}
	$\quad$ periodic Anderson model $\quad\qquad\qquad\qquad$ 2-band Hubbard model
	
	\subfloat[{\bf Non-interacting band-structure.} dispersion along the diagonal $(k,k,k)$-direction. Insets show a zoom to a low-energy window $\lbrack-150:150\rbrack$meV.
	]{
	{ \includegraphics[angle=0,width=.45\textwidth]{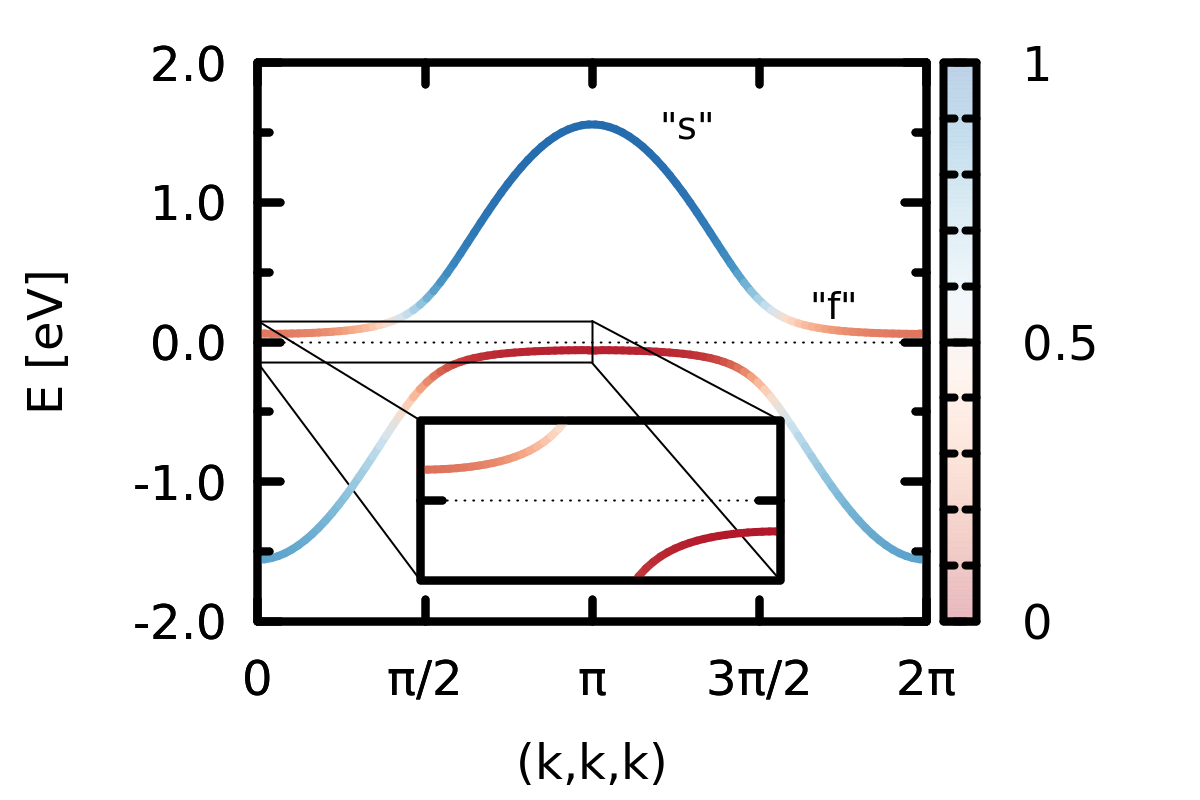}} 
	{ \includegraphics[angle=0,width=.45\textwidth]{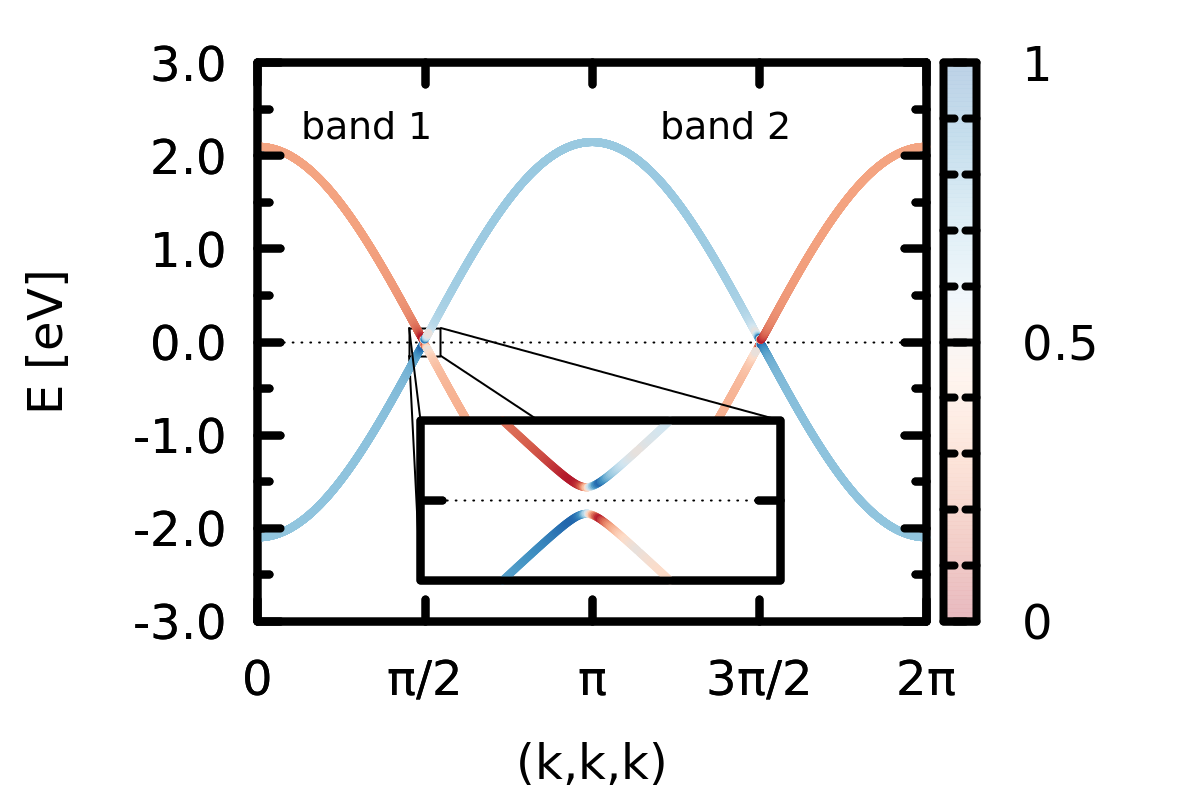}}
}

\subfloat[{\bf density of states.}]{
	{ \includegraphics[angle=0,width=.45\textwidth]{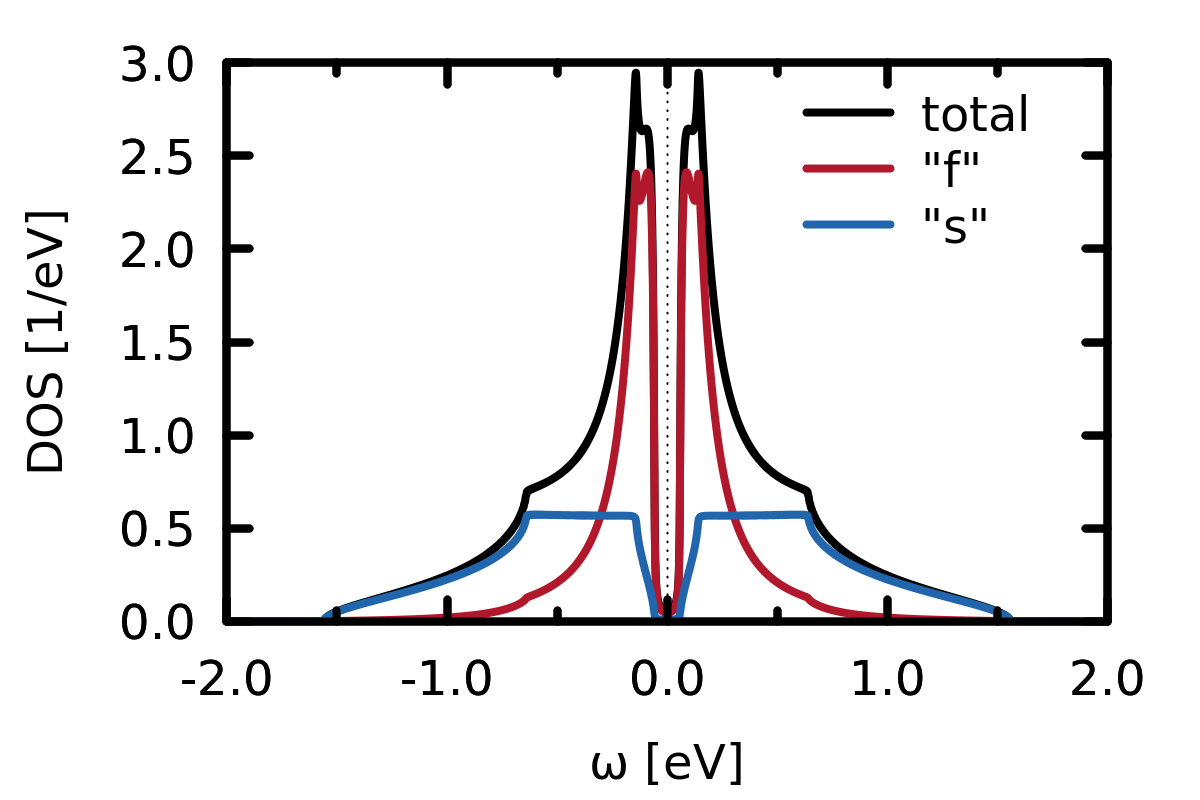}}
	{ \includegraphics[angle=0,width=.45\textwidth]{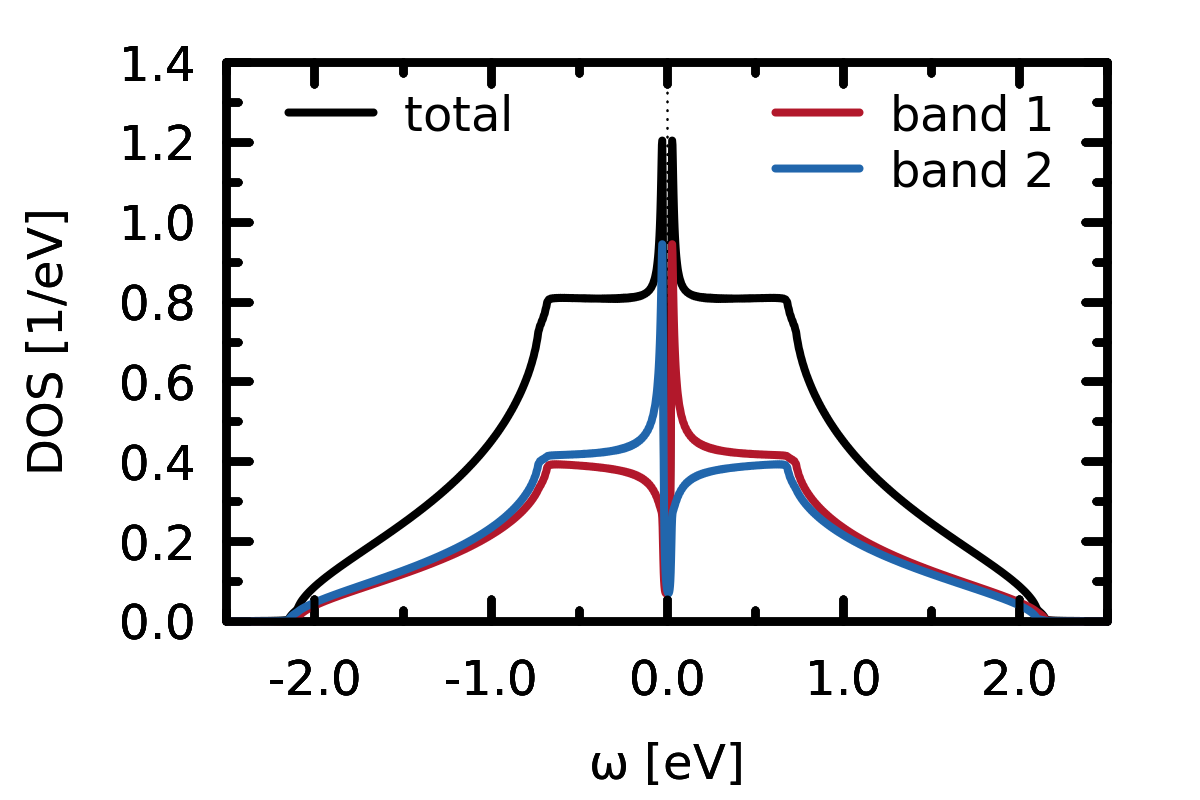}}
		}
				\caption{{\bf Non-interacting electronic structure of the models \eref{H2B}, \eref{HPAM2}.} Shown are (a) the dispersions and (b) the DOS
				of the 2-band Hubbard model (right), \eref{H2B}, and the periodic Anderson model (left), \eref{HPAM2}, in the non-interacting limit ($U=0$).
        Hubbard model: $t=V=0.25$eV ($\eta=1$), $\Delta^0_{dir}=50$meV. PAM: $t=0.25$eV, $V=0.3$eV. The Fermi level corresponds to zero energy.
				}
      \label{hubpam}
      \end{center}
\end{figure*}

The above discussed {\it ab initio} many-body results for prototype materials of correlated narrow-gap semiconductors
and Kondo insulators allowed for a quantitative description of diverse experimental observables, and helped elucidating the microscopic
origin of the observed anomalous behaviour, and identified crucial parameters that control the prominence of electronic correlation effects
in these system.
Here, we shall briefly return to the realm of reductionist models, with the goal to put the obtained realistic results
for individual compounds into a wider perspective.
We will focus the discussion on a single issue, namely the crossover temperature in the susceptibility $T^{max}_\chi$. 
We want to answer the question why, among correlated intermetallic semiconductors only FeSi, FeSb$_2$, and CrSb$_2$ display
a crossover from activated behaviour to a Curie-Weiss-like decay at experimentally accessible temperatures.
In \sref{band} we already surmised why closely related compounds such as FeAs$_2$ or RuSb$_2$ and RuSi display less or no
signatures of correlation effects. Here, we will put this analysis on a firmer footing, by studying the tendencies 
in both the crossover temperature $T^{max}_\chi$ and the peak value $\chi(T^{max}_\chi)$ of the susceptibility
for varying parameters. The models we consider are (i) a variant of a 2-band Hubbard model, and (ii) the periodic Anderson model.

\subsubsection{The model Hamiltonians}
\paragraph{(i) the 2-band Hubbard model.} We employ the Hamiltonian
\begin{widetext}
\begin{eqnarray}
H&=&\sum_{\svek{k},\sigma}(\epsilon_{\svek{k}}+\delta/2) (\cc_{\svek{k}1\sigma}\ca_{\svek{k}1\sigma} -  \cc_{\svek{k}2\sigma}\ca_{\svek{k}2\sigma})
+\sum_{\svek{k}\sigma}\left(V_{\svek{k}}\cc_{\svek{k}2\sigma}\ca_{\svek{k}1\pr\sigma}+ V^*_{\svek{k}}\cc_{\svek{k}1\sigma}\ca_{\svek{k}2\pr\sigma}\right)\nonumber\\
&&+\sum_{iLL\pr\sigma}U^{LL\pr} n_{iL\sigma}n_{iL\pr-\sigma} 
  +\sum_{iLL\pr\sigma}^{L\ne L\pr}  (U^{LL\pr}-J) n_{iL\sigma}n_{iL\pr\sigma}
\label{H2B}
\end{eqnarray}
\end{widetext}
at half-filling, using a cubic lattice with nearest neighbour hopping,  $\epsilon_{\svek{k}}=-2t\sum_{i=1}^{3} \cos(k_ia)$, a local crystal-field $\delta$, and a purely non-local inter-band hybridization
$V_{\svek{k}}=\eta\epsilon_{\svek{k}}$, $\eta\in\mathbb{R}$. The interaction is limited to density-density terms, and we use the cubic parametrization $U_{LL\pr}=U-2J(1-\delta_{LL\pr})$ with the Hubbard $U$ and the Hund's rule coupling $J$. 
In the non-interacting limit ($U=0$) the system has a direct charge gap $\Delta^0=\delta\frac{\eta}{\sqrt{1+\eta^2}}$: See \fref{hubpam}(a-b,right) for the 
dispersion and the density of states, respectively. We note that the orbital characters of the valence and conductions states at a given momentum are different,
as was found for FeSi and FeSb$_2$ within band-theory, see \sref{band}.%
\footnote{See, e.g., Refs.~\cite{PhysRevB.87.235104,PhysRevLett.114.185701} for related two-band Hubbard models that explore topological effects
via the inclusion of effective spin-orbit couplings.}%

\paragraph{(ii) the periodic Anderson model.}
We study the  PAM, \eref{HPAM}, in its particle-hole symmetric ($\epsilon_f=0$)%
\footnote{For the interesting physics of incoherence in the asymmetric PAM see, e.g., Ref.~\cite{PhysRevB.85.235110}.}
variant with local hybridization, $V_\svek{k}=V$:
\begin{equation}
H=\sum_{\svek{k}\sigma}\epsilon_\svek{k}\cc_{\svek{k}\sigma}\ca_{\svek{k}\sigma}+
V\sum_{\svek{k}\sigma}(\cc_{\svek{k}\sigma}\fa_{\svek{k}\sigma}+\fc_{\svek{k}\sigma}\ca_{\svek{k}\sigma})
+U\sum_{i}n^f_{i\sigma}n^f_{i-\sigma}
\label{HPAM2}
\end{equation}
on the 3d cubic lattice, $\epsilon_\svek{k}=-2t\sum_{i=1}^{3} \cos(k_ia)$, at half-filling. 
The non-interacting dispersion and density of states are
displayed in \fref{hubpam}(a-b, left), for a choice of parameters as indicated in the caption.
While the bare direct gap is given by $\Delta^0_{dir}=2V$, the bare indirect gap is much smaller, $\Delta^0_{indir}=\sqrt{D^2-4V^2}-D$, with $D=6t$ the half-bandwidth of the (unhybridized) $c$-electrons.

\medskip

Note that the orbitally resolved DOS of the above defined Hubbard (periodic Anderson) model neatly captures the qualitative aspects of the gap
in FeSi (\cbp). As in \cbp, both, valence and conduction states in the PAM are dominated by the $f$-contribution. 
In the 2-band Hubbard model, valence and conduction bands are instead of different orbital character, akin to FeSi---compare \fref{hubpam}(a) to \fref{DMFTCe32} (a,b).

\subsubsection{Dynamical mean-field theory results.} 
\begin{figure*}[!t]
  \begin{center}
	$\quad$ periodic Anderson model $\quad\qquad\qquad\qquad\qquad$ 2-band Hubbard model
	
	\subfloat[{dependence of $T^{max}_{\chi}$ on the Hubbard $U$.}]
	{
	{ \includegraphics[angle=0,width=.45\textwidth]{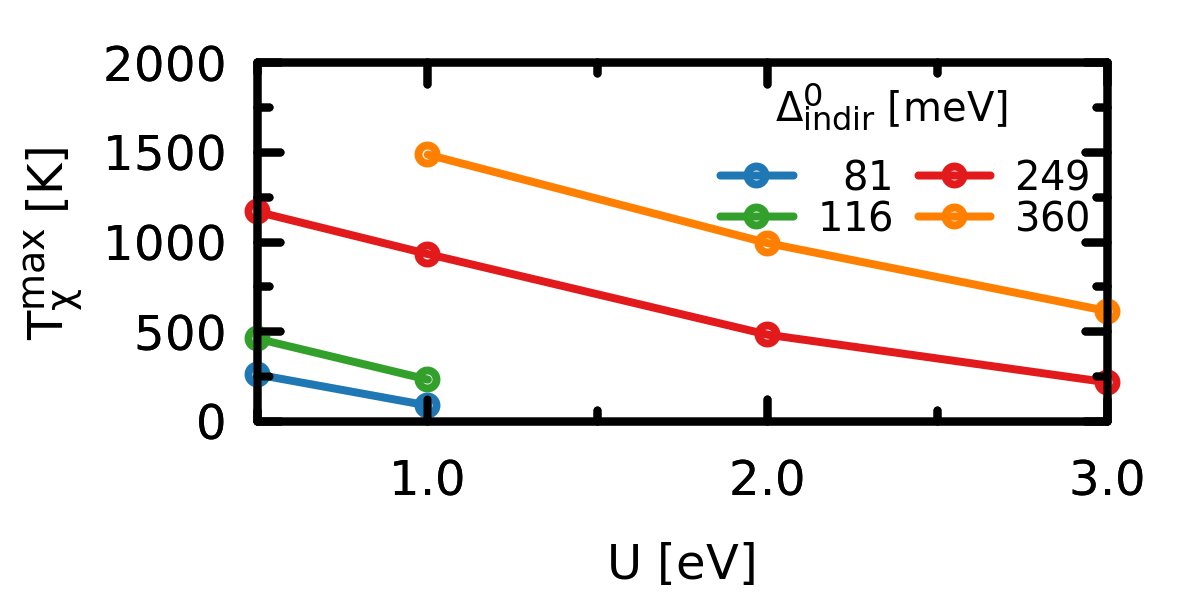}}
	{ \includegraphics[angle=0,width=.45\textwidth]{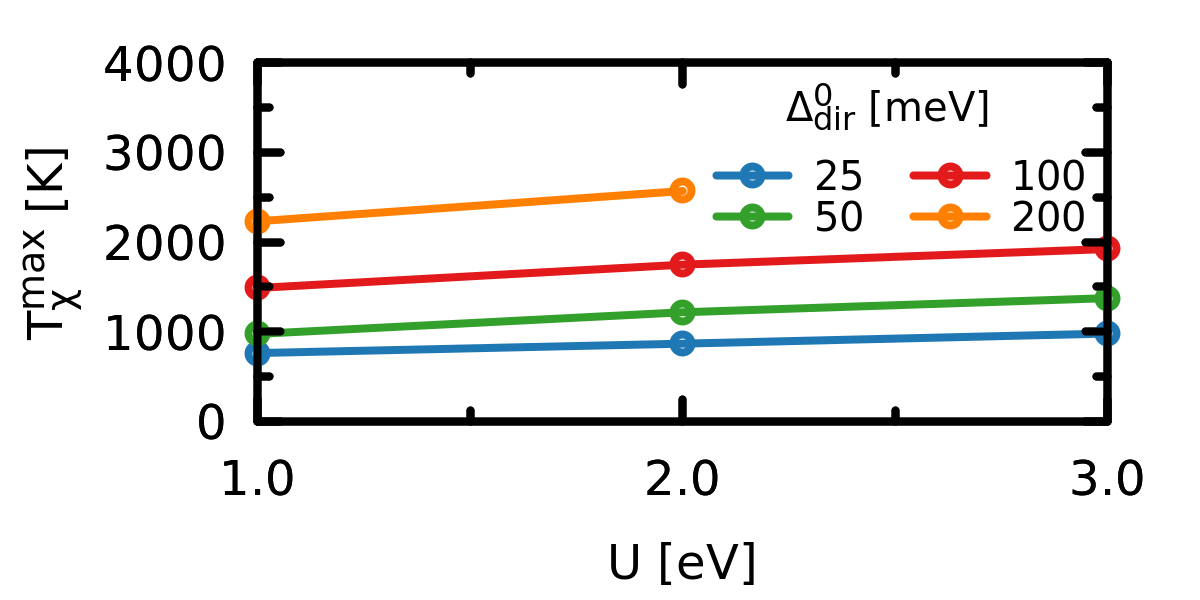}}
	}\\ 
	\subfloat[{dependence of $\chi(T^{max}_{\chi})$ on the Hubbard $U$.}]
	{
	{ \includegraphics[angle=0,width=.45\textwidth]{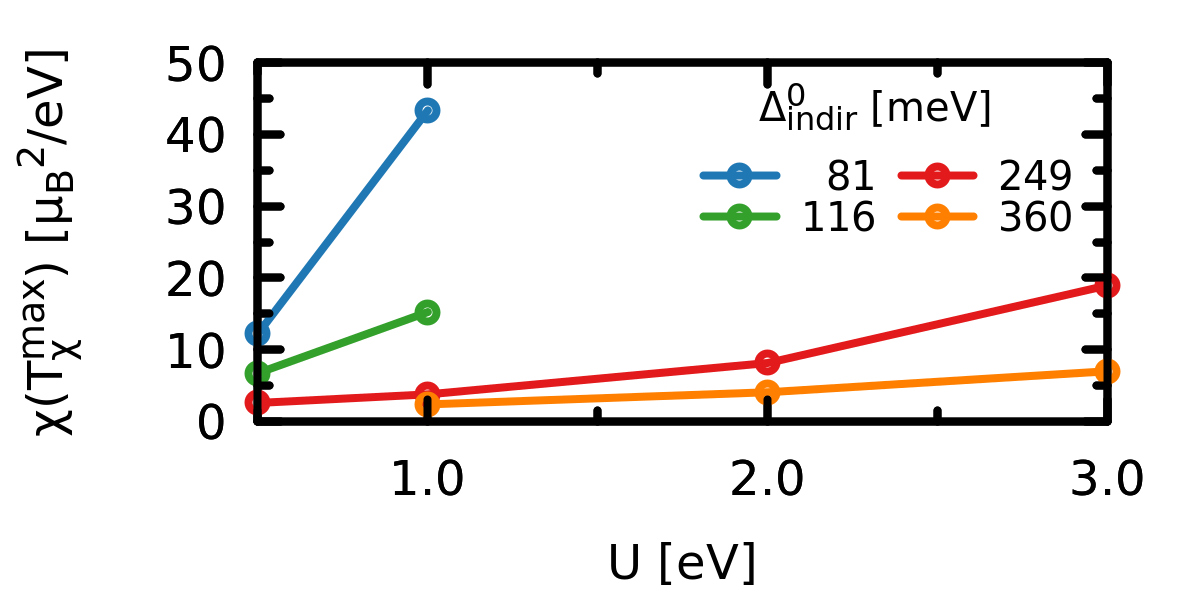}} 
	{ \includegraphics[angle=0,width=.45\textwidth]{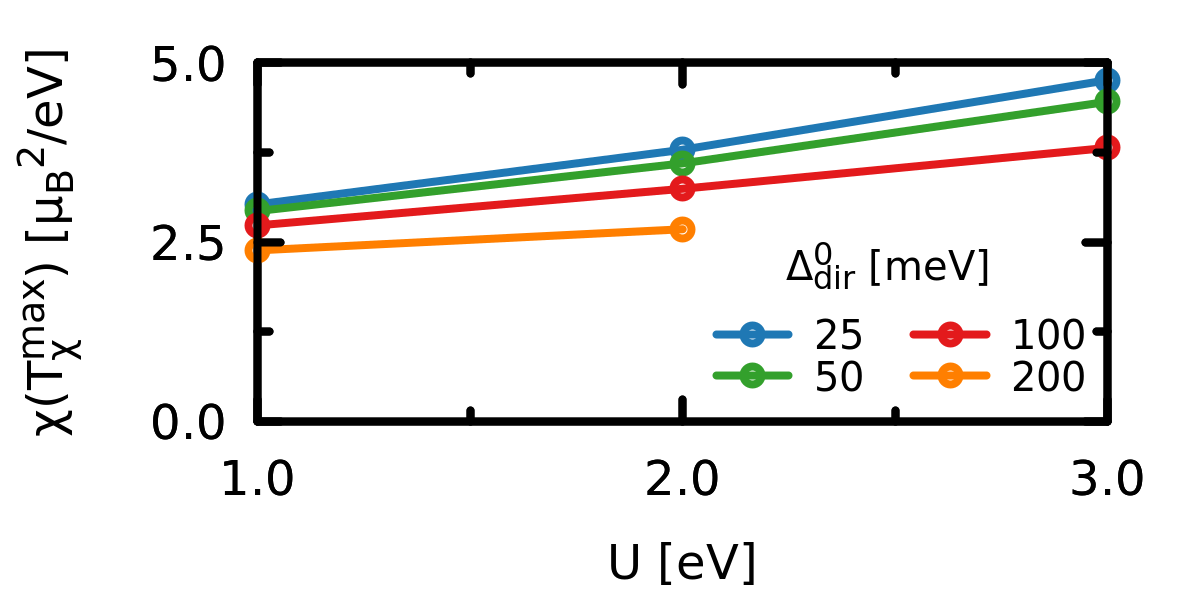}}
}	
				\caption{{\bf Dependence of magnetic fluctuations on the (bare) gap $\Delta^0$ and the strength of interactions $U$.} 
				$T^{max}_{\chi}$ and $\chi(T^{max}_{\chi})$ for the periodic Anderson model (left) and the 2-band Hubbard model (right).
				In all cases $t=0.25$eV. Results for the 2-band Hubbard model use a Hund's coupling $J=0$.
				}
      \label{hubpam2}
      \end{center}
\end{figure*}

We solve both models within DMFT, using the w2dynamics package\cite{w2dynamics}. 
Here, we do not allow for any long-range order, and leave the study of ordered phase for future work.
As a proxy for the experimental crossover in the spin response, we consider the maximum, $T^{max}_\chi$, in the {\it local} (DMFT) susceptibility
$\chi_{loc}^{\omega=0}(T)=g^2\int_0^\beta d\tau \sum_{LL\pr}\left \langle S_z^L(\tau)S_z^{L\pr}(0)  \right\rangle$%
\footnote{
In the PAM the uniform and the local magnetic susceptibility practically coincide down to $T^{max}_{\chi}$\cite{PhysRevB.85.165114}.
For a discussion of local vs.\ uniform susceptibility for the Hubbard model, see \sref{model} and \sref{pnas}.}.
We limit the sum over indices $L$, $L\pr$ to correlated orbitals and use $g=2$.
\Fref{hubpam2}
displays (a) $T^{max}_\chi$ and (b) $\chi(T^{max}_\chi)$ as a function of the Hubbard $U$ interaction for the PAM (left) and the Hubbard model (right),
for different values of the bare gap $\Delta^0$.
\begin{figure*}[!t]
\begin{center}
\subfloat{
	{ \includegraphics[angle=0,width=.45\textwidth]{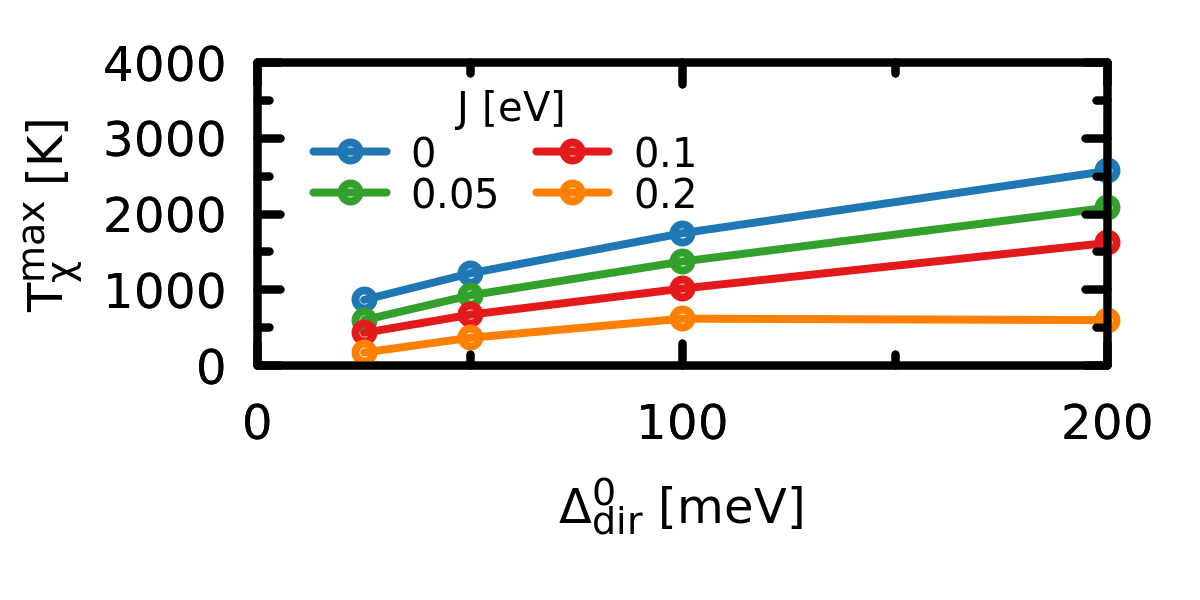}}
	{ \includegraphics[angle=0,width=.45\textwidth]{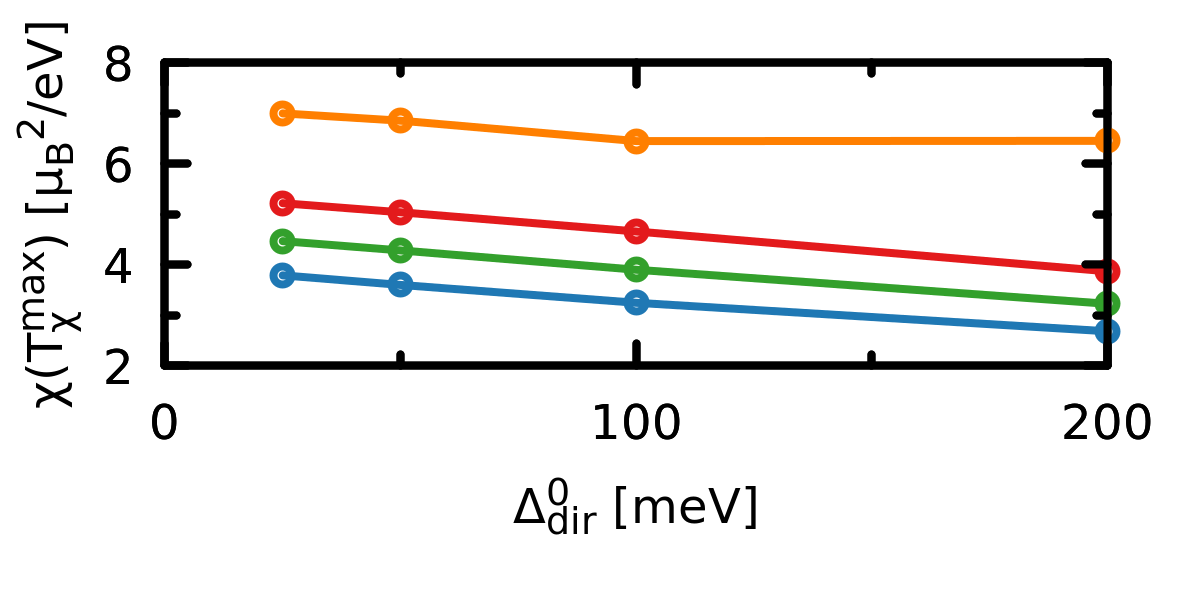}}
}
				\caption{{\bf Impact of the Hund's rule coupling $J$ on the magnetic response of the 2-band Hubbard model \eref{H2B}.} 
				$t=0.25$eV and $U=2$eV.
				}
      \label{hub2}
      \end{center}
\end{figure*}
For both models, $T^{max}_\chi$ quickly becomes inaccessibly high to experiments as the gap size grows.
This rationalizes, why enhanced susceptibilities cannot be seen in experiments for, e.g., FeAs$_2$ 
($\Delta_{\hbox{\tiny FeAs$_2$}}\sim 200$meV $\gg$ $\Delta_{\hbox{\tiny FeSb$_2$}}\sim 30$meV).

As to the influence of an increased Hubbard $U$ onto the crossover temperature, it grows slightly for the Hubbard model,
while it decreases strongly for the PAM.
In the Hubbard model, also the peak values of the susceptibility are quite insensitive to the Hubbard $U$, as well as to the gap size.
In the PAM, on the other hand, $\chi(T^{max}_{\chi})$ strongly depends on both $U$ and $\Delta^0_{indir}$.
This finding is congruent with the observation that the physics of Kondo insulators is more sensitive to external pressures than that of
correlated narrow-gap semiconductors (see \fref{deltaP}). Further, as was the case in the comparison between FeSi and \cbp\ in \fref{FeSiCe3trans}(b),
the susceptibility of the $f$-levels of the PAM is much larger than for the Hubbard model. Note, however, the
propensity of correlated narrow-gap semiconductors to exhibit ferromagnetic fluctuations rather than local-moment behaviour
implying a uniform susceptibility larger than the local one (see, e.g., the discussion in \sref{pnas}).

For the case of FeSi, realistic many-body calculations\cite{jmt_fesi} evidenced, as discussed in \sref{micro}, a strong dependence of the effective mass
and the lifetime of excitations on the Hund's rule coupling $J$. Here we study---in the Hubbard model context---the dependence of the magnetic response on $J$.
As displayed in \fref{hub2}, both the crossover temperature $T^{max}_\chi$ and the peak value of the (local) susceptibility strongly depend on the Hund's coupling.
Indeed the sensitivity on $J$ is notably more pronounced than the dependence on $U$ (see above).
The role of the Hund's coupling is to strongly enhance the susceptibility and to push the Curie-Weiss-like decay to much lower temperatures.
This tendency explains that while strong signatures of correlation effects are visible in the 3$d$-compound FeSi, the 4$d$-compounds RuSi and RuSb$_2$
with their much smaller Hund's rule $J$ (see \tref{tmarcs}), exhibit neither enhanced paramagnetism, nor Curie-Weiss-like behaviour.

Interestingly, in the Hubbard model, the magnitude (but not the crossover temperature) depends strongly on the strength of the hybridization:
For a constant gap $\Delta^0=\delta\cdot{\eta}/{\sqrt{1+\eta^2}}$, $\chi(T^{max}_{\chi})$ is larger for {\it smaller} $\eta$ (and thus larger $\delta$), i.e., for
weaker (and less dispersive) hybridizations.
This is congruent with the observation that $\chi_{loc}$ is much larger in the case of the PAM than for the 2-band Hubbard model.

In all, strong signatures of correlation effects in the magnetic response of narrow-gap semiconductors is favoured by a small gap, weak hybridizations, and
a large Hund's rule coupling. These prerequisites may explain why the discussed anomalous paramagnetic response is relatively rare. Indeed, contrary
to, e.g., ubiquitous mass enhancements, the occurrence of Curie-Weiss-like susceptibilities 
in $3d$-electron based semiconductors has so far only been established for
FeSi, FeSb$_2$, and CrSb$_2$.

\subsection{Electron-lattice effects}
\label{lattice}

\begin{figure}%
\includegraphics[width=0.45\textwidth]{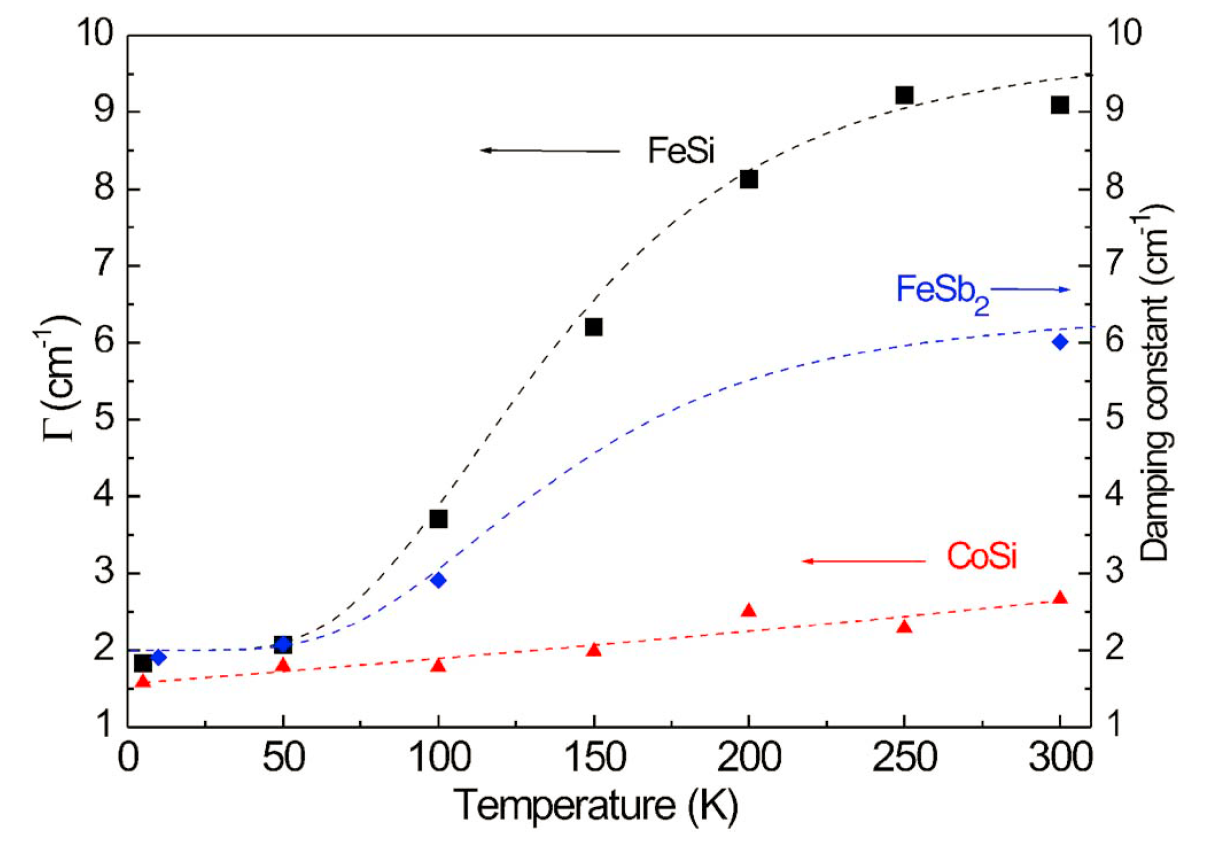}%
\caption{{\bf Temperature dependence of Raman/optical linewidths.} The symbols indicate the linewidth of the ``E'' mode (at $\sim22$meV) for FeSi and CoSi
and the damping of the infrared $B_{1u}$ mode (at $\sim24$meV) of FeSb$_2$; note that 10cm$^{-1}\approx1.2$meV. From Racu \etal\ \cite{racu:07C912}.}%
\label{phononwidth}%
\end{figure}

The realistic many-body calculations reviewed in the preceding sections were based on the Born-Oppenheimer approximation,
i.e., there was no interplay between electronic and phononic degrees of freedom.
Computations were even done for a fixed, temperature-independent lattice constant.
Given the sensitivity of correlated materials to minute changes in conditions, accounting for the volume expansion in FeSi
\cite{Vocadlo2002} might have noticeable results. In fact, its omission was blamed for slight discrepancies between the experimental and theoretical optical conductivity (see \sref{pnas}). 
This section is devoted to such electron-lattice effects: We will review experimental evidence for
 large electron-phonon couplings  in FeSi and FeSb$_2$,
discuss proposals for the consequences thereof for the electronic and phononic subsystems, as well as parse related {\it ab initio} calculations.
The influence of phonons onto the thermopower via the phonon-drag effect will be discussed in \sref{dianti} for the case of FeSb$_2$.
Let us also note in passing that phononic properties in their own right are quite interesting for some of the systems considered here\cite{PhysRevLett.120.016401}.

\subsubsection{Thermal conductivity.}
Let us briefly discuss the thermal conductivity $\kappa$ of FeSi\cite{PhysRevB.50.14933,PhysRevB.83.125209,Buschinger1997784,OUYANG201792}
and FeSb$_2$\cite{0295-5075-80-1-17008,sun_dalton,APEX.2.091102,doi:10.1063/1.4731251}, both of which---according to a Wiedemann-Franz analysis---are dominated by the lattice contribution for all measured temperatures ($T<300$K).
Interestingly, {\it ab initio} lattice dynamics calculations of the stoichiometric materials largely {\it overestimate} the thermal conductivity for, both, FeSb$_2$\cite{PhysRevB.89.035108} and FeSi, 
while simulations for semimetallic CoSi\cite{Pshenay-Severin2017} compare well to experiments\cite{PhysRevB.69.125111,Pshenay-Severin2017}.
Even more striking is the observation that in FeSi the thermal conductivity is larger in polycrystalline samples than for single crystals by a factor of two below 100K.
This is rather counter-intuitive, considering that boundary scattering is expected to suppress phonon-driven heat-conduction.
Finding smaller carrier densities in polycrystalline samples,
Sales \etal\ \cite{PhysRevB.83.125209} suggested that 
strong electron-phonon scattering is at the origin of the suppression of thermal conduction in single crystals.
In the case of FeSb$_2$, the behaviour is more conventional with thermal conduction significantly increasing 
for larger granularities\cite{Zhao_nano_fesb2,MRC:8871060} or sample sizes\cite{Takahashi2016}, see also \fref{Takahashi}.

\begin{figure}%
\includegraphics[width=0.45\textwidth]{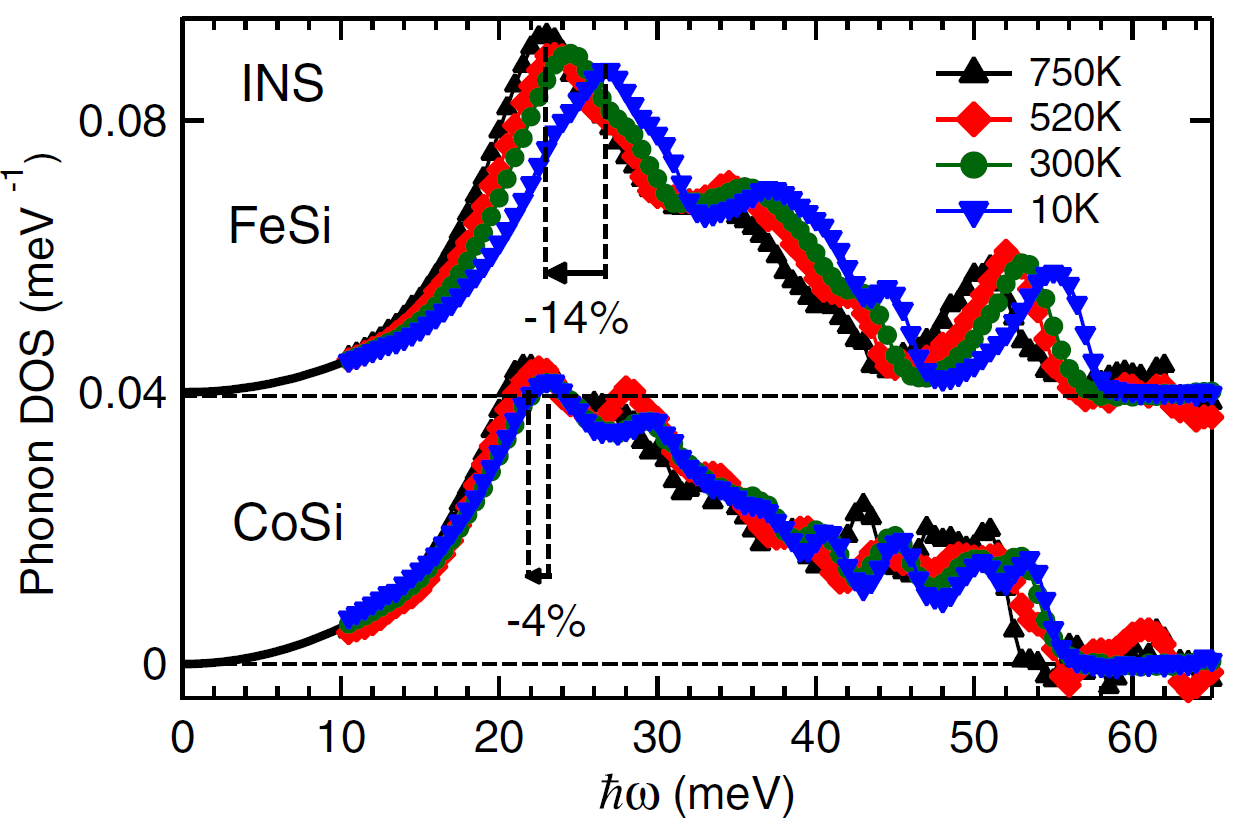}%
\caption{{\bf Phonon density of states.} The phonon DOS   for FeSi and CoSi were extracted from
inelastic neutron spectra on powder samples. Note the large softening of phonons in FeSi. From Delaire \etal\ \cite{Delaire22032011}.}
\label{delaire}%
\end{figure}

\subsubsection{Phonon structure and electron-phonon coupling.}

Some lattice properties such as the bulk modulus have already been discussed in conjunction with failures of band-theory for FeSi in \sref{fail}.
In that compound, unusual behaviour is also seen in other static lattice properties, such as elastic constants that exhibit an anomalous softening above $\sim100K$\cite{PhysRevB.82.155124}.

We now turn to dynamical lattice properties. Individual phonon modes can be studied by Raman and infrared spectroscopy.
Results for FeSi\cite{PhysRevB.55.R4863,Menzel2006718,PhysRevB.76.115103,racu:07C912,PhysRevB.79.165111,Ponosov2016}
and FeSb$_2$\cite{racu:07C912} indicate a notable softening of phonon modes and a strongly increasing phonon-linewidth (shown in \fref{phononwidth})
upon heating above  $\sim100$K.
These effects were interpreted as arising---via the electron-phonon coupling---from the metallization of FeSi above $\sim100$K.
This crossover provides for electron-phonon scattering events---absent in an insulator---and reduce the lifetime of phonons.
An analysis of the phonon lineshapes in Raman spectra by Racu \etal\ \cite{PhysRevB.76.115103} indeed pinpointed electron-phonon and not phonon-phonon interactions
to be at the origin of the temperature-induced broadening.
 
The phonon density of states can be accessed from inelastic neutron scattering (INS) and nuclear resonant inelastic scattering (NRIS) experiments.
In a seminal work\cite{Delaire22032011}, Delaire \etal\  presented a comprehensive INS study for FeSi in comparison to CoSi:
They found that the phonon softening in the metallic CoSi, which is of moderate 4\% (in the range of 10-750K), can be fully accounted for by the temperature-induced volume expansion.
Indeed, using experimental thermal expansion data, the experimental softening is congruent with computations in the quasi-harmonic approximation, and
more sophisticated {\it ab initio} molecular dynamics simulations give similar results.
In the case of FeSi, an inspection of the thermal expansion would suggest a phonon softening of a magnitude, 5\%, comparable to CoSi.
Instead, Delaire \etal  \cite{Delaire22032011} found in FeSi a softening three times as large: 14\%.
At low temperatures, the phonon dispersions of FeSi are in reasonable agreement with DFT calculations \cite{Delaire22032011,PhysRevB.83.125209}.
Above $\sim100$K, molecular dynamics simulations still give a good description, while the quasi-harmonic approximation now fails to account for the observed softening
of phononic dispersions, strongly suggesting a mechanism linked to the electronic degrees of freedom.

Parshin \etal\ \cite{Parshin2014} used the NRIS technique and resolved the fine structure of the partial phonon DOS associated with
the vibration of iron atoms. Moreover, they were able to extract phonon linewidths: Congruent with the Raman and optical spectroscopy, these
increase notably above 150K.

The empirical connection between phonon energies and temperature-induced changes in the electronic dispersion
has been illustrated by Delaire \etal, see \fref{delaire2}: 
\begin{enumerate}
	\item 
If there is no notable rearrangement of the electronic density of states at the Fermi level, phonon energies are modified only
through the thermal volume expansion, and the quasi-harmonic (QH) approximation captures the associated mode softening.
\item If the system evolves from a gapped electronic spectrum at low temperature to a state with finite weight at the Fermi level at high $T$,
an anomalous softening beyond the QH approximation occurs, in conjunction with a broadening of phonon linewidths.
The former is caused by electronic screening of inter-atomic potentials, while the latter originates from electron-phonon scattering.
 This behaviour is seen in FeSi (see above) and FeSb$_2$\cite{zaliznyak2014fesb2}, as well as Nb$_3$Sn where the low-temperature gap
is owing to superconductivity\cite{PhysRevLett.30.214}. Similar effects are also seen when metallizing FeSi through doping\cite{PhysRevB.91.094307}.
\item Conversely, if the electronic density at the Fermi level decreases with temperature, e.g., as a consequence of a
(heavy) Fermi liquid losing coherence, a stiffening of phonons is expected as force constants get unscreened. This phenomenon was indeed observed
in V-based A15 compounds \cite{PhysRevLett.101.105504}, as well as bcc V-based alloys\cite{PhysRevB.77.214112}.
\end{enumerate}

\subsubsection{Influence of vibrations onto the electronic structure.}

While these scenarios suggest a large influence of the electronic structure onto the phonons, lattice vibrations may conversely
modify the electronic structure. Indeed, via the electron-phonon coupling, the thermal activation of phonons provides scattering channels that may decrease the 
lifetime of electronic excitations. Given that FeSi metallizes through the occurrence of incoherent spectral weight at the Fermi level, lattice vibrations can at least accelerate the  crossover to the metallic state.

Looking at the literature of FeSi, the interplay of electrons and phonons could be interpreted as a chicken-and-egg questions:
Does the phonon system merely react to the (correlation-induced) evolution of the electronic and magnetic degrees of freedom, or are lattice vibrations themselves a
major driver of the metallization? Both standpoints have been forwarded on the basis of experimental and theoretical studies:

\begin{figure}%
\includegraphics[width=0.45\textwidth]{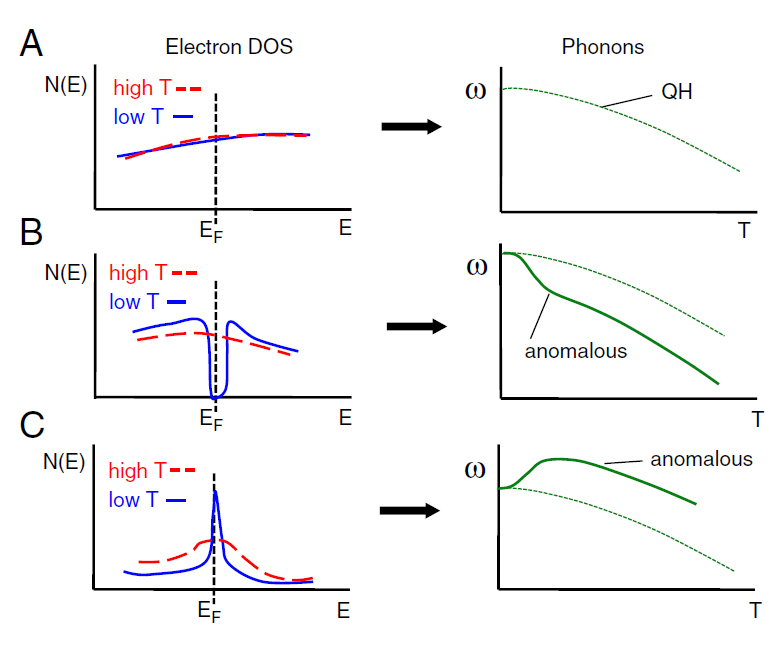}%
\caption{{\bf Impact of changes in the electronic structure onto phonon dispersions.} 
On the left three possible situations for temperature-induced changes in the electronic structure are shown: (A) no notable changes between low and high temperatures,
(B) crossover from a narrow-gap semiconductor at low temperature to a metallic state at high temperature, (C) a crossover from a coherent metal to an incoherent metal with rising temperature. On the right are shown the expected implications for the phonon dispersions in the presence of electron-phonon coupling: (A) the softening of modes is governed by thermal expansion of the lattice and is well described by the quasi-harmonic (QH) approximation. (B) the increased coupling between electrons and phonons induces an anomalous softening, (C) the decreased coupling between electrons and phonons leads to a hardening of modes. From Delaire \etal~\cite{Delaire22032011}.}%
\label{delaire2}%
\end{figure}

\paragraph{Thermal disorder.}

Jarlborg \cite{PhysRevB.59.15002,JARLBORG1997143,PhysRevB.76.205105,0034-4885-60-11-003} pioneered the view that 
lattice properties are predominantly responsible for the 
insulator-to-metal crossover in FeSi.%
\footnote{The metallization of a Kondo insulator via the electron-phonon coupling was studied theoretically e.g., in the Holstein-Kondo lattice model by Nourafkan and Nafari\cite{PhysRevB.79.075122}.}
Using a static supercell approach with a Gaussian distribution of atomic displacements,
he evidenced a gap closure at around 300K, found an increasing magnetic susceptibility of reasonable magnitude (at least when including
exchange enhancement factors of 3.5-6), a sensible specific heat, as well as spectral weight transfers in the optical conductivity over a range of more than 3.5eV.
This thermal disorder picture was later supported by the molecular dynamics (MD) simulations of Delaire \etal\ \cite{Delaire22032011}.
Moreover, investigating the effect of thermal disorder on inter-atomic force constants,
Stern and Madsen \cite{PhysRevB.94.144304} concluded that thermal disorder in conjunction with the thermal expansion can explain most of the anomalous softening of acoustic phonon branches. 

A few comments are in order: (i) in the MD calculations\cite{Delaire22032011} FeSi metallizes only for temperatures reaching the order of $k_BT\approx\Delta_{DFT}/2$, congruent
with the mean-square average of atomic displacements being controlled by $k_BT$. Experimentally, FeSi already metallizes for temperatures corresponding to a much smaller fraction of the gap.
(ii) the thermal disorder induces changes in the electronic structure on all energy scales, explaining why, in this scenario, optical weight is redistributed over large energy scales. However, the optical spectral weight redistribution $\Delta n$ (cf.\ \sref{pnas}) is, in the thermal disorder picture, a monotonously increasing function of energy up to 3.5eV\cite{PhysRevB.76.205105}, contrary to experimental findings (cf.\ \fref{FeSiDMFT1}).%
\footnote{See also Ref.~\cite{optic_prb} for a system with large electron-phonon coupling, where optical spectral weight transfers have been quantitatively described without accounting for thermal disorder.}
 (iii) the thermal disorder picture is unlikely to explain the occurrence of the Curie-Weiss-like decay in the magnetic susceptibility
at high temperatures.

\paragraph{Magneto-elastic coupling.}
An interesting proposal comes from Krannich \etal\ \cite{Krannich2015}: Using inelastic neutron and neutron resonant spin-echo spectroscopy, the authors evidence that
in FeSi the relative change with temperature of both the phonon linewidth and the phonon intensity neatly tracks the relative evolution of $3k_BT\cdot\chi(T)$, with $\chi(T)$ being
the uniform magnetic susceptibility.
The latter can be interpreted---in a local moment picture---as the square of an effective fluctuating magnetic moment $\mu_{eff}(T)$ (cf.\ the discussion in \sref{sus} and \sref{spinmodel}). This empirical observation strongly suggests a direct link between the emergence of (ferro)magnetic fluctuations and the phonon self-energy.%
\footnote{Note that also in the Kondo insulator YbB$_{12}$ a coupling between magnetic and phononic properties was suggested\cite{Rybina2007,0953-8984-24-20-205601}: While no perceptible
phonon softening occurs above the system's coherence temperature $T^*\approx50$K, the phonon intensity of modes (whose symmetry is thought to be commensurate with that of
relevant magnetic excitations\cite{0953-8984-24-20-205601}) significantly drops around $T^*$.
}
Using density functional perturbation theory in conjunction with a phenomenological broadening of the electronic DOS to mimic an incoherence-driven metallization
the authors computed phonon linewidths. Finding a significantly too small scattering rates (1meV vs.\ the experimental 3-5meV), they concluded that a metallization in the absence of magnetic fluctuations cannot account
for the observed phonon line-broadening.
As far as the phonon energies are concerned, Krannich \etal\ \cite{Krannich2015} found that the softening (of phonons at the R point in the phonon Brillouin zone) 
can be grouped into three temperature regimes: (i) at low temperatures the quasi-harmonic approximation applies, (ii) above 100K a regime of substantially increased softening
occurs, (iii) above 300K the softening resumes a rate similar to regime (i). This behaviour is shown schematically in \fref{delaire2} (B, right) and suggestively coincides
with the insulating, crossover, and metallic regime witnessed in spectroscopies probing charge degrees of freedom as well as transport properties.

\paragraph{The phonon spectator picture.}
Parshin \etal\ \cite{PhysRevB.93.081102,Parshin2016} performed nuclear inelastic scattering experiments of FeSi as a function of temperature and under pressure.
The rational behind this experiment is to disentangle the effects of (i) an increasing amplitude of atomic vibrations with raising temperature from (ii)
the thermally induced volume expansion.
Loosely speaking, pressure is used to reduce the volume to a value that is otherwise realized at lower temperatures.

As previously evidenced \cite{Parshin2016} the lowest peak in the phonon DOS has a sudden increase when heating above $T_c\sim 180$K.
Parshin \etal\ interpret $T_c$ as the critical temperature of the electronic subsystem, i.e., the characteristic scale for which the metallization occurs.
Applying pressure to FeSi at room temperature, a phonon DOS comparable to that at $T_c\sim 180$K (and ambient pressure)
is obtained at $P_c\sim 2$GPa\cite{PhysRevB.93.081102}. Comparing the two cases---(i) $T=180$K, $P\sim0$GPa and (ii) $T=300$K, $P\sim2$GPa---the crucial observation is that
what is similar at both conditions is the unit-cell volume, while the mean-square amplitude of atomic vibrations, $\left\langle u\right\rangle$, is largely different.
This finding has led Parshin \etal\ to dismiss the thermal disorder picture (which is driven by $\left\langle u\right\rangle$) and to assign
the change in unit-cell volume (instead of the phonons) as the dominant lattice contribution to the insulator-to-metal crossover in FeSi.
In that sense, the influence of the lattice is a static one, while phonons are mere spectators, not actors, to the metallization crossover.
It should be noted, however, that the electronic degrees of freedom in the cases (i) and (ii) are not quite identical: 
While pressure reduces the magnetic susceptibility, it is unlikely that the value of $\chi$ at $T=180$K and ambient pressure is reached by applying 2GPa at $T=300$K\cite{doi:10.1143/JPSJ.68.1693}. Moreover, while pressure increases the gap in activation-law fits\cite{Bauer1997794,PhysRevB.63.115103}, 
pressure does not undo the effects of increasing temperature from 180 to 300K in the resistivity\cite{PhysRevB.63.115103}.
Also, it should be noted that from the electronic point of view (see the optical conductivity in \fref{FeSiCe3spec}(b)), as well as the Raman linewidth (see \fref{phononwidth}), the reference temperature $T_C=180$K is already far above the metallization threshold.

\paragraph{Synthesis.}

In most of this chapter we have described theories that focus on the aspect of electronic correlation effects to describe the anomalous electronic, optical and magnetic properties of certain narrow-gap semiconductors. In this section we have instead reviewed experimental evidence, as well as some theoretical investigations that point to 
strong signatures of the electronic crossover in the phonon subsystem. To which degree the changes in the lattice degrees of freedom are a consequence of the electronic crossover,
or vice versa, is an unresolved issue. 
It is the author's opinion that effects of electronic correlations and the electron-phonon coupling, to an extent, conspire in causing the insulator-to-metal crossover.
Yet, the anomalous magnetic behaviour in FeSi and related systems may signal that the prevailing role is played by electronic correlation effects.
A full description of the coupled electron-phonon system is, however, still lacking. 
Indeed the interplay of electronic correlations and lattice degrees of freedom beyond the Born-Oppenheimer approximation is a challenging avenue for future research.

\begin{framed}
	\noindent
	{\bf Theories of correlated narrow-gap semiconductors: key points}
	\begin{itemize}
		\item signatures of correlation effects 
		in diverse observables can be quantitatively captured with realistic many-body simulations.
		\item insulator-to-metal crossovers are linked to incoherence induced by (ferromagnetic) spin fluctuations.
		\item lifetimes and effective masses in $d$-electron-based insulators are propelled by Hund's physics. 
		\item Kondo-insulating (covalent) behaviour can coexist with the suppression of quantum fluctuations by local crystal-field (ionic) physics.
		Indeed, we propose an orbital-selective Kondo insulator scenario for FeSi.
		\item important interplay of electronic and lattice degrees of freedom.
	\end{itemize}
\end{framed}

\section{Thermoelectricity}
\label{thermo}

The topic of this chapter is thermoelectricity in correlated narrow-gap semiconductors.
We begin \sref{thermointro} with a brief general introduction into the field of thermoelectrics:
We introduce relevant quantities, and review in \sref{optimize} current efforts to improve thermoelectric performance.
More details can be found in excellent books, e.g., Refs.~\cite{Zlatic,behnia}, or review articles\cite{PICHANUSAKORN201019,Shakouri2011,0034-4885-79-4-046502,doi:10.1080/09506608.2016.1183075,doi:10.1021/acs.chemrev.6b00255,ADMA:ADMA201605884,Heeaak9997}.
\Sref{correlS} then gives an overview over potential impacts of electronic correlations  onto thermoelectricity.
To explore the effects of simple many-body renormalizations onto the thermopower of a semiconductors, \sref{limit} discusses
an instructive analytical model. As a by product of the
latter, it is recapitulated that the thermopower owing to electronic diffusion has an upper bound in a coherent semiconductor.
Including effects of {\it in}coherence---ubiquitous in correlated materials---is shown to dramatically change transport
properties, to the effect that semi-classical Boltzmann approaches fail to describe them, as discussed in \sref{KubBoltz}.

This prelude sets the stage for the discussion of thermoelectric effects in the material classes that are the subject of this review.
As in the previous chapters, we shall pick representative materials for each class and review experimental findings and
theoretical efforts to describe them. 
Each subsection will discuss and highlight different aspects of electronic correlations and their impact on thermoelectricity:

\begin{framed}
	\begin{itemize}
	\item Silicides, \sref{silicides}: effects of incoherence and effective masses onto the thermopower.
	\item Marcasites, \sref{dianti}: limits of a purely electronic picture of thermoelectricity, the phonon-drag effect, couplings
	of phonons to charge carriers with large effective masses.
  \item Skutterudites, \sref{skutts}: large consequences of unreliable electronic structure calculations.
\item Heuslers, \sref{heuslers}: materials on the verge of thermoelectric particle-hole symmetry, impact of defects and disorder.
\end{itemize}
\end{framed}

Besides the materials discussed here, there are numerous other classes of compounds that are of interest from the thermoelectric
point of view, while at the same time probably hosting at least moderate correlation effects.
Among these materials are half-Heuslers\cite{0953-8984-26-43-433201}, 
other 4$d$ transition-metal intermetallics, pnictides, and chalcogenides\cite{Gonsalves2014} not covered here, complex oxides, 
sulfides, and selenides\cite{Herbert2016}.

\subsection{General considerations}
\label{thermointro}

\begin{figure}[!t]
  \begin{center}
	{\includegraphics[angle=0,width=.45\textwidth]{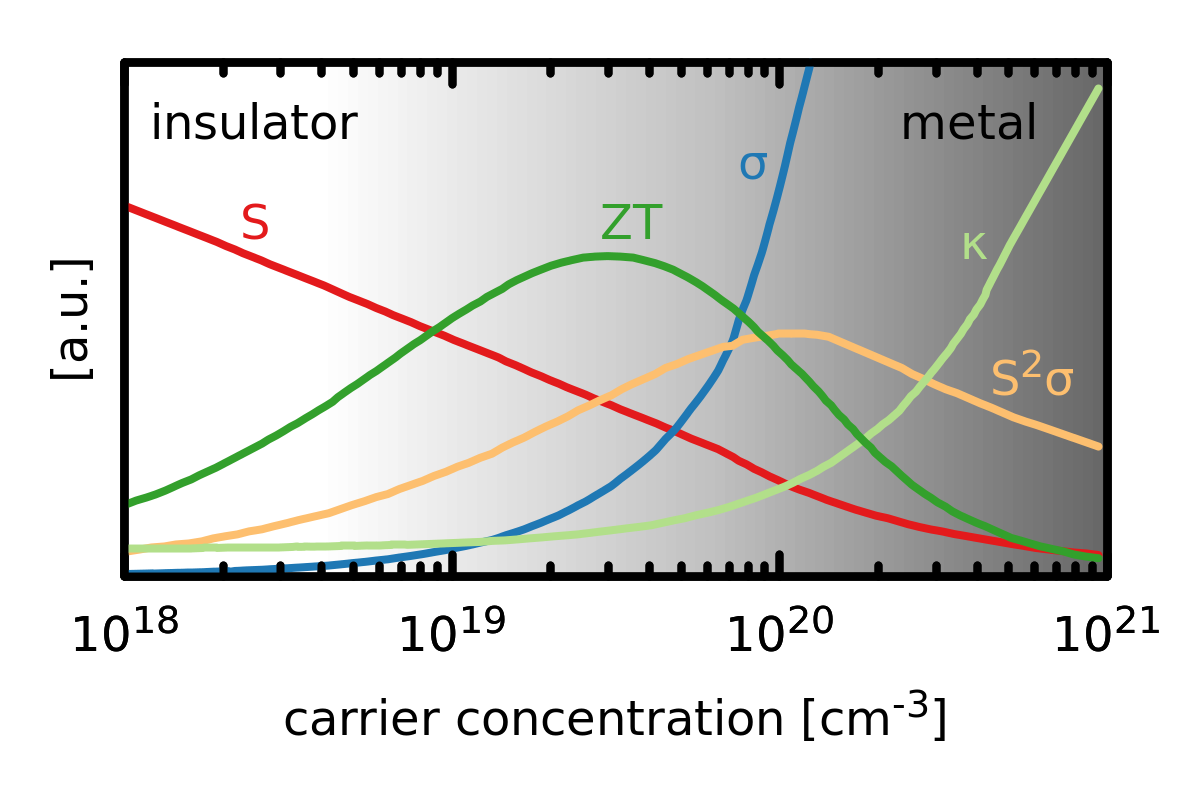}}
	    \caption{{\bf Optimizing thermoelectric performance.} 
			Shown is the typical carrier -density dependence of quantities relevant for thermoelectric devices at a given temperature: the conductivity $\sigma$, the thermopower $S$, the powerfactor $S^2\sigma$, the thermal conductivity $\kappa$, and the thermoelectric figure of merit $ZT$.}
      \label{ZT}
      \end{center}
\end{figure}

Thermoelectricity is the effect of a voltage drop $\Delta V$ occurring across a sample that is subjected to a temperature gradient $\Delta T$---or {\it vice versa}.
The most common mechanism of thermoelectricity is the net diffusion of electrons (or holes) from the hot toward the cold regions of the sample.
The displaced charges build up an electrical field that counteracts the diffusion, which defines, at equilibrium,  the  Seebeck coefficient $S=\Delta V / \Delta T$, also known as the thermopower.
The materials' properties that govern the efficiency of a thermoelectric conversion process---leading to electricity generation, or refrigeration---can be gathered into the so-called dimensionless figure of merit $ZT=S^2\sigma T/\kappa$. Besides the thermopower $S$, relevant quantities are the conductivity $\sigma=1/\rho$ and the thermal conductivity $\kappa$.%
\footnote{The minimal realization of an actual thermoelectric device consists of a thermocouple made of two materials. Then the overall figure of merit becomes $Z\bar{T}=\frac{(S_1-S_2)^2\bar{T}}{\left( \sqrt{\rho_1\kappa_1}+\sqrt{\rho_2\kappa_2}\right)^2}$ for the average temperature $\bar{T}$\cite{ioffe}.
Nonetheless, an individual material's $ZT$ still gives a good indication as to its potential thermoelectric performance.
}
As a rough guide, $ZT$ needs to be above unity---roughly corresponding to a conversion efficiency of 10-15\% at high temperatures---to be of any technological interest. 
The combination $S^2\sigma$ that enters $ZT$ is referred to as the power factor.
It is apparent that an optimization of the latter requires satisfying
conflicting demands. This dilemma is illustrated in \fref{ZT} where we display the typical dependency of relevant quantities on the carrier concentration:
The thermopower is typically large in insulators ($S\propto\Delta/T$ with the gap $\Delta$, see also below, \eref{eq:Ssc}) and small in metals (where, at low T, $S\propto T$). The conductivity, on the other hand, is large in
metals and exponentially suppressed in insulators. 
Hence, the power factor $S^2\sigma$ is largest in between, at densities corresponding to degenerate semiconductors.
The thermal conductivity, $\kappa=\kappa_e+\kappa_{ph}$, consists of an electronic and a phononic part.
While in metals the electronic contribution is preponderant, in the doped semiconductor regime the lattice  contribution by far dominates the detrimental thermal conduction
and severely limits $ZT$.

The conductivity and the thermopower can be expressed, e.g., in the Kubo linear response formalism. One finds\cite{PhysRevLett.80.4775,pruschke_review,PhysRevB.65.075102,jmt_fesb2}
\begin{equation}
S_{\alpha\beta}=-\frac{k_B}{\left|e\right|}\frac{A_1}{A_0}, \qquad\qquad\qquad \sigma_{\alpha\beta}=\frac{2\pi e^2}{\hbar}A_0
\label{Kubo}
\end{equation}
where%
\footnote{if neglecting electron-hole interactions, so-called vertex-corrections.}
 the response functions are
\begin{equation}
A_n=\int d\omega \beta^n (\omega-\mu)^n \left(-\frac{\partial f}{\partial \omega}\right) \sum_\svek{k}Tr \left[ v^\alpha_\svek{k}A_\svek{k}(\omega)v^\beta_\svek{k}A_\svek{k}(\omega)\right]
\label{KuboA}
\end{equation}
Here, $k_B$ is the Boltzmann constant, $\beta=1/(k_BT)$ the inverse temperature, $f(\omega)$ the Fermi function, and $\mu$ the chemical potential.
The trace goes over orbital indices $L$ and the Fermi velocities, $v_\svek{k}^\alpha$, and the spectral functions, $A_\svek{k}(\omega)$, are matrices in orbital space.
Further, $v^\alpha_\svek{k}$ has an orientational $\alpha$-dependence, which turns both $\sigma$ and $S$ into tensors, whose elements depend on the Cartesian directions $\alpha$, $\beta$ of the temperature
gradient and the electrical field, respectively.

\subsubsection{Optimizing thermoelectric performance.}
\label{optimize}
To increase thermoelectric efficiencies two main avenues are currently pursued\cite{ADMA:ADMA200600527,Snyder2008,Yang2016,GAYNER2016330}:%
\footnote{For interesting proposals that are beyond this classification see, e.g., the recent Refs.~\cite{Sun_CoSb3,Zhao2017}.} 

\paragraph{Phonon engineering.} Following the ``phonon-glass, electron crystal'' paradigm\cite{Slack1995},
the dominant lattice thermal conduction is sought to be limited by disordered unit-cells\cite{Snyder2008}, 
structuring samples on the nano\cite{ADMA:ADMA200600527,PICHANUSAKORN201019} and the mesoscale\cite{Biswas2012}, introducing
geometric constraints via superlattices \cite{PhysRevB.47.12727,doi:10.1063/1.1374458,BEYER2002965,Dresselhaus2013} or nanowires\cite{Hochbaum_Si_nano,Bokai_Si_nano},
or including loosely bound ``rattling'' atoms/modes\cite{doi:10.1063/1.357750,doi:10.1063/1.361828,PhysRevB.56.15081,PhysRevB.58.164,Voneshen2013}.
The global rationale is to introduce phonon scattering channels on length scales that are
long compared to the electronic mean-free path. Therewith, while reducing detrimental thermal conduction, the electronic powerfactor ideally is unhampered.
This strategy of phonon engineering is bounded by the physics of the amorphous limit, yet it has led to significant progress, yielding $ZT$ values in excess of 2.
A further route to reduce $\kappa$ are enhanced phonon-phonon scatterings owing to anharmonicities\cite{PhysRevB.85.155203,Zhao2014} linked to, e.g., ferroelectric lattice
instabilities\cite{Delaire2011,Li2015,Hong2016,PhysRevB.97.014306}.

\paragraph{Band engineering.}
The second major strategy is the enhancement of the power factor, i.e., an optimization of the electronic degrees of freedom.%
\footnote{See, however, \sref{dianti} and e.g., Refs.~\cite{FeSb2_Marco, Zhou01122015} for a phonon-mediated contribution to the thermopower.}
Besides reaching an optimal compromise between the thermopower and the conductivity (see \fref{ZT}), the crucial observation is that
the thermopower (somewhat akin to the Hall coefficient) is sensitive to the type of charge carriers. Indeed $S$ vanishes for particle-hole symmetric systems
and increases the more asymmetric the electronic spectrum is.
A large particle-hole asymmetry can be achieved by a variety of mechanisms, e.g.,

\begin{itemize}
	\item impurity levels that introduce sharp (=high-mass) density-of-states peaks on one side of the Fermi level\cite{Heremans554,C1EE02612G}.
\item band convergence, i.e., a high degeneracy of selective excitations \cite{Pei2011,Tang2015,PhysRevB.92.085205,Zhang2016}.
\item low-dimensional bands in three-dimensional crystals\cite{PhysRevLett.110.146601,PhysRevLett.114.136601} that lead to large effective masses.
This paradigm achieves the ideas of Hicks and Dresselhaus\cite{PhysRevB.47.16631} without the need for electronic confinement via quantum-well structures.
\item ``pudding mold'' band-structures as proposed by Kuroki and Arita \cite{doi:10.1143/JPSJ.76.083707} (see also Refs.~\cite{PhysRevB.78.115121,Usui2014}).
Here, the thermopower is enhanced by a large particle-hole asymmetry in the Fermi velocities.
\end{itemize}

The first three points advocate a large asymmetry of the excitation spectrum $A_\svek{k}(\omega)$, while the latter emphasizes the role of
transition matrix elements $v_\svek{k}$. 
In the simplest picture there is, however, a compensation between the spectral and the velocity route towards large $S^2\sigma$:
Commonly the Fermi velocity is approximated by the group velocity, $v(\svek{k})=1/
\hbar\partial_\svek{k}\epsilon(\svek{k})$%
\footnote{this is known as the Peierls approximation, see, e.g., Refs.~\cite{PhysRevB.67.115131,optic_prb}.}.
Then it is apparent that flat bands, that create sharp spectral features, have a small velocity (at least in some direction), which in turn hampers the response functions
(in that direction).%
\footnote{
As a nice illustration let us mention the red pigments $\alpha$-HgS (vermillion) and CeSF. While the good absorption properties
of the former owe to transitions of large amplitude between dispersive bands, in the latter relevant transitions have low transition matrix elements
but large non-dispersive spectral weight to transition into\cite{jmt_cesf}.
}

\medskip

The above mechanisms have been evidenced and/or studied for current state-of-the-art thermoelectrics,
all of which are \udl{doped intermetallic narrow-gap semiconductors}, such as Pb(Se,Te), Sn(Se,Te), and
Bi$_2$Te$_3$.
These materials combine \udl{good powerfactors} $S^2\sigma$ with reasonably low
thermal conductivities $\kappa$ to reach a favourable thermoelectric conversion efficiency.
Moreover, the intrinsically low thermal conduction in, e.g., 
lead telluride\cite{Delaire2011} alloys\cite{LaLonde2011526} could be combined with
 the above mentioned band-convergence \cite{Pei2011}, mesoscale phonon engineering \cite{Biswas2012},
as well as with impurity states \cite{Heremans554}.


\subsubsection{Electronic correlations and thermoelectricity: friends or foes?}
\label{correlS}

The propensity of correlated materials to exhibit large response functions begs the question whether electronic correlation effects
could provide means to improve upon the conventional thermoelectric materials just mentioned---all of which are {\it uncorrelated} narrow-gap semiconductors.
Several scenarios suggest an affirmative answer\cite{Zlatic,Herbert2016}:

On a model
level \cite{PhysRevLett.67.3724,PhysRevLett.80.4775,PhysRevB.65.075102,grenzebach_2006,PhysRevB.87.035126,PhysRevLett.109.266601},
as well as for realistic calculations of metallic
compounds\cite{oudovenko:035120,haule_thermo,PSSA:PSSA201300197,PhysRevLett.117.036401}, correlation effects
were shown to enhance the thermopower.
In fact, the low-temperature thermopower in correlated metals is boosted by $1/Z$, with the quasi-particle weight $Z<1$\cite{Behnia2004}.
However, since the conductivity scales with $Z^2$\cite{jmt_fesb2}, the powerfactor is actually independent of $Z$.%
\footnote{In fact, this is different from the semi-classical picture, where $\sigma\propto {m^*}^{-1}$, i.e., $\sigma\propto Z$ for a momentum-independent self-energy.}

Noting that the thermopower is a measure for the entropy per carrier\cite{ioffe,PhysRevLett.80.4775,PhysRevB.72.195109,PhysRevB.76.085122,PhysRevLett.117.036401,PhysRevB.87.035126},
there are further sources of a large thermoelectric response:
Indeed, in  the limit of large temperatures, the thermopower is entirely determined by the number of possible microstates (degeneracy) $g$,
\begin{equation}
S \xrightarrow{T\rightarrow\infty} \frac{1}{e}\left(\frac{\partial s}{\partial N}\right)_{E,V}=-\frac{k_B}{|e|}\frac{\partial \log g}{\partial N}
\label{Heikes}
\end{equation}
where $s$ is the entropy of the system, $N$ the number of particles, and $E$ and $V$ indicate that the derivative is taken at constant energy and volume.
\Eref{Heikes} is called Heikes formula \cite{PhysRevB.13.647}.
In correlated materials, the degeneracy is enhanced by the typically large number of spin- and orbital degrees of freedom.
In fact, the large thermopower in cobaltates has been proposed to originate from such entropic contributions\cite{PhysRevB.62.6869,wang_spinentropy}. 
Heikes-type formulae were demonstrated to also hold  for other oxides \cite{PhysRevB.83.165127}.

Large effects of entropy are also expected near phase transitions.
Indeed, large thermopower enhancements have been evidenced near structural instabilities.
The prime example is copper selenide, Cu$_2$Se,  a compound whose 
 transition from a low-temperature semi-conducting $\alpha$-phase 
 to a superionic $\beta$-phase is accompanied by a lambda-shaped thermopower anomaly \cite{ADMA:ADMA201302660,doi:10.1063/1.4827595}.
Similar effects have been evidenced also in other semiconductors with mixed ionic and electronic conduction, e.g., 
Ag$_{10}$Te$_4$Br$_3$\cite{Nilges2009} or AgCuS \cite{doi:10.1021/ja5059185}, where structural phase transitions induce sudden changes in the {\it ionic}
mobility.
In this context let us highlight the recent work of Sun \etal\ \cite{Sun_CoSb3}, who evidenced that a large temperature-dependence in
  {\it electronic} mobilities
induces pronounced features in the thermopower as well as in the Nernst coefficient.
A strong variation of the 
electronic
mobility could be engineered by fabricating a junction between materials of vastly different mobilities.
Large mobility gradients could, however, also be realized  in individual correlated materials in the vicinity of ordering instabilities, metal-insulator transitions 
or other phenomena that cause abrupt changes in electronic relaxation rates.
This mechanism of mobility gradients was suggested to be active for Ni-doped skutterudite CoSb$_3$ \cite{Sun_CoSb3} (cf.\ \sref{skutts}), as well as heavy-fermion compounds such
as CeRu$_2$Al$_{10}$\cite{Sun_CoSb3} and CeCu$_2$Si$_2$\cite{PhysRevLett.110.216408}.
Also the recent observation of colossal thermopower anomalies in doped double perovskites\cite{0022-3727-51-6-065104} could be interpreted in this vein.
Whether critical phenomena in the multifarious structural, magnetic, and electronic phase transitions in correlated materials may lead more generally to 
useful behaviours in the thermopower is yet an open question.

Further, large thermopowers occur when the excitation spectrum is \udl{strongly peaked}. Indeed as suggested by Mahan and Sofo \cite{Mahan07231996},
``the best thermoelectric'' has a very narrow peak within a couple of $k_BT$ around the Fermi level.
 Besides the paths of band-structure engineering on the one-particle level mentioned above, weakly dispersive, narrow features
can be generated by many-body enhancements
of effective masses. Particle-hole selective renormalizations can moreover result in an increased asymmetry, benefiting thermoelectricity, as suggested by Haule and Kotliar \cite{haule_thermo}. See, e.g., \tref{marcasitesqsgw} for the notable particle-hole asymmetric scattering rate in FeSb$_2$ and FeAs$_2$.

Many correlated narrow-gap semiconductors do indeed exhibit large thermoelectric effects: FeSi, FeSb$_2$, CrSb$_2$, and Fe$_2$VAl all have intrinsically large thermopowers
and powerfactors (see \tref{table1}). Yet, being insulators, this large response occurs at rather low temperatures (cf.\ the next section). The same applies to Kondo insulators, (cf.\ \sref{KI}). 
Although the undoped materials are hence unsuited for standard thermoelectric applications, such as waste-heat recovery at elevated temperatures or Peltier refrigeration down from ambient conditions, they do have potential for cooling applications or sensors at cryogenic temperatures\cite{Heremans2016}, and could significantly reduce the need for liquid helium. 
The figure of merit of correlated narrow-gap semiconductors and Kondo insulators, are however minuscules, because of their strong lattice thermal conductivity.
To address this issue, it has been suggested\cite{doi:10.1021/nl104090j} to combine the aspect of correlation effects with that of geometric
constraints mentioned earlier.

Conventional thermoelectrics are degenerate narrow-gap semiconductors. For correlated narrow-gap semiconductors, however,
doping typically does not help achieving high powerfactors at larger, more generally useful temperatures (see however sections \ref{skutts}, \ref{heuslers}). 
A likely reason for this drawback are the reduced electronic lifetimes that accompany the beneficial mass enhancements.
In fact, the scattering rate $\Gamma=-\Im\Sigma(\omega=0)$ is often not small compared to the gap, $\Gamma/\Delta\sim 1$.
Ss discussed in \sref{pnas} for FeSi, this leads to an incoherence-induced metallization at intermediate temperatures, and doping yields a full 
metal rather than a degenerate semiconductors.

\subsubsection{Limits of a purely electronic picture of thermoelectricity.}
\label{limit}

We find it instructive to discuss here a simple analytical model\cite{jmt_fesb2} of a correlated but coherent narrow-gap semiconductor.
By this we mean a system in which effective masses are renormalized and quasi-particle weights reduced, while
excitations remain coherent (scattering rate $\Gamma\ll 1$). 
The model serves several purposes: 
(i) it provides a minimal understanding of the coherent low-temperature regimes of systems such as FeSi,
(ii) it shows the limits of applicability of band-like transport calculations, and
(iii) it establishes an upper bound for the purely electronic thermopower in insulators.

Our starting point is \eref{Kubo} and \eref{KuboA}. We consider a half-filled system of two bands, whose non-interacting dispersions
are electron gas-like (quadratic in three dimensions), $\epsilon^k_{c,v}=\pm\Delta_0/2\pm\frac{\hbar^2k^2}{2m^0_{c,v}}$, for both (v)alence and (c)onduction states,
and separated by a crystal-field gap $\Delta_0$.
Using the Peierls approximation\cite{PhysRevB.67.115131,optic_prb} yields Fermi velocities $v_\svek{k}=\frac{1}{\hbar}\partial_\svek{k}\epsilon_\svek{k}$.
Correlation effects shall be encoded in a momentum-independent self-energy of the form $\Sigma^{c,v}(\omega)=\Re\Sigma^{v,c}(0)+(1-1/Z^{c,v})\omega-i\Gamma^{c,v}$.
Then the interacting dispersions will be given by
\begin{eqnarray}
\xi^\svek{k}_{c,v}&=&\pm\Delta/2\pm\frac{\hbar^2\svek{k}^2}{2m^*_{c,v}} 
\label{eqxi}
\end{eqnarray}
where $\Delta=Z\Delta_0$ is the renormalized charge gap, $m^*/m^0=1/Z$ the effective mass of the carriers, and the origin of the chemical potential is chosen as the mid-gap point.
Each excitation carries the reduced spectral weight $Z^{c,v}$, and has an inverse lifetime/scattering rate $\Gamma^{c,v}$.

In the limit of a coherent system ($\Gamma\ll 1$) and a large gap ($\beta\left|\Delta/2-\mu\right|\gg1$), \eref{KuboA} can be solved analytically to give
\begin{eqnarray}
A_n^{c/v}&=& (\pm 1)^n \frac{3\lambda}{\sqrt{2\pi^5\beta^3}}  e^{-\beta\Delta/2} \left\{ [\beta(\pm\mu-\Delta/2)]^n  - \frac{5}{2} n \right\}  
\end{eqnarray}
where all carrier specific parameters have been collected in
\begin{eqnarray}
\lambda_{c,v}&=&\frac{Z^2}{\Gamma}\frac{m_*^{5/2}}{m_0^2} e^{\pm \beta\mu}
\end{eqnarray}
Then, the thermopower acquires the form
\begin{eqnarray}
S&=&\oo{ \left| e \right| T} \left( \mu-\frac{\Delta}{2}\delta\lambda\right) -\frac{5}{2}\frac{k_B}{ \left| e \right|}\delta\lambda 
\label{eq:Ssc}
\end{eqnarray}
where we have defined the asymmetry parameter $\delta\lambda$ by
\begin{eqnarray}
\delta\lambda&=&\frac{\lambda^c-\lambda^v}{\lambda^c+\lambda^v}
\label{dl}
\end{eqnarray}
which itself depends on $T$ and $\mu$.
Hence, a large thermopower is achieved by an interplay of the gap, $\Delta$, the chemical potential, $\mu$, and
the particle-hole asymmetry, $\delta\lambda$. The latter is itself influenced by the bare masses ($m^0_{c,v}$), the chemical potential ($\mu$), and by the many-body effects of encoded in mass enhancements ($m^*_{c,v}$), scattering amplitudes
($\Gamma_{c,v}$), and quasiparticle weights ($Z_{c,v}$).

\begin{figure}[!t]
  \begin{center}
	{\includegraphics[angle=0,width=.45\textwidth]{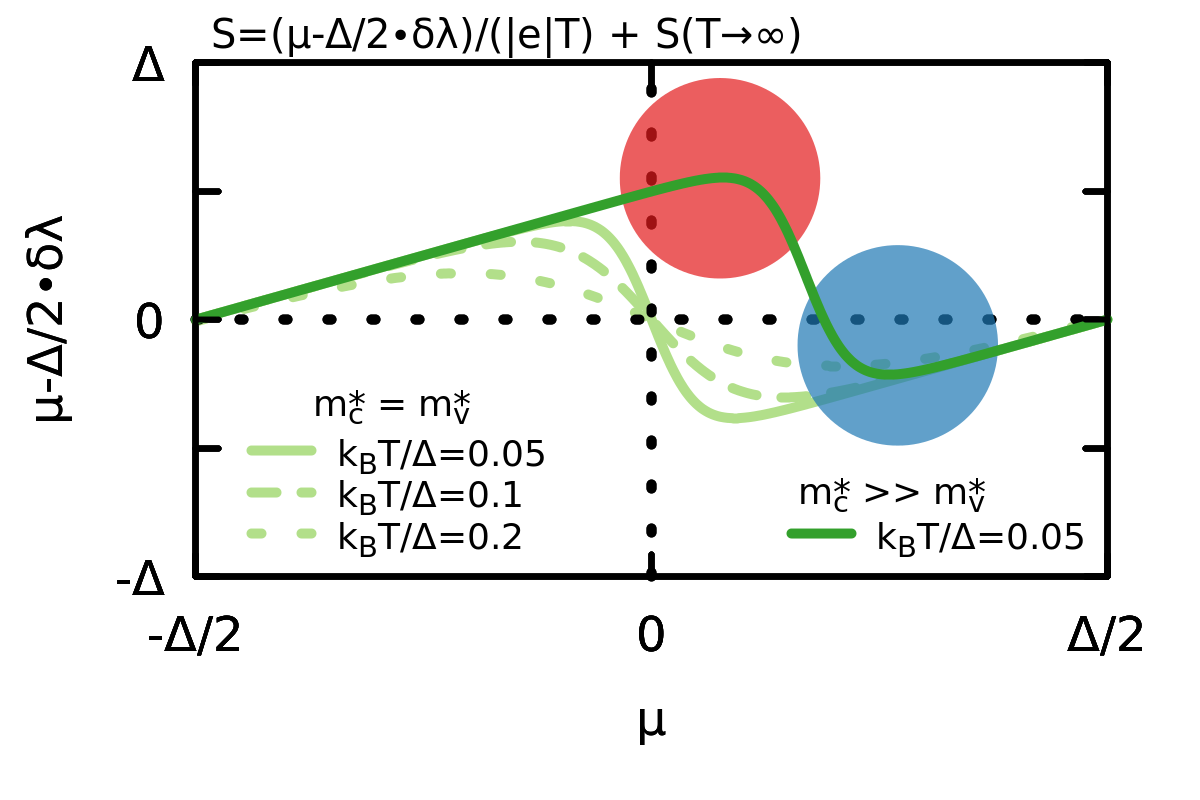}}
	    \caption{{\bf Understanding the thermopower of large gap coherent semi-conductors.} 
			Shown is the coefficient $\mu-\Delta\delta\lambda/2$ of the $1/T$ order of the thermopower of a coherent semiconductor as a function of the
			chemical potential within the gap, $-\Delta/2\le\mu\le\Delta/2$, for different temperatures. The coefficient vanishes in the particle-hole symmetric case (light green) at $\mu=0$. Particle-hole asymmetry, here created by different effective masses for valence $m^*_v$ and conduction $m^*_c$ carriers, lead to enhanced thermopowers.
			Note that since $|\delta\lambda|\le 1$ is bounded, the maximally possible thermopower is $|S|<\Delta/(|e|T)+5/2k_B$, see \eref{Smax}.}
      \label{modelphi}
      \end{center}
\end{figure}

\paragraph{Low-temperature behaviour.}
Let us have a look at the typical behaviour at low temperatures, where in systems such as FeSi or FeSb$_2$, coherent transport is expected.
In \fref{modelphi} we display the coefficient of the $1/(|e|T)$ term, namely $\mu-\Delta\delta\lambda/2$, as a function of the chemical potential inside the gap, $-\Delta/2\le\mu\le\Delta/2$, for different temperatures $T$.

In the particle-hole symmetric case, parameters for the valence and the conduction band are equal, i.e., $m^*_c=m^*_v$, and the asymmetry parameter becomes $\delta\lambda=\tanh(\beta\mu)$. Then, the thermopower vanishes for all temperatures, since $\mu=0$. However, small deviations from particle-hole symmetry, that could, e.g., be caused by impurity states pinning the chemical potential, can result in large thermopowers. In fact, the thermopower in narrow-gap semiconductors is quite sensitive to defects or off-stoichiometries, as also indicated by oftentimes notable sample dependencies (cf., e.g., sections \ref{skutts} and \ref{heuslers}).

Larger thermopowers are possible in case the electronic structure is strongly particle-hole asymmetric.
\Fref{modelphi} displays a case, where the effective mass of the conduction electrons is much higher than that of holes in the valence band.
Then, a large positive thermopower is attainable. This situation is for example realized in FeSi.
Doping the system with electrons---i.e., moving $\mu$ to the right in \fref{modelphi}---will drive the system through a sign change in $S$, to realize a negative thermopower albeit of smaller absolute magnitude. Experimentally this is seen, e.g., for Co-doped FeSi, see \fref{Sakai}. There, undoped ($x=0$, red) FeSi
has a large positive thermopower at low temperatures. Adding electrons ($x=0.05$, blue) switches the sign and yields an overall smaller $S$.

As seen in \fref{FeSiS} and \fref{Sakai}(right), the thermopower falls off quicker than $1/T$ above its peak value. This is potentially indicative
of a breakdown of the simplifications made in the current, simple model. Most notably, assuming the scattering rate to be vanishingly small might no longer
be a viable approximation. This issue will be discussed more in the next \sref{KubBoltz}, as well as for FeSi in \sref{silicides}. 

\medskip

\paragraph{An upper bound for the diffusive thermopower in semiconductors.}
Within the framework considered here, the thermopower cannot be arbitrarily large.
Indeed, the asymmetry parameter in \eref{dl} is bounded: $|\delta\lambda|\le 1$.
With the further constraint of the chemical potential remaining inside the gap, $|\mu|<\Delta/2$,
the thermopower of the coherent semiconductor thus obeys
\begin{equation}
\left|S(T)e\right|\le \Delta/T + 5/2k_B.
\label{Smax}
\end{equation}

In the coherent regime, correlation effects can hence enhance the thermopower by selective renormalizations (of, e.g., the effective mass).
The fundamental limit \eref{Smax}, which simply corresponds to fully eliminating either the electron or hole contribution to thermoelectric transport,
can, however, not be surmounted. While derived for a two-band model, also effects of degeneracies cannot change this fact.
Through similar considerations, Goldschmid and Sharp\cite{Goldsmid1999} actually suggested that the charge gap of an insulating system could be
read off from the peak value $S_{max}=S(T_{max})$ of the thermopower for which they assessed: $S_{max}\approx\Delta/(2eT_{max})$.

In \sref{dianti} (see in particular \fref{Sscaled}) we will encounter systems that violate the upper bound \eref{Smax}. 
As discussed there, it is believed that in these materials a major part of the thermopower comes from the so-called phonon-drag effect
that allows to circumvent the bound of the purely electronic thermopower considered here.

\subsubsection{Electronic correlations: a challenge for {\it ab initio} transport methodologies.}
\label{KubBoltz}

In currently used thermoelectric materials electronic correlations do not play a significant role.%
\footnote{This does not mean that all these materials are well-described by DFT. Indeed band-gaps in the undoped parent compounds are consistently underestimated.}
This allowed the successful use of band-structure methods to describe their electronic structure, and Boltzmann-derived methodologies\cite{PhysRevB.95.235137} to compute their
transport properties.
Applications of Boltzmann codes---such as those of Refs.~\cite{Madsen200667,Pizzi2014422}---are diverse and include theoretical studies of individual materials\cite{PhysRevB.68.125210,PhysRevB.76.085110,PhysRevB.81.195217}, theoretical support for experimental studies\cite{PhysRevB.79.075120,Rhyee2009}, and even high-throughput studies for thermoelectrics\cite{Madsen2006,PhysRevX.1.021012,C2CP41826F,PhysRevB.92.085205,PhysRevB.94.045122,Gorai2017,PhysRevX.6.041061}.
These simulations indeed reach predictive character: It was, e.g., suggested that PbSe could be a low-cost alternative to the commonly used, but expensive PbTe\cite{PhysRevB.82.035204}. This proposal was supported by subsequent experiments\cite{ADMA:ADMA201004200}.

In materials with sizable electronic correlation effects, however, both band-theory and the standard Boltzmann approach become inaccurate.
While we discussed the implications of many-body effects onto the electronic structure in \sref{theo}, we focus here on the challenges
that correlation effects pose for transport methodologies.
For this, we compare the Kubo approach (see \eref{Kubo}) with the commonly used semi-classical Boltzmann approach in the relaxation time approximation.
The latter can be obtained from the Kubo formalism in the limit of a vanishing scattering rate, $\Gamma\longrightarrow 0$.
Then, e.g., the conductivity and the thermopower of \eref{Kubo} simplify to
\begin{equation}
\sigma_{\alpha\beta}=\frac{e^2}{\Gamma} \sum_{\svek{k}L} \left(-\frac{\partial f}{\partial \omega}\right)_{\omega=\epsilon_{\svek{k}L}} \,  v^\alpha_{\svek{k}L}v^\beta_{\svek{k}L}
\label{Boltz}
\end{equation}
and
\begin{equation}
S_{\alpha\beta}= \frac{e}{T} \frac {\sum_{\svek{k}L} \left( \epsilon_{\svek{k}L}-\mu  \right)\left(-\frac{\partial f}{\partial \omega}\right)_{\omega=\epsilon_{\svek{k}L}} \,  v^\alpha_{\svek{k}L}v^\beta_{\svek{k}L} }  { \sum_{\svek{k}L} \left(-\frac{\partial f}{\partial \omega}\right)_{\omega=\epsilon_{\svek{k}L}} \,  v^\alpha_{\svek{k}L}v^\beta_{\svek{k}L}}
\label{BoltzS}
\end{equation}
where, for simplicity, we assumed a momentum- and band-independent scattering rate $\Gamma_{\svek{k}L}=\Gamma$ and have omitted inter-band contributions.
Then, the Boltzmann thermopower becomes independent of the scattering rate $\Gamma$.

\begin{figure*}[!t]
  \begin{center}
	{\includegraphics[angle=0,width=.32\textwidth]{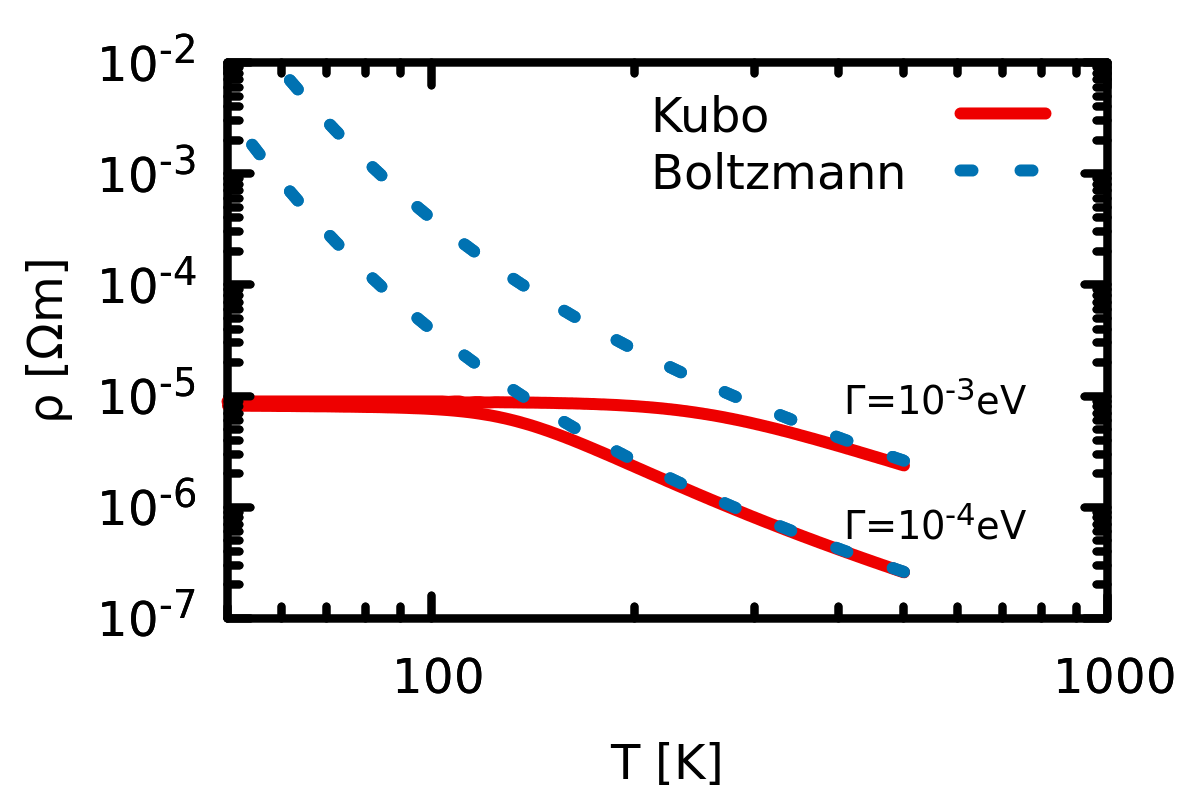}}
	{\includegraphics[angle=0,width=.32\textwidth]{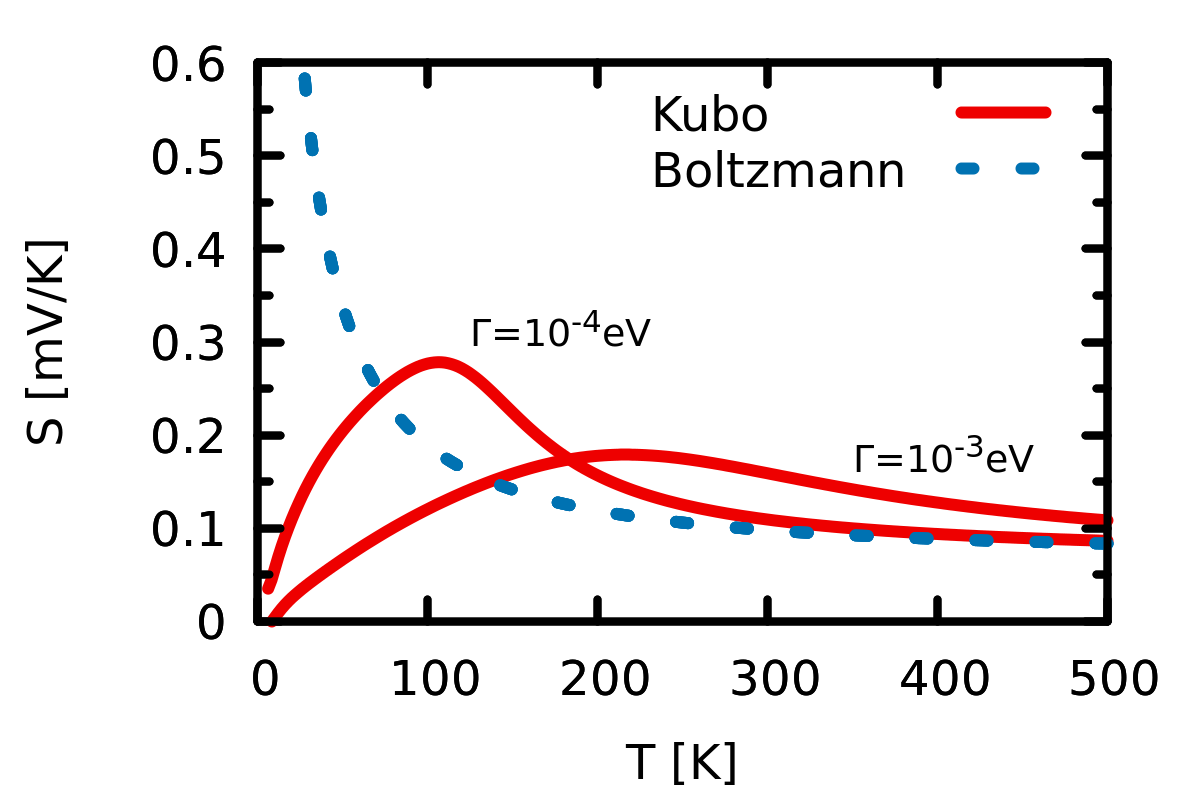}}
  {\includegraphics[angle=0,width=.32\textwidth]{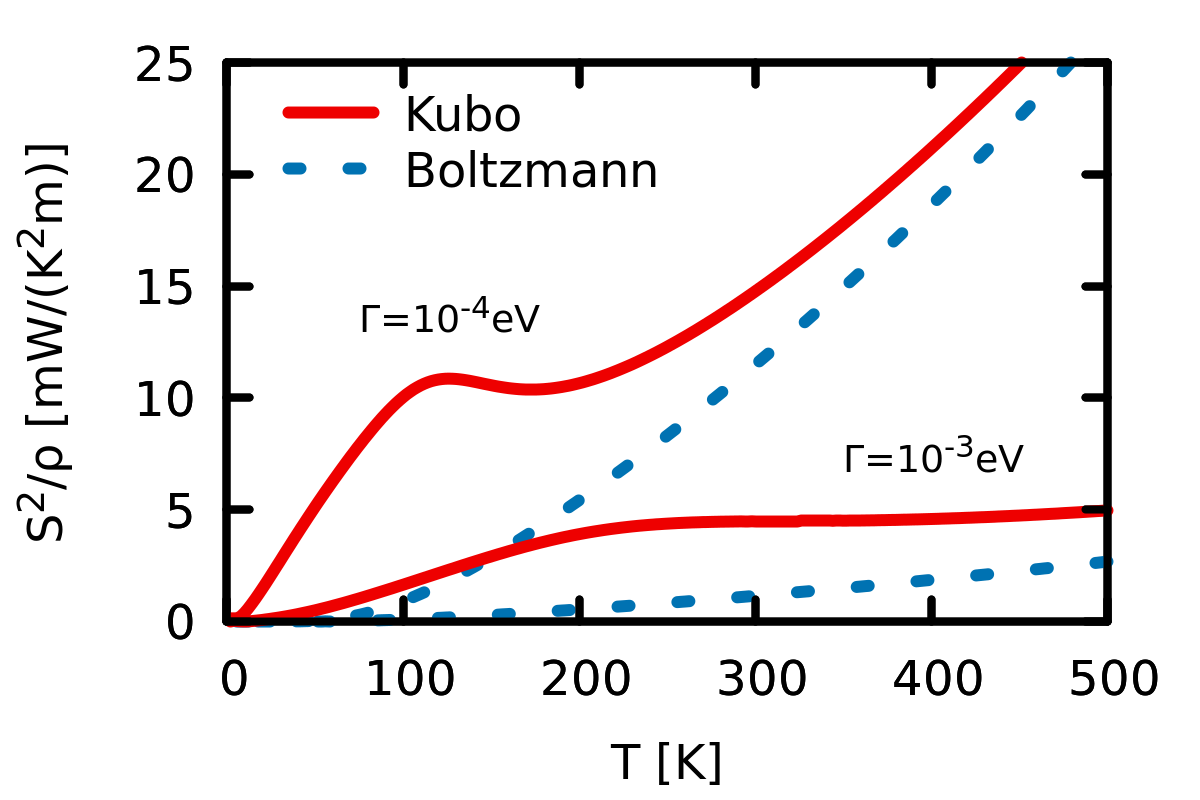}}
      \caption{{\bf Transport properties of a narrow-gap semiconductor model.} 
			Resistivity (left), thermopower (middle) and powerfactor (middle) of a model of a narrow-gap semiconductor consisting of two bands on a cubic lattice with nearest neighbour hopping $t_v=0.5$eV, $t_c=0.167$eV for (v)alence and (c)conduction states, respectively (i.e., an effective mass ratio $m_v/m_c=3$), separated by a crystal-field gap $\Delta=60$meV$=700$K$\times k_B$, with two electrons per site (half-filling), and two different scattering rates $\Gamma$, solved with the Kubo (\eref{Kubo}, red solid) and the semi-classical Boltzmann (\eref{Boltz}, \eref{BoltzS}, blue dashed) formalism. 
The only many-body effect here is the scattering rate $\Gamma$; no renormalization of the quasi-particle weight is accounted for ($Z=1$).
$\Gamma=10^{-4}$eV corresponds to the theoretical scattering rate of FeSi at 70K, see \fref{FeSiImS}.
}
      \label{BoKu}
      \end{center}
\end{figure*}

As an illustration of the importance of finite lifetime effects, we show in \fref{BoKu} a comparison for a generic case of a renormalized narrow-gap semiconductor
supplemented with a temperature-independent scattering rate.
At large temperature the Kubo and the Boltzmann approach yield similar resistivities, that follow activation laws, characteristic of coherent insulators.
However, already below 200-400K (for the given parameters), strong deviations between the two approaches occur: While the Boltzmann resistivities continue to grow
exponentially, the rise in the Kubo result slows down and $\rho$ tends to level off.
Trends toward resistivity saturation are commonly witnessed experimentally in correlated narrow-gap semiconductors (see \fref{overviewrho} for some examples).
Usually this situation is interpreted as indicative of impurity or defect derived in-gap states. Here, we propose that a finite (residual) scattering rate
for intrinsic charge carriers provides an alternative scenario for resistivity saturation in semiconductors.%
\footnote{In Kondo insulators, resistivity saturation has been suggested to be linked to topological effects\cite{Chang2017}.}
This suggestion is supported by experimental results for Ru-substituted FeSi: Indeed, Paschen \etal\ \cite{Paschen1999864} found the resistivity of 
Fe$_{0.9}$Ru$_{0.1}$Si to be similar to that of pristine FeSi\cite{PhysRevB.56.12916}, saturating, in particular, to approximately the same value at low temperatures.
The authors interpret this finding as strong evidence for an intrinsic origin of resistivity saturation in FeSi.
Moreover, in FeSi, nuclear magnetic resonance measurements\cite{doi:10.1063/1.372679} do not show (at least down to 50K) any clear characteristics that would be expected from in-gap states
(contrary to the cases of FeSb$_2$ and FeGa$_3$, cf.\ \sref{ingap}).%
\footnote{See, however, the discussion of possible donor levels in Ref.~\cite{PhysRevB.56.12916}, as well as their interpretation
in terms of spin polarons\cite{0295-5075-51-5-557,PhysRevB.84.073108}.}

Turning to the thermopowers of the model, \fref{BoKu}(middle),  we see large differences between the two approaches: The Boltzmann result is independent of the scattering rate, see \eref{BoltzS}, and diverges inverse proportional to $T$, as expected for a fully coherent semiconductor (see also \sref{limit}).
The Kubo result, on the other hand, largely depends on the scattering rate. Towards absolute zero, $S$ vanishes as required by the third law of thermodynamics.
At intermediate temperatures, the thermopower is non-monotonous in temperature and assumes a shape common to that seen in experiments, see \fref{overviewS}.
Interestingly, for some temperatures, the Kubo thermopower can be larger than the Boltzmann result. Peak values of the thermopower, however, increase with growing coherence (smaller scattering rate $\Gamma$).

The resulting powerfactors are shown in \fref{BoKu}(right). While $S^2\sigma$ is larger for small scattering rates, it is interesting to note that
properly accounting for effects of finite lifetimes via the Kubo formalism yields larger values than the semi-classical Boltzmann approach.

This model calculations thus show: (i) a quantitative description of transport properties in regimes of non-negligible lifetime effect requires going beyond
the Boltzmann approach (for metallic systems in which accounting for correlation effects proved crucial, see, e.g., Refs.~\cite{Held_thermo,arita:115121,PSSA:PSSA201300197,PhysRevLett.117.036401}),
(ii) finite lifetime effects can actually yield a better thermoelectric performance than would be anticipated from semi-classical approaches.
Consequently, high-throughput screenings\cite{Madsen2006,PhysRevX.1.021012,C2CP41826F,PhysRevB.94.045122,Gorai2017} based on DFT and Boltzmann approaches might 
overlook promising thermoelectric materials.

\begin{framed}
\noindent
{\bf Impact of many-body renormalizations on transport quantities}
	\begin{itemize}
		\item (selective) mass enhancements can boost thermopowers.
		\item in a coherent regime, the thermopower of a semiconductor has $\Delta/T$ as upper bound.
		\item describing incoherent transport is beyond semi-classical Boltzmann theory.
		\item residual scattering of intrinsic charge carriers can cause a saturation of the resistivity at low $T$,
		and causes a decay of the thermopower for $T\rightarrow 0$.
	\end{itemize}
\end{framed}

\subsection{Silicides: FeSi, RuSi and their alloys.}
\label{silicides}

The electronic structure of the prototypical correlated narrow-gap semiconductor FeSi was discussed in \sref{theo}, and its
properties compared to Kondo insulators in \sref{KI}. Here, we focus on the thermopower of the material, that was already shown
in \fref{overviewS} and \fref{FeSiCe3trans}(c).
In \fref{FeSiS} we again display the experimental results, now in comparison to theoretical simulations.

\subsubsection{Simulations of the thermopower.}
\begin{figure*}[!t]
  \begin{center}
	{\includegraphics[angle=0,width=.6\textwidth]{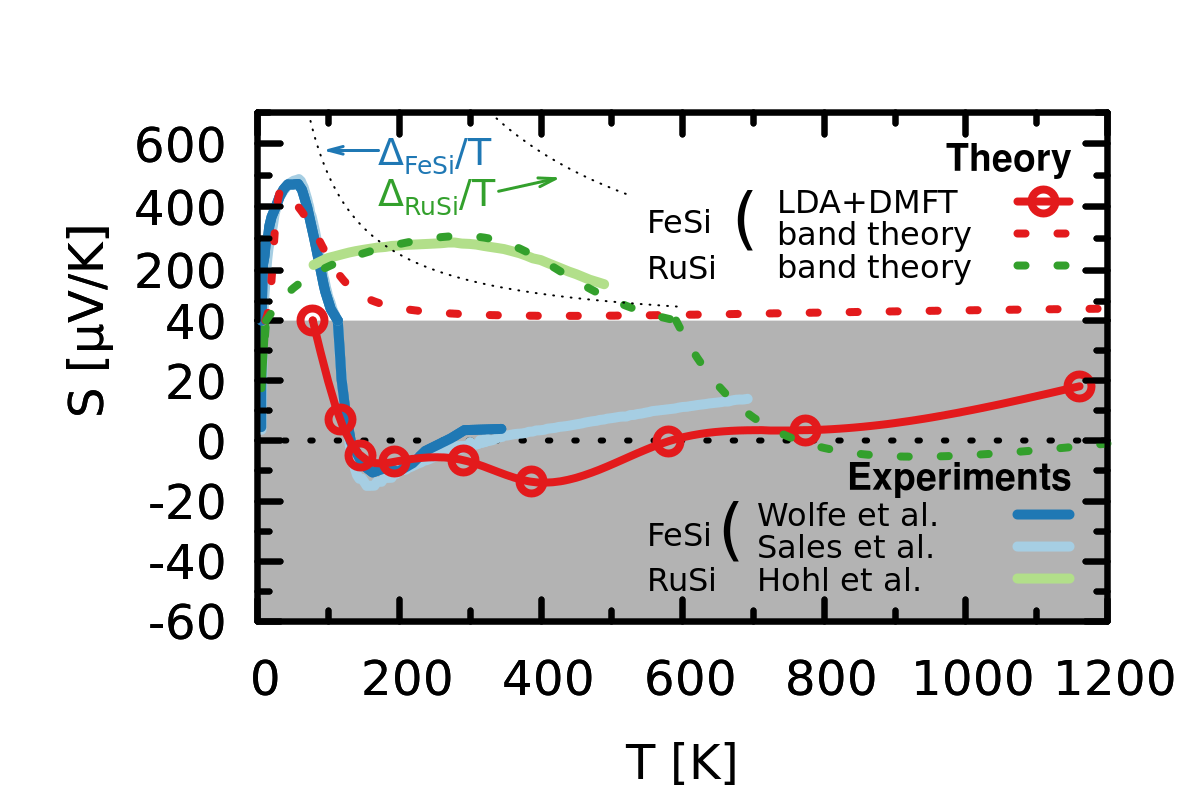}}
      \caption{{\bf Thermopower of FeSi and RuSi.} 
			Shown are theoretical simulations of the thermopower: The ``band-theory'' results (FeSi: red dashed, RuSi: green dashed) correspond
			to DFT results that have been slightly hole-doped (FeSi: 0.1\% holes per iron, RuSi: 0.25\% holes per Ru), and,
			in the case of FeSi, supplemented with an effective mass of two, in order to narrow the band-gap (cf.\ \sref{pnas}).
			In both cases a scattering rate of 1meV was used.
			Many-body results for FeSi \cite{jmt_fesi} (red circles, connecting line is a mere guide to the eye) agree well with experimental results.
			Also indicated are the fundamental limits (cf.\ \sref{limit}) of the thermopower in a purely electronic picture: $\Delta_{FeSi}/T$ and
			$\Delta_{RuSi}/T$.			
			Experimental data from Wolfe \etal\ \cite{Wolfe1965449}, Sales \etal\ \cite{PhysRevB.83.125209}, and Hohl \etal\ \cite{Hohl199839}.
			Note the change in scale above 40$\mu$V/K. Figure adapted from Ref.~\cite{jmt_hvar}.}
      \label{FeSiS}
      \end{center}
\end{figure*}

\begin{figure*}[!t]
  \begin{center}
	\subfloat{
	{\includegraphics[clip=true,trim=0 0 0 0,angle=0,width=.45\textwidth]{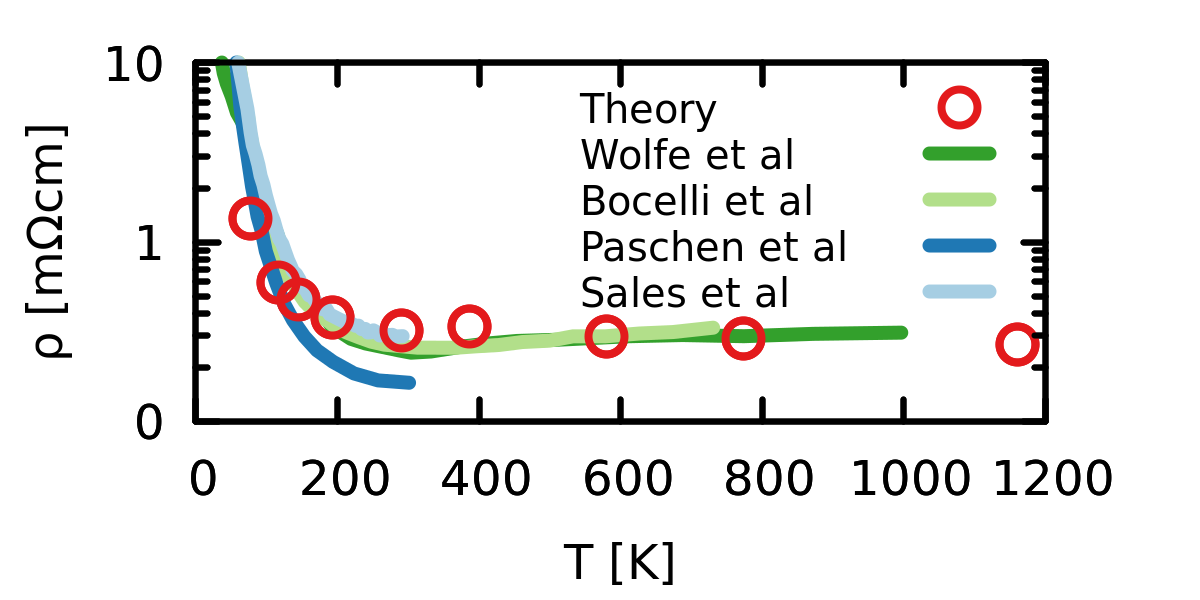}}
	$\quad$
	{\includegraphics[clip=true,trim=0 0 0 0,angle=0,width=.45\textwidth]{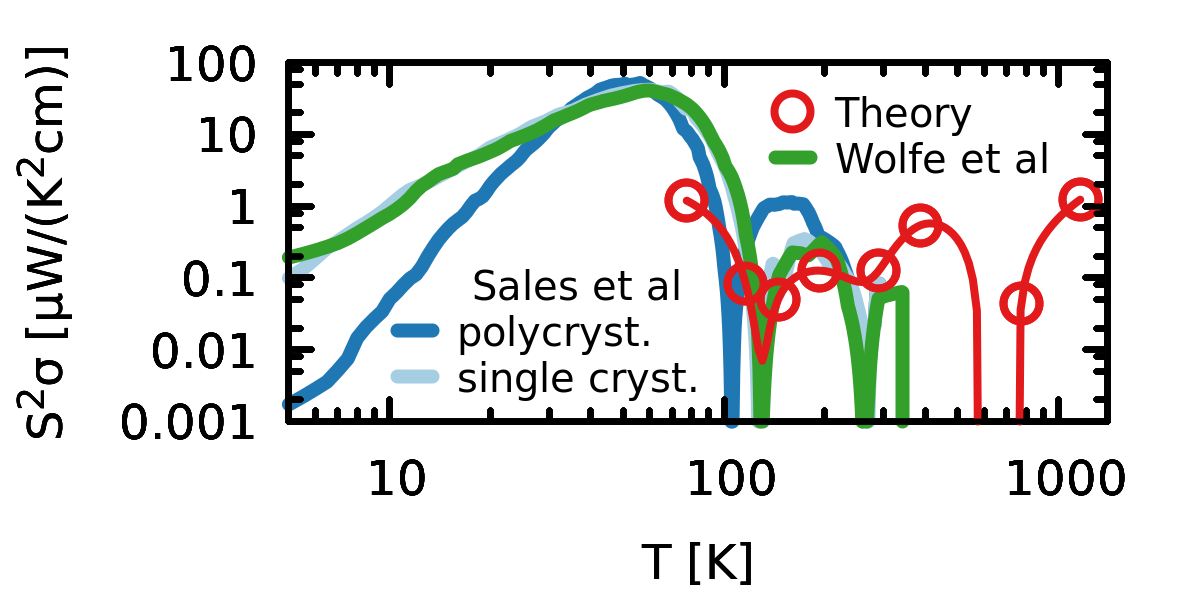}}
		}
	      \caption{{\bf Resistivity and powerfactor of FeSi.}  Many-body DFT+DMFT resistivity (left) and powerfactor (right) \cite{jmt_fesi} in comparison to experimental
				results from Wolfe \etal\ \cite{Wolfe1965449}, Bocelli \etal\ \cite{Bocelli_FeSi}, Paschen \etal\ \cite{PhysRevB.56.12916}, and Sales \etal\ \cite{PhysRevB.83.125209}.
								Adapted from Ref.~\cite{jmt_fesi}.}
      \label{FeSirhoPF}
      \end{center}
\end{figure*}

At temperatures below 100K, i.e., when FeSi is in its coherent regime (cf.\ \sref{pnas}), its thermopower can be reproduced by
a slightly hole-doped and renormalized band-structure. This is illustrated in \fref{FeSiS}, where we plot a thermopower obtained from band-theory
doped  with 0.1\% holes per iron (see also the work of Jarlborg \cite{PhysRevB.59.15002}),%
\footnote{That tiny amounts of holes improves congruence might not be a question of introducing charge, but simply of changing the particle-hole asymmetry.}
 and supplementing the dispersions with an effective mass of two so as to narrow the band-gap to the experimental value (cf.\ \sref{pnas}).
The agreement of this band-theory-derived thermopower with experiment, however, worsens significantly when temperature rises.
Indeed, above 100K the experimental thermopower decreases quickly in absolute magnitude and changes sign twice.
Good agreement with experimental results is found,
when computing the thermopower of FeSi within Kubo's linear response formalism (see \sref{KubBoltz}) from the
realistic many-body electronic structure discussed in \sref{pnas}. 
In particular, the sign changes as a function of temperature, which indicate the transition between
hole ($S>0$) and electron ($S<0$) dominated transport, are captured. This non-monotonousness in the thermopower was already heralded by
the moving of the chemical potential, as seen in the right inset of \fref{FeSiDMFT1} (top left): Starting from low temperatures, the chemical potential moves down,
therewith reducing the hole contributions to the thermopower, before passing, at around 120K, the point of thermoelectric particle-hole symmetry,
below which the thermopower becomes negative. At higher temperatures yet, the trend reverses and the thermopower changes sign again.
Besides the thermopower, also for the resistivity, and the powerfactor derived from both of them, there is good agreement between experiment
and the many-body simulations, see \fref{FeSirhoPF}.
Whether the many-body picture for FeSi\cite{jmt_fesi} also accounts for the compound's Hall\cite{springerlink:10.1134/1.567951,PhysRevB.58.10288,PhysRevB.88.245203} and its intriguing Nernst coefficient\cite{PhysRevB.88.245203}, is an open question.

\subsubsection{Influence of correlation-induced incoherence.}
\begin{figure}[!h]
  \begin{center}
	{\includegraphics[angle=0,width=.45\textwidth]{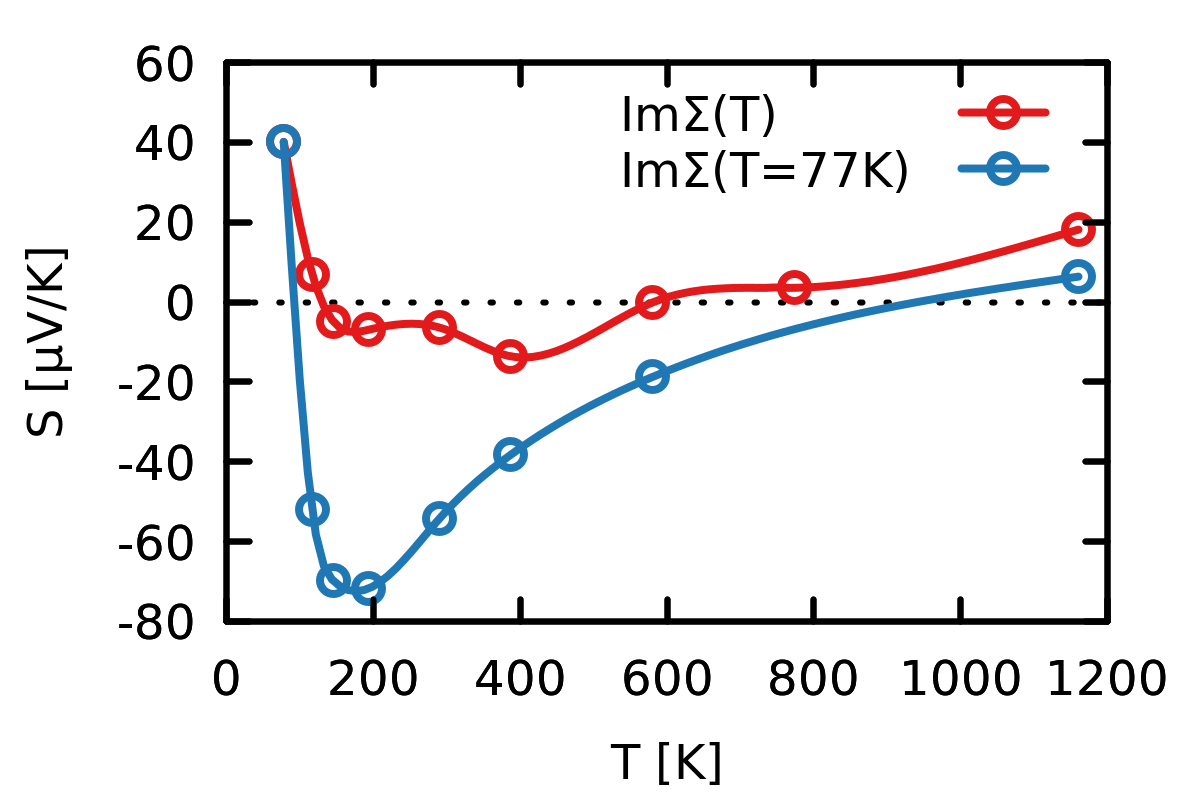}}
      \caption{{\bf Thermopower of FeSi: influence of incoherence.} Shown is the thermopower of FeSi within DFT+DMFT (red, see also \fref{FeSiS}),
			and the thermopower obtained from the same calculation if fixing the degree of incoherence, $\Im\Sigma(\omega)$, to that
			corresponding to 77K (blue). The comparison shows that the increase in lifetime effects (see also \fref{FeSiImS}) quenches the thermopower.}
      \label{FeSicoh}
      \end{center}
\end{figure}

Having validated the accuracy of transport simulation in the many-body framework, we can
elucidate the origin of the quenching of the thermopower above 100K. Is it a consequence of the chemical potential
moving, or does the onset of incoherent conduction destroy the voltage drop across the sample?
In \fref{FeSicoh} we therefore compare the original DFT+DMFT thermopower (from \fref{FeSiS}) with a calculation in which lifetime
effects, i.e., $\Im\Sigma(\omega)$, are frozen to their values at $T=77$K. We see that the resulting thermopower 
still switches sign, but remains much larger on an absolute scale.
Hence, the moving of the chemical potential has a large influence on what is the dominant type of carriers (electrons or holes),
but it is
the onset of incoherent spectral weight at the Fermi level above 100K (cf.\ the spectral function in \fref{FeSiCe3spec}(a))
that is the main culprit for the substantial decrease in the thermoelectric signal. In that sense, short lifetimes, i.e., a substantial $\Im\Sigma(\omega)$,
are antagonistic to thermoelectric performance in semiconductors.

\subsubsection{Doped FeSi.}
If incoherent spectral weight at the Fermi level kills the thermopower, than so does doping.
Indeed, while the thermopower of FeSi is robust with respect to different samples and preparation techniques,
any disturbance away from stoichiometry vastly decreases the response.

\begin{figure*}[!t]
  \begin{center}
	{\includegraphics[clip=true,trim=0 0 0 0,angle=0,width=.525\textwidth]{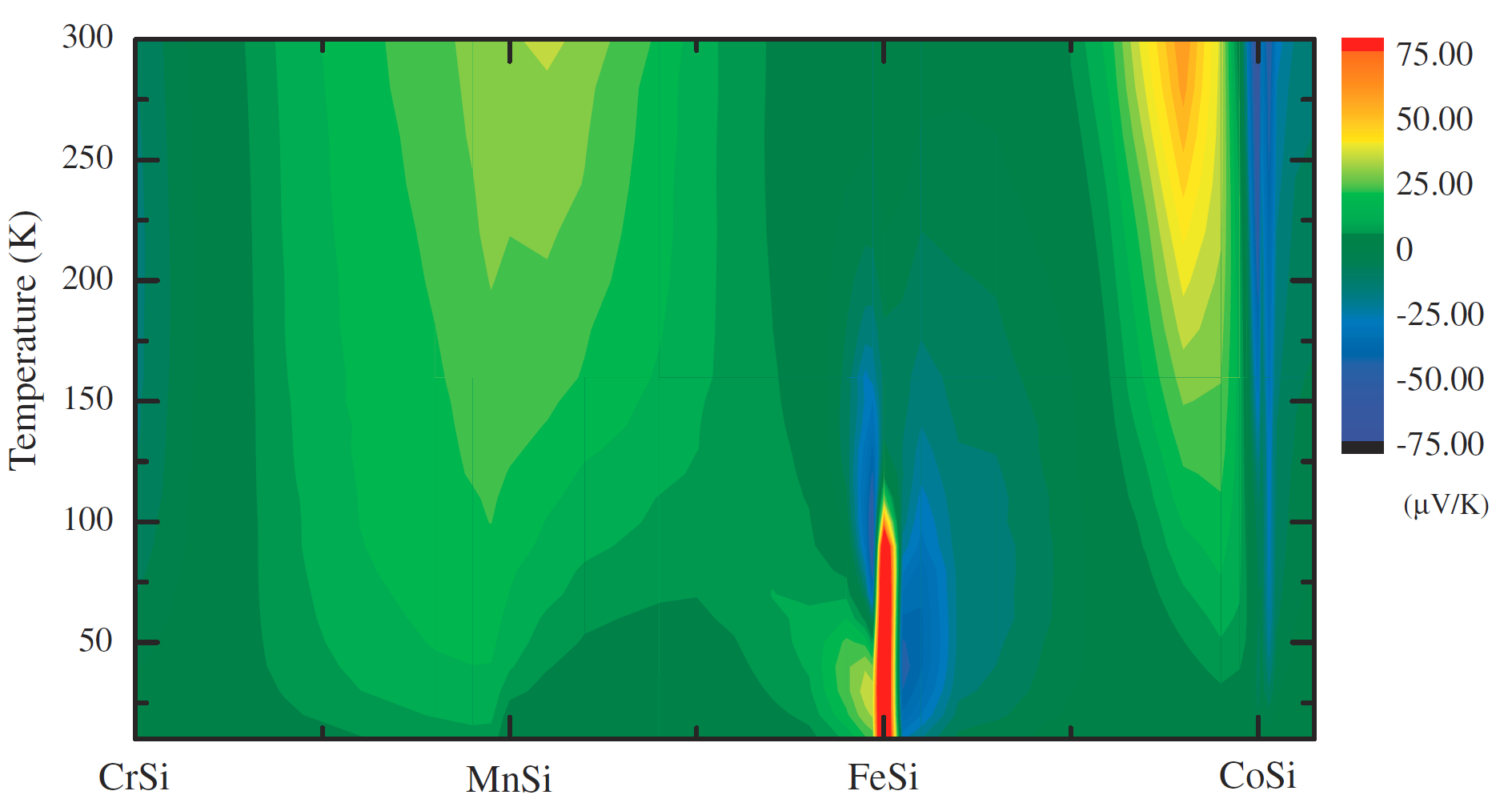}}
	{\includegraphics[clip=true,trim=0 0 0 0,angle=0,width=.4\textwidth]{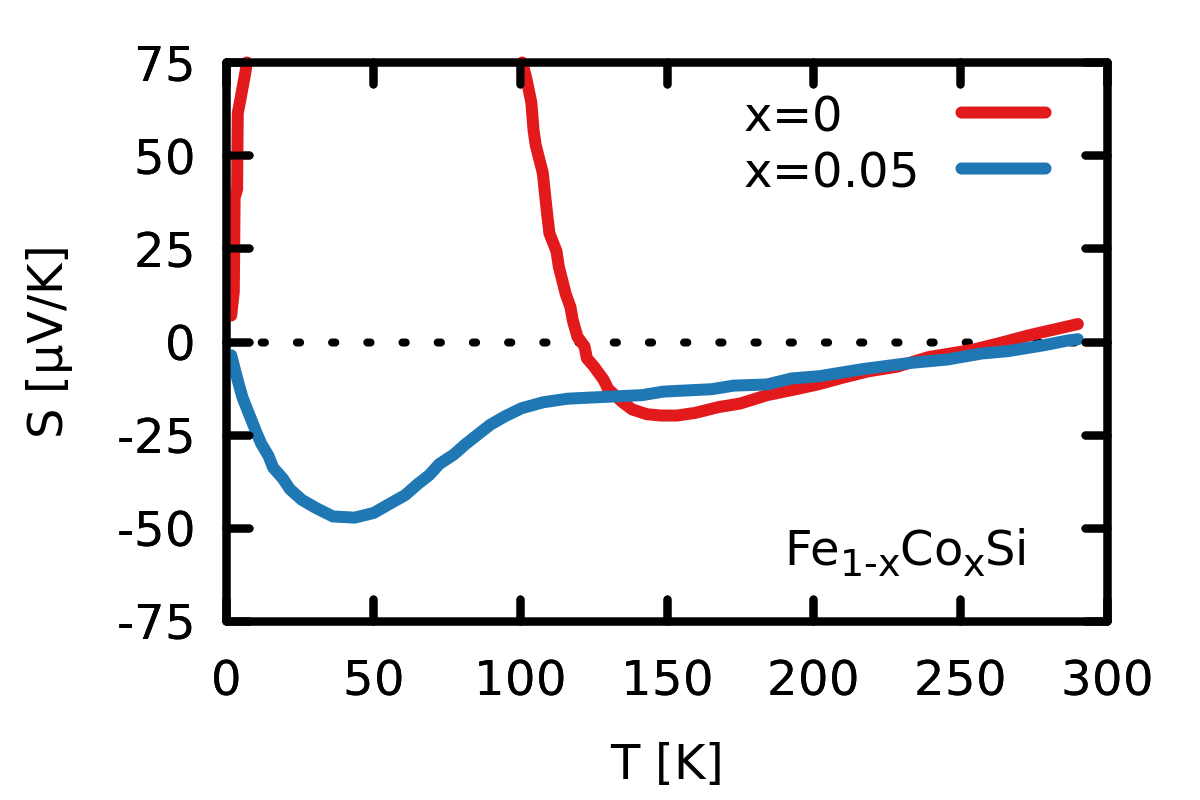}}
		\caption{{\bf Thermopower of monosilicides.} The thermopower of the 3$d$ monosilicides and their alloys is shown as colour code
		in the phase-diagram spanned by composition and temperature (left). For two vertical cuts, corresponding to stoichiometric FeSi and
		Fe$_{0.95}$Co$_{0.05}$Si, the thermopower is shown as a function of temperature (right). The behaviour at low-temperatures is congruent
		with the simple model shown in \fref{modelphi}. Used with permission from Sakai \etal\ \cite{JPSJ.76.093601}.
		\textcircled{c} 2007 The Physical Society of Japan (J. Phys. Soc. Jpn. 76, 093601).
		}
      \label{Sakai}
      \end{center}
\end{figure*}

The uniqueness of pure FeSi was impressively illustrated by Sakai \etal\ \cite{JPSJ.76.093601} who studied
the thermopower of monosilicides and their alloys: In the continuous phase-diagram of temperature vs.\ composition, see \fref{Sakai},
FeSi sticks out as an island of largely enhanced thermoelectricity. 
The above discussed effects of incoherence explain why the narrow (red) strip of large and positive thermopower is
restricted to low temperatures. Its confinement with respect to doping (the chemical potential) can be understood
from the simple model, presented in \sref{limit}:
Indeed, in an insulator with conduction electrons heavier than valence holes, large positive thermopowers are possible, as
indicated in \fref{modelphi} that shows the coefficient of the $1/T$ behaviour.
Moving the chemical potential to the left (hole doping) continuously decreases the thermopower
which, however, remains positive. 
A trend similar to that shown here for Fe$_{1-x}$Mn$_x$Si, is realized in ligand-substituted FeSi$_{1-x}$Al$_x$ \cite{PhysRevB.58.10288,PhysRevB.78.075123}.

Increasing $\mu$ (electron doping) beyond the maximum in \fref{modelphi}
causes the thermopower to decrease abruptly, change sign, and realize negative values of smaller magnitude.
This is indeed, what is found experimentally, as can be inferred from \fref{Sakai} and Refs.~\cite{OUYANG201792} for Co-doping, and is seen in Ref.~\cite{PhysRevB.50.8207}
also for Ir-doping.

\subsubsection{A comparison of FeSi with RuSi.}
\label{FeRuSi}

\begin{figure}[!t]
  \begin{center}
	{\includegraphics[angle=0,width=.45\textwidth]{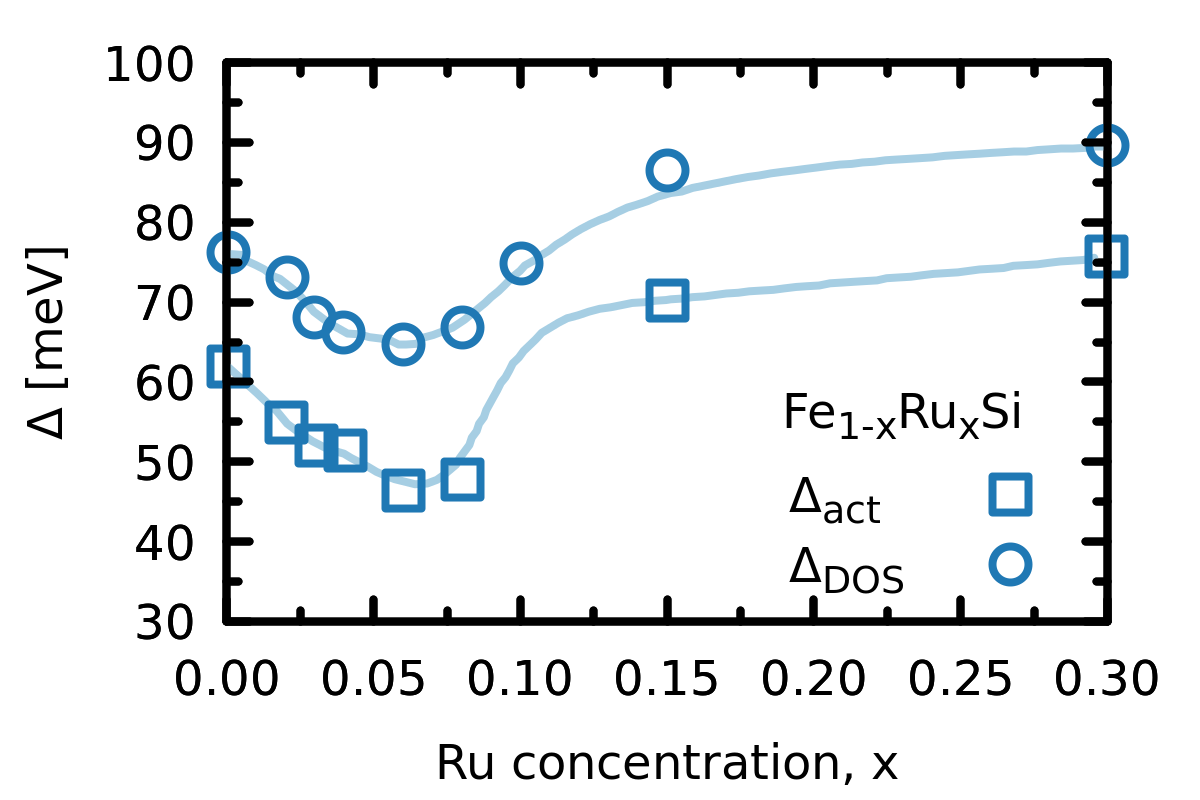}}
      \caption{{\bf Gap in the alloy Fe$_{1-x}$Ru$_x$Si.} Shown is the evolution of the indirect gap 
			as extracted from the resistivity  via an activation-law fit, $\Delta_{\hbox{\tiny act}}$ (squares), and a fit 
      using a density-of-states {\`a} la \fref{FeSiDOSscheme}(middle), $\Delta_{\hbox{\tiny DOS}}$ (circles).			
			Adapted from Mani \etal\ \cite{PhysRevB.65.245206}.
			}
      \label{FeRuSiDelta}
      \end{center}
\end{figure}

We find it instructive to compare FeSi to its isostructural and isoelectronic 4$d$ homologue RuSi, and
discuss the series Fe$_{1-x}$Ru$_{x}$Si in view of its thermopower%
\footnote{For a comparison of the related couple FeSb$_2$ and RuSb$_2$, see Refs.~\cite{APEX.2.091102,herzog_fesb2,cava2013fesb2},
and for the pair FeGa$_3$ and RuGa$_3$, see Ref.~\cite{monika_fega3}.}.
As discussed in \sref{bandFeSi} in conjunction with band-structure results,
ruthenium silicide, RuSi, is a semiconductor with an indirect gap of $260-310$meV (obtained from resistivity measurements\cite{Buschinger1997238,Hohl199839})
and an optical gap of $\sim 400$meV\cite{Buschinger1997238,Vescoli1998367}.
Interestingly, the trend in the size of the gap in the alloy series Fe$_{1-x}$Ru$_{x}$Si is not monotonous with $x$ \cite{PhysRevB.65.245206}, see \fref{FeRuSiDelta}:
In fact up to a ruthenium concentration of 6\% the charge gap is found to decrease with respect to FeSi. At higher concentrations the gap then starts to grow, passes
the initial value of FeSi at around 15\% ruthenium, and augments further up to the value of stoichiometric RuSi.

With the insight into FeSi (\sref{pnas}, \fref{FeSicoh}, and \sref{PAM}), this behaviour was understood as follows\cite{jmt_hvar}. In fact, there are, in Fe$_{1-x}$Ru$_{x}$Si,
two opposing tendencies:
(i) ruthenium has a larger atomic radius than iron.  Thus, with increasing Ru content, the hybridization gap will shrink as the lattice expands.
This effect is immediate\cite{PhysRevB.65.245206}, and wins for low ruthenium concentrations.
(ii) with $x$ there is a crossover in the dominant orbital character of low-energy excitations, namely from a 3$d$ to a 4$d$ radial distribution
(cf.\ also the discussion in Ce$_3$Bi$_4$Pt$_{3-x}$Pd$_x$ in \sref{Ce3dft}).
As a consequence of the greater extension of the latter, interaction matrix elements, such as the Hubbard $U$ and the Hund's $J$, will have smaller values
(see Refs.~\cite{jmt_wannier,jmt_mno}, as well as the comparison of $U$ and $J$ of FeSb$_2$ and RuSb$_2$ in \tref{tmarcs} in \sref{band}).
As discussed in \sref{pnas}, and elaborated on in the model calculations of \sref{PAM}, the Hund's rule coupling drives the strong correlation physics in FeSi.
This is the reason why correlation effects are near absent in RuSi, in congruence with the fact that band-structure theory underestimates its gap---instead of the large overestimation for FeSi, cf.\ \sref{bandFeSi}.
The effect of diluting the Hund's rule coupling in Fe$_{1-x}$Ru$_{x}$Si with growing $x$ becomes preponderant for larger ruthenium concentrations,
diminishes effects of band-narrowings, and, thus, causes the gap to increase.

The thermopower of RuSi, see \fref{FeSiS}, while not achieving the very large values of FeSi at low temperatures, is noteworthy in size over an
extended temperature regime, with $S\sim250\mu$V/K from 100 to 500K:
As discussed in \sref{limit}, the thermopower of a coherent insulator is controlled by the size of the gap $\Delta$ and the particle-hole asymmetry,
while it cannot exceed $\Delta/T$. RuSi, having a larger gap than FeSi, exhibits a  thermopower that indeed surpasses the envelope function, $\Delta_{\hbox{\small FeSi}}/T$, of FeSi, for $200\hbox{K}\le T\le 500$K, as indicated in \fref{FeSiS}. 
However, RuSi obeys its own boundary, $\Delta_{\hbox{\small RuSi}}/T$, and the distance between the largest possible and the actual thermopower is in fact much larger than is the case for FeSi.
Hence, the particle-hole asymmetry (called $\delta\lambda$ in \sref{limit}) is smaller in the 4$d$ compound.

The dilution of the  Hund's rule coupling upon replacing Fe with Ru will, besides the decrease in effective masses, also diminish effects of incoherence.
This should push the quenching of the thermopower (cf.\ \fref{FeSicoh}) to higher temperatures.
Therefore, isovalent, yet non-isoazimuthal and non-isovolume substitutions in correlated narrow-gap semiconductors might produce larger thermopowers 
in the temperature regime of 100 to 300K.
Besides the greater coherence, the loss of thermoelectric particle-hole asymmetry ($\delta\lambda$) when departing from pure FeSi, is partly compensated by the enlarged gap $\Delta$ (for $x\ge 0.15$). While the power factor $S^2\sigma$ of stoichiometric FeSi reaches favourable
40 $\mu$W/(K$^2$cm) at around 60K, see \fref{FeSirhoPF}, a notable improvement beyond 100K can be expected in Fe$_{1-x}$Ru$_x$Si for $x>0$.
Finally, the substitution will also reduce the thermal lattice conductivity $\kappa$, yielding a better figure of merit. 
Nonetheless, it is unlikely that Ru-substitution alone will result in a useful $ZT$, let alone a good price-performance ratio.

\begin{framed}
	\noindent
	{\bf FeSi: lessons for thermoelectricity}
	\begin{itemize}
		\item effective masses enhance peak thermopowers at low $T$: correlations beneficial
		\item incoherence (finite lifetimes) quenches the thermopower: correlations detrimental
		\item in the presence of lifetime effects transport properties cannot be captured with semi-classical Boltzmann approaches
		\item potential path to extend regime of large thermopowers: partial isoelectronic substitution of $4d$ for $3d$ transition metals (dilution of Hund's rule coupling)
	\end{itemize}
\end{framed}


\subsection{Marcasite di-Antimonides: FeSb$_2$, CrSb$_2$ \& Co.} 
\label{dianti}

The material that has particularly fuelled the interest in thermoelectric effects in correlated narrow-gap semiconductors is FeSb$_2$.
Indeed, its thermopower has rightly been called ``colossal'' by Bentien \etal\ \cite{0295-5075-80-1-17008}, who found values up to -45mV/K at around 10K, and the
powerfactor is the largest ever measured%
\footnote{With the possible exception of elemental bismuth, where, when combining results from different experiments (see data compiled in Ref.~\cite{ISSI196613}), $S^2\sigma$ can reach spectacular 81mW/K$^2$cm at 3K.}
with up to
8mW/K$^2$cm at 28K\cite{PhysRevB.86.115121}. 
While basic transport, optical and magnetic properties of FeSb$_2$ have been discussed in \sref{overview} and its electronic structure in \sref{marcasites},
we here make the connection to its thermoelectric properties, and review works that try to explain the unusual behaviour.
Here, one has to make a clear distinction between polycrystalline samples and single crystals.
Indeed, while the temperature dependence of the thermopower is qualitatively similar for both cases,
the colossal magnitude of $S$ occurs only in single crystals. The thermopower measured for polycrystalline samples\cite{bentien:205105,sun:033710,0022-3727-43-20-205402,10.1063/1.3556645} instead never exceeds 1mV/K, see, e.g., \fref{Takahashi} (left). 

\subsubsection{Polycrystalline samples---physics akin to FeSi?}

Where it peaks, the thermopower of FeSb$_2$ is negative, indicative of preponderant contributions from electrons. This is opposite to the hole-like, positive thermopower of FeSi, discussed above. This finding is in line with the particle-hole asymmetry evidenced in electronic structure calculations (see, e.g., the DOS in \fref{marcasitesqsgw}). Indeed
spectral features are much sharper and of higher density for states below the Fermi level than above it.
When described in terms of an effective model in the spirit of \sref{limit}, this asymmetry translates into effective masses that are larger for valence than for conduction bands. 
Assuming long electronic lifetimes at low temperatures, the qualitative behaviour of the thermopower of FeSb$_2$ can be discussed in terms of \fref{modelphi}.
There, to account for the hierarchy of masses, $m_v^*\gg m_c^*$ instead of the shown $m_v^*\ll m_c^*$, the figure needs to be mirrored with respect to both, the x- and the y-axis.
Then it is expected that electron-doping leads to a gradual decrease of the thermopower that however remains negative.
This is indeed what is found experimentally, see, e.g., the case of FeSb$_{2-x}$Te$_x$ in Refs.~\cite{10.1063/1.3556645,doi:10.1063/1.4731251}. 
Hole-doping, on the other hand, quenches the thermopower more rapidly and leads to a sign change already at very low doping, see, e.g., Ref.~\cite{bentien:205105}
for the case of FeSb$_{2-x}$Sn$_x$. 
A quantitative simulation of thermoelectric effects in polycrystalline samples of FeSb$_2$ is still outstanding. However, it can be expected  that once
the obstacles to describe the electronic structure of the compound (discussed in \sref{marcasites}) are overcome, a computation of (intrinsic) transport properties will 
come into reach.

\begin{figure}[!t]
  \begin{center}
	{\includegraphics[angle=0,width=.45\textwidth]{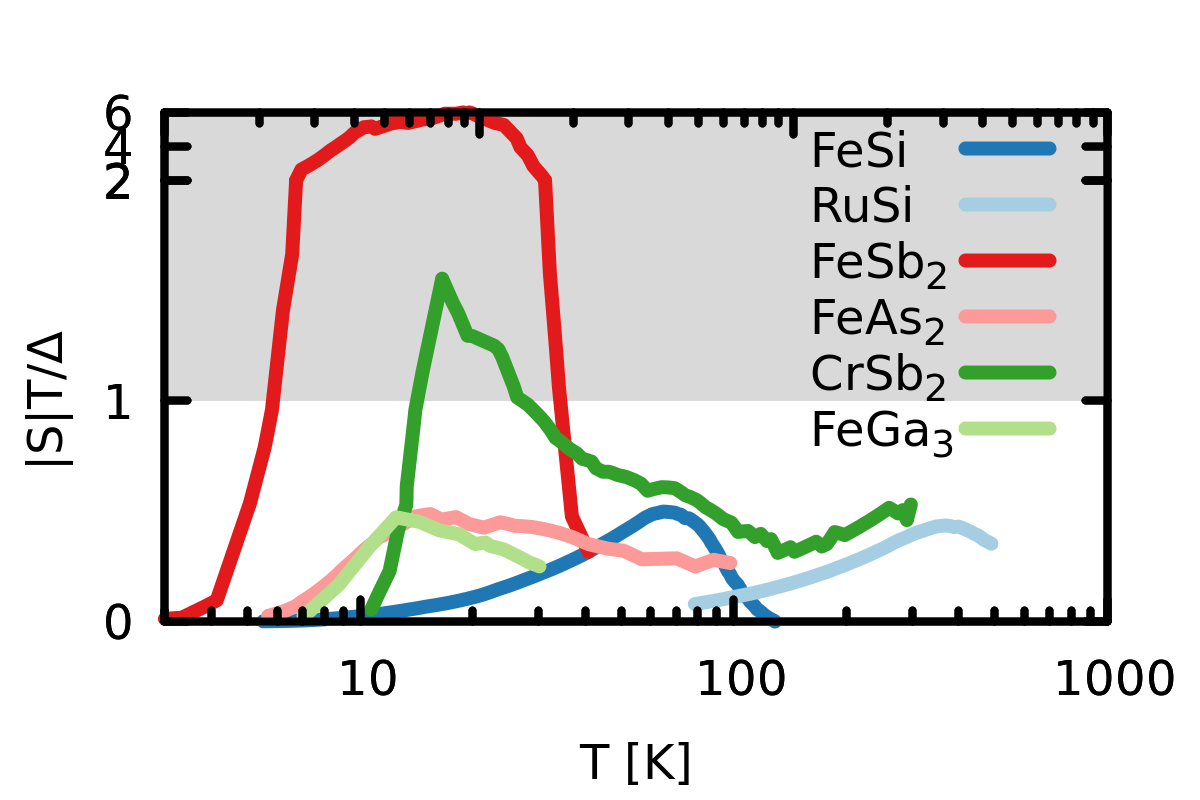}}
      \caption{{\bf Thermopower of binary intermetallics scaled with their charge gap $\Delta$.}
			FeSi ($\Delta=60$meV): Sales \etal\ \cite{PhysRevB.83.125209}, RuSi ($\Delta=215$meV): Hohl \etal\ \cite{Hohl199839}, FeSb$_2$ ($\Delta=35$meV), FeAs$_2$ ($\Delta=200$meV): Sun \etal\ \cite{PhysRevB.88.245203}, CrSb$_2$ ($\Delta=50$meV): Sales \etal\ \cite{PhysRevB.86.235136}, FeGa$_3$ ($\Delta=400$meV): Wagner \etal\ \cite{PhysRevB.90.195206}.
			The (grey) shaded area indicates the regime in which $S$ violates the fundamental limit of the thermopower of purely electronic origin (cf.\ \sref{limit}).
			}
      \label{Sscaled}
      \end{center}
\end{figure}

\subsubsection{Single crystals---do correlations cause the large thermopower?}
It is in single crystals\cite{0295-5075-80-1-17008,sun_dalton,PhysRevB.86.115121,Takahashi2016} that the thermopower of FeSb$_2$ was found to be unusually large, indeed ``colossal'', with, however, notable variations on the quality\cite{doi:10.1143/JPSJ.80.054708} and size\cite{MRC:8871060,Takahashi2016} of samples (see also below).
Given the pronounced effects of electronic correlation effects in this system (discussed in \sref{overview}), it was proposed that also the thermoelectric response
owes to many-body effects. 

Indeed, extending the comparison of FeSb$_2$ with its isostructural homologues FeAs$_2$ and RuSb$_2$ (see the discussion in \sref{marcasites} and also \sref{PAM}) to thermoelectric properties,
one notes that only the strongly correlated FeSb$_2$ boosts an unusually large thermopower\cite{sun:153308,APEX.2.091102,sun_dalton}.
Also, the thermoelectric properties of doped FeSb$_2$, e.g., the mentioned  FeSb$_{2-x}$Te$_x$
are indeed that of a correlated metal, i.e., the low-temperature thermopower is linear in $T$, 
with an enhancement factor, via the effective mass, of the order of 15~\cite{sun_dalton}.
Moreover, the {\it shape}  of the thermopower of FeSb$_2$ can actually be modelled\cite{sun:153308,PhysRevB.86.115121} with phenomenological
expressions that include many-body renormalizations\cite{jmt_fesb2}.
However, the modelling failed to account for the colossal {\it magnitude} of the thermopower. 
Here, it is instructive to examine the experimental results in terms of the model that was discussed in \sref{limit}.
Since, at low temperatures, FeSb$_2$ is a coherent semiconductor (cf.\ \sref{overview}), the conditions to apply the band-like model are met.
In this coherent, purely electronic picture---which, on top, neglects so-called vertex corrections---the thermopower cannot exceed $\Delta/T$, where $\Delta$
is the charge gap. In \fref{Sscaled} we plot the experimental thermopowers of several narrow-gap semiconductors, scaled with $T/\Delta$. Any value of $|S|T/\Delta$
in excess of unity indicates a violation of the theoretical upper bound of the thermopower. The antimonides FeSb$_2$ and CrSb$_2$, subject of this section, indeed defy the constraint. In the case of FeSb$_2$ the violation amounts to almost one order of magnitude.%
\footnote{Note that in \fref{Sscaled} we have used the thermopower from Ref.~\cite{PhysRevB.88.245203}, where the peak-value is -18meV/K, which is less than half of the record -45meV/K found in Ref.~\cite{0295-5075-80-1-17008}.}
Hence, to explain the large thermopower in the antimonides, either the omitted vertex corrections (describing, e.g., excitonic effects) somehow strongly boost the response,
or, alternatively, an additional not electron-diffusive mechanism is at work in these materials.

\begin{figure*}[!t]
  \begin{center}
	{\includegraphics[angle=0,width=.65\textwidth]{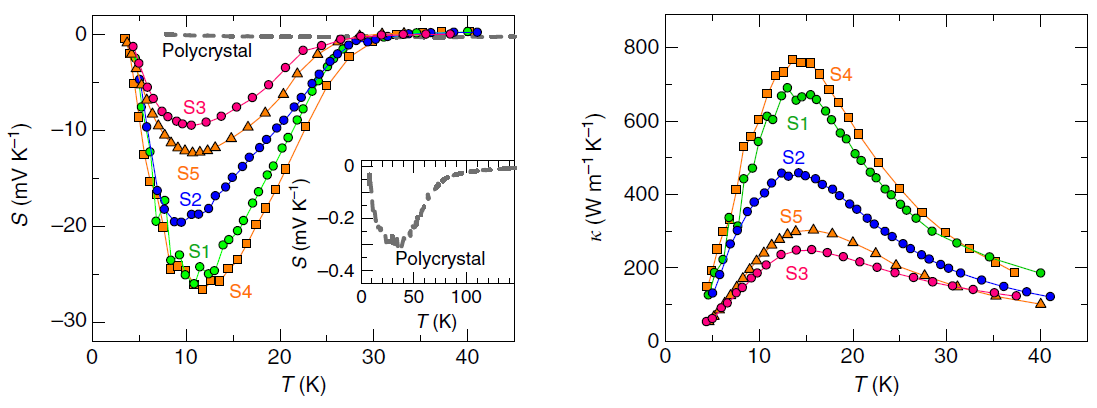}}
      \caption{{\bf Thermopower and thermal conductivity of FeSb$_2$: Influence of sample dimensions.}
			Shown are thermopower $S$ (left) and thermal conductivity $\kappa$ (right) of FeSb$_2$ as obtained
			by Takahashi \etal\ \cite{Takahashi2016} for single crystalline samples of different dimensions:
			S1: $250\mu$m$\times 245\mu$m$\times2.5$mm
			S2: $210\mu$m$\times 205\mu$m$\times2.3$mm
			S3: $80\mu$m$\times 160\mu$m$\times1.8$mm
			S4: $270\mu$m$\times 250\mu$m$\times3.3$mm
			S5: $130\mu$m$\times 150\mu$m$\times2.9$mm. The inset on the left displays the thermopower of polycrystalline samples. From Takahashi \etal\ \cite{Takahashi2016}.
			}
      \label{Takahashi}
      \end{center}
\end{figure*}

\subsubsection{Empirical evidence for a phonon-drag thermopower.}

As a potential candidate the so-called phonon-drag effect\cite{PhysRev.92.248,PhysRev.96.1163} has been suggested for FeSb$_2$\cite{jmt_fesb2}. 
This electron-phonon scattering mediated enhanced electron-drift has been evidenced in pure, conventional semiconductors, such as p-type Germanium\cite{PhysRev.94.1134},
or Indium Antimonide\cite{PhysRev.99.1889}, but also in pure metals and dilute alloys\cite{Blatt}.
In the following we will summarize empirical evidence that motivates the presence of the phonon-drag in FeSb$_2$ and hints at why it is negligible in FeAs$_2$ which obeys the $\Delta/T$-rule.

When cooling FeSb$_2$ below its insulator-to-metal crossover, optical spectroscopy witnesses a large change in phonon lifetimes, suggesting an important electron-lattice coupling~\cite{perucci_optics,PhysRevB.82.245205} (cf.\ the discussion for FeSi in \sref{lattice}).  
Also polarized Raman scattering experiments indicate notable electron-phonon couplings, that are strongly temperature dependent below 40K\cite{PhysRevB.81.144302} (see \fref{phononwidth}).
Further, we note that, as can be expected for a non-electronic mechanism of the thermopower, the magneto-thermopower of \fesb, $MT=[S(B)-S(0)]/S(0)$ is very low for samples with large $S(B=0)$~\cite{0295-5075-80-1-17008}. 
Interestingly, the large low-temperature thermopower in FeSb$_2$ is also accompanied by a notable increase in the nuclear spin-lattice relaxation rate\cite{1742-6596-150-4-042040}.

Also, the mentioned dependence of the thermopower onto the type of samples can be consistently interpreted in favour of the phonon-drag scenario.
Indeed, a decrease in the phonon mean-free path by scatterings of non-electronic origin is expected to lower phonon-drag contributions to the thermopower, as well as  lattice contributions to the thermal conductivity.
This naturally explains why the low-temperature thermopower found in polycrystalline samples~\cite{bentien:205105}, polycrystalline thin films~\cite{sun:033710} and nano-particular samples\cite{C3DT51535D}
is significantly lower than in single crystals. That all types of samples yield a comparable response at higher temperatures\cite{1742-6596-150-4-042040}
suggests that the phonon-drag contribution to the thermopower decreases with the emergence of phonon-phonon or umklapp scattering, as does the lattice thermal conductivity (cf.\ \fref{Takahashi}).

The trend of the thermopower with the phonon mean-free path has been put onto a much firmer footing 
by studies in which the granularity or size of samples was varied in a controlled fashion:
Pokharel \etal\ \cite{MRC:8871060} found that the thermopower of nano-composites roughly grows logarithmically with the average grain size.
Takahashi \etal\ \cite{Takahashi2016} studied rod-shaped single crystals of varying dimension, see \fref{Takahashi}. They established a clear proportionality
between the phonon mean-free path extracted from the thermal conductivity and that 
determined from the thermopower---assuming a simple phonon-drag model\cite{PhysRev.96.1163}.

\subsubsection{Comparison to FeAs$_2$}

At first glance, phonon related properties, including the phonon-drag effect, might be expected to be important also for the isoelectronic and isostructural FeAs$_2$.
Indeed, the thermal conductivity is found to be even larger in the arsenide\cite{APEX.2.091102}.
Yet, the maximal thermopower of FeAs$_2$ is much smaller than that of Fesb$_2$, and, in particular, compatible with a purely electronic picture (cf.\ \fref{Sscaled}).
The following might explain this disparity:
The thermopower is a measure for the entropy per carrier. Hence the phonon contribution $S_{ph}$ to the thermopower will roughly scale with the lattice specific 
heat $C_V$ times the electron-phonon coupling constant $g$, divided by the electron density: $S_{ph}\propto gC_V/n$.
With Debye temperatures of $348$~K for \fesb\cite{bentien:205105,APEX.2.091102}, and $510$~K for \feas\cite{APEX.2.091102}, the specific heat of \fesb\ will be 
larger than that of \feas.
The charge carrier concentration at temperatures where the thermopower is maximal, on the other hand, is larger for \fesb, $n\sim 8\cdot 10^{14}$/cm$^3$ for the sample
with largest $S$ in Ref.~\cite{sun_dalton}, whereas $n\sim 5\cdot 10^{14}$/cm$^3$ for \feas.
Also the electron-phonon coupling---the least accessible quantity (see Diakhate \etal\ \cite{PhysRevB.84.125210} for a computational study) is likely to be larger in the antimonide: 
The electronic contribution to the specific heat of \fesb\  exhibits a notable feature below 50K\cite{APEX.2.091102}.%
\footnote{However, the unlocking of spins in \fesb\ becomes appreciable only beyond this regime at around 150K, where the entropy reaches $R\log2$, owing to a second and larger hump in the specific heat,
and in congruity with the susceptibility\cite{koyama:073203,PhysRevB.67.155205}. 
Indeed \fesb\ becomes paramagnetic above  100K~\cite{PhysRevB.67.155205}, 
and a Curie-like downturn appears at temperatures above 350K~\cite{koyama:073203}, whereas the susceptibility of \feas\ is flat up to 350K\cite{APEX.2.091102}.
} 
That there is no analogue in the spin response, may indicate that the features's contribution to the entropy is associated with either the charge degrees of freedom (excitonic or charge ordering effects) or is, indeed, an electron-phonon effect.
No such feature appears in the specific heat of FeAs$_2$.\footnote{P.\ Sun, private communication.} 
Together, this suggests a much smaller phonon-drag effect in FeAs$_2$.

Comparing the thermopower of the antimonide and the arsenide (see Fig. 1 in Ref.~\cite{APEX.2.091102}) one notes, however, that
the Seebeck coefficient of \feas\ becomes larger than that of \fesb\ 
above 35K. This might indicate---if the phonon-drag picture holds---that the {\it effective} electron-phonon coupling in \fesb\ has
sufficiently decreased (by umklapp and phonon-phonon scattering) so that the thermopower is now dominated by the electronic degrees of freedom, i.e.,
the larger gap in \feas\ causes a larger response.
Indeed, an approximate simulation\cite{jmt_fesb2} of purely electron-diffusive contributions to the thermopower yielded good results for FeAs$_2$
above 15K, and, for FeSb$_2$, in the interval 35-50K which is high enough for phonon-contributions to have decayed and low enough to
not cross the insulator-to-metal transition of the compound.

\begin{figure}[!t]
  \begin{center}
	{\includegraphics[angle=0,width=.4\textwidth]{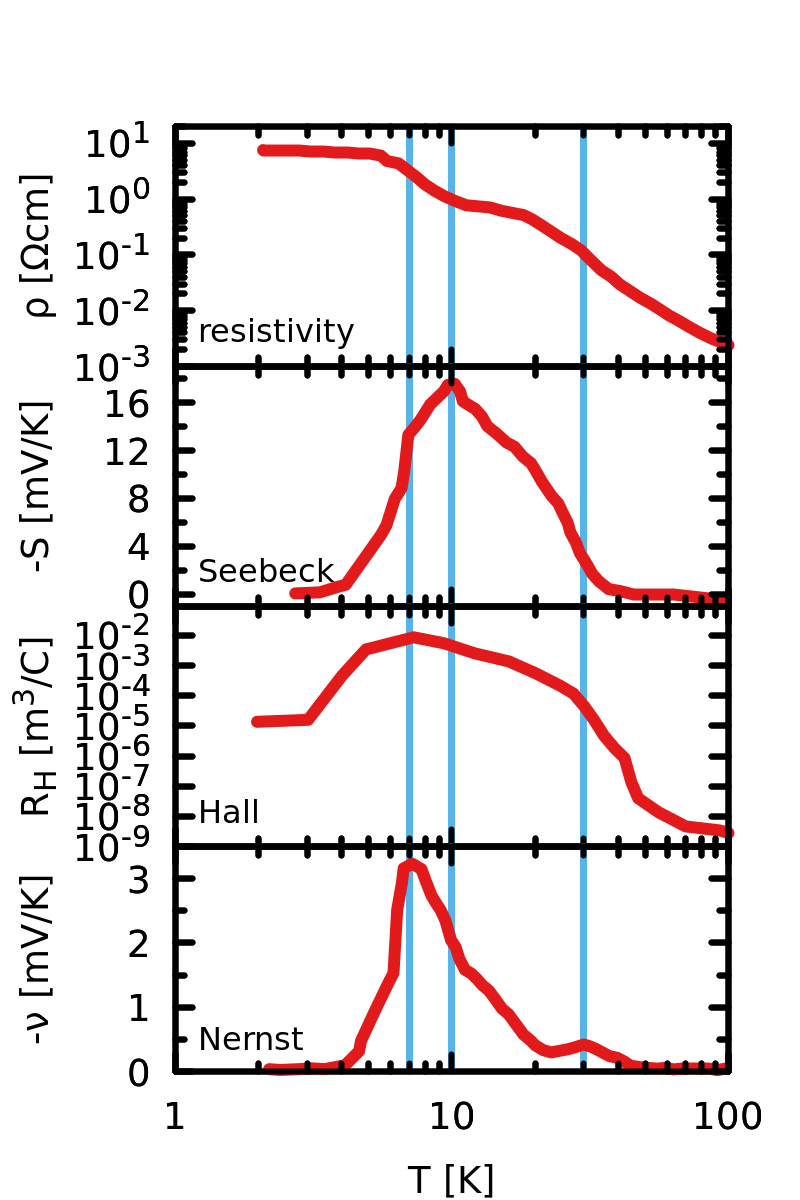}}
      \caption{{\bf Correlation of features in transport observables of FeSb$_2$.}
			From top to bottom: resistivity $\rho$, thermopower $S$, Hall coefficient $R_H$, and Nernst coefficient $\nu$ of FeSb$_2$.
			Vertical lines indicate characteristic temperatures that are common to all shown observables.
			Data compiled from Sun \etal\ \cite{PhysRevB.88.245203}.
			}
      \label{FeSb2_correl_feature}
      \end{center}
\end{figure}

\subsubsection{Constraints to a theory of FeSb$_2$: correlation of features in transport observables.}

For the development of a microscopic understanding, it is paramount to establish
as many empirical constraints as possible, against which a candidate theory can be tested.
In the case of FeSb$_2$ the comprehensive work of Sun \etal\ \cite{sun:153308,PhysRevB.88.245203}  demonstrated that there is
a striking inter-dependence of diverse transport observables. As depicted in \fref{FeSb2_correl_feature},
there are (at least) three temperatures at which the resistivity, the thermopower, and the Hall and Nernst coefficient
display distinct features:
Inflection points in the thermopower coincide with extrema in the Nernst coefficient, while the thermopower is maximal
for inflection points in both the resistivity and the Nernst coefficient.
Likewise, at these characteristic temperatures features appear in the Hall coefficient, as well as the magnetoresistance (see \fref{fig:threelevels}(d)).
This intimate linkage between electrical, magnetic, and thermoelectric quantities is a harbinger for a common microscopic origin,
posing severe constraints on possible explanations of the colossal thermopower.

\subsubsection{Modelling the phonon-drag in FeSb$_2$: evidence for important role of in-gap states.}
\label{pdrag}

Realizing that the characteristic temperatures just mentioned all lie far below the insulator to metal crossover in FeSb$_2$,
it is reasonable to assume electronic lifetimes to be long. In that case, one may attempt to describe transport properties
within the semi-classical Boltzmann approach (cf.\ the discussion in \sref{KubBoltz}).

\begin{figure*}[!t]
 \includegraphics[width=0.99\textwidth]{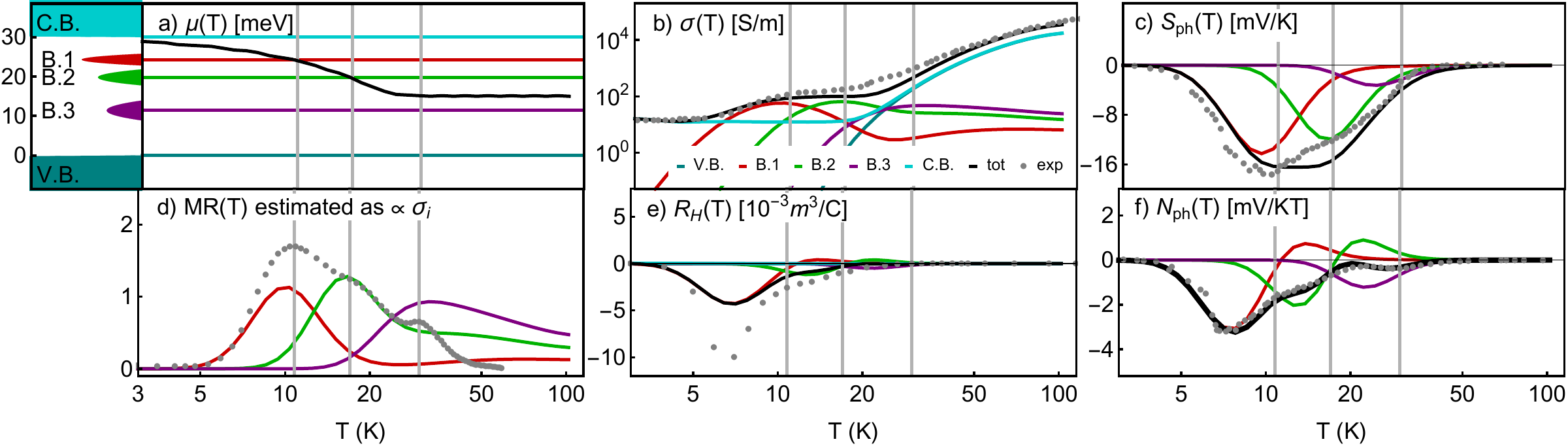}
 \caption{{\bf Boltzmann modelling of transport functions of FeSb$_2$.}
(a) electronic structure consisting of a valence band (VB), a conduction band (CB), and three in-gap states (B.1-B.3) at energies $E_i=11.4$, $19.8$, $24.3$ meV. The black line
indicates the temperature dependence of the chemical potential. (b) conductivity $\sigma$ including contributions from all states. (c)
phonon-drag thermopower owing to an injection of momentum into the in-gap states. (d) conductivity contributions of the in-gap states in comparison
to the magneto resistance MR. (e) the Hall coefficient, (f) the phonon-drag Nernst coefficient.
The grey vertical lines are guides to the eye indicating maxima in the MR.
From Battiato \etal\ \cite{FeSb2_Marco}, with experimental data from Sun \etal\ \cite{PhysRevB.88.245203}.
}
 \label{fig:threelevels}
\end{figure*}

Battiato \etal\ \cite{FeSb2_Marco} recently investigated transport properties of FeSb$_2$
by fitting experimental data to Boltzmann expressions for different microscopic scenarios:
Consistent with earlier findings\cite{jmt_fesb2}, a purely electronic setting was found unable to account for the large amplitude of thermoelectric quantities,
and, moreover, did not produce the characteristic temperature scales. 
Indeed, considering that both the Hall coefficient and the thermopower are anti-symmetric under particle-hole inversion,
one would expect the two quantities to exhibit a somewhat similar behaviour. Remarkably, the Hall coefficient's profile is instead more similar to the Nernst coefficient.
Also, a conventional phonon-drag mechanism---in which momentum is transferred to valence and/or conduction state carriers---did not result in the correct temperature profiles of observables.
Finally, Battiato \etal\ \cite{FeSb2_Marco} proposed a new, unconventional phonon-drag scenario: Assuming the presence of in-gap excitations and a phonon-drag that couples to their density of states, a quantitative description of all major transport functions was achieved, see \fref{fig:threelevels}.

The congruence of the temperature profile can be explained by the fact that in this scenario, the thermopower and the Nernst coefficient 
have the opposite symmetry under a particle-hole transformation when compared to their usual purely electronic counterparts: In the phonon-drag scenario
the Seebeck coefficient is even, while the Nernst coefficient is odd, see \tref{tab:f_functions}.%
\footnote{It should further be stressed that in the modelling of Ref.~\cite{FeSb2_Marco} the positions and linewidths of features
in the transport observables could not (within reason) be tuned with parameters such as the lifetimes and effective masses, but are determined
solely by the positions $E_i$ of the in-gap states.
Further, it was shown that capturing the large magnitude of the thermopower and the Nernst coefficient, required a momentum transfer of phonons
into the electronic system that is small enough to have no apparent signature onto the thermal conductivity. 
Indeed, the experimental thermal conductivity is dominated by phonon-crystal boundary scatterings \cite{0295-5075-80-1-17008}.
} 

In the Nernst coefficient, the proposed phonon-coupling to the density of in-gap states circumvents Sondheimer cancellation\cite{PhysRevB.64.224519}, 
without requiring multi-band effects.  Previously, it had been proposed that a strongly energy-dependent scattering rate $\Gamma(\epsilon)$ causes the colossal Nernst signal\cite{PhysRevB.88.245203}. 
For a scattering rate of the form $\Gamma(\omega)\propto\omega^r$, the exponent can be determined by $r=-e/k_B\times \nu/(R_H\sigma)$. 
The phonon-drag formalism formally yields a large exponent $r=-229.2$, in reasonable agreement with Ref.~\cite{PhysRevB.88.245203}, albeit in the absence of strongly dynamical electronic correlation effects.

{
\begin{table}
  \begin{tabular}{ | r | c | c | c | c | c |}
  \hline
\rowcolor{blue} transport kernel& $\sigma$ & R$_H$ & S & N & MR \\
  \hline 
\cellcolor{lblue}  electronic &\includegraphics[width=0.03\textwidth]{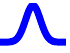} &  \includegraphics[width=0.03\textwidth]{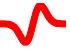}  &  \includegraphics[width=0.03\textwidth]{Figures/odd.pdf}  &  \includegraphics[width=0.03\textwidth]{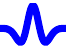} & \includegraphics[width=0.03\textwidth]{Figures/even.pdf} \\ \hline
  \cellcolor{lblue}phonon-drag & - & - &  \includegraphics[width=0.03\textwidth]{Figures/even.pdf} &  \includegraphics[width=0.03\textwidth]{Figures/odd.pdf} & - \\ \hline
  \end{tabular}
   \caption{{\bf Symmetry of transport kernels.} 
	The curves show the schematic shape (even: blue, odd: red) of the kernel (=integrands) of the transport functions: conductivity, Hall coefficient, thermopower, Nernst coefficient and magnetoresistance. E.g., for the conductivity, $\sigma\propto
	\int d\omega \left(-\frac{\partial f}{\partial w}\right) \sum_\svek{k}Tr \left[ v^\alpha_\svek{k}A_\svek{k}(\omega)v^\beta_\svek{k}A_\svek{k}(\omega)\right]$, the integrand is
	even, i.e., the sign of contributions is positive for both electrons ($\omega>0$) and holes ($\omega<0$), cf.\ \eref{KuboA}. 
	The even/odd character of the thermopower and the Nernst coefficient swap when going from the electron diffusive to the phonon-drag scenario.
	Reproduced from Battiato \etal\ \cite{FeSb2_Marco}. 
	} 
    \label{tab:f_functions}
\end{table}
}

\subsubsection{Existence and possible origin of in-gap states.}
\label{ingap}

To strengthen the unconventional phonon-drag picture, we here review experimental findings regarding in-gap states and their nature in FeSb$_2$, as well as mention
attempts to pinpoint their origin using realistic electronic structure calculations.

\begin{figure}[!th]
 \includegraphics[width=0.49\textwidth]{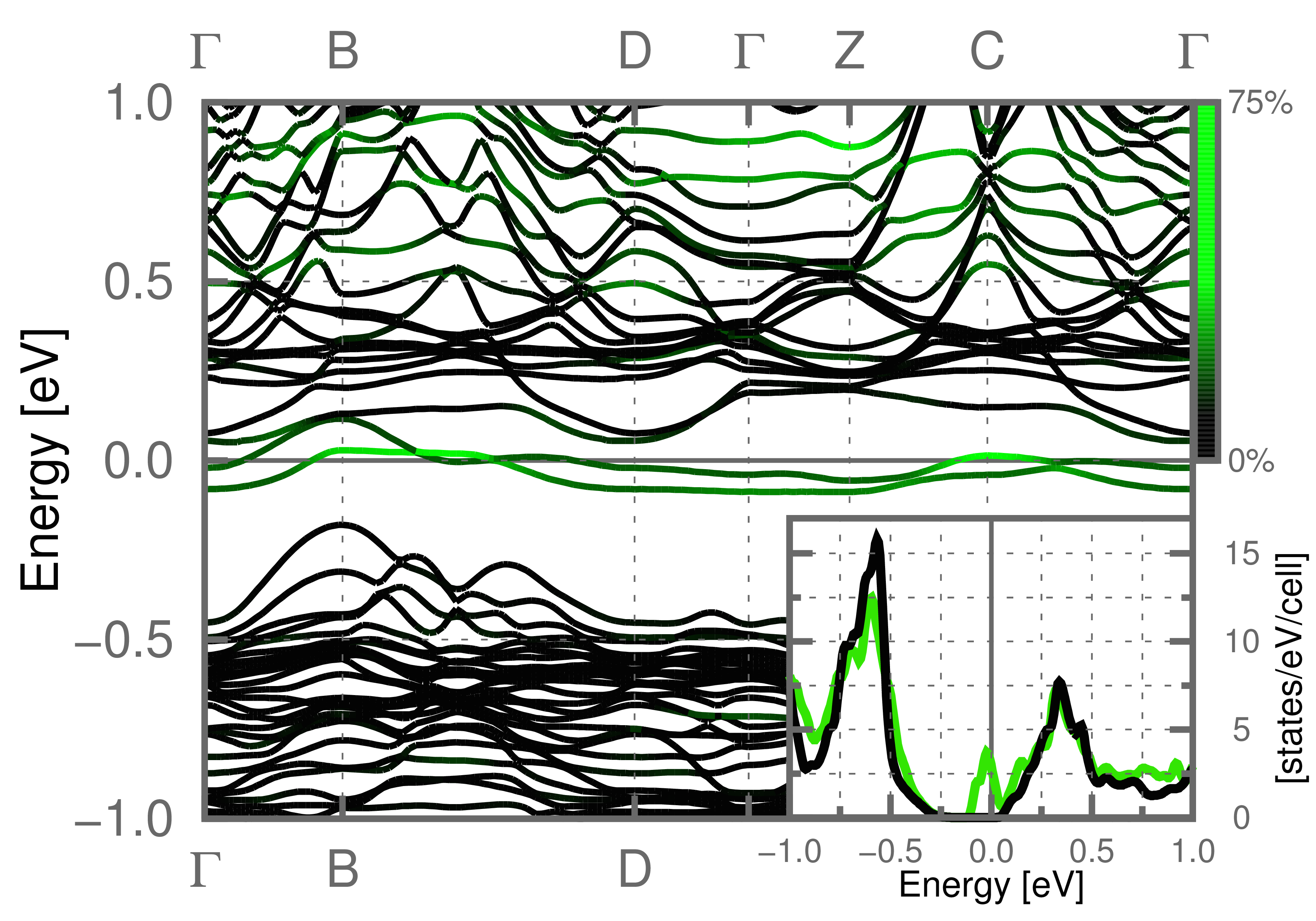}
 \caption{{\bf Antisite defects in FeSb$_2$.}
Band-structure of Fe$_{25}$Sb$_{47}$, mimicking Fe$_{1+x}$Sb$_{2-x}$.
Hues of green indicate orbital admixtures originating from the extra iron atom.
Inset: density of states of pure (black) and Fe-rich (green) FeSb$_2$.
This calculation uses the modified Becke-Johnson (mBJ)\cite{mBJ} exchange potential to account for the insulating nature of FeSb$_2$. From Ref.~\cite{FeSb2_Marco}.
}
 \label{excessFe}
\end{figure}

In-gap states in FeSb$_2$ have been held responsible for activation-law regimes with distinct energies in the material's resistivity\cite{Fan1972136,0295-5075-80-1-17008},
as well as the low-temperature upturn in the magnetic susceptibility\cite{koyama:073203}.
Using pressure-dependent magneto-transport measurements, Takahashi \etal\ \cite{Takahashi2013}, disentangled the band-gap energies and found strong evidence for 
the smallest gap to originate from an impurity band.

Contrary to FeSi\cite{doi:10.1063/1.372679}, also low-temperature nuclear magnetic resonance measurements of both FeSb$_2$ and RuSb$_2$ (as well as FeGa$_3$\cite{PhysRevB.89.104426}) by Gippius \etal\ \cite{Gippius_AMR,1742-6596-150-4-042040}  show clear indications of localized in-gap states,
suggested to be located slightly below the conduction band and to be of $S=1/2$ character. This finding is consistent with the strong reaction of the in-gap states
to magnetic fields (see the magnetoresistance in \fref{fig:threelevels}(d)). Also, Storchak \etal\ \cite{0953-8984-24-18-185601} interpreted their muon spin rotation spectra as showing the presence of in-gap states, proposed to be of spin-polaron nature.

Modelling several defect and impurity scenarios from first principles, Battiato \etal\ \cite{FeSb2_Marco} found the off-stoichiometry Fe$_{1+x}$Sb$_{2-x}$ to be a likely candidate for the emergence of in-gap states. Only for this substitution, electronic states were found to emerge inside the fundamental gap.
\Fref{excessFe} shows a simulation of a single anti-site defect: Fe$_{25}$Sb$_{47}$. Indeed two impurity bands appear close to the conduction states. 
The size of the supercell corresponds to $x\approx 0.04$ in Fe$_{1+x}$Sb$_{2-x}$. This concentration is, however, much larger than the estimated number of defect,
 $x\approx 4.5\cdot 10^{-5}$, that is needed to account for the phonon-drag amplitude.
Using Curie's law, it was shown\cite{FeSb2_Marco} that the expected magnetic signature of the required excess iron is consistent with 
the experimental magnetic susceptibility, which shows a low-temperature upturn of less than
$5\cdot 10^{-5}$emu/mol\cite{PhysRevB.67.155205,PhysRevB.72.045103,PhysRevB.74.195130,koyama:073203,APEX.2.091102,sun_dalton,PhysRevB.83.184414} for all accessed temperatures.
The proposed origin of in-gap states is further supported by the binary Fe-Sb phase diagram\cite{Richter1997247} that may suggest Fe-richness.
Moreover, Fuccillo \etal\ \cite{cava2013fesb2} demonstrated that even the smallest iron concentrations in Ru$_{1-x}$Fe$_x$Sb$_2$ causes the appearance of the low temperature characteristics of FeSb$_2$, suggesting a possible link to the transition-metal stoichiometry.
Finally, recent transport measurements of Sanchala \etal\ \cite{Sanchela2015205} showed that excess iron (or, equivalently, Sb-deficiency) has a large impact onto the peak value
of the thermopower.

\subsubsection{Outlook.}
\label{pdrago}

At first glance, the phonon-drag scenario for FeSb$_2$ appears a disappointment on several levels. However, a closer inspection
actually reveals new paths for future research:

(i) {\it phonon engineering.}
	The phonon-drag contribution to the thermopower $S_{ph}$ and the thermal lattice conductivity $\kappa$ are inter-linked, see, e.g., \fref{Takahashi}.
	Maximizing the first while minimizing the latter hence means satisfying antagonistic effects.
	The usual electron-crystal--phonon-glass paradigm is therefore likely to be unsuitable: An undifferentiated decrease in
	the phonon mean-free path to lower $\kappa$ will equally diminish the phonon-drag effect. 
	Any attempt at phonon engineering has therefore to be selective.
	It is {\it a priori} uncertain whether a separation between the phonon modes that dominate the phonon-drag and those propelling thermal conduction
	is possible, let alone whether they can be independently manipulated.
	In a pioneering work, Zhou \etal\ \cite{Zhou01122015} demonstrated the feasibility of a phonon filter that achieves just that.
	The authors showed that in the case of silicon
	the phonon-drag induced figure of merit, $ZT$, could be boosted by more than one order of magnitude. The key observation is that phonon modes contributing to the
	phonon-drag have typically a larger mean-free path, as well as notably lower energies than modes dominating thermal conduction.
	It is then proposed that nano-clusters can be used as impurities to preferentially scatter high-energy phonons, thus acting as a low-pass filter\cite{Zhou01122015}.
	These findings provide a new path for optimizing the phonon-drag mechanism in
	thermoelectric materials for applications at temperatures up to some fraction of their Debye temperatures.

(ii) {\it occurrence in other materials.}
  The phonon-drag scenario proposed for FeSb$_2$ relies on an unconventional mechanism, in which momentum is transferred to in-gap states.
  However, defect states are notoriously difficult to control (cf.\ the large sample dependence of $S$). Therefore it could be expected that technical applications are cumbersome to
	realize for this system and that the microscopic mechanism might be non-universal, i.e., not found in other materials.
  Yet, despite the large variation in the thermopower, the signatures of a narrow band inside the fundamental gap is consistently
	found around the same energy, not just in other transport and spectroscopic probes, but also in the electronic specific heat. This may indicate
	that the impurity band is an intrinsic feature of the FeSb$_2$ compound or crystal structure.
	The magnitude of the phonon-drag thermopower will, however, depend on minute details, and the interplay of position of in-gap states, the fundamental gap, the electron-phonon coupling, as well as phonon energies. It seems plausible that the inter-period substitutions Ru for Fe, or As for Sb, thus sufficiently perturb the phonon system
	(in addition to changes in the electronic structure, see \sref{marcasites}), so as to destroy the conditions for a large phonon drag in FeAs$_2$ and RuSb$_2$.
  A substitution that changes the mass-balances much less is replacing Fe with Cr. Interestingly,	the isostructural CrSb$_2$---despite its quite different electronic structure
	(d$^4$ configuration in antiferromagnetic CrSb$_2$ vs.\ d$^6$ in paramagnetic FeSb$_2$)---also violates the upper bound of electronic diffusive thermoelectricity\cite{PhysRevB.86.235136},  see \fref{Sscaled}.
	Therefore it is tempting to propose that the material realizes the same phonon-drag mechanism as in FeSb$_2$.
	Indeed, also in CrSb$_2$ the existence of a donor impurity band below the conduction states has been suggested\cite{PhysRevB.86.235136}.
	A further system could be FeGa$_3$\cite{PhysRevB.90.195206}. As mentioned before, nuclear magnetic resonance measurements evidenced in-gap states there as well\cite{PhysRevB.89.104426}.
	Also, more recent thermopower data\cite{PhysRevB.90.195206}, while not violating $\Delta/T$, do exhibit a characteristically sharp peak of up to -15meV/K at 12K, as well
	as a pronounced sample, and interestingly, polarization-dependence in the thermoelectric response.
	In conclusion, the phonon-drag effect in intermetallic narrow-gap semiconductors with open $d$-electron shells might be more prevalent than initially thought.
	Further experimental investigations, in particular magnetoresistance and Nernst effect measurements for CrSb$_2$ and FeGa$_3$ are called for.

(iii) {\it correlation effects.} In the scenario for FeSb$_2$ electronic correlation effects do not seem to play any role. Indeed, the phonon-drag mechanism
does not inject momentum into the correlation-effects hosting valence and conduction carriers.
Instead phonons couple to in-gap states.
As remarked by Takahashi \etal\ \cite{Takahashi2016}, this, in fact, makes
the phonon-drag effect particularly large:
Indeed, according to the phenomenological treatment due to Herring \cite{PhysRev.96.1163}, the phonon-drag contribution to the thermopower
reads $S_{ph}=\pm\alpha \frac{\tau_{ph}v_{ph}^2}{\mu T}$. Here, $\tau_{ph}$ and $v_{ph}$ are the phonon lifetime and velocity, respectively,
and $\alpha$ a parameter related to the electron-phonon coupling strength that determines the fraction of the phonon momentum-loss/transfer.
Crucially, the $S_{ph}$ is inversely proportional to the mobility of the electrons ($-$) or holes ($+$) that dominate the regular electron diffusive thermopower. In the semi-classical picture, $\mu=e\tau/m^*$, where $\tau$ is the electronic lifetime and $m^*$ the charge carriers' effective mass. 
Hence, {\it large effective masses enhance the phonon-drag}. From cyclotron resonance spectra,  a mass $m^*=5.4m_e$ was indeed extracted for 
the impurity states in FeSb$_2$\cite{Takahashi2016}.
Masses of this order of magnitude are, however, not at all uncommon for intrinsic carriers in correlated materials.
This suggests that strongly correlated electrons may actually lead to enhanced thermopowers and powerfactors also in semiconductors---with a little help of phonons.

\medskip

Hence, instead of wrapping up and shelving the FeSb$_2$ puzzle, the reviewed recent theoretical and experimental insights actually open up new directions in the field of thermoelectrics. While most likely not relevant to thermoelectricity in FeSb$_2$, studying its colossal thermopower may have led to 
the unravelling of a new paradigm: the combination of the time-honoured phonon-drag effect and the physics of strongly correlated
electrons.

\begin{framed}
	\noindent
	{\bf Lessons from FeSb$_2$: colossal thermopower caused by unconventional phonon-drag}
	\begin{itemize}
		\item FeSb$_2$ and CrSb$_2$ violate upper bound \eref{Smax} of purely diffusive thermopower.
		\item correlation of features (extrema, inflection points) in diverse transport observables constrain microscopic modelling.
		\item phonon-drag coupling to extrinsic in-gap states gives a consistent description of all transport quantities.
		\item potential path for large thermopowers: coupling phonons to correlated electrons with large effective masses: enhanced phonon-drag thermopowers; 
		advanced phonon-engineering to reduce detrimental thermal conduction while leaving beneficial phonon-drag intact.
	\end{itemize}
\end{framed}

\subsection{Skutterudite Antimonides: a focus on CoSb$_3$.}
\label{skutts}

Skutterudites are the prototypical compounds in which the phonon-glass, electron-crystal paradigm\cite{Slack1995} can be realized\cite{PhysRevB.56.15081}.
These materials crystallize in the CoAs$_3$-structure (cubic space-group $Im\bar{3}$) harbouring four formula units.
Crucially, the structure has voids at the 2a Wyckoff positions, that can be filled with
rare earth, alkali metal, alkaline earth, or other elements, that are only weakly bound to the skutterudite host.
As a consequence, the localized fillers exhibit large anharmonic vibrations that diminish the propagation of
phonons, and thus heat.%
\footnote{However, it has been noted that even some of the unfilled skutterudites have intrinsically low thermal conductivities,
see, e.g., Ref.~\cite{PhysRevB.94.075122} for a discussion of FeSb$_3$.}
Besides the possibility to reach low thermal conductivities, skutterudite systems exhibit notably large powerfactors.
For example, $S^2\sigma$ of CoSb$_3$ reaches $9.5\mu$W/(K$^2$cm) at 625K\cite{doi:10.1063/1.3557068} (see \tref{table1}).
Over the years, this led to a continuous optimization of the thermoelectric figure of merit.
$ZT\sim 1$ at 800K was achieved in the year 2000\cite{Sales1325}, multiple fillings boosted the value to
$\sim1.6$ in 2011\cite{doi:10.1021/ja111199y}, and $\sim1.9$ was reached in 2016\cite{ROGL201430}.

Providing an all-encompassing overview of the research in skutterudites is far beyond the scope of this work.
Instead we will focus on a single (unfilled) compound, CoSb$_3$, to highlight challenges in the {\it ab initio} modelling of
its thermoelectric properties.

\subsubsection{Experimental and theoretical knowledge.} 

\begin{figure*}[th]
  \begin{center}
	\subfloat[CoSb$_3$: thermopower]
	{\includegraphics[clip=true,trim=0 0 0 0,angle=0,width=.45\textwidth]{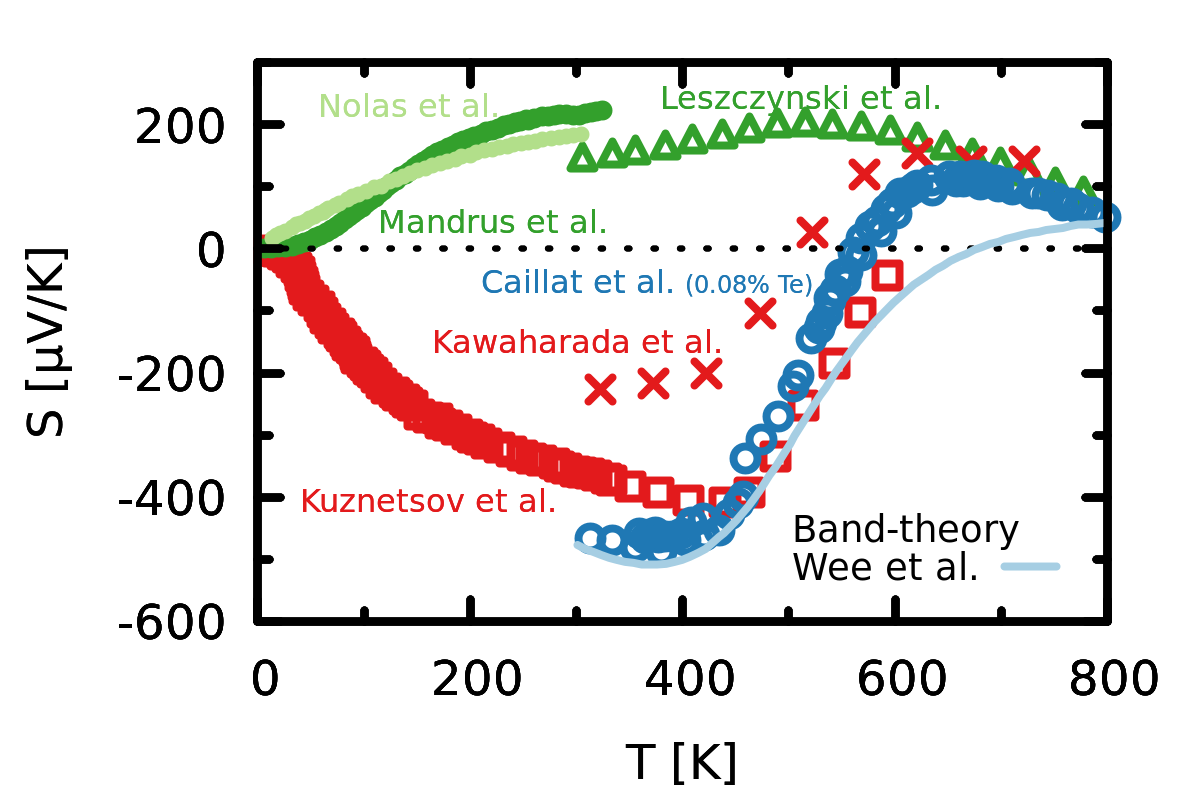}}
	\subfloat[Co$_{0.999}$Ni$_{0.001}$Sb$_3$: thermopower]
	{\includegraphics[clip=true,trim=0 0 0 0,angle=0,width=.45\textwidth]{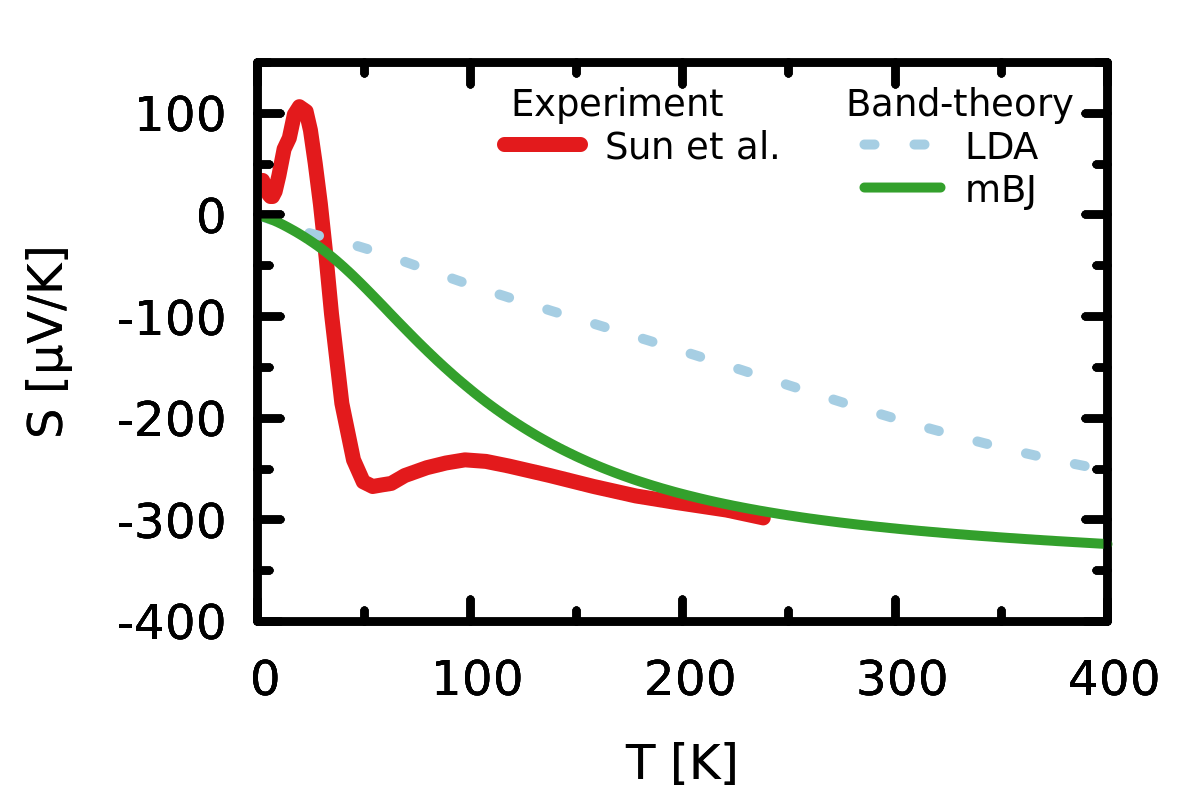}}
	
	\subfloat[CoSb$_3$: band-structure]
	{\includegraphics[clip=true,trim=0 0 0 0,angle=0,width=.45\textwidth]{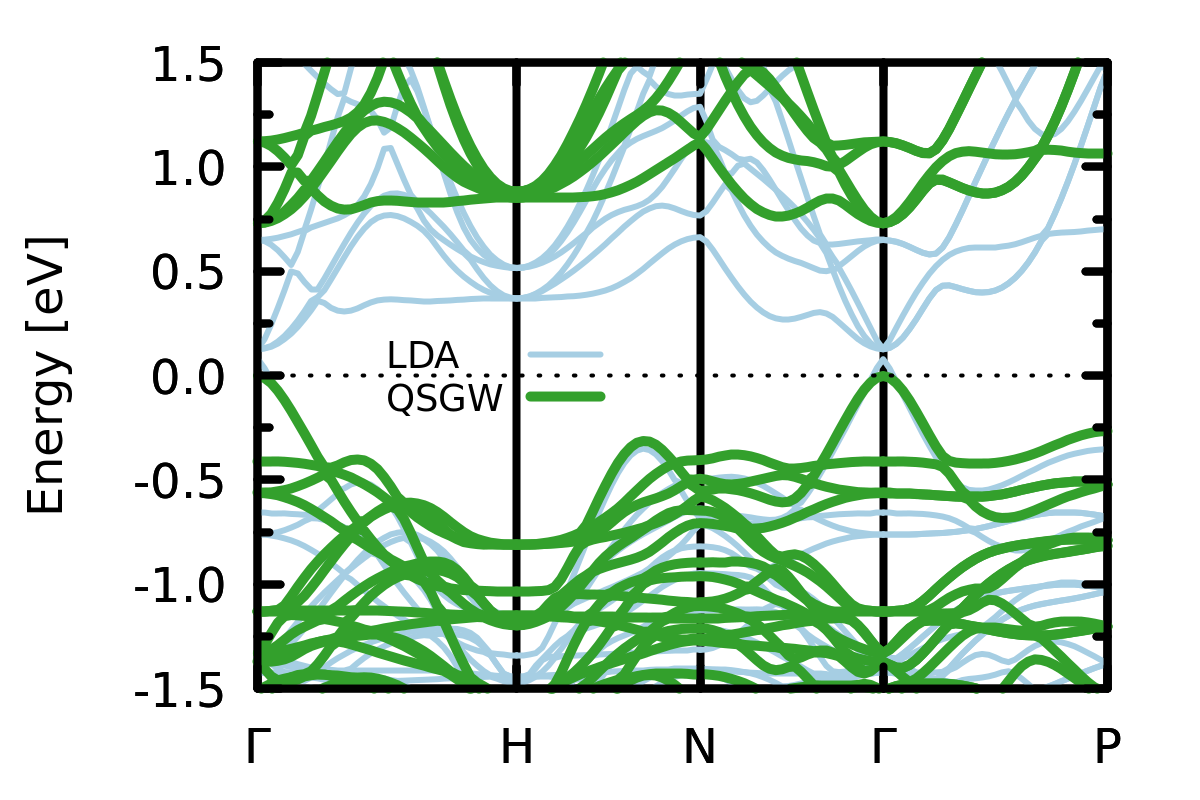}}
	\subfloat[Co$_{0.999}$Ni$_{0.001}$Sb$_3$: band-structure]
	{\includegraphics[clip=true,trim=0 0 0 0,angle=0,width=.45\textwidth]{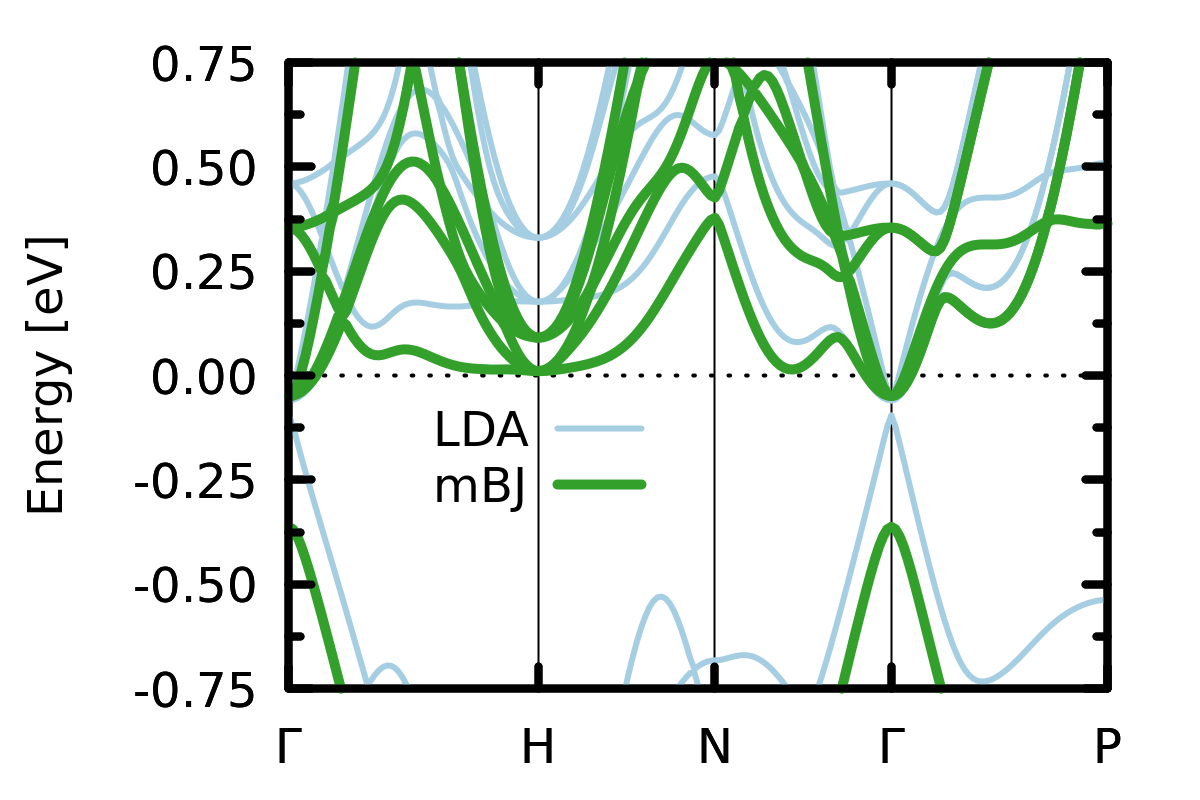}}
		\caption{{\bf Thermopower and band-structure of CoSb$_3$.} (a) thermopower of CoSb$_3$ from  
		Mandrus \etal\ \cite{PhysRevB.52.4926},
		Nolas \etal\ \cite{PhysRevB.58.164},
		Leszczynski \etal\ \cite{0022-3727-46-49-495106},
		Kawaharada \etal\ \cite{KAWAHARADA2001193},
		Kuznetsov \etal\ \cite{0953-8984-15-29-315}, and of a
   		lightly electron-doped (0.08\% Te) sample from Caillat \etal\ \cite{doi:10.1063/1.363405}.
			A comparison is shown to Boltzmann transport calculations of Wee \etal\ \cite{PhysRevB.81.045204} based on an LDA band-structure with the carrier density adjusted 
			to the value extracted from the Hall coefficient of the n-type sample of Caillat \etal\ \cite{doi:10.1063/1.363405}.
		(b) thermopower of lightly Ni-doped CoSb$_3$. Experimental data from Sun \etal\ \cite{Sun_CoSb3}. The simulated thermopowers 
		use the Kubo code from Ref.~\cite{jmt_fesi} with a constant scattering rate $\Gamma=8$meV and foot on the LDA and mBJ band-structures shown in panel (d). There the doping is simulated in the virtual crystal approximation (VCA).
		(c) LDA and QS{\it GW} band-structure of stoichiometric CoSb$_3$.
		(b) \& (d) adapted from Sun \etal\ \cite{Sun_CoSb3}.}
      \label{CoSb3}
      \end{center}
\end{figure*}

To start, experimental results for this compound scatter notably\cite{doi:10.1063/1.360402}. For example, values for the charge gap, as extracted
from activation-law fits to the resistivity fall into the large range of 230-700meV\cite{doi:10.1063/1.360402,Tang2015,KAWAHARADA2001193,PhysRevB.52.4926}.
Optical absorption experiments find, depending on temperature, a gap from 250 to 320meV, while
Shubnikov-de Haas measurements\cite{RAKOTO199913}  and photoemission experiments\cite{doi:10.1143/JPSJ.71.2271} suggest gap values as low as 30-40meV.
Also the thermopower is highly dependent on the sample preparation\cite{ZAAC:ZAAC201500179}, to the extend that both $p$ and $n$-type samples have been measured\cite{doi:10.1063/1.360402,PhysRevB.52.4926,PhysRevB.56.1911,PhysRevB.58.164,doi:10.1063/1.363405,0022-3727-46-49-495106,KAWAHARADA2001193,0953-8984-15-29-315},
see \fref{CoSb3}(a).
From the Hall coefficient ($R_H>0$\cite{PhysRevB.56.1911,PhysRevB.65.115204}) as well as electronic structure calculations (DOS larger for conduction states, see \fref{CoSb3}(c))
one would naively expect a positive thermopower for the stoichiometric compound.
One might surmise that the origin of these variations lies in the loosely packed crystal structure which makes skutterudite compounds prone to diverse types of defects.
Therefore, also the very small gap values extracted from Shubnikov-de Haas and photoemission experiments might be linked to impurity bands.

These experimental uncertainties are met with substantial variations in theoretical results for the perfectly stoichiometric compound.
Using experimental atomic positions
yields a direct gap in the Kohn-Sham spectrum of CoSb$_3$ that varies from 40-60meV (LDA)\cite{PhysRevB.50.11235,PhysRevB.58.15620}, over 170meV (ACBN0)\cite{PhysRevB.94.165166} to 310meV (mBJ)\cite{Sun_CoSb3}.%
\footnote{The valence count, however, varies little, there are $\sim 7.6$ $d$-electrons on each Co-site (wien2k with muffin-tin radius of $2.5a_0$).}
For the band-structure of slightly electron-doped CoSb$_3$ within LDA and mBJ, see \fref{CoSb3}(d) and its discussion below.
On top of this, also the equilibrium volume largely depends on the functional used in the relaxation: One finds 
difference of about 7\% when relaxing the structure within LDA or GGA \cite{PhysRevB.58.15620}.
Using these theoretical structures, large variations in the gap size are found for a given functional\cite{PhysRevB.81.045204}.
The strong link between electronic and structural properties is also congruent with quantum molecular dynamics simulations\cite{C7TC03603E}.
The latter evidence a substantial temperature-dependence in the electronic structure in CoSb$_3$, and might explain the temperature-induced band-convergence
seen in experiments\cite{Tang2015}.

Within DFT it is the treatment of exchange contributions that largely affects the size of the gap. 
Using quasi-particle self-consistent QS{\it GW}, we find an even larger gap of
730meV, see \fref{CoSb3}(c). This increase in the size of the gap with respect to DFT-LDA occurs despite the presence of significant
band-narrowings. Indeed, within QS{\it GW} the valence states are renormalized towards the Fermi with a factor of $m^*/m_{LDA}\sim1.8$.
Due to the perturbative nature of the {\it GW} approach, it tends to underestimate effective masses in compounds with open 3$d$-shells
 (see also the discussion on di-antimonides in \sref{marcasites}).
This finding thus indirectly advocates the presence of substantially enhanced effective masses in CoSb$_3$.
Experimentally, masses of 2-5$m_e$ were indeed extracted from the Hall coefficient\cite{doi:10.1063/1.363405}, or from the thermopower when doping with Ni \cite{doi:10.1063/1.371287}, and 1-3$m_e$ when adding Yb fillers \cite{Tang2015}.
Given the competition\cite{jmt_pnict,jmt_svo_extended} of exchange self-energies (widen gaps) and mass enhancements (shrink gaps),
as well as moderately large values of effective masses,
an accurate modelling of the electronic structure of skutterudites requires a methodology that treats exchange and correlations beyond DFT,
and correlations beyond the perturbative {\it GW}. Viable approaches will be mentioned in \sref{conclusions}.

Despite issues related to defect-derived in-gap states and the putative presence of electronic correlation effects, several transport observables have been successfully modelled
from first principles: Guo \etal\ \cite{Guo2015} computed the thermal lattice conductivity of unfilled and filled CoSb$_3$ and found good agreement with experiments.
Also thermoelectric simulations have been performed with some success: 
Using the Boltzmann response formalism on top of band-structure calculations, several groups\cite{PhysRevB.58.15620,PhysRevB.71.155119,PhysRevB.81.045204,PhysRevB.72.085126,PSSA:PSSA201532609} reproduced experimental thermopowers when adjusting the carrier density to values extracted from the experimental Hall coefficients, see \fref{CoSb3}(a) for an example.

\subsubsection{Doped CoSb$_3$.}
While for thermoelectric applications, doping CoSb$_3$ by introducing filler atoms is the way to go, let us here nonetheless highlight
the intriguing case of lightly Ni-doped CoSb$_3$\cite{doi:10.1063/1.371287,PhysRevB.65.115204,Sun_CoSb3}. 
Indeed introducing 3$d$-transition-metal-derived carriers to the skutterudite has pronounced effects on transport and magnetic properties even for very low concentrations\cite{PhysRevB.65.115204,Sun_CoSb3}:
Up to 1\% Ni, the resistivity remains insulating, while the magnetic susceptibility develops a Curie-tail at low temperatures.%
\footnote{At larger dopings the system becomes metallic with mobile charge carriers having effective masses enhanced by a factor of 2-5\cite{doi:10.1063/1.371287,PhysRevB.65.115204}, similar to CoSb$_3$ with Yb-fillings\cite{Tang2015}.
 }
The Hall coefficient becomes negative
already for the tiniest amounts of Ni. Concomitantly, as seen in \fref{CoSb3}(c) for 0.1\% Ni, also the thermopower is negative, with the exception of a pronounced positive peak below 25K, that is accompanied by a large peak in the Nernst coefficient\cite{Sun_CoSb3}.

To first approximation, one can try to model the electronic structure of Co$_{1-x}$Ni$_x$Sb$_3$ within the virtual crystal approximation (VCA). The resulting LDA and mBJ band-structures are shown in \fref{CoSb3}(c) for $x=0.001$.
These bands are very close to those without doping, albeit shifted rigidly downwards.
Crucially the mBJ electronic structure is notably more particle-hole asymmetric than the LDA one, owing especially to a
rearrangement of conduction bands, as well as the increased gap below the conduction states.
The corresponding thermopower has been computed within the Kubo formalism, using a constant scattering rate and fixing the temperature dependence of  the chemical potential by the 
requirement of charge neutrality (instead of adjusting the carrier density to experimental values\cite{PhysRevB.58.15620,PhysRevB.81.045204,PhysRevB.72.085126}).
With the chemical potential residing inside the conduction bands, the LDA and mBJ-based thermopowers shown in \fref{CoSb3}(c) are metallic.
We note that---as expected---the visibly more particle-hole-asymmetric mBJ-band-structure translates into a larger thermopower.
Vis-{\`a}-vis experiment, the mBJ-based thermopower approaches the experimental result above 100K.
In fact---if the electronic structure is reliable and lifetime effects are minor (cf.\ \sref{KubBoltz})---congruence between simulation and experiment is
expected at high enough temperatures when putative donor-band contributions can be neglected.
This quantitative congruence suggests that the mBJ approach gives a qualitatively correct description of the intrinsic electronic structure 
of (lightly doped) CoSb$_3$.
The qualitative discrepancy below 100K, however, point to sizable contributions to the thermopower from
effects not contained in the (intrinsic) band-structure. 
Indeed, using a phenomenological band-structure that includes a spin-orbit splitting (see also Ref.~\cite{PhysRevB.71.155119}) and several impurity bands, Kajikawa \cite{Kajikawa2016} 
successfully fitted the transport observables of Sun \etal\ \cite{Sun_CoSb3}.
In the grander scheme of things, the onset of the extrinsic low-temperature region can be viewed as a transition into a regime of vastly different electronic mobility\cite{Sun_CoSb3}.
As discussed in \sref{correlS}, such mobility gradients can be a source of enhanced thermoelectric effects.
Indeed, the large Nernst coefficient in Co$_{0.999}$Ni$_{0.001}$Sb$_3$ can be quantitatively explained in this scenario\cite{Sun_CoSb3}.

\begin{framed}
	\noindent
	{\bf Lessons for the skutterudite CoSb$_3$: }
	\begin{itemize}
		\item unreliable electronic structures: large dependence on DFT functional.
		\item indications for competition of exchange and correlation effects beyond DFT.
		\item need for electronic structure calculations to include defects.
		\item potential path for large thermoelectric effects: engineering of mobility gradients.
	\end{itemize}
\end{framed}

\subsection{Heuslers compounds: a focus on Fe$_2$VAl}
\label{heuslers}

The Heusler class of compounds
displays an enormous richness in physical phenomena, with applications in spintronics, solar cells, magneto-calorics, and thermoelectrics%
\cite{Graf20111,Galanakis2016}.
Here, we will limit the discussion to the so-called full-Heusler structure X$_2$YZ
which crystallizes in the cubic L2$_1$ structure (space group $Fm\bar{3}m$).
We again pick a representative of this subclass, that is both of fundamental and of potential technological interest: Fe$_2$VAl.

\begin{figure*}[!th]
  \begin{center}
{\includegraphics[clip=true,trim=0 0 0 0,angle=0,width=.45\textwidth]{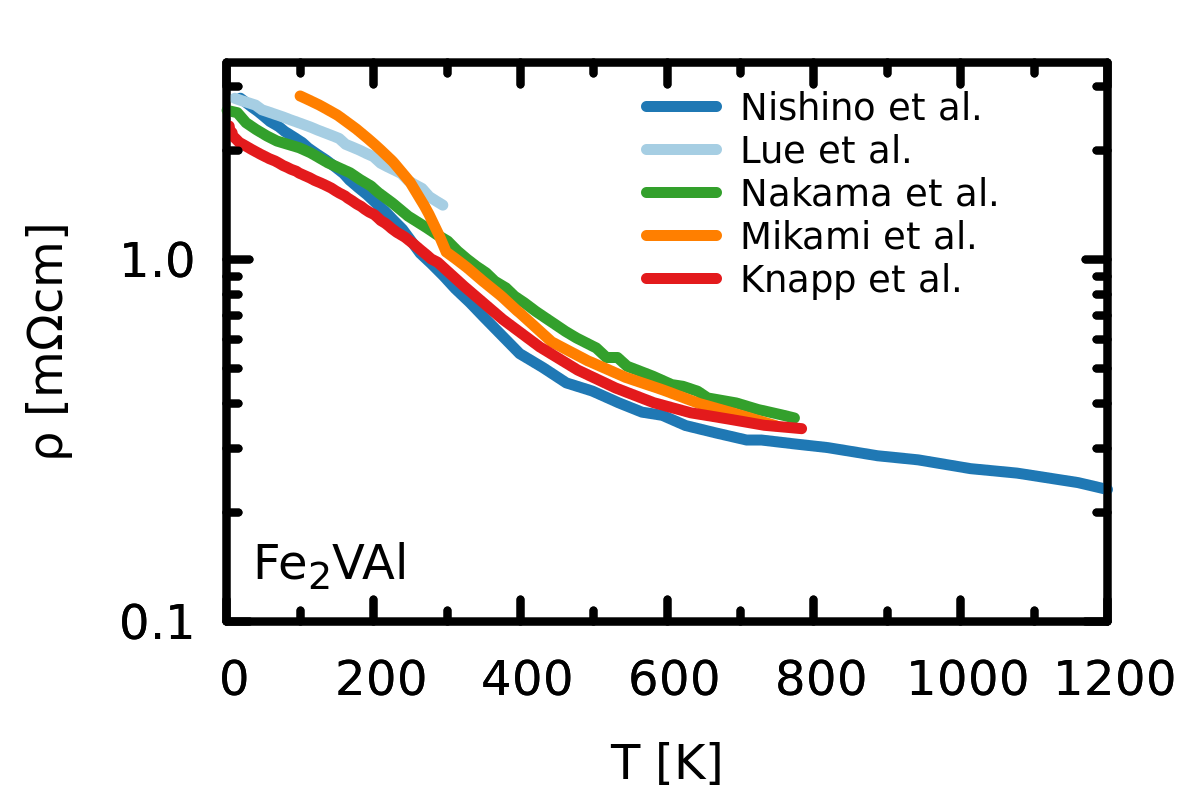}}
{\includegraphics[clip=true,trim=0 0 0 0,angle=0,width=.45\textwidth]{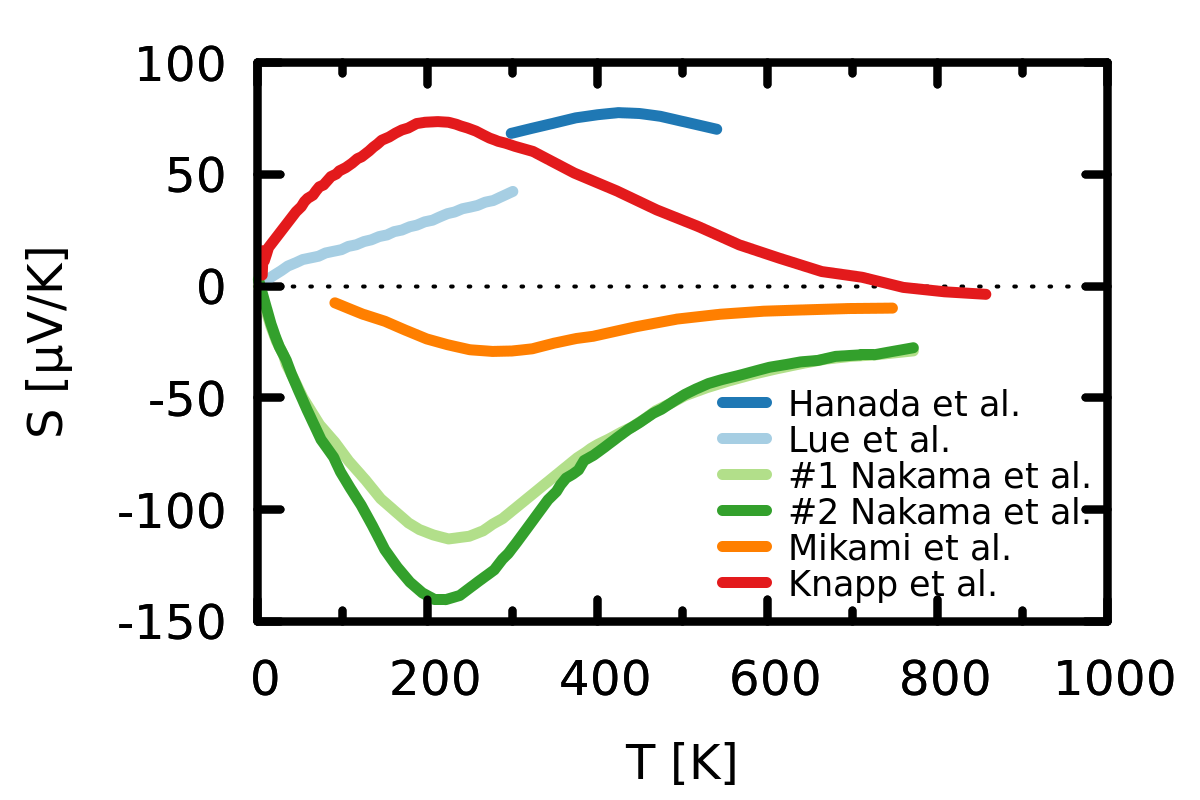}}
		\caption{{\bf Resistivity and thermopower of Fe$_2$VAl.} (a) resistivity from
		Nishino \etal\ \cite{PhysRevLett.79.1909}, 
		Lue \etal\ \cite{PhysRevB.66.085121},  
		Nakama \etal\ \cite{doi:10.1143/JPSJ.74.1378}, 
		Mikami \etal\ \cite{MIKAMI2009444}, 
		Knapp \etal\ \cite{PhysRevB.96.045204}, 
		(b) thermopower  from  
		Hanada \etal\ \cite{HANADA200163}, 
		Lue \etal\ \cite{PhysRevB.66.085121},  
		Nakama \etal\ \cite{doi:10.1143/JPSJ.74.1378} (2 different samples), 
		Mikami \etal\ \cite{MIKAMI2009444}, and 
		Knapp \etal\ \cite{PhysRevB.96.045204}. 
		}
      \label{Fe2VAlexp}
      \end{center}
\end{figure*}

\subsubsection{Experimental knowledge.} 
The material is a semiconductor with a narrow gap of 90-130meV\cite{0953-8984-12-8-318,doi:10.1143/JPSJ.74.1378}, 
100-200meV \cite{PhysRevLett.84.3674,Shreder2015}, or  210-280meV\cite{PhysRevB.58.9763} as extracted from transport, optical, and nuclear magnetic resonance (NMR) measurements, respectively. Indeed, the resistivity of \fval, \fref{Fe2VAlexp}(a) has a shape that seems typical for the materials investigated in this review.
A closer inspection, see \fref{overviewrho}, reveals, however, that the resistivity increase upon lowering temperature is orders of magnitude smaller than
in other materials with similar gap sizes.
Indeed, despite insulating characteristics in the resistivity, photoemission spectroscopy\cite{PhysRevLett.79.1909,doi:10.7566/JPSCP.3.017036,2053-1591-1-1-015901} sees a metallic edge at the Fermi level, and---the activated behaviour of the spin-lattice relaxation rate notwithstanding---also NMR\cite{PhysRevB.58.9763} suggests the presence of a residual density of states. 
This disparity between transport (and thermodynamic) quantities on the one hand, and spectroscopic probes on the other has
been suggested to derive 
from off-stoichiometry\cite{PhysRevB.63.233303,0953-8984-28-28-285601}, anti-site disorder\cite{PhysRevB.63.165109} or weak localization\cite{doi:10.7566/JPSCP.3.017036}.

In the spin sector, \fval\ has been characterized as paramagnetic down to lowest temperatures. Early 
specific heat measurements suggested metallic contributions from carriers with effective electronic masses of the order of 20-70 times the band mass\cite{0953-8984-10-8-002,PhysRevB.57.14352,PhysRevB.58.6855}.
Together this prompted the suggestion of possible $3d$-orbital-based heavy fermion behaviour\cite{0953-8984-10-8-002}. Subsequently, however,
 Lue \etal\ \cite{PhysRevB.60.R13941} established that these finding were owing to the presence of extrinsic magnetic clusters\cite{PhysRevB.78.064401}. 
Extracting the ``intrinsic'' contribution to the electronic specific heat from measurements in a varying magnetic field, however, still yielded masses enhanced by a factor of 5\cite{PhysRevB.60.R13941}.
The intrinsic magnetic susceptibility is much flatter with no signs of towards activated behaviour up to 300K\cite{ISHIKAWA2007e616,Shreder2008,0953-8984-28-28-285601}. NMR relaxation rates, however, do grow exponentially above 300K\cite{PhysRevB.58.9763}, yet without any signs towards a crossover to Curie-Weiss-like behaviour is seen up to 600K, see \fref{overviewchi}.

\begin{figure}[!t]
  \begin{center}
{\includegraphics[clip=true,trim=0 0 0 0,angle=0,width=.45\textwidth]{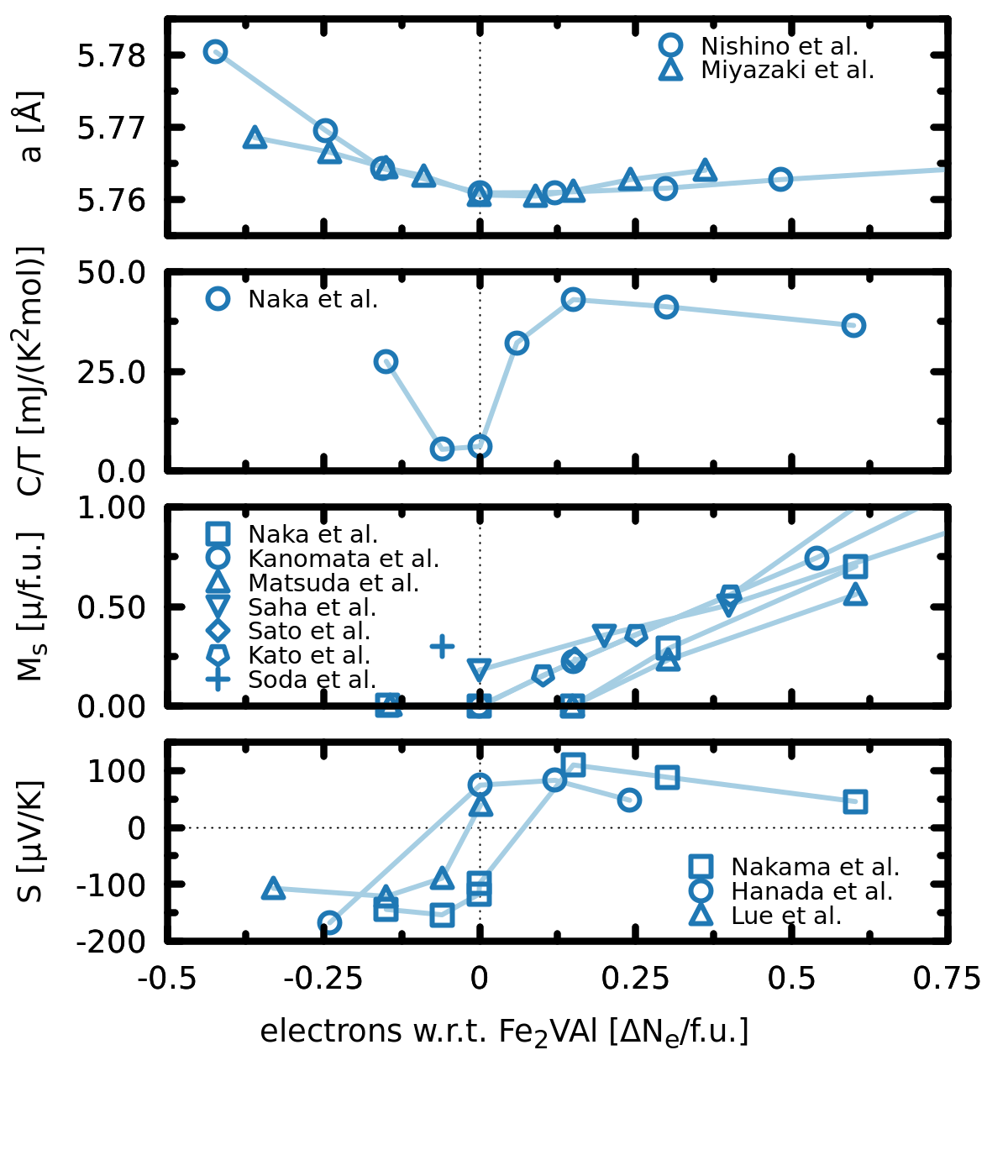}}
		\caption{{\bf Singular nature of Fe$_2$VAl.} From top to bottom: Composition dependence of (a) the lattice constant $a$, (b) the saturation ordered moment $M_s$, (c)
		the specific heat divided by temperature $C/T$  at $T=2$K, and (d) the thermopower $S$ at $T=300$K with respect to stoichiometric Fe$_2$VAl
		measured in electrons $\Delta N_e$ per formula unit.
		Data for
		(a) Fe$_{2+x}$V$_{1-x}$Al from Nishino \etal\ \cite{PhysRevLett.79.1909}, 
		Fe$_2$V$_{1+x}$Al$_{1-x}$ from Miyazaki \etal\ \cite{2053-1591-1-1-015901},
		(b) Fe$_{2+x}$V$_{1-x}$Al from Naka \etal\ \cite{PhysRevB.85.085130},
		(c) Fe$_{2+x}$V$_{1-x}$Al from Naka \etal\ \cite{PhysRevB.85.085130},
		Kanomata \etal\ \cite{KANOMATA2001390},
		Matsuda \etal\ \cite{doi:10.1143/JPSJ.69.1004},
		Kato \etal\ \cite{0953-8984-12-8-318},
		Fe$_2$V$_{1-x}$Cr$_x$Al from Saha \etal\ \cite{PhysRevB.79.174423},
		Fe$_2$VAl$_{1-\delta}$ from Sato \etal\ \cite{PhysRevB.82.104408},
		Fe$_{2−x}$V$_{1+x}$Al from Soda \etal\ \cite{SODA2004338}
		(partly extracted from Naka \etal\ \cite{0953-8984-25-27-275603}),
		(d) Fe$_{2+x}$V$_{1-x}$Al from Nakama \etal\ \cite{doi:10.1143/JPSJ.74.1378},
	   Hanada \etal\ \cite{HANADA200163},
		Lue \etal\ \cite{PhysRevB.66.085121}
		(extracted from Naka \etal\ \cite{PhysRevB.85.085130}).
		}
      \label{Fe2VAlprox}
      \end{center}
\end{figure}

In the chemical phase-space---given, e.g., by the pseudo-binary alloy series (Fe$_{1-x}$V$_x$)$_3$Al or the off-stoichiometry Fe$_2$V$_{1+x}$Al$_{1-x}$---the full-Heusler
composition \fval\ is in many respects distinguished.
This is illustrated in \fref{Fe2VAlprox}: As a function of the number of valence electrons, stoichiometric \fval\ realizes (a) the minimal equilibrium volume, (b)
the minimal specific heat, is (c) in direct proximity to ferromagnetic long range order, and (d) is placed close to thermoelectric particle-hole symmetry.

The property that both electron and hole doping increases the volume of \fval\ is reminiscent of the Kondo insulator \cbp\ (cf.\ the discussion in \sref{Ce3dope}).
However, the physical origin behind this phenomenon is very different: In \cbp\ the volume is indeed linked to the valence count\cite{FISK1995798}.
In the case of \fval\ the counter-intuitive volume decrease upon substituting iron with larger vanadium atoms has been suggested to originate
from a gain in cohesive energy. Indeed, Kato \etal\ \cite{Kato1998} have shown that long range structural order in (Fe$_{1-x}$V$_x$)$_3$Al consistently increases
when going from $DO_3$-Fe$_3$Al to the $L2_1$-Fe$_2$VAl ternary Heusler composition. 
This finding roots in the specifics of the Heusler crystal-structure: There are two inequivalent iron sites in Fe$_3$Al that, among others, have different nearest neighbour distances.
Vanadium, being larger than iron, preferentially substitutes the larger site, which gets completely filled when reaching the $L2_1$ composition\cite{NISHINO1997461}.
For excess vanadium, the smaller iron site gets populated and the lattice necessarily expands.%
\footnote{In this respect the substitution with vanadium is unique. Indeed, substituting with the larger Ti the volume increases continuously, while for
Cr---which has a radius only slightly larger than V---realizes a minimal volume at $x\approx0.2$ in (Fe$_{1-x}$Cr$_x$)$_3$Al\cite{NISHINO1997461}.}

That also the specific heat is minimal for \fval, indicates that the ternary Heusler compound is an insulating island immersed in a metallic sea of adjacent alloy compositions.
This finding is corroborated by high-temperature extrapolations of the magnetic susceptibility that find Pauli contributions to be minimal
for stoichiometric \fval\cite{0953-8984-28-28-285601}.
On top of that, \fval\ is in proximity to ferromagnetic long-range order that is reached by,
both, excess iron (e.g., for $x\ge 0.05$ in Fe$_{2+x}$V$_{1-x}$Al\cite{PhysRevB.85.085130}), and excess vanadium\cite{SODA2004338}. 
Indeed, stoichiometric \fval\ represents an exact zero of the Slater-Pauling function, $M=(N-24)\mu_B$, for the spontaneous ferromagnetic moment\cite{Graf20111,Galanakis2016}.
Here, $N$ is the total number of valence electrons per formula unit. Since here $N=24$, it follows that \fval\ is non-magnetic.

\begin{figure}[!th]
  \begin{center}
{\includegraphics[clip=true,trim=0 0 0 0,angle=0,width=.45\textwidth]{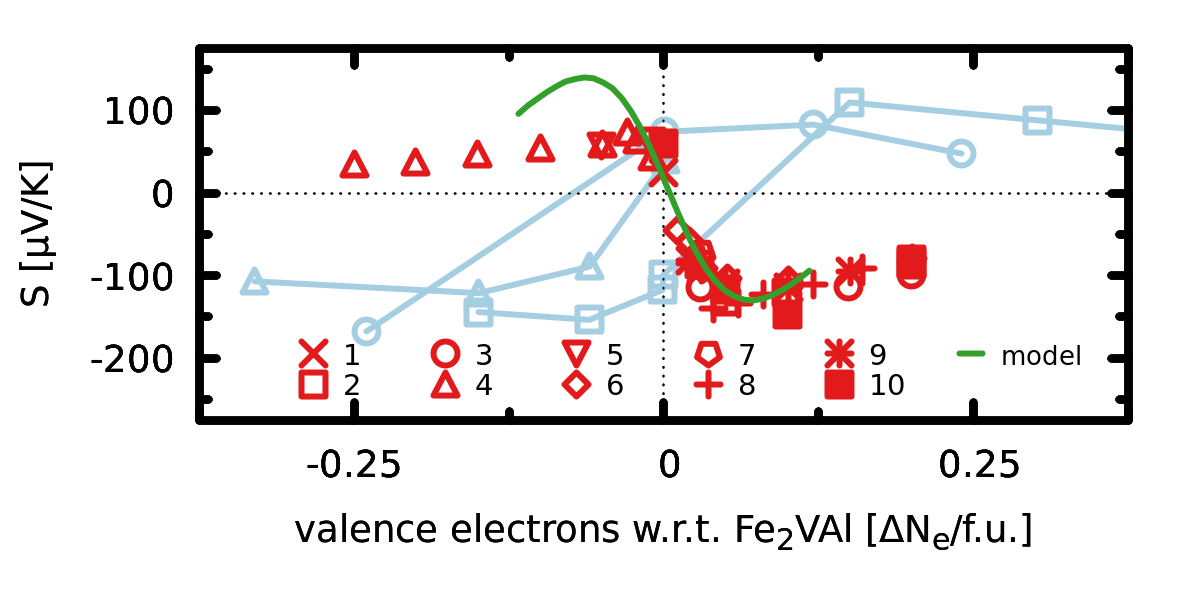}}
		\caption{{\bf Thermopower of Fe$_2$VAl: doping with fourth element vs.\ off-stoichiometry.} Thermopower at $T=300$K when doping with a fourth element: 
		(1) stoichiometric Fe$_2$VAl,
		(2) Fe$_2$V(Al$_{1-x}$Si$_x$),
		(3) Fe$_2$V(Al$_{1-x}$Ge$_x$),
		(4) Fe$_2$(V$_{1-x}$Ti$_x$)Al,
		(5) Fe$_2$(V$_{1-x}$Zr$_x$)Al,
		(6) Fe$_2$(V$_{1-x}$Mo$_x$)Al,
		(7) Fe$_2$(V$_{1-x}$W$_x$)Al,
		(8) (Fe$_{1-x}$Co$_x$)$_2$VAl,
		(9) (Fe$_{1-x}$Pt$_x$)$_2$VAl
		as extracted from Nishino \cite{1757-899X-18-14-142001} (see there for references),
		(10) Fe$_2$VAl$_{1-x}$Sn$_x$ \cite{Skoug2009}.
		The solid (green) line is a fit using \eref{eq:Ssc} applicable to semiconductors for $k_BT<\Delta$, cf.\ the typical shape in \fref{modelphi}.
		For comparison, results from off-stoichiometric samples is reproduced (in light blue) from \fref{Fe2VAlprox}. See Ref.~\cite{Soda2016MF201603} for a similar collection.
		}
      \label{Fe2VAlS}
      \end{center}
\end{figure}

Also thermopower measurements place \fval\ into a tight spot. Indeed, there is a tremendous sample dependence in $S$, as shown in \fref{Fe2VAlexp}(b) and \fref{Fe2VAlprox}.
While the thermopower most of the time reaches large values, the sign of the response varies from experiment to experiment.
Note that the thermopower is much more sample dependent than the resistivity shown in \fref{Fe2VAlexp}(a), as it is very sensitive to the balance of electron and hole contributions.
As discussed in \sref{limit} in the model context, this sensitivity suggest that intrinsic \fval\ is placed close to thermoelectric particle-hole symmetry,
for which $S$ exactly cancels but acquires very large values in the immediate vicinity.%
\footnote{
We specified {\it thermoelectric} particle-hole symmetry, because different observables may experience compensation at different points.
For example, in (Fe$_2$V)$_{1+x}$Al$_{1-3x}$, the Hall coefficient vanishes for $x=2.7\cdot 10^{-3}$, while $S=0$ already for $x=4\cdot 10^{-4}$\cite{PhysRevB.63.233303}.}
Moreover, the trend in the  off-stoichiometry of the composition series Fe$_{2+x}$V$_{1-x}$Al is unconventional, see \fref{Fe2VAlprox}: Adding in this way electrons(holes)
to the system causes a more hole-like(electron-like) response, in even qualitative defiance of the rigid band picture of carrier doping.
Similarly in the Hall coefficient, $R_H>0$ for Fe$_2$VAl$_{1.05 }$, while $R_H<0$ for Fe$_{2}$VAl$_{0.95}$\cite{PhysRevB.65.075204}.
These counter-intuitive findings may be linked again to the unconventional effects of off-stoichiometry onto the crystal structure, cf.\ \fref{Fe2VAlprox}(a). 
On the other hand, doping the Heusler compound by substitutions with a fourth element, yields (at small enough doping) the standard form of \eref{eq:Ssc}, as indicated by 
the model curve in \fref{Fe2VAlS}.

The large thermopower in conjunction with the low resistivity values in \fval-based systems are indicative of large thermoelectric powerfactors.
Indeed, Fe$_2$VAl$_{1-x}$Si$_x$ with $x = 0.10$ shows a large (n-type) power factor of $S^2/\rho=5.4$mW/(K$^2$m) in the range of $T=300-400$K\cite{Kato2001}.
Also Fe$_2$V$_{1+x}$Al$_{1-x}$ reaches a very favourable (n-type) powerfactor of $6.8$mW/(K$^2$m) for $x=0.05$ and p-type $4.2$mW/(K$^2$m) for $x=-0.03$\cite{2053-1591-1-1-015901}. These value fall into the same performance class as standard Bi$_2$Te$_3$-based thermoelectrics for which $S^2/\rho\sim 4-5$mW/(K$^2$m) in the same temperature range.
However, the corresponding figures of merit are small: at 300K, e.g.,
$ZT\approx0.13$ for Fe$_2$VAl$_{0.9}$Ge$_{0.1}$\cite{PhysRevB.74.115115}, or
$ZT\approx0.14$ for Fe$_2$VAl$_{0.9}$Si$_{0.07}$Sb$_{0.03}$\cite{MIKAMI2009444}. Indeed the
thermal conductivity of stoichiometric \fval\ reaches of 28W/(Km) at 300K\cite{PhysRevB.74.115115}, which is an order of magnitude worse than in Bi$_2$Te$_3$.

\subsubsection{Theoretical results.}
\paragraph{Electronic structure.}

\begin{figure*}[!th]
  \begin{center}
{\includegraphics[clip=true,trim=0 0 0 0,angle=0,width=.45\textwidth]{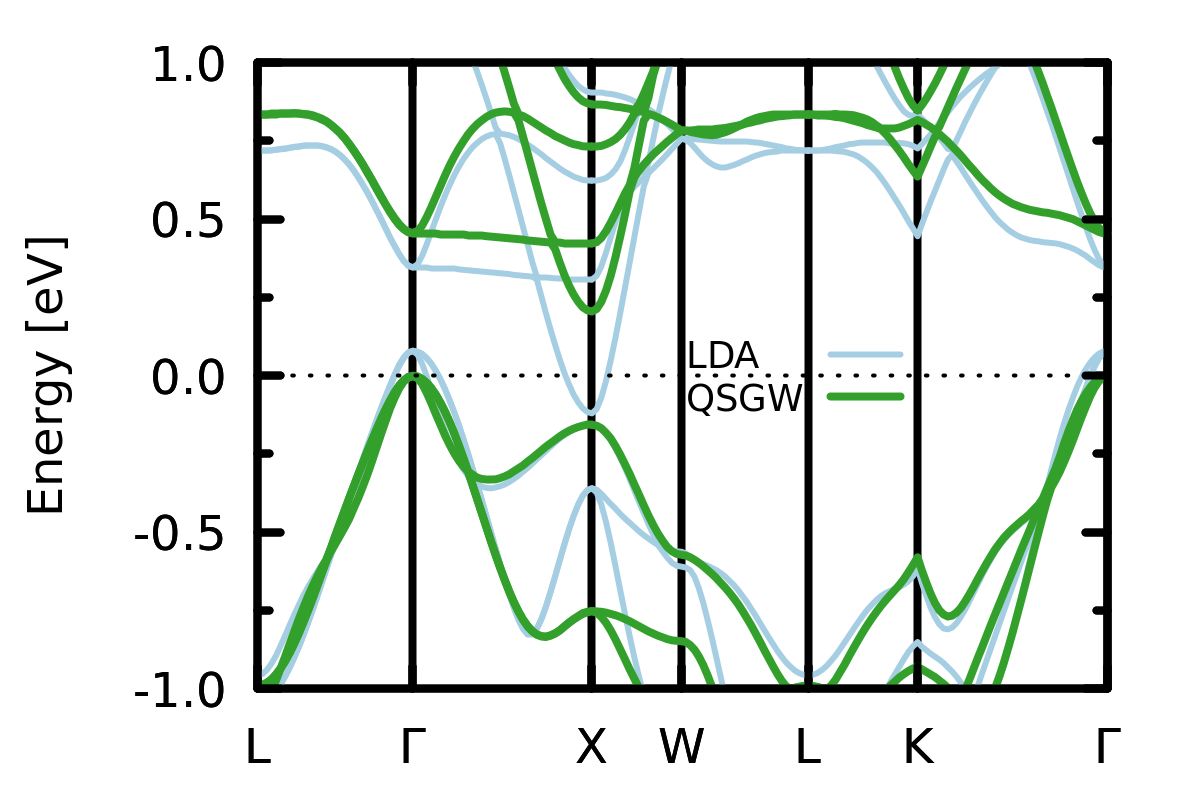}}
{\includegraphics[clip=true,trim=0 0 0 0,angle=0,width=.45\textwidth]{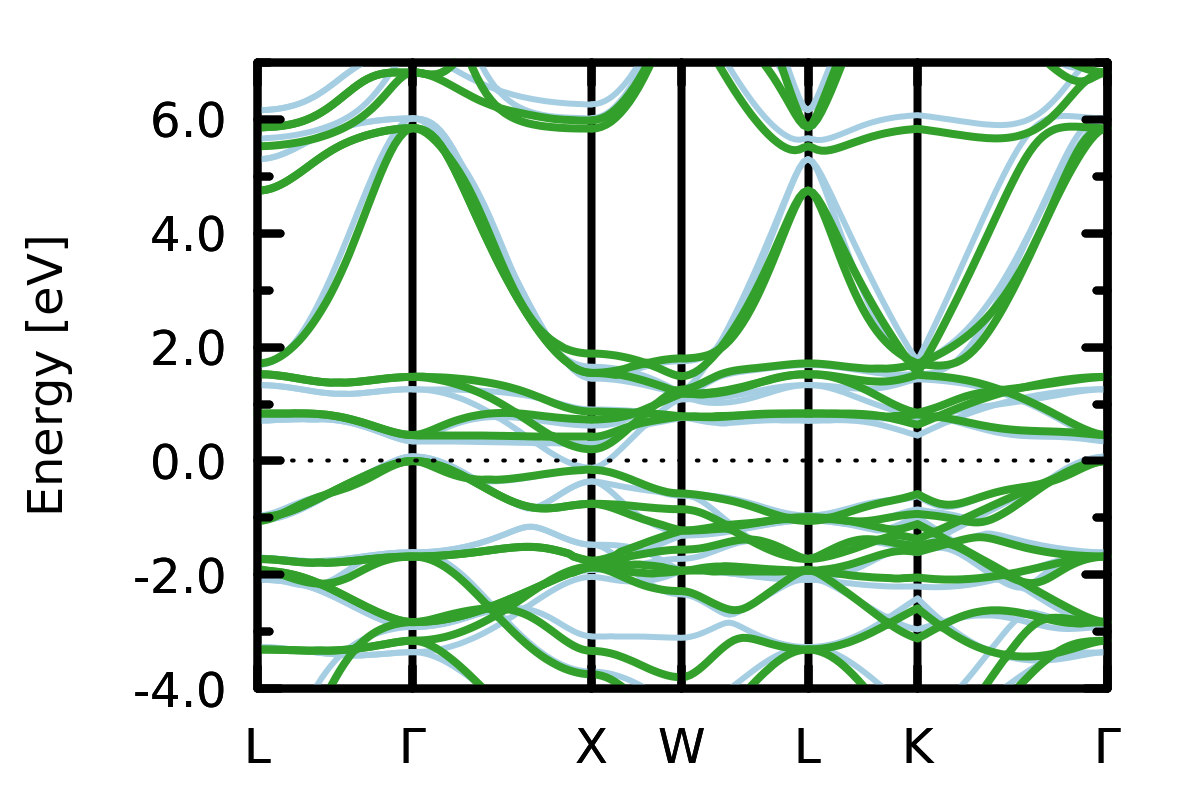}}
		\caption{{\bf Band-structure of Fe$_2$VAl.} Compared are metallic DFT-LDA (light blue) with QS{\it GW} (green) that yields a gap of 210meV. Left and right show different energy windows.
		}
      \label{Fe2VAlBND}
      \end{center}
\end{figure*}

Standard band-structure methods characterize \fval\ as a non-magnetic semi-metal,
with a hole-pocket at $\Gamma$ and an electron-pocket at $X$ and a low-density pseudogap of $100-200$meV in the DOS\cite{PhysRevB.57.14352,PhysRevB.58.6855,PhysRevB.58.9732,0953-8984-10-8-002}, see the LDA results in \fref{Fe2VAlBND}.
Improving on the exchange part in DFT, \fval\ becomes a semiconductor with an indirect gap between states of dominant Fe-t$_{2g}$ at $\Gamma$ and
V-e$_g$ states at the $X$-point. The size of this gap varies wildly in size depending
on which functional and which admixture parameter is used, e.g., 200meV (mBJ)\cite{PhysRevB.83.205204,PhysRevB.84.125104,2053-1591-3-7-075022}, 580 meV (PBE0)\cite{PhysRevB.84.125104}, or 340-620meV (B1-WC)\cite{PhysRevB.83.205204,PhysRevB.84.125104}. 
Therefore, we compare in \fref{Fe2VAlBND} results of LDA (metallic) to that of the many-body perturbation theory QS{\it GW}\footnote{which in particular does not depend on empirical parameters such as
the fraction of Hartree-Fock admixtures.}, which yields a  gap of $\sim210$meV.
Contrary to exchange-improved DFT (see, e.g., Refs.~\cite{PhysRevB.83.205204,2053-1591-3-7-075022}), the conduction states are not merely shifted upwards rigidly in QS{\it GW},
but experience a narrowing of their band-width, see \fref{Fe2VAlBND}(b). Interestingly, the effect is largest for the V-$t_{2g}$-derived band that disperses from  below 2eV at the $L$ or $X$-point to around 6eV  at $\Gamma$.

\begin{figure}[!th]
  \begin{center}
{\includegraphics[clip=true,trim=0 0 0 0,angle=0,width=.45\textwidth]{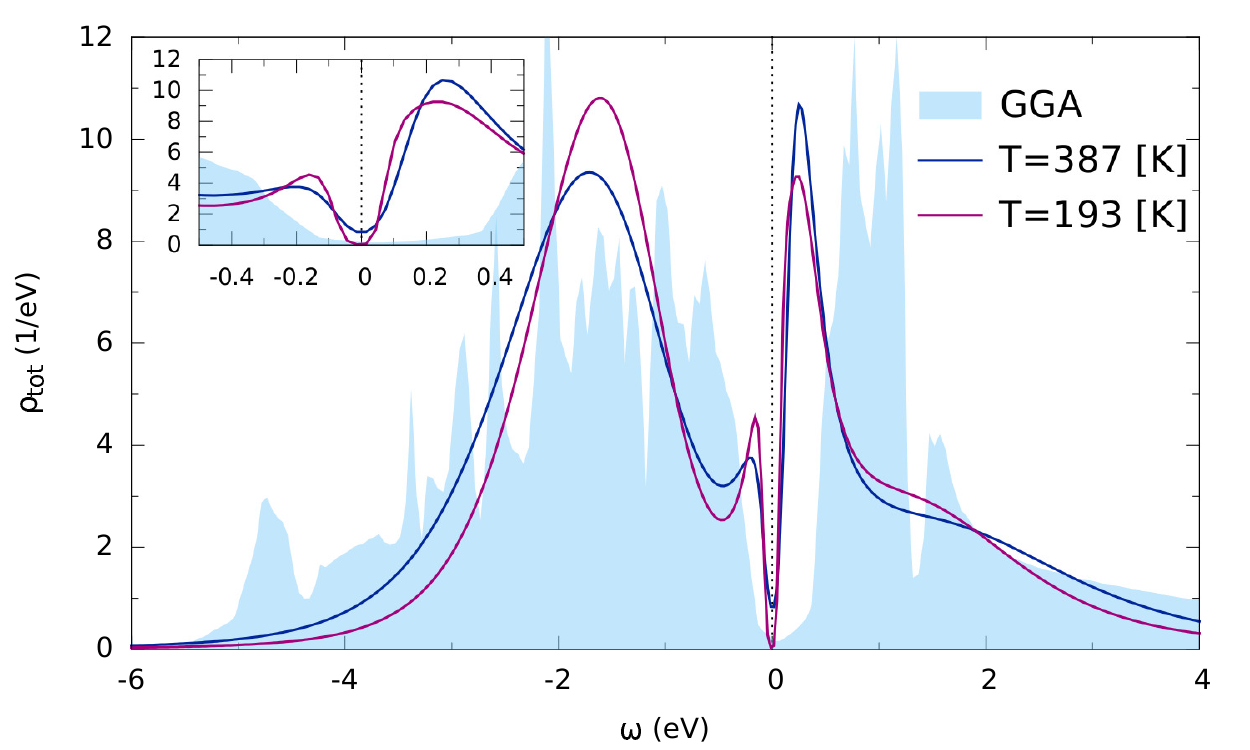}}
		\caption{{\bf Fe$_2$VAl within GGA+DMFT.} Local spectral functions at $T=387$K and $T=193$K within GGA+DMFT in comparison to the GGA DOS.
		From Kristanovski \etal\ \cite{PhysRevB.95.045114}.%
		}
      \label{Fe2VAldmft}
      \end{center}
\end{figure}

Such band-width narrowings are a consequence of dynamical correlation effects.
Treating the latter within DFT+dynamical mean-field theory, Kristanovski \etal\ \cite{PhysRevB.95.045114} recently investigated the end member compounds ($x=0$, $x=1$) of
the Fe$_{2+x}$V$_{1-x}$Al series.%
\footnote{
For DMFT calculations for another compounds in the binary Fe-Al system, see Galler \etal\ \cite{PhysRevB.92.205132};
for magnetic properties of half-metallic Heusler compounds within DMFT, see, e.g., Refs.~\cite{RevModPhys.80.315,0953-8984-23-25-253201} and references therein.
}
For Fe$_2$VAl they found, see \fref{Fe2VAldmft},  a strongly renormalized pseudogap of about 150meV at low temperatures and a mass-enhancement factor of 1.4 for vanadium states that 
were identified to dominate correlation-induced changes near the Fermi level.
In particular, the spectral weight at the Fermi level was shown to be strongly dependent on temperature, causing
a metallization with increasing temperature,
reminiscent of what happens in FeSi (cf.\ \sref{pnas}).%
\footnote{Ref.~\cite{PhysRevB.95.045114}
uses $U=3.36$eV and $J=0.71$eV for both Fe and V atoms, while
constrained DFT yields $U_{\hbox{\tiny Fe}}=4$eV, $U_{\hbox{\tiny V}}=1.5$eV\cite{PhysRevB.84.125104}, and
our constrained RPA for a maximally localized Wannier setup consisting of Fe-$3d$ and V-$3d$ orbitals gives
$U_{\hbox{\tiny Fe}}=2.5-3.3$eV, $J_{\hbox{\tiny Fe}}=0.4-0.6$eV, 
$U_{\hbox{\tiny V}}=1.3-1.8$eV, $J_{\hbox{\tiny V}}=0.15-0.3$eV, where the ranges owe to an intra-shell orbital dependence.
}

\paragraph{Disorder and magnetism.}
An alternative suggestion to explain that insulator-like transport properties coinciding with metallic photoemission spectrum resorts to issues of defects and disorder.
Indeed, the discussed DMFT calculations build on a metallic GGA band-structure and incorporate correlation effects originating from local interactions. 
Longer range interaction lead, however, to significant exchange contributions in the self-energy that open a charge gap as discussed above for hybrid functionals and {\it GW}
(cf.\ also the cases of FeSb$_2$ in \sref{marcasites} and CoSb$_3$ in \sref{skutts}).
Starting from an insulating host, diverse types of defects and disorder introduce states inside the gap\cite{PhysRevB.83.205204}.

Besides changes in the density-of-states at low-energies, deviations from perfect \fval\ also strongly affect magnetic properties:
As mentioned above, stoichiometric \fval\ verifies the Slater-Pauling rule for the magnetic moment of full-Heusler compounds\cite{Graf20111,Galanakis2016},
namely $M=(N-24)\mu_B=0$, with $N=24$ the total number of valence electrons per formula unit.
Deviations from the stoichiometric valence count induces a spontaneous moment in experiments (see above).
This is reproduced by
band-theories---based on supercell setups or the coherent potential approximation---that see ferromagnetic order induced by excess vanadium\cite{PhysRevB.77.134444,doi:10.1143/JPSJ.72.698}, excess iron\cite{doi:10.1143/JPSJ.72.698}, as well as anti-site defects (V for Fe, or Fe for Al)\cite{PhysRevB.83.205204}.
Among point defects, anti-site substitutions were shown to have the lowest formation enthalpies\cite{PSSB:PSSB201600441}.
In this picture, \fval\ is an intrinsic semiconductor whose puzzling transport and spectral properties are dominated by extrinsic effects.

In all, the electronic-structure of \fval\ is not yet properly understood.
Indeed, as was the case for skutterudites (cf.\ \sref{skutts}) the current Heusler compound presents a very challenging case, in which correlation and exchange effects beyond DFT-LDA need to be considered. Viable approaches will be mentioned in \sref{conclusions}.
On top of this, the interplay of correlation effects and defects need to be considered, 
as pioneered in different contexts, e.g.,\ by Lechermann \etal\ \cite{1367-2630-17-4-043026,PhysRevB.95.195159}.

\paragraph{Thermoelectricity.}
From the theory side, the large powerfactors in Fe$_2$YZ Heusler compounds has been rationalized to originate from
an effective low-dimensionality in the electronic structure\cite{PhysRevLett.114.136601}.
By this mechanism, it was shown that bulk semiconductors can realize highly directional conduction otherwise only found
in actual low-dimensional systems.

Further, computing transport properties of \fval\ from first principles, using the semi-classical Boltzmann approach in the relaxation time approximation
helped shedding some light onto the controversy of the compounds electronic structure.
Do \etal \cite{PhysRevB.84.125104} computed transport properties  on top of 
their DFT (metallic with pseudogap) and DFT+U (insulating) electronic structure.
A comparison with experimental results might elucidate whether from a thermoelectric point of view the pseudogap or the fully gapped scenario
is more adequate.
Using the pseudogap electronic structure, no thermopower values above $\pm50\mu$V/K can be realized at 300K, even when considering (rigid band) doping
\cite{PhysRevB.84.125104,XU201322,Al-Yamani2016}.
This finding is at odds with experiments for ``stoichiometric'' \fval, see \fref{Fe2VAlexp}(b).
The gapped electronic structure can qualitatively reproduce both the magnitude and the shape  of the experimental thermopower\cite{PhysRevB.83.205204,PhysRevB.84.125104},
however only for densities that seem to be at odds with those extracted from the experimental Hall coefficient.
Do \etal \cite{PhysRevB.84.125104} suggested that the impact of defects needs to be included for a consistent description of \fval.
Indeed, it was shown by Bilc and Ghosez \cite{PhysRevB.83.205204} and subsequently by Bandaru \etal\ \cite{doi:10.1080/15567265.2017.1355948} that antisite defects largely impact the thermopower, and can easily flip its sign. While the powerfactor in the presence of such defects slightly decreases\cite{PhysRevB.83.205204}, $ZT$ gets significantly
enhanced in the simulations\cite{doi:10.1080/15567265.2017.1355948} due to a notable reduction of thermal conduction.

\begin{framed}
\noindent
	{\bf Lessons for the Heusler compound Fe$_2$VAl: }
	\begin{itemize}
		\item large, qualitative uncertainties in electronic structure.
		\item need for treatment of both exchange and correlations beyond DFT.
		\item interplay of intrinsic and defect, disorder derived states crucial
		for thermoelectric properties.
	\end{itemize}
\end{framed}

\section{Conclusions and outlook}
\label{conclusions}

In this article we have reviewed signatures of electronic correlation effects,
their microscopic origin, and their impact onto thermoelectricity in correlated narrow-gap semiconductors.

The salient characteristic of these systems is the large and unconventional temperature dependence in physical
observables. In defiance of mere thermal activation across a fundamental gap, the most prominent members of these
materials display an insulator-to-metal crossover at temperatures corresponding to energies of only a fraction of the
gap size. Concomitantly, the magnetic susceptibility first grows to large values, before fluctuations disembogue into
a Curie-Weiss-like decay.

In the microscopic picture emerging from realistic many-body calculations, the evolution of charge and spin degrees of freedom
are intimately linked: The unlocking of spin fluctuations on transition-metal atoms cause the Brillouin zone momentum
to seize being a good quantum number. The widening decomposition onto momentum-states corresponds to a diminishing lifetime
of electronic excitations at a given momentum, causing incoherent spectral weight to spill into the gap, eventually filling it
to realize a bad metal phase. Contrary to Mott insulators, the magnetic susceptibility is not dominated by local
spin fluctuation. In fact, the main driving force of many-body effects in these systems was identified to be the Hund's rule coupling%
\footnote{In the genealogy of PhD advisors, Friedrich Hund is incidentally the author's ``great-grandfather''.}
entailing fluctuations to have ferromagnetic characteristics. The interplay of the Hund's rule coupling and the hybridization gap
explains the large variation of correlation signatures witnessed among various materials.

New realistic many-body simulations for \cbp\ and the study of the Kondo-insulating nature of its $J=5/2$ states allowed for an 
in-depth comparison of heavy-fermion and $d$-electron-based semiconductors.
Based on a quite general classification of paramagnetic insulators from the dynamical mean-field perspective, we suggest 
an orbital-selective scenario for FeSi: While some iron-$3d$ orbitals are Kondo-insulating, in others
quantum fluctuations are suppressed by a more conventional gapping of non-local hybridizations.

As to the influence of correlation effects onto thermoelectricity, details matter, but simplistically one could summarize
that  (selective) mass enhancements (via the real-parts of the self-energy) are beneficial for a large thermopower,
while (global) incoherence (described by the imaginary parts of the self-energy) is detrimental.
We pointed hence out that many-body renormalizations that increase the particle-hole asymmetry are to be looked for, 
while coherence (excitation lifetimes) could be increased by diluting the atoms that host the dominant Hund's rule coupling.

We further elaborated on new paths to exploit electronic correlations in thermoelectrics: In one proposal 
many-body effects have been suggested for the design
of mobility gradients that were shown to enhance both the thermopower and the Nernst coefficient\cite{Sun_CoSb3}.
Further, the time-honoured phonon-drag effect might find a revival and technical relevance. Indeed, it
has been proposed that its effect is enhanced by coupling lattice vibrations to charge carriers with large effective masses\cite{Takahashi2016}.
Further, it was shown\cite{Zhou01122015} that phonon modes contributing to the phonon-drag can be separated from those dominating
unwanted thermal conduction, heralding a new direction in the phonon-engineering route towards high-performance thermoelectrics.

The faithful simulation of thermoelectric properties of the silicides, antimonides, skutterudites, Heusler compounds discussed here,
exposed several challenges for the future:

\begin{enumerate}
	\item Electronic structure: in all materials discussed here dynamical correlation effects beyond DFT\cite{RevModPhys.71.1253,RevModPhys.61.689} or {\it GW}\cite{hedin,ferdi_gw,RevModPhys.74.601} have to be accounted for. In some, however, also
	exchange contributions to the self-energy beyond DFT and DFT+DMFT have been shown to be crucial for the electronic structure, let alone the computation
	of transport properties. This calls for the application of 
	higher-level theories, such as the recent {\it GW}+DMFT\cite{PhysRevLett.90.086402,0953-8984-26-17-173202,0953-8984-28-38-383001,Tomczak2017review}, QS{\it GW}+DMFT\cite{jmt_sces14,Choi2016}, SEX+DMFT\cite{paris_sex}, or  AbinitioD$\Gamma$A\cite{Anna_ADGA,JPSJ_ADGA,CPC_ADGA}.
	\item Defects and disorder: as in more conventional semiconductors, effects of defects and disorder are a particular concern, especially when doping the stoichiometric
	parent compounds. Accounting for these effects in the presence of electronic correlation effects is a challenging, yet necessary endeavour for the future.
\item	Electron-phonon effects: future studies must address the impact of electronic correlations onto the phonon-structure, 
 the interplay of electronic incoherence and thermal-disorder, as well as the phonon-drag effect.
\end{enumerate}

\bigskip

\begin{acknowledgments}

The author gratefully acknowledges discussions on topics related to this review with
H.\ Aoki, M.\ Aronson,
M.\ Battiato, S.\ B{\"u}hler-Paschen, O.\ Delaire, A.\ Galler, M.\ Gam{\.z}a, A.\ Georges, K.\ Haule, K.\ Held, G.\ Kotliar, J.\ Kune{\v{s}}, G.\ K.\ H.\ Madsen, E.\ Maggio, J.\ Mravlje, Hyowon Park, C.\ Petrovic, J.\ Pflaum, A.\ Prokofiev, G.\ Sangiovanni, F.\ Steglich, Peijie Sun, P.\ Tomes, A.\ Toschi, Wenhu Xu, V. Zlati{\'c}, Z.\ Zhong,
as well as M.\ Battiato, S.\ B{\"u}hler-Paschen, K.\ Held, J.\ Kune{\v{s}} for comments on parts of the manuscript.
We thank A.\ Hausoel for providing pre-release capabilities and support of the w2dynamics package.
This work has been supported by the Austrian Science Fund (FWF) through project ``LinReTraCe'' P~30213-N36. 
Some calculations were performed on the Vienna Scientific Cluster (VSC).

\end{acknowledgments}

\newpage




\end{document}